\documentclass[12pt,english,a4paper,twoside]{report}
\newlength{\defbaselineskip}
\newcommand{\setlinespacing}[1]%
           {\setlength{\baselineskip}{#1 \defbaselineskip}}

\usepackage{babel}
\usepackage{amsmath,exscale,amssymb,latexsym}
\usepackage[mathscr]{euscript}
\catcode`æ=13 \def^^e6{\ae} \catcode`Æ=13 \def^^c6{\AE}
\catcode`ø=13 \def^^f8{\o} \catcode`Ø=13 \def^^d8{\O}
\catcode`å=13 \def^^e5{\aa} \catcode`Å=13 \def^^c5{\AA}

\usepackage{varioref}%\usepackage{feynmp}
\usepackage[dvips]{graphicx,color}
\begin{document}%\begin{fmffile}{thesis21}
\title{\thispagestyle{empty}{\huge Quantum gravity,
effective fields and string theory}\vspace{1cm}\\}
\author{\thispagestyle{empty}{\bf Niels Emil Jannik Bjerrum-Bohr}\vspace{0.5cm} \\
\large The Niels Bohr Institute\\
\large University of Copenhagen\vspace*{2cm}\\
\includegraphics[scale=1]{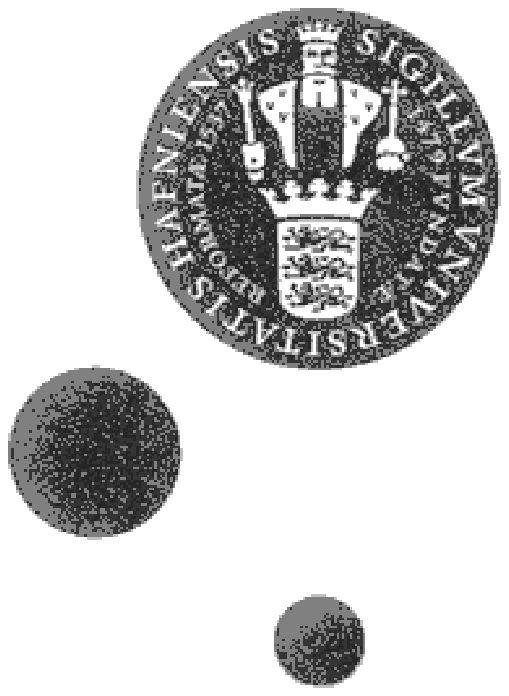}
\vspace{2cm}\\ \\
Thesis submitted for the degree of Doctor of Philosophy in Physics\\
at the Niels Bohr Institute, University of Copenhagen.\date{28th
July}} \thispagestyle{empty}\maketitle{\thispagestyle{empty}}
\thispagestyle{empty}
\begin{abstract}
In this thesis we will look into some of the various aspects of
treating general relativity as a quantum theory. The thesis falls
in three parts. First we briefly study how general relativity can
be consistently quantized as an effective field theory, and we
focus on the concrete results of such a treatment. As a key
achievement of the investigations we present our calculations of
the long-range low-energy leading quantum corrections to both the
Schwarzschild and Kerr metrics. The leading quantum corrections to
the pure gravitational potential between two sources are also
calculated, both in the mixed theory of scalar QED and quantum
gravity and in the pure gravitational theory. Another part of the
thesis deals with the (Kawai-Lewellen-Tye) string theory
gauge/gravity relations. Both theories are treated as effective
field theories, and we investigate if the KLT operator mapping is
extendable to the case of higher derivative operators. The
constraints, imposed by the KLT-mapping on the effective coupling
constants, are also investigated. The KLT relations are
generalized, taking the effective field theory viewpoint, and it
is noticed that some remarkable tree-level amplitude relations
exist between the field theory operators. Finally we look at
effective quantum gravity treated from the perspective of taking
the limit of infinitely many spatial dimensions. The results are
here mostly phenomenological, but have some practical and
theoretical implications as it is verified that only a certain
class of planar graphs will in fact contribute to the $n$-point
functions at $D=\infty$. This limit is somewhat an analogy to the
large-$N$ limit of gauge theories although the interpretation of
such a graph limit in a gravitational framework is quite
different. We will end the thesis with a summary of what have been
achieved and look into the perspectives of further investigations
of quantum gravity.
\end{abstract}\newpage

\pagestyle{headings}
\newpage \thispagestyle{empty} \vspace*{1cm}\newpage
\tableofcontents
\newpage

\section*{Acknowledgements}
Especially I would like to thank my advisor Prof. Poul Henrik
Damgaard for stimulating discussions, and for his help and advise
during my graduate studies.

I would also like to thank Prof. John F. Donoghue and Prof. Barry
R. Holstein for our mutual collaborations, for their support, and
for nice and enlightening discussions on quantum gravity and effective
field theory.

I am very grateful to the Department of Physics and Astronomy at
the University of California at Los Angeles, and to Prof. Zvi Bern for
good and interesting discussions on the KLT-relations and on
quantum gravity. Also I would like to thank Gilda Reyes for help and
support during my stay at UCLA.

I would like to thank all of my colleges at the Niels Bohr
Institute for generating a stimulating academic environment and
for creating a good social atmosphere.

I would like to thank the staff at the Niels Bohr Institute
and especially Hanne and Marianne for their help and support. Also
Lisbeth and the staff at the Library deserve much thanks.

I am also very thankful for the financial support from the Leon Rosenfeld
Scholarship Foundation.

Finally I would like to {\it emphasize} that I wish to thank
everyone who have helped me though the process of the making
this thesis.
\\
\\
\\
\\
\\
\\
\\
Niels Bohr Institutet den 28. april,
\\ \\

Niels Emil Jannik Bjerrum-Bohr
\newpage
\section*{List of publications}
\begin{itemize}
\item N.~E.~J.~Bjerrum-Bohr,
{\bf Leading quantum gravitational corrections to scalar QED},
Phys.\ Rev.\ D {\bf 66} (2002) 084023 [arXiv:hep-th/0206236].
\item N.~E.~J.~Bjerrum-Bohr, J.~F.~Donoghue and B.~R.~Holstein,
{\bf Quantum corrections to the Schwarzschild and Kerr metrics},
Phys.\ Rev.\ D {\bf 68}, 084005 (2003) [arXiv:hep-th/0211071].
\item N.~E.~J.~Bjerrum-Bohr, J.~F.~Donoghue and B.~R.~Holstein,
{\bf Quantum gravitational corrections to the nonrelativistic scattering  potential of two masses},
Phys.\ Rev.\ D {\bf 67}, 084033 (2003) [arXiv:hep-th/0211072].
\item N.~E.~J.~Bjerrum-Bohr,
{\bf String theory and the mapping of gravity into gauge theory},
Phys.\ Lett.\ B {\bf 560}, 98 (2003) [arXiv:hep-th/0302131].
\item N.~E.~J.~Bjerrum Bohr,
{\bf Generalized string theory mapping relations between gravity and gauge theory},
Nucl.\ Phys.\ B {\bf 673}, 41 (2003) [arXiv:hep-th/0305062].
\item N.~E.~J.~Bjerrum-Bohr,
{\bf Quantum gravity at a large number of dimensions},
Nucl.\ Phys.\ B {\bf 684} (2004) 209 [arXiv:hep-th/0310263].
\end{itemize}

\chapter{Introduction}
Everywhere in Nature gravitational phenomena are present and daily
experiences teach us to recognize, appreciate and live with the
basic effects and constraints of gravity. A life completely
without earthly gravity would indeed be very different!

Newton realized that tides, falling apples, and the motion of the
planets all originate from the same force of Nature -- the
universal gravitational attraction. Everyone knows that things are
supposed to fall if you drop them. The great invention of Newton
was to put the experimental facts of Nature into a mathematical
formalism which predicts both the motion of a falling apple and
the orbit of the Moon. The fundamental equation is his
gravitational law, also known as the Newtonian potential, which
predicts the potential energy of the gravitational attraction
between two bodies:
\begin{equation}
\boxed{V(r) = -G \frac{m_1 m_2}{r}}
\end{equation}
here $\left(V(r)\right)$ is a measure for the potential energy,
$\left(m_1\right)$ and $\left(m_2\right)$ are the masses of the
two particles, $\left(r\right)$ is the distance between the masses
and $\left(G\right)$ is the gravitational constant. The
gravitational potential energy thus is predicted by Newton to be
proportional to masses and inverse proportional to distances.
Newton's law together with his mechanic principles
(ref.~\cite{Newton}) form the foundation of theoretical physics
and create the basis for understanding and predicting theoretical
effects of gravity.

Einstein realized that the gravitational attraction is in fact
really proportional to the total energy of matter and not only to
rest masses.
\begin{figure}[h]
\centering\parbox{7cm}{\includegraphics[height=1.5cm]{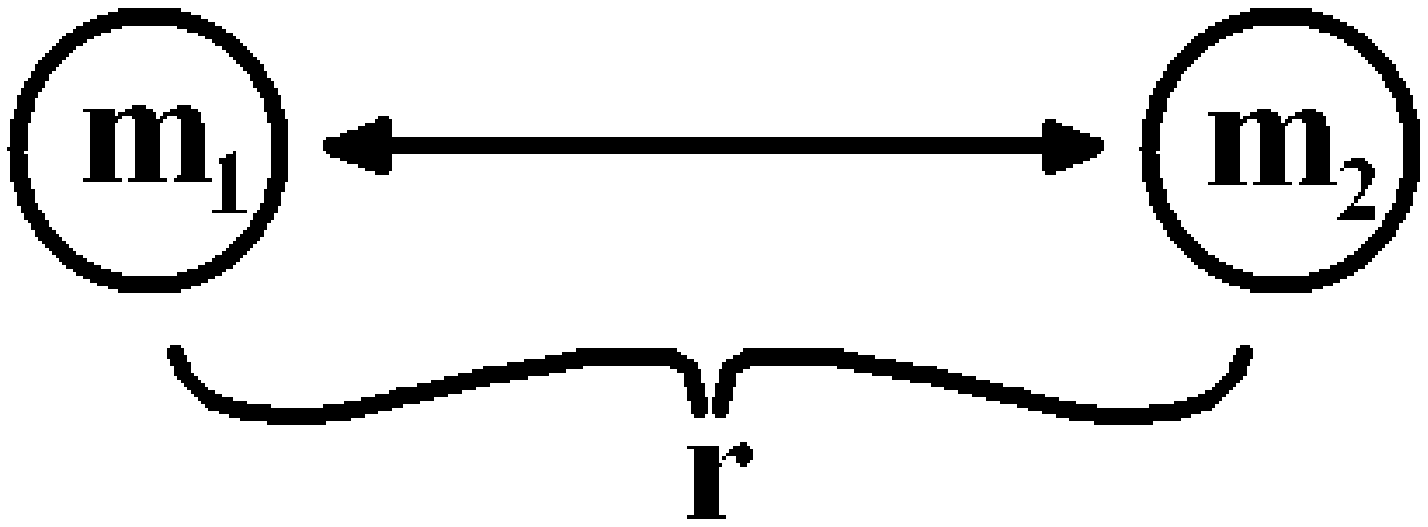}}\parbox{4cm}
{\includegraphics[height=2.5cm]{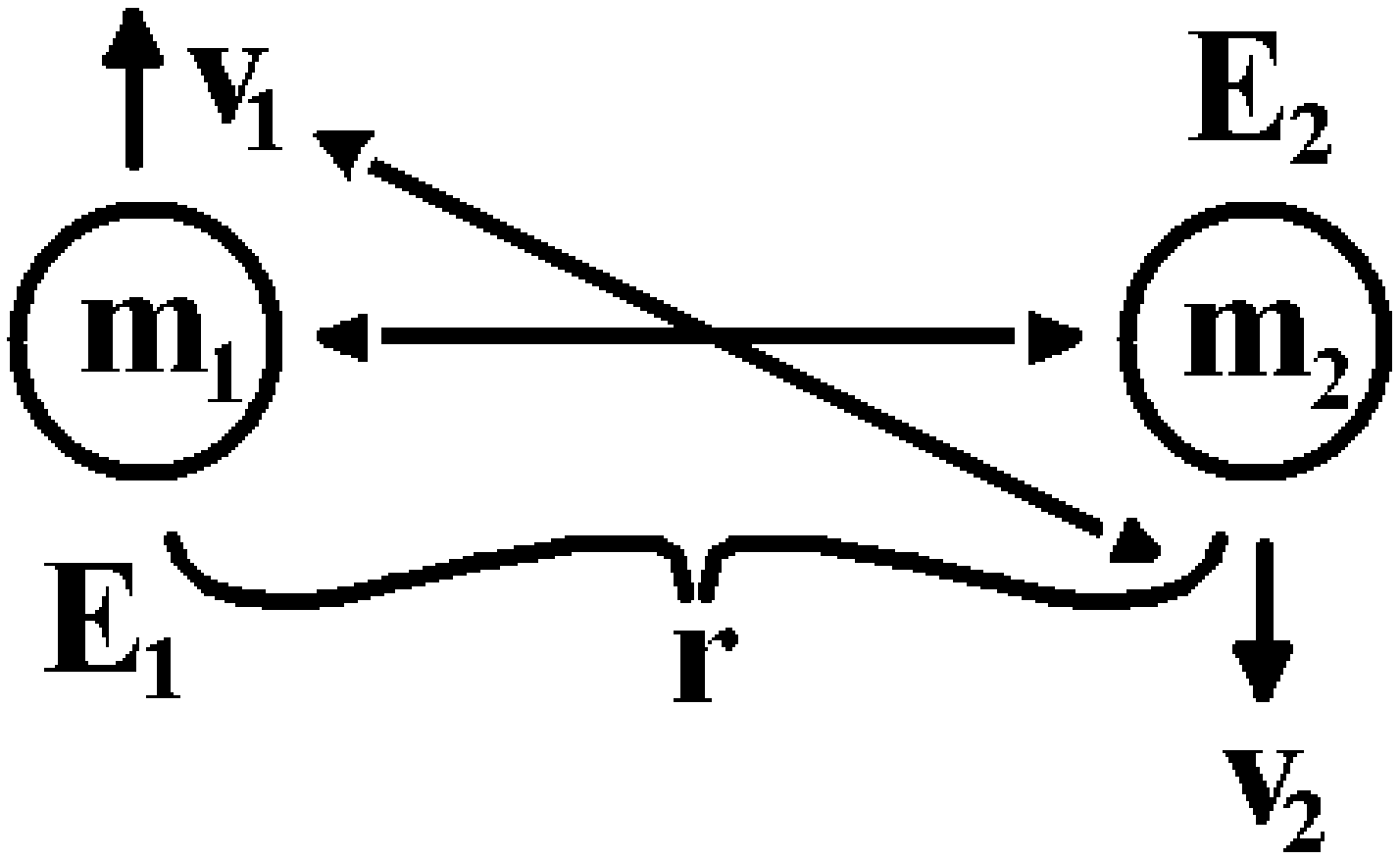}}
\caption{Illustration of the gravitational attraction between
two sources, with or without a kinetic energy.}
\end{figure}
At relativistic velocities this can be observed -- as an extra
gravitational attraction caused by the kinetic energy of
particles. Thus relativistic particles such as light rays are in
fact also sources of gravity. The first confirmation of Einstein's
proposal was through the direct observation of light rays bending
around the Sun.
\begin{figure}[h]
\centering\includegraphics[height=4cm]{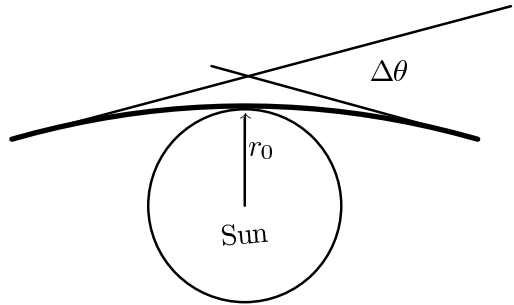}
\caption{Illustration of light rays bending around the Sun.}
\end{figure}

The theory of general relativity (ref.~\cite{Einstein}) provides a
framework for extending Newton's theory to objects with no rest
mass and relativistic velocities. In general relativity one solves
the basic field equation, the Einstein equation:
\begin{equation}\boxed{
R_{\mu\nu}(g_{\mu\nu}) - \frac12 R(g_{\mu\nu}) g_{\mu\nu} = 16\pi
G T_{\mu\nu} - \lambda g_{\mu\nu}}
\end{equation}
where $\left(g_{\mu\nu}\right)$ is the gravitational metric, $\left(R^{\alpha}_{\
\beta\mu\nu}\right)$ is a measure for the curvature of
space\footnote{\footnotesize{$\left(R_{\mu\nu} = R^\beta_{\
\mu\nu\beta}\right)$, $\left(R \equiv g^{\mu\nu}R_{\mu\nu}\right)$}} and $\left(T_{\mu\nu}\right)$
is the total energy-momentum tensor for the sources of gravity.
The cosmological constant $\left(\lambda\right)$ is needed on cosmological
scales, and is today believed to have a non-zero expectation value
in the Universe. Whenever we have a concrete gravitational problem
(a specified energy distribution, $i.e.$, a mathematical
expression for the energy-momentum tensor) we solve the Einstein
equation for the metric field. The metric will be a local object
and depend on the geometry of space. In this way a solution of the
gravitational problem is found. General relativity
relates the geometry of space and gravity. Einstein's description
holds in the fully relativistic regime, and its low-energy and
nonrelativistic predictions fit the expectations of Newtonian
mechanics.

The mathematical theories for gravitational phenomena have been
very successful and calculations of orbits for planets in the
solar system fit experimental data with great precision.
Classical gravitational models for the Universe are being
investigated and observations so far support the theories up to
the precision of the measurements. It must be concluded that
general relativity is a very successful theory for the classical
gravitational phenomena of the Universe.

Quantum mechanics and quantum field theory are the great other
fundamental physical theories. They essentially explain to us how
to interpret physics at the atomic and subatomic scales. Quantum
mechanics provide the theoretical basis for most solid state
physics. Without a proper knowledge of quantum mechanics we could
in fact neither build semi-conductor chips nor lasers. Most
chemistry and atomic physics rely heavily on quantum physics.
Quantum mechanics do not only give models for atoms, but also for
nuclear physics and for the interactions of the elementary
particles in the Standard Model. This is the theory uniting the
electro-weak and strong interactions of elementary particles in a
$$\boxed{{\rm U(1)} \times {\rm SU}(2) \times {\rm SU}(3)}$$ gauge theory. It
is believed to form the basis of particle physics and to be valid
until energies around $\sim 1$ TeV.

The crucial distinctions between classical and quantum theories,
$e.g.$, the replacement of $c$-numbers by operators etc, rule out
the possibility that the Universe fundamentally could have a
combined classical and quantum mechanical description. As strong
experimental evidence ($e.g.$, ref.~\cite{Aspect}) are in favor of
a quantum description of Nature at the fundamental Planckian
scale, $i.e.$, $\left(E_{\text {Planck}}\sim 10^{19}\ {\rm
GeV}\right)$, the logical consequence is that an elementary theory
for gravity will have to be a quantum theory.

So the problem can be stated in the following way. As general
relativity is a classical theory -- it cannot be the whole story
of gravity and there thus must exist a fundamental theory that in
the limit of classical energy scales provides us with the
classical expectations.
\begin{figure}[h]
\centering\includegraphics[height=4cm]{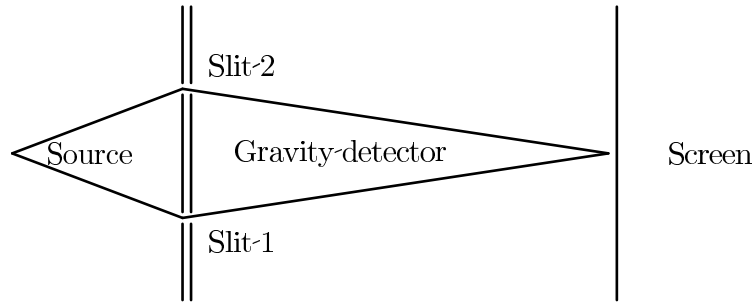}
\caption{Feynman's double slit experiment, see ref.~\cite{Feynman}. An
experiment to illustrate intuitively why the Universe
fundamentally must be of a classical or of a quantum nature,
$i.e.$, a combined description is not possible. If
gravitational waves from the quantum particle in the experiment
are {\it classical}, there is something wrong from a {\it quantum} mechanical
perspective, because when we in principle could detect which slit
the quantum particle went. This would disagree with the principles
of quantum mechanics. On the other hand if the wave-detector senses
gravitational waves, with a {\it quantum} nature, $i.e.$, wave
amplitudes, surely gravity self must be a {\it quantum} theory for the
experiment to be consistent! Thus a fundamental theory for gravity
has to be a {\it quantum} theory, if we assume that all other theories of
Nature are {\it quantum} theories.}
\end{figure}

It is not obvious in any way to understand general
relativity as a quantum theory -- and in fact general relativity and
quantum mechanics seem to be based on completely different
perceptions of physics -- nevertheless this question is one of the most
pressing questions of modern theoretical physics and has been the
subject of many authors, $e.g.$, see refs. De Witt~\cite{Dewitt},
Feynman~\cite{Feynman,Feynman2}, Gupta et al~\cite{Gupta}, Faddeev
and Popov~\cite{Faddeev}, Mandelstam~\cite{Mandelstam},
Schwinger~\cite{Schwinger2,Schwinger}, Weinberg~\cite{Weinberg},
Iwasaki~\cite{Iwasaki}, van Dam~\cite{Dam1,Dam2}, 't Hooft and
Veltman~\cite{Veltman1,Veltman2}, Zakharov~\cite{Zak}, Capper et
al~\cite{Capper:pv,Capper:vb,Capper:dp}, Duff~\cite{Duff1,duff3,Duff2},
Deser~\cite{Deser:cz,Deser:cy,Deser:1974xq}, Berends and
Gastmans~\cite{Berends:1975ah} and Goroff and
Sagnotti~\cite{Goroff:1985th}, just to mention a few.
All sorts of interpretational complications arise promoting
general relativity to be a quantum theory. The adequate starting point
for such a theory appears to be to interpret general relativity as a quantum
field theory, to let the metric be the basic gravitational field,
and to quantize the Einstein-Hilbert action:
\begin{equation}\boxed{
{\cal L }_{\text EH} = \int d^4 x \sqrt{-g}{R\over 16\pi G}}
\end{equation}
where $\left(g = \det (g_{\mu\nu})\right)$ and $\left(R\right)$ is the scalar curvature.
However the above action is not self contained as loop diagrams
will generate new terms not present in the original action,
$i.e.$, the action cannot be renormalized in a traditional
way, see refs.~\cite{Veltman1,Veltman2,Deser:cz,Deser:cy,Deser:1974xq,Goroff:1985th,
vandeVen:1991gw}.
This is the renormalization problem of general relativity.

We will here investigate quantum gravity from the point of view of
treating it as an effective field theory below the energies of the
Planck scale, see refs.~\cite{Weinberg:1978kz,{Donoghue:1993eb},{Donoghue:dn}}.
As we will observe as an effective field theory there
are no problems with the renormalization of the action. For a recent review of
general relativity as an effective field theory, see ref.~\cite{Burgess:2003jk}.
The thesis work can be divided into
three parts. First it was investigated how to make quantum
predictions for observables by treating gravity as an effective
field theory. This part of the thesis is covered by the research
papers, see refs.~\cite{B1,B2,B3}. These papers explain the effective
field theory treatment of gravity and show that below the Planck scale,
quantum and post-Newtonian corrections to the Schwarzschild and
Kerr metrics and to scattering matrix potentials can be
calculated. The thesis work covered by the research papers, see refs.~\cite{B4,B5}
is next discussed. The discussion is the Kawai-Lewellen-Tye string
relations, which is treated from the viewpoint of quantum gravity and
Yang-Mills theory as effective field theories. Generalized gauge-gravity
relations for tree amplitudes are derived from this investigation, and it
is observed that the KLT-mapping is extendable in the framework
of effective field theories.
Finally we look at the paper ref.~\cite{B6}. Here we take up the old idea by
Strominger, see ref.~\cite{Strominger:1981jg}, which discusses quantum gravity as
a theory in the limit of an infinite number of dimensions. We extend this idea
to quantum gravity seen as an effective field theory.

In the thesis we will everywhere work
in harmonic gauge, use units $\left(c=\hbar=1\right)$ and
metric $\left(+1,-1,-1,-1\right)$ when anything else is not explicitly stated.

\chapter{Quantum gravity as an effective field theory}
\section{Introduction}
We will in this section consider general relativity as an
effective quantum field theory.

In the concept of effective field theories, see
ref.~\cite{Weinberg:1978kz}, the couplings which should be
included in a particular Lagrangian are perturbatively determined
by the energy scale. It is traditional to require strict
renormalization conditions in a quantum field theory, however this
has no meaning for effective field theories. Every term consistent
with the underlying symmetries of the fields of the theory has to
be included in the action, and by construction any effective field
theory is hence trivially renormalizable and has an infinite
number of field couplings. The various terms of the action
correspond to different energy scales for the theory in a
perturbative manner. At each loop order only a finite number of
terms of the action need to be accounted for. In non-effective
field theories the Lagrangian is believed to be fundamental and
valid at every energy scale -- from this viewpoint the effective
action is less fundamental, $i.e.$, at sufficiently high energies
it has to be replaced by some other theory, perhaps a string
theory description etc. However at sufficiently low energies the
effective Lagrangian presents both an interesting path to avoid
the traditional renormalization problems of non-renormalizable
theories and a way to carry out explicit field theory
calculations.

It is a known fact that quantum field theories of pure general
relativity, as well as quantum theories for general relativity
including scalar, see refs.~\cite{Veltman1,Veltman2,Goroff:1985th,{vandeVen:1991gw}},
fermion or photon fields, see refs.~\cite{Deser:cz,Deser:cy,{Deser:1974xq}},
suffer from severe problems with renormalizability in the
traditional meaning of the word.

This apparent obstacle can be avoided if we treat general
relativity as an effective field theory. We consider then the
gravitational action as consisting not solely of the Einstein
curvature term plus the minimal couplings of the matter terms, but
as an action consisting of all higher derivative terms which are
generally covariant. As an effective field theory, general
relativity thus can be dealt with as any other quantum field
theory up to the limiting scale of the Planck energy $\sim
10^{19}$ GeV.

Actions for general relativity where a finite number of additional
derivative couplings have been allowed for are discussed in the
literature, see refs.~\cite{Pais:1950za,Stelle:1976gc,Stelle:1977ry,Simon:ic},
and various issues concerning general relativity as a classical or
a quantum theory with higher derivative terms have been dealt with.
Chiral perturbation theories have many uses for effective action
approaches, see ref.~\cite{gdh}. Treating general relativity as an effective
field theory in order to calculate leading 1-loop pure
gravitational corrections to the Newtonian potential was first
done in refs.~\cite{Donoghue:1993eb,Donoghue:dn}. This calculation has
since been followed by the various
calculations, see refs.~\cite{Muzinich:1995uj,Hamber:1995cq,akh,Thesis,ibk}.
In ref.~\cite{Donoghue:2001qc} effective action techniques were used to
calculate the quantum corrections to the Reissner-Nordstr\"om and
Kerr-Newman metrics. Here we will present the results of the three
thesis papers, see refs.~\cite{B1,B2,B3} which sort out the numerical
inconsistencies of previous calculations. The first ref.~\cite{B1}
considers the combined theory of general relativity and scalar
QED. In this paper the quantum and post-Newtonian corrections to
the mixed scattering matrix potential of these two theories are
calculated.\footnote{\footnotesize{In ref.~\cite{sharaz} we will
consider the corresponding mixed
calculation with QED and general relativity.}} In the other
two papers refs.~\cite{B2,B3} the quantum and
post-Newtonian corrections to the Schwarzschild and Kerr metric as
well as to the Newtonian potential are calculated correctly for
the first time.

The background field method first introduced in ref.~\cite{Dewitt}
will be employed here.

The results will be discussed in relation to, see refs.~\cite{Iwasaki,Donoghue:2001qc,
Barker:bx,Barker:ae,Okamura,Khriplovich:2004cx}. The used vertex rules
are presented in appendix~\ref{vertex}.

\section{General relativity as an effective field theory}
In this section we will consider in a more detailed manner how to
treat general relativity as an effective field theory.

The Einstein action for general relativity has the form:
\begin{equation}
{\cal L} = \sqrt{-g}\left[\frac{2R}{\kappa^2}+ {\cal
L}_{\text{matter}}\right],
\end{equation}
here $\left(\kappa^2 = 32\pi G\right)$ is defined as the gravitational
coupling, the curvature tensor is defined as $\left(R^\mu_{\
\nu\alpha\beta} \equiv
\partial_\alpha \Gamma_{\nu\beta}^\mu
-\partial_\beta \Gamma_{\nu\alpha}^\mu
+\Gamma^\mu_{\sigma\alpha}\Gamma_{\nu\beta}^\sigma
-\Gamma_{\sigma\beta}^\mu\Gamma_{\nu\alpha}^\sigma\right)$ and the
determinant of the metric field $\left(\det(g_{\mu\nu})\right)$ is denoted as $\left(g\right)$. The
term $\left(\sqrt{-g}{\cal L}_{\rm matter}\right)$ is a covariant expression
for the inclusion of matter into the theory. We can include any type
of matter. As a classical theory the above Lagrangian defines the
theory of general relativity.

Any effective field theory can be seen as an expansion in energies
of the light fields of the theory below a certain scale. Above the
scale transition energy there will be additional heavy fields that
will manifest themselves. Below the transition the heavy degrees
of freedom will be integrated out and will hence not contribute to
the physics. Any effective field theory is built up from terms
with higher and higher numbers of derivative couplings on the
light fields and obeying the gauge symmetries of the basic theory.
This gives us a precise description of how to construct effective
Lagrangians from the gauge invariants of the theory. We expand the
effective Lagrangian in the invariants ordered in magnitude of
their derivative contributions. Derivatives of light fields
$\left(\tilde\partial\right)$ will essentially go as powers of
momentum, while derivatives of massive fields
$\left(\partial\right)$ will generate powers of the interacting
masses. As the interacting masses usually are orders of magnitude
higher than the momentum terms --- the derivatives on the massive
fields will normally generate the leading contributions.

An effective treatment of general relativity results in the following Lagrangian:
\begin{equation}
{\cal L} =
\sqrt{-g}\left[\frac{2R}{\kappa^2}+c_{1}R^2+c_{2}R^{\mu\nu}R_{\mu\nu}+\ldots\right],
\end{equation}
the ellipses in the equation express that we are dealing with an
infinite series --- additional higher derivative terms have to be
included at each loop order. The coefficients ($c_1$, $c_2$, $\ldots$) represent
the appearance of the effective couplings in the action. Through cosmology we can bound
the effective coupling constants, we should have $\left(c_1,c_2 > 10^{70}\right)$ in order to generate
measurable deviations from the predictions of traditional general relativity, $e.g.$,
see ref.~\cite{Stelle:1977ry}.

As an example of how to construct an effective Lagrangian for a theory we can look at the
case of QED with charged scalars and photons. A general covariant version of the scalar
QED Lagrangian is:
\begin{multline}
{\cal L} = \sqrt{-g}\bigg[-\frac14
\left(g^{\alpha\mu}g^{\beta\nu}F_{\alpha\nu}F_{\mu\beta}\right)
+ (D_\mu\phi+ieA_\mu)^*(g^{\mu\nu})(D_\nu\phi+ieA_\nu) -
m^2|\phi|^2 \bigg],
\end{multline}
where $\left(F_{\mu\nu} \equiv  {D}_\mu A_\nu - {D}_\nu A_\mu=
\partial_\mu A_\nu - \partial_\nu A_\mu\right)$, and $\left({D}_\mu\right)$ denotes
the covariant derivative with respect to the gravitational field,
$\left(g_{\mu\nu}\right)$. As $\left(\phi\right)$ is a scalar $\left({D}_\mu \phi = \partial_\mu
\phi\right)$.

Counting the number of derivatives in each term of the above
Lagrangian we see that the term with
$\left(g^{\alpha\mu}g^{\beta\nu}F_{\alpha\nu}F_{\mu\beta}\right)$ goes as $\left(\sim
\tilde\partial\tilde\partial\right)$, while the scalar field terms go
as $\left(\sim \partial\partial\right)$ and $\left(\sim 1\right)$ respectively.
Thus seen in
the light of effective field theory, the above Lagrangian
represents the minimal derivative couplings of the gravitational
fields to the photon and complex scalar fields.

Typical 1-loop field singularities for the mixed graviton and
photon fields in the minimal theory are known to take the
form ref.~\cite{Deser:cz}:
\begin{eqnarray}
\sqrt{-g}T_{\mu\nu}^2, && \sqrt{-g}R_{\mu\nu}T^{\mu\nu},
\end{eqnarray}
where: $$T_{\mu\nu} = F_{\mu\alpha}F_{\nu}^{\ \alpha}
-\frac14g_{\mu\nu}F^{\alpha\beta}F_{\alpha\beta},$$ is the Maxwell
stress tensor, and where: $$R^\mu_{\ \nu\alpha\beta} \equiv
\partial_\alpha \Gamma^\mu_{\nu\beta}-
\partial_\beta \Gamma^\mu_{\nu\alpha}+\Gamma^\mu_{\sigma\alpha}
\Gamma^\sigma_{\nu\beta}-\Gamma^\mu_{\sigma\beta}\Gamma^\sigma_{\nu\alpha},$$
is the Einstein curvature tensor. Examples of similar 1-loop
divergences for the mixed graviton and scalar fields are:
\begin{eqnarray}\sqrt{-g}R^{\mu\nu} \partial_\mu \phi^*
\partial_\nu \phi, & \sqrt{-g} R |\partial_\mu \phi|^2,
& \sqrt{-g} R m^2 |\phi|^2.
\end{eqnarray}
The two photon contributions are seen to go as $\left(\sim
\tilde\partial\tilde\partial\tilde\partial\tilde\partial\right)$, while
the scalar contributions go as $\left(\sim
\tilde\partial\tilde\partial\partial\partial\right)$ and
$\left(\tilde\partial\tilde\partial\right)$ respectively. So clearly the 1-loop
singularities correspond to higher derivative couplings of the
fields.

Of course there will also be examples of mixed terms with both
photon, graviton and complex scalar fields. We will not consider
any of these terms explicitly.

As we calculate the 1-loop diagrams using the minimal theory,
singular terms with higher derivative couplings of the fields will
unavoidably appear. None of these singularities need to
be worried about explicitly, because the combined theory is
seen as an effective field theory.

To make the combined theory an effective field theory we thus need to
include into the minimal derivative coupled Lagrangian a piece as:
\begin{equation}
{\cal L}_{\text{photon}} = \sqrt{-g}\Big[c_1 T_{\mu\nu}^2 + c_2R_{\mu\nu}T^{\mu\nu} + \ldots \Big],
\end{equation}
for the photon field, and:
\begin{equation}\begin{split}
{\cal L}_{\text{scalar}} &= \sqrt{-g} \Big[d_1 R^{\mu\nu} \partial_\mu \phi^* \partial_\nu \phi + d_2 R |\partial_\mu \phi|^2
+ d_3 R m^2 |\phi|^2 + \ldots \Big],
\end{split}\end{equation}
for the scalar field. The ellipses symbolize other higher
derivative couplings at 1-loop order which are not included in the
above equations, $e.g.$, other higher derivative couplings and mixed
contributions with both photon, graviton and scalar couplings.

The coefficients $(c_1$, $c_2$, $d_1$, $d_2$ and $d_3, \ldots)$ in
the above equations are in the effective theory seen as energy
scale dependent couplings constants to be measured experimentally.
All singular field terms from the lowest order Lagrangian are thus
absorbed into the effective action, leaving us with a theory at
1-loop order, with a finite number of coupling coefficients to be
determined by experiment.

The effective combined theory of scalar QED and general relativity
is thus in some sense a traditional renormalizable theory at
1-loop order. At low energies the theory is determined only by the
minimal derivative coupled Lagrangian, however at very high
energies, higher derivative terms will manifest themselves in
measurable effects, and the unknown coefficients ($c_1$, $c_2,
\ldots$),  ($d_1$, $d_2$ and $d_3, \ldots$) will have to be
determined explicitly by experiment. This process of absorbing
generated singular field terms into the effective action will of
course have to continue at every loop order.

\subsection{Analytic and non-analytic contributions}
Computing the leading long-range and low-energy quantum
corrections of an effective field theory, a useful distinction is
between non-analytical and analytical contributions from the
diagrams. Non-analytical contributions are inherently non-local
effects not expandable in a power series in momentum. Such effects
come from the propagation of massless particle modes, $e.g.$,
gravitons and photons. The difference between massive and massless
particle modes originates from the impossibility of expanding a
massless propagator $\left (\sim \frac{1}{q^2}\right )$ while we
have the well known:
\begin{equation}
\frac{1}{q^2-m^2} = -\frac{1}{m^2}\Big(1+\frac{q^2}{m^2}+\ldots \Big),
\end{equation}
expansion for the massive propagator. No $\left(\sim
\frac{1}{q^2}\right)$-terms are generated in the above expansion of the
massive propagator, thus such terms all arise from the propagation
of massless modes.

As will be explicitly seen, to leading order the non-analytic
contributions are governed only by the minimally coupled Lagrangian.

The analytical contributions from the diagrams are local effects and
thus expandable in power series.

Non-analytical effects are typically originating from terms which in the
S-matrix go as, $e.g.$, $\left(\sim \ln(-q^2)\right)$ or
$\left(\sim\frac1{\sqrt{-q^2}}\right)$, while the generic
example of an analytical contribution is a power series
in momentum $\left(q\right)$. Our interest is
only in the non-local effects, thus we will only consider the
non-analytical contributions of the diagrams.

The high-energy renormalization of the effective field
theory thus does not concern us --- we focus only
on finding the low-energy leading non-analytical momentum
terms from the 1-loop diagrams. Singular analytical
momentum effects, which are to be absorbed into the coefficients
of the higher derivative couplings of the effective theory,
are of no particular interest to us as they have
no low-energy manifestation.

\subsection{The quantization of general relativity}
The procedure of the background field quantization is as follows.
We define the metric as the sum of a background
part $\left(\bar g_{\mu\nu}\right)$ and a
quantum contribution
$\left(\kappa h_{\mu\nu}\right)$
where $(\kappa)$ is defined from
$\left(\kappa^2 = 32G\pi\right)$.

The expansion of $(g_{\mu\nu})$ is done as follows, we define:
\begin{equation}
g_{\mu\nu}\equiv \bar g_{\mu\nu} + \kappa h_{\mu\nu}.
\end{equation}
From this equation we get the expansions for the upper metric
field $\left(g^{\mu\nu}\right)$ (it is defined to be the inverse matrix),
and for $\left(\sqrt{-g}\right)$ (where $\left(\det(g_{\mu\nu})=g\right)$):
\begin{equation}\begin{split}
g^{\mu\nu} &= \bar g^{\mu\nu} - \kappa h^{\mu\nu} + h_\alpha^\mu h^{\alpha\nu}+\ldots\\
\sqrt{-g} &= \sqrt{-\bar g}\left[1+ \frac12\kappa h -\frac14h_\beta^\alpha h_\alpha^\beta
+\frac18 h^2+ \ldots\right],
\end{split}\end{equation}
in the above equation we have made the definitions:
\begin{equation}
h^{\mu\nu}\equiv \bar g^{\mu\alpha} \bar g^{\nu\beta}
h_{\alpha\beta}
\end{equation}
and
\begin{equation}h  \equiv \bar g^{\mu\nu} h_{\mu\nu}.
\end{equation}

In the quantization, the Lagrangian is expanded in the gravitational
fields, separated in quantum and background parts, and the vertex factors
as well as the propagator are derived from the expanded
action. Details of these calculations can be found in the literature,
for a detailed derivation, see ref.~\cite{Thesis}.

The matter Lagrangian is also expanded. In the scalar QED case we expand the
action in the scalar field $(\phi)$ in the vector field $(A_\mu)$ as
well as in the gravitational field $(h_{\mu\nu})$.
The expanded result for the photon parts reads:
\begin{equation}\begin{split}
{\cal L} &= -\frac14\kappa h\left(\partial_\mu A_\alpha
\partial^\mu A^\alpha - \partial_\mu A_\alpha \partial^\alpha
A^\mu\right)\\& + \frac12\kappa h^{\mu\nu}\left(\partial_\mu
A_\alpha
\partial_\nu A^\alpha + \partial_\alpha A_\mu \partial^\alpha A_\nu
- \partial_\alpha A_\mu \partial_\nu A^\alpha -\partial_\alpha A_\nu \partial_\mu
A^\alpha\right),
\end{split}\end{equation}
while the complex scalar part can be quoted as:
\begin{multline}
{\cal L} = \frac12\kappa h\left(|\partial_\mu\phi|^2 -
m^2|\phi|^2\right) -\kappa
h^{\mu\nu}(\partial_\mu\phi^*\partial_\nu\phi)
+\left(ieA_\mu\partial^\mu \phi^*\phi - ieA_\mu \phi^*\partial^\mu
\phi\right) +e^2A_\mu A^\mu |\phi|^2\\  + \frac12\kappa
h\partial_\mu \phi^*(ieA^\mu)\phi-\frac12\kappa h(ieA^\mu)
\phi^*\partial_\mu \phi - \kappa
h^{\mu\nu}\partial_\mu\phi^*(ieA_\nu)\phi + \kappa h^{\mu\nu}
(ieA_\mu)\phi^*\partial_\nu\phi .
\end{multline}

One can immediately find the vertex rules for the lowest
order interaction vertices of photons, complex scalars and
gravitons from the above equations.

Another necessary issue to consider is the gravitational coupling of
fermions, an area which is not very investigated in the literature. See
ref.~\cite{Veltman2,Deser:cy} for some useful considerations and calculations.

To couple a fermion to a gravitational field an additional formalism is needed.
A coupling through a metric is not possible, see e.g.~ref.~\cite{cartan,weyl},
so a new object {\it the vierbein} thus has to be introduced. Resembling the matrix
square root of the metric, it allows the Dirac equation
in curved space-time to be written consistently in the following way:
\begin{equation}
\sqrt{e}{\cal L}_m=\sqrt{e}\bar{\psi}(i\gamma^a{e_a}^\mu D_\mu-m)\psi,
\end{equation}
here $\left({e_a}^\mu\right)$ is the vierbein field which
connects the global curved space-time with a space-time which is
locally flat.  The vierbein is connected to the metric through the
following set of relations:
\begin{equation}\begin{split}
{e^a}_\mu {e^b}_\nu\eta_{ab}&=g_{\mu\nu},\\
{e^a}_\mu e_{a\nu}&= g_{\mu\nu},\\
e^{a\mu}e_{b\mu}&=\delta^a_b,\\
e^{a\mu}{e_a}^\nu&=g^{\mu\nu},
\end{split}\end{equation}
and it is thus indeed seen to be the matrix square root of the metric.

We also have to define a covariant derivative which acts on fermion
fields:
\begin{equation}
D_\mu\psi=\partial_\mu\psi+{i\over 4}\sigma^{ab}\omega_{\mu ab},
\end{equation}
Insisting that (a torsion-free scenario):
\begin{equation}
e^a_{\ \mu;\nu} = \partial_\nu e^a_{\ \mu} - \Gamma^\sigma_{\mu\nu} e^a_{\ \sigma}
+ {\omega_{\nu}}_{\ b}^a e^b_{\ \mu} = 0,
\end{equation}
one can relate $(\omega_{\mu ab})$ to the vierbein field and derive that:
\begin{equation}
\omega_{\mu ab}={1\over 2}{e_a}^\nu(\partial_\mu e_{b\nu}-\partial_\nu
e_{b\mu})-{1\over
2}{e_b}^\nu(\partial_\mu
e_{a\nu}-\partial_\nu e_{a\mu})
+{1\over 2}{e_a}^\rho{e_b}^\sigma(\partial_\sigma
e_{c\rho}-\partial_\rho e_{c\sigma}){e_\mu}^c.
\end{equation}
In order to quantize the vierbein fields we make the expansion:
\begin{equation}
{e^a}_\mu=\delta^a_\mu+c^{(1)a}_\mu+c^{(2)a}_\mu+\ldots,
\end{equation}
the superscript indices in the above equation correspond to the number of powers
of the gravitational coupling $\left(\kappa\right)$. This notation will
useful later, dealing with the corrections to the metric.

The inverse vierbein matrix is:
\begin{equation}
{e_a}^\mu=\delta_a^\mu-c_a^{(1)\mu}-c_a^{(2)\mu}+{c^{(1)\mu}_b{c^{(1)b}_a+\ldots}},
\end{equation}
so that the metric becomes:
\begin{equation}\begin{split}
g_{\mu\nu}&=\eta_{\mu\nu}+c^{(1)}_{\mu\nu}+c^{(1)}_{\nu\mu}+c^{(2)}_{\mu\nu}
+c^{(2)}_{\nu\mu}
+{c^{(1)a}}_\mu c^{(1)}_{a\nu}+\ldots,\\
g^{\mu\nu}&=\eta^{\mu\nu}-c^{(1)\mu\nu}-c^{(1)\nu\mu}-c^{(2)\mu\nu}
-c^{(2)\nu\mu}+c^{(1)a\mu} c^{(1)\nu}_a+c^{(1)\mu a} {c^{(1)}_a}^\nu+ c^{(1)\mu a}
c^{(1)\nu}_a+\ldots,
\end{split}\end{equation}
in terms of the vierbein fields.
Only the symmetric part of the $c$-field is necessary to keep in expansions
and derivations of the vertex rules, this is because the non-analytic (long-range) effects
only arise from such contributions. Antisymmetric field components will not generate
non-analytic effects and are associated with the freedom of transforming among
local Lorentz frames.

To first order in $(\kappa)$ the metric is connected to the
vierbein fields in the following way:
$$c^{(1)}_{\mu\nu}\rightarrow {1\over
2}(c^{(1)}_{\mu\nu}+c^{(1)}_{\nu\mu})={1\over 2}h^{(1)}_{\mu\nu},$$
and we can rewrite all vierbein equations in terms of the metric fields:
\begin{eqnarray}
{\rm det}\,e&=&1+c+{1\over 2}c^2-{1\over
2}{c_a}^b{c_b}^a+\ldots\nonumber\\
&=&1+{1\over 2}h+{1\over 8}h^2-{1\over 8}{h_a}^b{h_b}^a+\ldots,
\end{eqnarray}
and so forth for the other equations.

Making all expansions we arrive at:
\begin{eqnarray}
\sqrt{e}{\cal L}_m^{(0)}&=&\bar{\psi}({i\over
2}\gamma^\alpha\delta^\mu_\alpha\partial_\mu^{LR}-m)\psi,\\
\sqrt{e}{\cal L}_m^{(1)}&=&-{1\over
2}h^{(1)\alpha\beta}\bar{\psi}i\gamma_\alpha
\partial_\beta^{LR}\psi-{1\over 2}h^{(1)}\bar{\psi}({i\over
2}\not\!{\partial}^{LR}-m)\psi,\\
\sqrt{e}{\cal L}_m^{(2)}&=&-{1\over
2}h^{(2)\alpha\beta}\bar{\psi}i\gamma_\alpha
\partial_\beta^{LR}\psi-{1\over 2}h^{(2)}\bar{\psi}({i\over
2}\not\!{\partial}^{LR}-m)\psi\\
&-&{1\over 8}h^{(1)}_{\alpha\beta}
h^{(1)\alpha\beta}\bar{\psi}i\gamma^\gamma\partial_\lambda^{LR}\psi+{1\over
16}(h^{(1)})^2\bar{\psi}i\gamma^\gamma\partial_\gamma^{LR}\psi\nonumber \\
&-&{1\over
8}h^{(1)}\bar{\psi}i\gamma^\alpha{h_\alpha}^\lambda\partial_\lambda^{LR}\psi
+{3\over
16}h_{\delta\alpha}^{(1)}h^{(1)\alpha\mu}\bar{\psi}i\gamma^\delta\partial_\mu^{LR}\psi\nonumber\\
&+&{1\over 4}h_{\alpha\beta}^{(1)}h^{(1)\alpha\beta}\bar{\psi}
m\psi-{1\over
8}(h^{(1)})^2\bar{\psi}m\psi\nonumber \\
&+&{i\over 16}h_{\delta\nu}^{(1)}(\partial_\beta
h^{(1)\nu}_\alpha-\partial_\alpha h^{(1)\nu}_\beta)
\epsilon^{\alpha\beta\delta\epsilon}\bar{\psi}\gamma_\epsilon\gamma_5\psi\nonumber ,
\end{eqnarray}
where
$$\bar{\psi}\partial_\alpha^{LR}\psi\equiv \bar{\psi}\partial_\alpha\psi-
(\partial_\alpha\bar{\psi})\psi.$$
as the the expanded Dirac Lagrangian.

From these equations vertex forms for fermions can be derived. A summary of the vertex
rules needed for our survey of the scattering amplitudes and the metric
is presented in appendix~\ref{vertex}.

\section{Quantum Corrections to the Schwarzschild and Kerr Metrics}
In this section we will discuss the results of the paper
ref.~\cite{B2}, which was devoted to a study of the quantum
and post-Newtonian corrections to the Schwarzschild and Kerr
metrics. Looking at the 1-loop lowest order radiative corrections to the
interactions of massive particles, it was shown that long-range
corrections to the Schwarzschild and Kerr metrics was possible to obtain,
through the calculation of the non-analytic gravitational
radiative corrections for a massive source. The results for the Kerr
metric was obtained using a massive fermion as the source, while the
Schwarzschild result was found using a massive scalar source.

The classical result for the Schwarzschild metric is refs.~\cite{wein,sch}:
\begin{eqnarray}
g_{00}&=&\left(1-{Gm\over r}\over 1+{Gm\over r}\right)=1-2{Gm\over r}
+2{G^2m^2\over r^2}+\ldots\nonumber,\\
g_{0i}&=&0\nonumber,\\
g_{ij}&=&-\delta_{ij}(1+{Gm\over r})^2-{G^2m^2\over r^2}\left(1+{Gm\over
r}
\over 1-{Gm\over r}\right){r_ir_j\over r^2}\nonumber\\
&=&-\delta_{ij}\left(1+2{Gm\over r}+{G^2m^2\over r^2}\right)
-{r_ir_j\over r^2}{G^2m^2\over
r^2}+\ldots,\label{eq:sch}
\end{eqnarray}
while the corresponding result for the Kerr metric is ref.~\cite{ker}:
\begin{eqnarray}
g_{00}&=&\left(1-{Gm\over r}\over 1+{Gm\over
r}\right)+\ldots=1-2{Gm\over r}
+2{G^2m^2\over r^2}+\ldots\nonumber,\\
g_{0i}&=&{2G\over r^2(r+mG)}(\vec{S}\times\vec{r})_i+\ldots=\left(
{2G\over r^3}-{2G^2m\over
r^4}\right)(\vec{S}\times\vec{r})_i+\ldots\nonumber,\\
g_{ij}&=&-\delta_{ij}(1+{Gm\over r})^2-{G^2m^2\over r^2}\left(1+{Gm\over
r}
\over 1-{Gm\over r}\right){r_ir_j\over r^2}+\ldots\nonumber\\
&=&-\delta_{ij}\left(1+2{Gm\over r}+{G^2m^2\over r^2}\right)
-{r_ir_j\over r^2}{G^2m^2\over
r^2}+\ldots.\label{eq:ker}
\end{eqnarray}
We note that in both the above two expressions for the metrics we have kept
only terms linear in momentum.

One of the key result of this paper was the reproduction
of the expressions for the classical parts of the metric
using Feynman diagrams. Together with these classical
post-Newtonian corrections
we produce some additional corrections not present
in general relativity. These are interpreted as the
long-distance quantum corrections to the metric.
The results for the metric including these corrections are in the
Schwarzschild case:
\begin{eqnarray}
g_{00}&=&1-2{Gm\over r}
+2{G^2m^2\over r^2}+{62G^2 m \hbar\over 15\pi r^3} +\ldots\nonumber,\\
g_{0i}&=& 0\nonumber,\\
g_{ij} &=&-\delta_{ij}\left(1+2{Gm\over r}+{G^2m^2\over r^2}+{14G^2 m
\hbar\over 15\pi r^3} -{76G^2m\hbar (1-\log r)\over 15 \pi r^3}\delta_{ij}\right)
\nonumber \\  &-&{r_ir_j\over
r^2}\left({G^2m^2\over r^2}+{76G^2 m \hbar\over 15\pi
r^3}+{76G^2m\hbar (1-\log r)\over 5 \pi r^3}\right)+\ldots,\label{eq:sh21}
\end{eqnarray}
while for the Kerr metric:
\begin{eqnarray}
g_{00}&=&1-2{Gm\over r}
+2{G^2m^2\over r^2}+{62G^2 m \hbar\over 15\pi r^3}+\ldots\nonumber,\\
g_{0i}&=&\left( {2G\over r^3}-{2G^2m\over r^4}+{36G^2\hbar\over 15\pi
r^5}\right)(\vec{S}\times\vec{r})_i+\ldots\nonumber,\\
g_{ij}&=&-\delta_{ij}\left(1+2{Gm\over r}+{G^2m^2\over r^2}+{14G^2 m
\hbar\over 15\pi r^3}-{76G^2m\hbar (1-\log r)\over 15 \pi r^3}\delta_{ij}
\right)\nonumber \\  &-&{r_ir_j\over
r^2}\left({G^2m^2\over r^2}+{76G^2 m \hbar\over 15\pi
r^3}+{76G^2m\hbar (1-\log r)\over 5 \pi r^3}\right)+\ldots.\label{eq:ker21}
\end{eqnarray}
The quantum corrections in the above expressions carry an
explicit factor of ($\hbar$) put in by dimensional analysis.
The spin independent classical parts are the same for both the
Kerr and the Schwarzschild metrics. This is as expected from
general relativity. That also the quantum contributions satisfy
this property of universality of fermions and bosons is another
interesting achievement of our calculations.

\section{How to derive a metric from a static source}
The metric for a static source can be extracted using
the Einstein equation. The metric will be a function
of the energy-momentum tensor, that will have to be derived
for the particular source. Next we solve for the metric order
by order in the gravitational coupling $(\kappa)$.

To lowest approximation a source is just a point particle, and
depending on the type of matter, with or without a spin.
To higher orders of approximation the picture
of a point particle is no longer correct, because there will be
a surrounding gravitational field which effects
have to be included in the calculations, $e.g.,$ the gravitational
self-coupling. In order to add such effects into the expression
for the energy-momentum tensor it has to be corrected so that
it includes the energy-momentum carried by the gravitational fields.

The masslessness of the propagating graviton assures us that there will be
long-ranged fields around the source, and that the gravitational fields
will dependent on the transverse momentum, $i.e.$, the exchanged momentum
between sources.

Calculating the 1-loop radiative corrections to the source, non-analytic
terms like
$\left(\sim\sqrt{-q^2}\right)$ and $\left(\sim q^2\ln-q^2\right)$,
together with analytic terms like $\left(q^2,q^4...\right)$ are generated.
Here $\left(q\right)$ defines
the transferred momentum. As we previously have discussed, the analytic
terms have a short-ranged nature, while the long-ranged terms always will
be non-analytic. The term $\left(\sim\sqrt{-q^2}\right)$ will
carry the long-ranged post-Newtonian effects, while the term
$\left(q^2\ln-q^2\right)$ will give rise
to a quantum contribution and be proportional to $\left(\hbar\right)$.

Including the 1-loop radiative corrections to the
energy-momentum tensor we can solve the equation of motion for the
post-Newtonian and quantum corrections to the metric.

The metric will take the generic form:
\begin{eqnarray}
{\rm metric} &\sim& Gm\int {d^3q \over (2\pi)^3 }
e^{i\vec{q}\cdot\vec{r}} {1 \over \vec{q}^2}
\left[ 1- a G
{\vec{q}^2 }\sqrt{m^2 \over \vec{q}^2} - b G
{\vec{q}^2 } \log(\vec{q}^2) - c G {\vec{q}^2 }+\ldots
\right] \nonumber \\
&\sim& Gm \left[ {1\over r} +{a G m\over  r^2} +{ b G  \hbar\over
r^3}
+{c G} \delta^3(x) +\ldots \right],
\end{eqnarray}
where $\left(a\right)$, $\left(b\right)$ and $\left(c\right)$ are
dimensionless factors.
The non-analytic terms are completely unrelated to the high-energy scale of gravity. The predictions from such terms are
only related to the low-energy scale of general relativity.
The effective terms of the Lagrangian do not provide
long-distance corrections of the metric at 1-loop
leading order and will hence be neglected in the calculations.

\section{The form factors}
In this section we will discuss the generic expressions
for the energy-momentum tensors of scalars and fermions.

The transition density:
\begin{equation}
\langle p_2|T_{\mu\nu}(x)|p_1\rangle,
\end{equation}
expresses the energy-momentum tensor in quantum
mechanics. If we make the requirement that
$\left(T_{\mu\nu}\right)$ is a second rank tensor and
that $\left(T_{\mu\nu}\right)$ is conserved, $i.e.$,
$\left(\partial^\mu T_{\mu\nu}=0\right)$, we arrive at:
\begin{equation}
\langle p_2|T_{\mu\nu}(x)|p_1\rangle ={e^{i(p_2-p_1)\cdot x}\over
\sqrt{4E_2E_1}}\left[2P_\mu P_\nu F_1(q^2)+(q_\mu
q_\nu-\eta_{\mu\nu} q^2)F_2(q^2)\right],
\end{equation}
as the most general scalar form for
$\left(T_{\mu\nu}\right)$. In the above expression the following
definitions have been made:
\begin{equation}
P_\mu={1\over 2}(p_1+p_2)_\mu,
\end{equation}
and
\begin{equation}
q_\mu=(p_1-p_2)_\mu,
\end{equation}
as wells as we have taken the traditional normalization:
\begin{equation}
\langle p_2|p_1\rangle =2E_1(2\pi)^3\delta^3(\vec{p}_2-\vec{p}_1).
\end{equation}

For fermions the corresponding expression for the
energy-momentum tensor is, see ref.~\cite{pag}:
\begin{equation}\begin{split}
\langle p_2|T_{\mu\nu}|p_1\rangle &=\bar{u}(p_2)\left[ F_1(q^2)P_\mu P_\nu{1\over
m}\right. -F_2(q^2)({i\over 4m}\sigma_{\mu\lambda}q^\lambda
P_\nu+{i\over 4m}\sigma_{\nu\lambda} q^\lambda P_\mu)\\
&+\left. F_3(q^2)(q_\mu q_\nu-\eta_{\mu\nu}q^2){1\over m}\right]
u(p_1).
\end{split}\end{equation}
In the above expression for the fermions the notation of
Bjorken and Drell, ref.~\cite{bjd}, is taken.
The condition $\left(F_1(q^2=0)=1\right)$ is related to the conservation of
energy and momentum as in the scalar case. Imposing that angular
momentum is conserved:
\begin{equation}\begin{split}
\hat{M}_{12}&=\int d^3x(T_{01}x_2-T_{02}x_1)\\
&\stackrel{q\rightarrow 0}{\longrightarrow}-i(\nabla_{q})_2\int d^3x
e^{i\vec{q}\cdot\vec{r}}T_{01}(\vec{r})
+i(\nabla_{q})_1\int d^3xe^{i\vec{q}\cdot\vec{r}}T_{02}(\vec{r}),
\end{split}\end{equation}
requires that:
\begin{equation}
\lim_{q\rightarrow 0}\langle p_2|\hat{M}_{12}|p_1\rangle ={1\over 2}={1\over 2}
\bar{u}_\uparrow(p)\sigma_3u_\uparrow(p)F_2(q^2).
\end{equation}
The constraint of angular momentum conservation thus implies:
$\left(F_2(q^2=0)=1\right)$.

Calculating the energy-momentum tensor
for massive scalars and fermions we have to lowest order the
expressions:
\begin{equation}
\langle p_2|T_{\mu\nu}^{(0)}(0)|p_1\rangle ={1\over \sqrt{4E_2E_1}}\left[2P_\mu P_\mu
-{1\over 2}(q_\mu
q_\nu-\eta_{\mu\nu}q^2)\right],\label{eq:emom}
\end{equation}
for the scalar, together with:
\begin{equation}\begin{split}
\langle p_2|T_{\mu\nu}^{(0)}(0)|p_1\rangle &=\bar{u}(p_2){1\over
2}\left(\gamma_\mu P_\nu
+\gamma_\nu P_\mu\right)u(p_1)\\
&=\bar{u}(p_2)\left[{1\over m}P_\mu P_\nu -{i\over
4m}(\sigma_{\mu\lambda}q^\lambda
P_\nu+\sigma_{\nu\lambda}q^\lambda P_\mu)\right]u(p_1),
\end{split}\end{equation}
for the fermion. These expressions can be read off from the vertex
rules.

\section{How to derive the metric from the equations of motion}
In this section we describe how to calculate the metric from the
Einstein equation of motion. The starting point is the basic
Einstein-Hilbert action which is given by:
\begin{equation}
S_g=\int d^4x \sqrt{-g}\left({1\over 16\pi G}R+{\cal
L}_m\right),
\end{equation}
where $\left({\cal L}_m\right)$ defines a Lagrangian density for
a specific type of matter. Here we will consider both fermionic
and bosonic matter.

Through the variational principle we can generate the
equation of motion:
\begin{equation}
R_{\mu\nu}-{1\over 2}g_{\mu\nu}R=-8\pi GT_{\mu\nu}.
\end{equation}
This is the Einstein equation. The energy-momentum tensor
$\left(T_{\mu\nu}\right)$ can formally be expressed as:
\begin{equation}
T_{\mu\nu}={2\over \sqrt{-g}}{\partial\over \partial
g^{\mu\nu}}(\sqrt{-g}{\cal L}_m).
\end{equation}

To solve the field equation we have to expand the Einstein equation
in terms of the metric. Hence we make a weak field limit expansion
of $\left(g_{\mu\nu}\right)$ in powers of the gravitational
coupling $\left(G\right)$:
\begin{eqnarray}
g_{\mu\nu}&\equiv&\eta_{\mu\nu}+
h_{\mu\nu}^{(1)}+h_{\mu\nu}^{(2)}+\ldots,\nonumber\\
g^{\mu\nu}&=&\eta^{\mu\nu}-h^{(1)\mu\nu}-h^{(2)\mu\nu}
+h^{(1)\mu\lambda} {h^{(1)}}_\lambda{}^\nu+\ldots,
\end{eqnarray}
the superscripts in these expressions count the number of
powers of $\left(\kappa\right)$ which go into the field. All indices in the
expressions are raised or lowered by $\left(\eta_{\mu\nu}\right)$.

The determinant can be expanded as:
\begin{equation}
\sqrt{-g}=\exp{1\over 2}{\rm tr}\log\,g=1+{1\over 2}(h^{(1)}+h^{(2)})
-{1\over
4}h^{(1)}_{\alpha\beta}h^{(1)\alpha\beta}+{1\over 8}h^{(1)2}+\ldots,
\end{equation}
together with the curvatures:
\begin{eqnarray}
R_{\mu\nu}^{(1)}&=&{1\over 2}\left[\partial_\mu\partial_\nu h^{(1)}+
\partial_\lambda\partial^\lambda
h_{\mu\nu}^{(1)}-\partial_\mu\partial_\lambda {h^{(1)}}^\lambda{}_\nu
-\partial_\nu\partial_\lambda {h^{(1)}}^\lambda{}_\mu\right],\nonumber\\
R^{(1)}&=&\Box h^{(1)}-\partial_\mu\partial_\nu
h^{(1)\mu\nu},\nonumber\\
R_{\mu\nu}^{(2)}&=&{1\over 2}\left[\partial_\mu\partial_\nu h^{(2)}+
\partial_\lambda\partial^\lambda
h_{\mu\nu}^{(2)}-\partial_\mu\partial^\lambda h^{(2)}_{\lambda\nu}
-\partial_\nu\partial^\lambda h^{(2)}_{\lambda\mu}\right]\nonumber\\
&-&{1\over 4}\partial_\mu h^{(1)}_{\alpha\beta}\partial_\nu
h^{(1)\alpha\beta} -{1\over 2}\partial_\alpha
h^{(1)}_{\mu\lambda}\partial^\alpha h^{(1)}_{\lambda\nu}+{1\over
2}\partial_\alpha h^{(1)}_{\mu\lambda}
\partial^\lambda h^{(1)\alpha}_\nu\nonumber\\
&+&{1\over 2}h^{(1)\lambda\alpha}\left[\partial_\lambda\partial_\nu
h^{(1)}_{\mu\alpha}+\partial_\lambda\partial_\mu h^{(1)}_{\nu\alpha}-
\partial_\mu\partial_\nu h^{(1)}_{\lambda\alpha}-\partial_\lambda
\partial_\alpha h^{(1)}_{\mu\nu}\right]\nonumber\\
&+&{1\over 2}\left(\partial_\beta h^{(1)\beta\alpha}-{1\over 2}
\partial^\alpha h^{(1)}\right)\left(
\partial_\mu h^{(1)}_{\nu\alpha}+\partial_\nu h^{(1)}_{\mu\alpha}
-\partial_\alpha h^{(1)}_{\mu\nu}\right),\nonumber\\
R^{(2)}&=&\Box h^{(2)} -\partial^\mu\partial^\nu
h^{(2)}_{\mu\nu}-{3\over 4}\partial_\mu
h^{(1)}_{\alpha\beta}\partial^\mu h^{(1)\alpha\beta}+{1\over
2}\partial_\alpha h^{(1)}_{\mu\lambda}
\partial^\lambda h^{(1)\mu\alpha}\nonumber\\
&+&{1\over 2}h^{(1)\lambda\alpha}\left(2\partial_\lambda\partial^\beta
h^{(1)}_{\alpha\beta}-\Box
h^{(1)}_{\lambda\alpha}-
\partial_\lambda\partial_\alpha h^{(1)}\right)\nonumber\\
&+&\left(\partial^\beta h^{(1)\alpha}_\beta-{1\over 2}
\partial^\alpha h^{(1)}\right)
\left(\partial^\sigma h^{(1)}_{\sigma\alpha}-{1\over 2}
\partial_\alpha h^{(1)}\right),
\end{eqnarray}
for a detailed calculation of these quantities, $e.g.$, see ref.~\cite{Thesis}.

The propagator is defined making a gauge choice. Harmonic gauge correspond to,
$\left(g^{\mu\nu} \Gamma^\lambda_{\mu\nu}=0\right)$. As a field expansion this
the same as:
\begin{eqnarray}
0&=&\partial^\beta h^{(1)}_{\beta\alpha}-{1\over 2}\partial_\alpha
h^{(1)}\nonumber\\
&=&\left(\partial^\beta h^{(2)}_{\beta\alpha}-{1\over 2}\partial_\alpha
h^{(2)}-{1\over 2}h^{(1)\lambda\sigma}
\left(\partial_\lambda h^{(1)}_{\sigma\alpha}+
\partial_\sigma h^{(1)}_{\lambda\alpha} -\partial_\alpha
h^{(1)}_{\sigma\lambda}\right)\right) \label{eq:gau}.
\end{eqnarray}

Using all these results we can write the Einstein equation as:
\begin{equation}
\Box h^{(1)}_{\mu\nu}-{1\over 2}\eta_{\mu\nu}\Box h^{(1)}
-\partial_\mu\left(\partial^\beta h^{(1)}_{\beta\nu}-{1\over
2}\partial_\nu h^{(1)}\right)-\partial_\nu\left(\partial^\beta
h^{(1)}_{\beta\mu}-{1\over 2}\partial_\mu
h^{(1)}\right)=-16\pi GT_{\mu\nu}^{\rm matt},
\end{equation}
or using the gauge condition:
\begin{equation}
\Box \left(h_{\mu\nu}^{(1)}-{1\over 2}\eta_{\mu\nu}h^{(1)}\right)
=-16\pi G T_{\mu\nu}^{\rm matt},
\end{equation}
which is equivalent to:
\begin{equation}
\Box h_{\mu\nu}^{(1)}=-16\pi G\left( T_{\mu\nu}^{\rm matt}-{1\over
2}\eta_{\mu\nu} T^{\rm matt}\right).
\end{equation}
In this form the Einstein equation relates the gravitational field
$\left(h_{\mu\nu}^{(1)}\right)$ to the energy-momentum tensor
$\left(T_{\mu\nu}\right)$, where the trace is expressed
as: $\left(T=\eta^{\mu\nu}T_{\mu\nu}\right)$.

We can then calculate the metric as:
\begin{equation}\begin{split}
h_{\mu\nu}(x) &= -16\pi G \int d^3y D(x-y)\Big (T_{\mu\nu}(y)-{1\over
2}\eta_{\mu\nu}
T(y)\Big)  \\
&= -16\pi G \int {d^3q \over (2\pi)^3} e^{i\vec{q}\cdot\vec{r}}{1\over
\vec{q}^2}\Big(T_{\mu\nu}(q)-{1\over
2}\eta_{\mu\nu}
T(q)\Big).
\end{split}\end{equation}

As an explicit example we have in limit where $\left(q^2 \lll
m^2\right)$ for both fermions and scalars:
\begin{equation}
\langle p_2|T_{\mu\nu}^{(0)}(0)|p_1\rangle \simeq m\delta_{\mu 0}\delta_{\nu 0}.
\end{equation}
Using the above form of the Einstein equation we end up with:
\begin{equation}
h_{\mu\nu}^{(1)}(\vec{q})=-{8\pi Gm\over \vec{q}^2}\times\left\{
\begin{array}{cc}
1& \mu=\nu=0\\ 0&\mu=0,\nu=i\\
\delta_{ij}&\mu=i,\nu=j\end{array}\right.+\ldots,
\end{equation}
which in coordinate space results in:
\begin{equation}
h_{\mu\nu}^{(1)}(\vec{r})=f(r)\times\left\{
\begin{array}{cc}1& \mu=\nu=0\\ 0&\mu=0,\nu=i\\
\delta_{ij}&\mu=i,\nu=j\end{array}\right.+\ldots\label{eq:sch1},
\end{equation}
with: $\left(f(r)=-{2Gm\over r}\right)$. This is the leading order Schwarzschild
result ref.~\cite{sch}. A list of the used Fourier transforms is given
in appendix~\ref{Fourier}.

The fermion will to lowest order also have a spin component:
\begin{equation}
\langle p_2|T_{0i}^{(0)}(0)|p_1\rangle \simeq
\chi_2^\dagger\vec{\sigma}\chi_1\times\vec{q},
\end{equation}
This generates the following off-diagonal result for the metric:
\begin{equation}
h_{0i}^{(1)}(\vec{q})=-8\pi i G{1\over \vec{q}^2}(\vec{S}\times\vec{q})_i,
\end{equation}
which in coordinate space is:
\begin{equation}
h_{0i}^{(1)}(\vec{r})={2G\over r^3}(\vec{S}\times\vec{r})_i,
\end{equation}
This result for the off-diagonal metric
is in complete agreement with the Kerr result ref.~\cite{ker}.

Thus to first order in the gravitational coupling we can solve for the
metric given a specific energy-momentum distribution. We will now extend
these results to second order in the gravitational coupling.

The second order Einstein equation has the form:
\begin{eqnarray}
R^{(2)}_{\mu\nu}-{1\over 2}\eta_{\mu\nu}R^{(2)}- {1\over
2}h^{(1)}_{\mu\nu}R^{(1)}=0.
\end{eqnarray}
Expanding the above expression we get:
\begin{equation}
\Box h^{(2)}_{\mu\nu}-{1\over 2}\eta_{\mu\nu}\Box h^{(2)}
-\partial_\mu\left(\partial^\beta h^{(2)}_{\beta\nu}-{1\over
2}\partial_\nu h^{(2)}\right)-\partial_\nu\left(\partial^\beta
h^{(2)}_{\beta\mu}-{1\over 2}\partial_\mu
h^{(2)}\right)\equiv-16\pi GT_{\mu\nu}^{\rm grav}\label{eq:soe}.
\end{equation}
The gravitational field itself generates an energy-momentum distribution which
we will write as $\left(T^{\rm grav}_{\mu\nu}\right)$. It takes the form:
\begin{eqnarray}
8\pi GT_{\mu\nu}^{\rm grav}&=&-{1\over 2}h^{(1)\lambda\kappa}\left[
\partial_\mu\partial_\nu
h^{(1)}_{\lambda\kappa}+\partial_\lambda\partial_\kappa h^{(1)}_{\mu\nu}
-\partial_\kappa\left(\partial_\nu
h^{(1)}_{\mu\lambda}+\partial_\mu
h^{(1)}_{\nu\lambda}\right)\right]\nonumber\\
&-&{1\over 2}\partial_\lambda h^{(1)}_{\sigma\nu}\partial^\lambda
h^{(1)\sigma}{}_\mu +{1\over 2}\partial_\lambda
h^{(1)}_{\sigma\nu}\partial^\sigma h^{(1)\lambda}{}_\mu -{1\over
4}\partial_\nu
h^{(1)}_{\sigma\lambda}\partial_\mu
h^{(1)\sigma\lambda}\nonumber\\
&-&{1\over 4}\eta_{\mu\nu}(\partial_\lambda h^{(1)}_{\sigma\chi}
\partial^\sigma h^{(1)\lambda\chi}
-{3\over 2}\partial_\lambda h^{(1)}_{\sigma\chi}\partial^\lambda
h^{(1)\sigma\chi})
-{1\over 4}h^{(1)}_{\mu\nu}\Box h^{(1)}\nonumber\\
&+&{1\over 2}\eta_{\mu\nu}h^{(1)\alpha\beta}\Box h^{(1)}
_{\alpha\beta}\label{eq:gmn}.
\end{eqnarray}
Employing the gauge condition in the above expression, and isolating for
$\left(h^{(2)}_{\mu\nu}\right)$ we end up with:
\begin{eqnarray}
\Box \bigg(h_{\mu\nu}^{(2)}&-&\left.{1\over
2}\eta_{\mu\nu}h^{(2)}\right)
=-16\pi G T_{\mu\nu}^{\rm grav}\nonumber\\
&+&\partial_\mu \left( h^{(1)\lambda\sigma}\left[\partial_\lambda
h^{(1)}_{\sigma\nu}-
{1\over 2}\partial_\nu h^{(1)}_{\lambda\sigma}\right]\right)
+\partial_\nu\left(
 h^{(1)\lambda\sigma}\left[\partial_\lambda h^{(1)}_{\sigma\mu}-
{1\over 2}\partial_\mu
h^{(1)}_{\lambda\sigma}\right]\right)\\ \nonumber
&-&\eta_{\mu\nu}\partial^\alpha\left(h^{(1)\lambda\sigma}\left[\partial_\lambda
h_{\alpha\sigma}^{(1)}-{1\over 2}\partial_\alpha
h_{\lambda\sigma}^{(1)}\right]\right),
\end{eqnarray}
or using the solution we found for $\left(h^{(1)}_{\mu\nu}\right)$:
\begin{equation}
\Box h_{\mu\nu}^{(2)} =-16\pi G \left(T_{\mu\nu}^{\rm grav}-{1\over
2}\eta_{\mu\nu}T^{\rm grav}\right)
-\partial_\mu\left(f(r)\partial_\nu f(r)\right)
-\partial_\nu\left(f(r)\partial_\mu f(r)\right)\label{eq:eet}.
\end{equation}
For a fermion there will also be a corresponding off-diagonal equation:
\begin{eqnarray}
\Box h_{0i}^{(2)}&=&-16\pi GT_{0i}^{\rm grav}-\nabla_i(h^{(1)}_{0j}
\nabla_jh_{00}^{(1)})
+\nabla_i(h^{(1)}_{jk}\nabla_jh^{(1)}_{0k}) \label{eq:eeu},
\end{eqnarray}
As it is easily verified that:
\begin{equation}
\nabla_i(h^{(1)}_{0j}\nabla_jh_{00}^{(1)})=\nabla_i(h^{(1)}_{jk}
\nabla_jh^{(1)}_{0k})=0,
\end{equation}
the second order off-diagonal Einstein equation has the form:
\begin{equation}
\Box h_{0i}^{(2)} =-16\pi G T_{0i}^{\rm grav}.
\end{equation}
This together with the above equation concludes our investigations
of how to calculate the metric from the energy-momentum tensor. What has
to be done in order to derive the metric is to use the above equations
and to Fourier transformation to get the coordinate results.
The used Fourier transforms are given in appendix~\ref{Fourier}.

\section{Loop corrections and the energy-momentum tensor}
\begin{figure}[h]
\begin{center}
\begin{tabular}{c}
\includegraphics[scale=1.3]{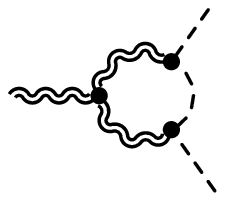}\\ (a)
\end{tabular}
\hspace{1cm}
\begin{tabular}{c}
\includegraphics[scale=1.3]{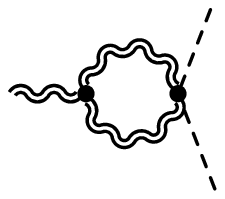}\\ (b)
\end{tabular}
 \caption{The only gravitational radiative
diagrams which carry non-analytic contributions.\label{scalar}}
\end{center}
\end{figure}

In the case of scalars the 1-loop non-analytic radiative vertex corrections
to the energy-momentum tensor were first calculated in refs.~\cite{Donoghue:1993eb,Donoghue:dn}, and an
error in these results was corrected in ref.~\cite{Thesis}. In order to determine the radiative corrections
to the vertex form factors we have to calculate the diagrams, see figure~\ref{scalar}.

These diagrams have the following expressions:
\begin{equation}
A^{\mu\nu}_{(a)}=iP^{\alpha,\lambda\kappa}iP^{\gamma\delta,\rho\sigma}
i\int {d^4\ell\over
(2\pi)^4}{\tau_{6\alpha\beta,m}(p,p'-\ell) \tau_{6\gamma\delta}(p'-\ell,p',m')
\tau^{\mu\nu}_{10\rho\sigma,\lambda\kappa}(\ell,q)\over \ell^2(\ell-q)^2
((\ell-p')^2-m^2)},
\end{equation}
and
\begin{equation}
A^{\mu\nu}_{(b)}={i\over 2}
P^{\alpha\beta,\lambda\kappa}iP^{\gamma\delta,\rho\sigma}
\tau_{8\alpha\beta,\gamma\delta}(p,p')\int{d^4\ell\over (2\pi)^4}
{\tau^{\mu\nu}_{10\lambda\kappa,\rho\sigma}(\ell,q)\over
\ell^2(\ell-q)^2}.
\end{equation}
The formal expressions for the vertex factors can be found in
appendix~\ref{vertex}.

In terms of the form factors the results for the diagrams are:
\begin{equation}\begin{split}
F_1(q^2)&=1+{Gq^2\over \pi}(-{3\over 4}\log{-q^2\over m^2}+
{1\over 16}{\pi^2 m\over \sqrt{-q^2}})+\ldots\nonumber\\
F_2(q^2)&=-{1\over 2}+{Gm^2\over \pi}(-2\log{-q^2\over m^2}+{7\over
8}{\pi^2m\over \sqrt{-q^2}})+\ldots.
\end{split}\end{equation}

These corrections lead to the results for the energy-momentum tensor,
using the integrals listed in appendix~\ref{Fourier}.
\begin{eqnarray}
T_{00}(\vec{r})&=&\int{d^3q\over
(2\pi)^3}e^{i\vec{q}\cdot\vec{r}}\left(mF_1({q}^2)+
{\vec{q}^2\over 2m}F_2(q^2)\right)\nonumber\\
&=&\int{d^3q\over (2\pi)^3}e^{i\vec{q}\cdot\vec{r}}\left[m+\pi
Gm^2(-{1\over 16}+{7\over 16})|\vec{q}|+{Gm\over
\pi}\vec{q}^2\log{\vec{q}^2}({3\over
4}-1)\right]\nonumber\\
&=&m \delta^3(r)-{3Gm^2\over 8\pi r^4}-{3Gm\hbar \over 4\pi^2
r^5},\nonumber\\
T_{0i}(\vec{r})&=&0,\nonumber\\
T_{ij}(\vec{r})&=&{1\over 2m}\int{d^3q\over
(2\pi)^3}e^{i\vec{q}\cdot\vec{r}}(q_iq_j-\delta_{ij}\vec{q}^2)F_2
(q^2)\nonumber\\
&=&\int{d^3q\over (2\pi)^3}e^{i\vec{q}\cdot\vec{r}} \left[{7\pi
Gm^2\over
16|\vec{q}|}(q_iq_j-\delta_{ij}\vec{q}^2)-
(q_iq_j-\delta_{ij}\vec{q}^2) {Gm\over \pi}\log \vec{q}^2
\right]\nonumber\\
&=& -{7Gm^2\over 4\pi r^4}\left({r_ir_j\over r^2}-{1\over
2}\delta_{ij}\right)+ {Gm\hbar \over 2\pi^2 r^5}\left(9\delta_{ij}-15{r_i r_j\over r^2}\right).\nonumber\\
\end{eqnarray}

\begin{figure}[h]
\begin{center}
\begin{tabular}{c}
\includegraphics[scale=1.3]{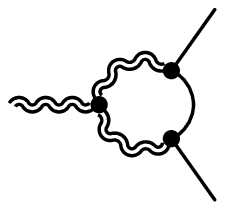}\\ (a)
\end{tabular}
\hspace{1cm}
\begin{tabular}{c}
\includegraphics[scale=1.3]{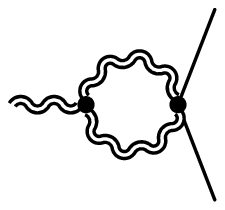}\\ (b)
\end{tabular}
 \caption{The only gravitational radiative
diagrams which carry non-analytic contributions.\label{fermions}}
\end{center}
\end{figure}

For fermions we have to calculate the diagrams, see figure~\ref{fermions}.

They lead to the results in terms of form factors:
\begin{equation}\begin{split}
F_1(q^2)&=1+{Gq^2\over \pi}\left({\pi^2m\over 16\sqrt{-q^2}}-
{3\over 4}\log{-q^2\over m^2}\right) +\ldots,\\
F_2(q^2)&=1+{Gq^2\over \pi}\left({\pi^2m\over 4\sqrt{-q^2}}+
{1\over 4}\log{-q^2\over m^2}\right) +\ldots,\\
F_3(q^2)&={Gm^2\over \pi}\left({7\pi^2m\over
16\sqrt{-q^2}}-\log{-q^2\over m^2}\right) +\ldots,
\end{split}\end{equation}
which were calculated in the paper, ref.~\cite{B2}.

Again these results are used to calculate the energy-momentum tensor:
\begin{eqnarray}
T_{00}(\vec{r})&=&\int{d^3q\over(2\pi)^3}e^{i\vec{q}\cdot\vec{r}}\left(mF_1(-\vec{q}^2)+
{\vec{q}^2\over m}F_3(-\vec{q}^2)\right),\nonumber\\
T_{0i}(\vec{r})&=&i\int{d^3q\over(2\pi)^3}e^{i\vec{q}\cdot\vec{r}}{1\over
2}(\vec{S}\times\vec{q})_iF_2(-\vec{q}^2),\nonumber\\
T_{ij}(\vec{r})&=&{1\over m}\int{d^3q\over(2\pi)^3}e^{i\vec{q}\cdot\vec{r}}
(q_iq_j-\delta_{ij}\vec{q}^2)F_3(-\vec{q}^2),
\end{eqnarray}
where we have defined ($\vec{S}= \vec{\sigma}/2$).

The form factors then become:
\begin{eqnarray}
T_{00}(\vec{r})&=&\int{d^3q\over(2\pi)^3}e^{i\vec{q}\cdot\vec{r}}\left( m +{3Gm^2\pi \over 8}|\vec{q}|
-{Gm\over 4\pi }\vec{q}^2\log \vec{q}^2\right)+\ldots \nonumber \\
&=&m\delta^3(\vec{r}) -{3Gm^2\over 8\pi r^4}-{3Gm\hbar\over 4\pi r^5}+\ldots,\nonumber\\
T_{0i}(\vec{r})&=&{i\over 2}\int{d^3q\over(2\pi)^3}e^{i\vec{q}\cdot\vec{r}}(\vec{S}\times\vec{q})_i
\left(1-{Gm\pi\over 4}|\vec{q}| - {G\over 4\pi}\vec{q}^2\log \vec{q}^2 \right)+\ldots,\nonumber \\
&=& {1\over 2}(\vec{S}\times\vec{\nabla})_i \delta^3(\vec{r})+\left(-{Gm\over 2\pi r^6} +{15G\hbar\over
4\pi^2r^7} \right)(\vec{S}\times\vec{r})_i+\ldots \nonumber\\
T_{ij}(\vec{r})&=&\int{d^3q\over(2\pi)^3}e^{i\vec{q}\cdot\vec{r}}\left({7Gm^2\pi\over16|\vec{q}|}-{Gm\over \pi}\log \vec{q}^2 \right)
\left(q_iq_j-\delta_{ij}\vec{q}^2\right)+\ldots\nonumber \\
&=&-{7Gm^2\over 4\pi r^4}\left({r_ir_j\over r^2}-{1\over 2}\delta_{ij}\right) +
{Gm\hbar \over 2\pi^2 r^5}\left(9\delta_{ij}-15{r_i r_j\over r^2}\right)+\ldots.\quad
\end{eqnarray}

\begin{figure}[h]
\begin{center}
\includegraphics[scale=1.3]{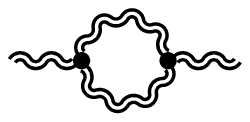}\hspace{1cm}\includegraphics[scale=1.3]{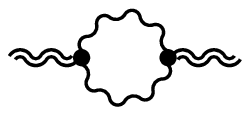}
\caption{The vacuum polarization diagrams including the ghost diagram.\label{vacuumq}}
\end{center}
\end{figure}

Together with the vertex correction there is an additional effect due
to the vacuum polarization of the graviton.
These additional correction terms come from the vacuum polarization corrections
to the propagator and can be graphically depicted, see figure~\ref{vacuumq}.

To understand this better we have to take a look at the effective action.
At 1-loop level it can be written in the following way:
\begin{equation}
Z[h] = -\int d^4x d^4y{1\over 2} \left[ h_{\mu\nu}(x) \right.
\left.\Delta^{\mu\nu,\alpha\beta}(x-y)h_{\alpha\beta}(y)+\textit{O}(h^3)\right]
+Z_{\rm matter}[h,\phi].
\end{equation}
Properly renormalized, the operator: ($\Delta^{\mu\nu,\alpha\beta}(x-y)$), will
contain a contribution from the vacuum polarization.

This can be mathematically expressed as:
\begin{equation}
\Delta^{\mu\nu,\alpha\beta}(x-y) = \delta^4(x-y)
D^{\mu\nu,\alpha\beta}_2 +\hat{\Pi}^{\mu\nu,\alpha\beta}(x-y)
+\textit{O}(\partial^4),
\end{equation}
where the differential operator $\left(D_2^{\mu\nu,\alpha\beta}\right)$ will
give the traditional harmonic gauge propagator and
$\left(\hat{\Pi}^{\mu\nu,\alpha\beta}(x-y)\right)$ contains the vacuum
polarization correction. To generate the effect on the metric we
solve the equation of motion once again:
\begin{eqnarray}
 \Box h_{\mu\nu}(x)&+& {\cal P}_{\mu\nu\alpha\beta}\int
d^4y\hat{\Pi}^{\alpha\beta,\gamma\delta}(x-y)h_{\gamma\delta}(y)\nonumber\\
 &=&-16\pi G(T^{\rm grav}_{\mu\nu} -{1\over 2}\eta_{\mu\nu} T^{\rm grav})
-\partial_\mu(f(r)\partial_\nu f(r))
-\partial_\nu(f(r)\partial_\mu f(r)),\nonumber\\
\quad
\end{eqnarray}
here $\left(P_{\mu\nu,\alpha\beta}\right)$ and
$\left({\cal I}_{\mu\nu\alpha\beta}\right)$ are defined as:
\begin{eqnarray}
{\cal P}_{\mu\nu\alpha\beta} &=& {\cal I}_{\mu\nu\alpha\beta} -{1\over 2}
\eta_{\mu\nu} \eta_{\alpha\beta} \nonumber \\
{\cal I}_{\mu\nu\alpha\beta} &=& {1\over 2}
(\eta_{\mu\alpha}\eta_{\nu\beta}+\eta_{\nu\alpha}\eta_{\mu\beta}).
\end{eqnarray}

Using the harmonic gauge condition, we arrive at:
\begin{eqnarray}
\Box h_{\mu\nu} &=& -16\pi G(T_{\mu\nu}^{\rm grav}-{1\over 2}\eta_{\mu\nu}
T^{\rm grav})-\partial_\mu(f(r)
\partial_\nu f(r)) -\partial_\nu(f(r)\partial_\mu
f(r))\nonumber \\
&+&16\pi G\int d^4y
d^4z~{\cal P}_{\mu\nu\alpha\beta}\hat{\Pi}^{\alpha\beta,\gamma\delta}
(x-y)D(y-z)(T_{\gamma\delta}^{\rm matt}(z)-{1\over
2}\eta_{\gamma\delta} T^{\rm matt}(z)).\nonumber\\
\quad
\end{eqnarray}

\begin{figure}[h]
\begin{center}
\parbox{3.5cm}{\includegraphics[scale=1.3]{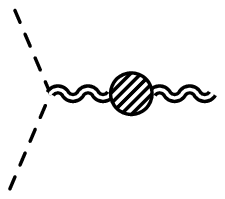}}
\hspace{2cm}
\parbox{3.5cm}{\includegraphics[scale=1.3]{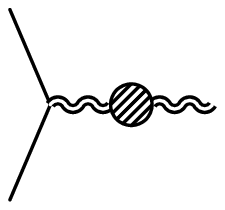}}
\caption{Vacuum polarization modification of the energy-momentum
tensor for a scalar and a fermion.}
\end{center}
\end{figure}

't Hooft and Veltman ref.~\cite{{Veltman1}} were the first to calculate
the vacuum polarization diagram in gravity. The source of the non-locality of the
diagram is the factor of $\left(q^4\log(-q^2)\right)$ and the specific form is:
\begin{eqnarray}
\hat{\Pi}_{\alpha\beta,\gamma\delta}&=&-{2G\over
\pi}\log(-q^2)\left[{21\over 120}q^4
I_{\alpha\beta,\gamma\delta}+{23\over
120}q^4\eta_{\alpha\beta}\eta_{\gamma\delta} \right.\nonumber
-\left.{23\over 120}q^2(\eta_{\alpha\beta}q_\gamma q_\delta
+\eta_{\gamma\delta}q_\alpha
q_\beta)\right.\\
&-&\left.{21\over 240}q^2(q_\alpha q_\delta\eta_{\beta\gamma}+q_\beta
q_\delta\eta_{\alpha\gamma} +q_\alpha
q_\gamma\eta_{\beta\delta}+q_\beta q_\gamma\eta_{\alpha\delta})
+{11\over 30}q_\alpha q_\beta q_\gamma
q_\delta\right].
\end{eqnarray}
Using this result we see that the vacuum polarization introduces a
shift in the metric by:
\begin{equation} \delta h^{(2){\rm vac pol}}_{\mu\nu}(x)=  32 G^2 \int
{d^3q
\over (2\pi)^3}
e^{i\vec{q}\cdot\vec{r}}\log (\vec{q}^2) \left[{21\over
120}T_{\mu\nu}^{\rm matt}(q) + \left({1\over 240}\eta_{\mu\nu}
-{11\over 60}{q^\mu q^\nu\over \vec{q}^2}\right)
   T^{\rm matt}(q)\right].
\end{equation}
This is in terms of components:
\begin{eqnarray}
\delta h_{00}^{(2){\rm vac pol}} &=& - { 43G^2m\hbar  \over 15\pi r^3},
\nonumber \\
\delta h_{ij}^{(2){\rm vac pol}} &=& {G^2 m \hbar\over 15\pi r^3}
(\delta_{ij} +44 {r_i r_j\over r^2})-{44G^2 m(1-\log r) \hbar\over
15 \pi r^3}\left(\delta_{ij}-3\frac{r_i r_j}{r^2}\right).
\end{eqnarray}

Adding all logarithmic corrections together we find:
\begin{eqnarray}
\delta h_{00}^{(2){\rm vertex}} (r)&=&-16\pi G\int{d^3q\over
(2\pi)^3}e^{i\vec{q}\cdot\vec{r}} {1\over
\vec{q}^2}\left({m\over 2}F_1(-\vec{q}^2)
-{\vec{q}^2\over 4m}F_2(-\vec{q}^2)\right)\nonumber\\
&=&-16\pi G\int{d^3q\over (2\pi)^3}e^{i\vec{q}\cdot\vec{r}} {Gm\over
\pi}({3\over 8}+{1\over
2})\log\vec{q}^2={7G^2m\hbar\over \pi r^3},
\nonumber\\
\delta h_{0i}^{(2){\rm vertex}}(r)&=&0,\nonumber\\
\delta h_{ij}^{(2){\rm vertex}}(r)&=&-16\pi G\int{d^3q\over
(2\pi)^3}e^{i\vec{q}\cdot\vec{r}} {1\over
\vec{q}^2}\log\vec{q}^2\left({m\over 2}
F_1(-\vec{q}^2))\delta_{ij}\right.\nonumber\\
&+&\left.{1\over 2m}(q_iq_j+{1\over 2}\delta_{ij}\vec{q}^2)
F_2(-\vec{q}^2)
\right)\nonumber\\
&=&-16\pi G\int{d^3q\over (2\pi)^3}e^{i\vec{q}\cdot\vec{r}}
{Gm\over \pi}\left(\delta_{ij}({3\over 8}-{1\over
2})-{q_iq_j\over \vec{q}^2}\right)\nonumber\\
&=&-{G^2m\hbar\over \pi r^3}\left(\delta_{ij}+8{r_ir_j\over r^2}\right)
+{8G^2m\hbar(1-\log r)\over \pi r^3}\left(\delta_{ij}-3\frac{r_i r_j}{r^2}\right)
.
\end{eqnarray}

These results lead to the metric:
\begin{eqnarray}
g_{00}&=&1-2{Gm\over r}
+2{G^2m^2\over r^2}+{62G^2 m \hbar\over 15\pi r^3} +\ldots.\nonumber\\
g_{0i}&=& 0.\nonumber\\
g_{ij} &=&-\delta_{ij}\left(1+2{Gm\over r}+{G^2m^2\over r^2}+{14G^2 m
\hbar\over 15\pi r^3}-{76G^2m\hbar (1-\log r)\over 15 \pi r^3}\right) \nonumber\\&-&{r_ir_j\over
r^2}\left({G^2m^2\over r^2}+{76G^2 m \hbar\over 15\pi
r^3}+{76G^2m\hbar (1-\log r)\over 5 \pi r^3}\right)+\ldots.
\end{eqnarray}
which clearly is identical to the classical Schwarzschild metric plus
a unique set of quantum corrections, caused by the quantum nature of the
theory of general relativity.

For fermions we make the observation that all diagonal components of
the vacuum polarization are completely identical to the bosonic case.
For the off-diagonal terms there is an extra component associated with the spin.
It can be written:
\begin{eqnarray}
h_{0i}^{(2){\rm vac pol}} &=&  32 G^2\int{d^3q\over
   (2\pi)^3}e^{i\vec{q}\cdot\vec{r}} \log \vec{q}^2 ~{21\over 240}
iF_2(q^2)(\vec{S}\times\vec{q})_i \nonumber \\
&=& {21G^2\hbar \over 5\pi r^5} (\vec{S}\times\vec{r})_i.
\end{eqnarray}
Adding all corrections together we end up with:
\begin{eqnarray}
g_{00}&=&1-2{Gm\over r}
+2{G^2m^2\over r^2}+{62G^2 m \hbar\over 15\pi r^3}+\ldots.\nonumber\\
g_{0i}&=&\left( {2G\over r^3}-{2G^2m\over r^4}+{36G^2\hbar\over 15\pi
r^5}\right)(\vec{S}\times\vec{r})_i+\ldots.\nonumber\\
g_{ij}&=&-\delta_{ij}\left(1+2{Gm\over r}+{G^2m^2\over r^2}+{14G^2 m
\hbar\over 15\pi r^3}-{76G^2m\hbar (1-\log r)\over 15 \pi r^3}\right)\nonumber\\
&-&{r_ir_j\over
r^2}\left({G^2m^2\over r^2}+{76G^2 m \hbar\over 15\pi
r^3}+{76G^2m\hbar (1-\log r)\over 5 \pi r^3}\right)+\ldots.
\end{eqnarray}
This is the result for the classical Kerr metric together with
unique quantum corrections.

The reproduced classical metric and the results for the quantum corrections
are essentially the main results of the paper ref.~\cite{B2}.
In the next section we will discuss the results of the papers
refs.~\cite{B1,B3}, here the diagrammatic 1-loop quantum
corrections will be discussed from the point of view of defining
a potential.

\section{Calculations of the scattering matrix potential}
In this section we will discuss on more general grounds how to
define a potential from the scattering matrix and how to derive
the 1-loop scattering matrix potential for a pure gravitational interaction
and in the mixed theory of scalar QED and gravity.

The general form for any diagram contributing to the scattering
matrix is, $e.g.$, in the mixed theory of gravity scalar QED:
\begin{equation}\begin{split}
{\cal M} &\sim \Big(A+B q^2 + \ldots + (\alpha_1 \kappa^2 +
\alpha_2 e^2) \frac{1}{q^2} +\beta_1 e^2\kappa^2\ln(-q^2) +
\beta_2 e^2\kappa^2 \frac{m}{\sqrt{-q^2}} + \ldots\Big),
\end{split}\end{equation}
where $A, B, \ldots$ correspond to the local analytical
interactions and $\alpha_1, \alpha_2$ and $\beta_1, \beta_2,
\ldots$ correspond to the leading non-analytical, non-local, long-range interactions.

The space parts of the non-analytical terms Fourier transform as:
\begin{equation}\begin{split}
\int \frac {d^3 q}{(2\pi)^3} e^{i\vec q \cdot \vec r}\frac 1{|{\bf q}|^2}& = \frac {1}{4\pi r},\\
\int \frac {d^3 q}{(2\pi)^3} e^{i\vec q \cdot \vec r}\frac 1{|{\bf q}|}& = \frac {1}{2\pi^2 r^2},\\
\int \frac {d^3 q}{(2\pi)^3} e^{i\vec q \cdot \vec r}\ln({\bf q}^2)& = \frac {-1}{2\pi r^3},\\
\end{split}\end{equation}
so clearly these terms will contribute to the long-range
corrections.

The non-analytical contribution, corresponding to the
$\left(\frac{1}{q^2}\right)$-part, gives as seen the Newtonian and Coulomb
potentials respectively. The other non-analytical contributions
generate the leading quantum and classical corrections to the
Coulomb and Newtonian potentials in powers of $\left(\frac1{r}\right)$. It is
necessary to have non-analytic contributions in the matrix
element, to ensure that the S-matrix is unitary.

The analytic contributions will not be considered. As
noted previously these corrections correspond to local
interactions, and are thus only needed for the high-energy
manifestation of the theory. Many of the analytical corrections
will be divergent, and hence have to be carefully absorbed into
the appropriate terms of the effective action of the theory.

We will not consider the radiative corrections due to soft
bremsstrahlung in this approach. In some of the diagrams in gravity,
as well as in QED, there is a need for introducing soft
bremsstrahlung radiative corrections to the sum of the diagrams
constituting the vertex corrections. We will not consider this
aspect of the theory in this approach, as we are not computing the
full amplitude of the S-matrix. Furthermore certain effects have
been included in the recent work of ref.~\cite{Donoghue:2001qc}, where
the gravitational vertex corrections are treated. This issue
should be dealt with at some stage refining the theories, however
here we will carry on and simply concentrate on the leading
post-Newtonian and quantum corrections to the scattering matrix,
leaving this concern for future investigations.

\subsubsection{The definition of the potential}
The various definitions of the potential have been discussed at
length in the literature. We will here define the potential
directly from the scattering matrix amplitude.

In the quantization of general relativity the definition of a
potential is certainly not obvious. One can choose between several
definitions of the potential depending on, $e.g.$, the physical
situation, how to define the energy of the fields, the diagrams
included etc.

Clearly a valid choice of potential should be gauge invariant to
be physically reasonable, but while other gauge theories like QCD
allow a gauge invariant Wilson loop definition --- $e.g.$, for a
quark-anti-quark potential, this is not directly possible in
general relativity.

It has however been attempted to make the equivalent of a Wilson loop
potential for quantum gravity. A Wilson-like potential
seems to be possible to construct in general relativity using the
Arnowitt-Deser-Misner formula for the total energy of the
system ref.~\cite{Modanese:1994bk}. This choice of potential has been discussed
in ref.~\cite{Muzinich:1995uj} in the case of pure gravity coupled to
scalar fields.

A recent suggestion ref.~\cite{Kazakov:2000mu} is that one should use
the full set of diagrams constituting the scattering
matrix, and decide the nonrelativistic potential from the total sum
of the 1-loop diagrams. As the full 1-loop scattering matrix
is involved, this choice of potential gives a gauge invariant
definition.

This choice of potential is equivalent to that
of ref.~\cite{Hamber:1995cq}, where the scalar source pure
gravitational potential was treated.

This choice of potential, which includes all 1-loop diagrams, seems
to be the simplest, gauge invariant definition of the
potential.

We will calculate the nonrelativistic potential using the the
full amplitude. We will simply relate the expectation value for
the ($iT$) matrix to the Fourier transform of the potential
($\tilde
V({\bf q})$) in the nonrelativistic limit in the following way:
\begin{equation}
\langle k_1,k_2 | i T | k_1', k_2' \rangle = -i\tilde {V}({\bf
q})(2\pi)\delta(E - E'),
\end{equation}
where $\left(k_1\right)$, $\left( k_2\right)$ and $\left(k_1'\right)$, $\left(k_2'\right)$ are the incoming and
outgoing momentum respectively, and $\left(E-E'\right)$ is the energy
difference between the incoming and outgoing
states. Comparing this to the definition of the
invariant matrix element $\left(i{\cal M}\right)$ we get from the diagrams:
\begin{equation}
\langle k_1,k_2 | i T | k_1', k_2' \rangle =
(2\pi)^4\delta^{(4)}(k_1-k_1'+k_2-k_2')(i{\cal M}),
\end{equation}
we see that (we have divided the above equation with $\left(2m_1 2m_2\right)$
to obtain the nonrelativistic limit):
\begin{equation}
\tilde V({\bf q}) = -\frac{1}{2m_1}\frac{1}{2m_2}{\cal M},
\end{equation}
so that
\begin{equation}
V({\bf x}) = -\frac{1}{2m_1}\frac{1}{2m_2}\int
\frac{d^3k}{(2\pi)^3}e^{i{\bf k} \cdot {\bf x}}{\cal M}.
\end{equation}
This is how we define the nonrelativistic potential
generated by the considered non-analytic parts. In the above equation
$\left({\cal M}\right)$
is the non-analytical part of the amplitude of the scattering
process to a given loop order. This definition of the potential is
also used in ref.~\cite{Hamber:1995cq}.

We will note here that the above definition of a potential $\left( V({\bf
q})\right)$ using the scattering amplitude is not the only possibility. The
nonrelativistic bound state quantum mechanics potential is arrived
at, if we subtract off the second order Born contribution. This can
be done using the following prescription:
\begin{eqnarray}
i\langle f |T| i\rangle &=& -2\pi i\delta(E-E')\nonumber\\
&\times&\bigg[\langle f| \tilde V_{bs}({\bf q})| i\rangle +\sum_n
\frac{\langle f |\tilde V_{bs}({\bf q})| n\rangle \langle n|
\tilde V_{bs}({\bf q})| i\rangle}{E-E_n+i\epsilon} + \ldots\bigg].
\end{eqnarray}
For a detailed definition of the bound state potential,
see ref.~\cite{Iwasaki}. In a treatment involving the Hamiltonian there
will be terms such as $\left(\frac{Gp^2}r\right)$ contributing to the potential at
the same order. To lowest order in Einstein-Infeld-Hoffmann
coordinates the the bound state potential $\left(\tilde V_{bs}({\bf q})\right)$
will be:
\begin{equation}
\tilde V_{bs} (r)= V(r) + {7 Gm_1m_2 (m_1+m_2)\over 2c^2 r^2}.
\end{equation}

\section{The results for the Feynman diagrams}
\subsection{The diagrams contribution to the non-analytical parts of the scattering matrix
in the combined theory of scalar QED and general relativity}
Of the diagrams contributing to the scattering matrix only a
certain class of diagrams will actually contribute to the sum of
the non-analytical terms considered here --- the logarithmic and
square-root parts. In this treatment we will only in detail consider the
diagrams which contribute with non-analytical contributions.
Diagrams with many massive propagators will usually only
contribute with analytical terms. Due to the involved vertex rules, some of
the diagrams will have a somewhat complicated algebraic structure.
To do the diagrams we developed an algebraic program
for Maple 7(TM)\footnote{\footnotesize{Maple
and Maple V are registered trademarks of Waterloo Maple Inc.}}.
The program contracts the various
indices and performs the loop integrations. In the following we
will go through the diagrams and discuss how they are calculated
in detail. We will begin with the tree diagrams.

\subsubsection{The tree diagrams}
\begin{figure}[h]\vspace{0.5cm}
\begin{minipage}{0.43\linewidth}
\begin{center}
\includegraphics[scale=1]{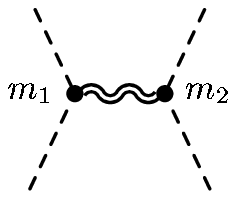}
\end{center}
{\begin{center}1(a)\end{center}}
\end{minipage}
\begin{minipage}{0.43\linewidth}
\begin{center}
\includegraphics[scale=1]{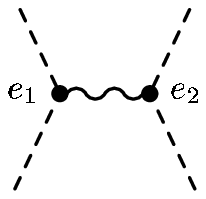}
\end{center}
{\begin{center} 1(b)\end{center}}
\end{minipage}
\caption{The set of tree diagrams contributing to the potential.}
\label{tree}\end{figure}

The set of tree diagrams contributing to the scattering matrix are
those of figure \ref{tree}.
 The formal expression for these diagrams
are:
\begin{equation}
{i \cal M}_{\rm 1(a)} = \tau_2^{\mu\nu}(k_1,k_2,m_1)\bigg[ \frac{i{\cal P}_{\mu\nu\alpha\beta}}{q^2}\bigg] \tau_2^{\alpha\beta}(k_3,k_4,m_2),
\end{equation}
and
\begin{equation}
{i \cal M}_{\rm 1(b)} = \tau_1^{\mu}(k_1,k_2,e_1)\bigg[ \frac{-i{\eta}_{\mu\nu}}{q^2}\bigg] \tau_1^{\nu}(k_3,k_4,e_2).
\end{equation}
These diagrams yield no complications. Contracting all indices and
preforming the Fourier transforms one ends up with:
\begin{equation}
V_{\rm 1(a)}(r) = -\frac{G m_1 m_2}{r},
\end{equation}
\begin{equation}
V_{\rm 1(b)}(r) = \frac{e_1 e_2}{4\pi r},
\end{equation}
where $\left(e_1,m_1\right)$ and $\left(e_2,m_2\right)$ are the two charges and masses of
the system respectively. This is of course the expected results
for these diagrams. One gets the Newtonian and Coulomb terms for
the potential of two charged scalars.

The next class of diagrams we will consider is that of the box diagrams.
\subsubsection{The box diagrams and crossed box diagrams }
\begin{figure}[h]\vspace{0.5cm}
\begin{minipage}{0.5\linewidth}
\begin{center}
\includegraphics[scale=1.15]{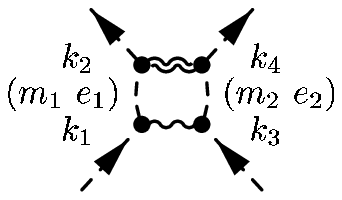}
\end{center}
{\begin{center} 2(a)\end{center}}
\end{minipage}
\begin{minipage}{0.5\linewidth}
\begin{center}
\includegraphics[scale=1.15]{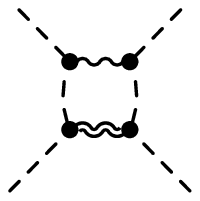}
\end{center}
{\begin{center} 2(b)\end{center}}
\end{minipage}\vspace{0.5cm}
\begin{minipage}{0.5\linewidth}
\begin{center}
\includegraphics[scale=1.15]{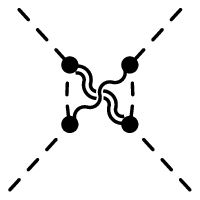}
\end{center}
{\begin{center} 2(c)\end{center}}
\end{minipage}
\begin{minipage}{0.5\linewidth}
\begin{center}
\includegraphics[scale=1.15]{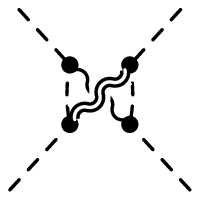}
\end{center}{\begin{center} 2(d)\end{center}}
\end{minipage}
\caption{The set of box and crossed box diagrams contributing to
the non-analytical parts of the potential.}\label{box1}
\end{figure}
There are four distinct diagrams. See figure \ref{box1}. Two
crossed box and two box diagrams. We will not treat all diagrams
separately but rather discuss one of the diagrams in detail and
then present the total result. The diagram 2(a)
is defined in the following way:
\begin{equation}\begin{split}
{i \cal M}_{\rm 2(a)} & =
\int\frac{d^4l}{(2\pi)^4}\frac{i}{(l+k_1)^2-m_1^2}\frac{i}{(l-k_3)^2-m_2^2}\\&
\times \tau_1^{\gamma}(k_1,k_1+l,e_1)\bigg[
\frac{-i\eta_{\gamma\delta}}{l^2}\bigg]
\tau_1^{\delta}(k_3,-l+k_3,e_2)\\& \times
\tau_2^{\mu\nu}(l+k_1,k_2,m_1)\bigg[ \frac{i{\cal
P}_{\mu\nu\sigma\rho}}{(l+q)^2}\bigg]
\tau_2^{\sigma\rho}(k_3-l,k_4,m_2).
\end{split}\end{equation}
where we have chosen a certain parametrization for the momenta in
the diagram, the side with mass $\left(m_1\right)$ and charge $\left(e_1\right)$ has
$\left(k_1\right)$, $\left(k_2\right)$ as incoming and outgoing momentum respectively.
Correspondingly the other side with mass $\left(m_2\right)$ and charge
$\left(e_2\right)$ has $\left(k_3\right)$, $\left(k_4\right)$ as incoming and
outgoing momentum respectively. These diagrams are quite hard to do. In
appendix we will present all the needed integrals as well as we will discuss
some calculational simplifications which are useful in doing the
box-integrals, see appendix~\ref{integrals}.

The final sum for these diagrams is:
\begin{equation}
V_{\rm 2(a)+2(b)+2(c)+2(d)}(r) = \frac{10Ge_1e_2}{3\pi^2 r^3}.
\end{equation}

\subsubsection{The triangular diagrams}
\begin{figure}[h]
\begin{minipage}{0.5\linewidth}
\begin{center}
\includegraphics[scale=1.1]{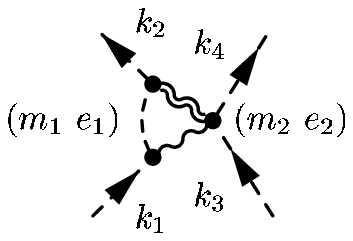}
\end{center}
{\begin{center}3(a)\end{center}}
\end{minipage}
\begin{minipage}{0.5\linewidth}\vspace{0.4cm}
\begin{center}
\includegraphics[scale=1.1]{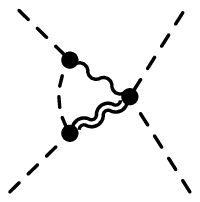}
\end{center}
{\begin{center}3(b)\end{center}}
\end{minipage}\vspace{0.3cm}
\begin{minipage}{0.5\linewidth}
\begin{center}
\includegraphics[scale=1.1]{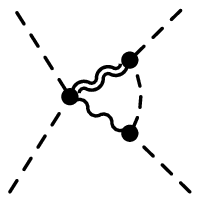}
\end{center}
{\begin{center}3(c)\end{center}}
\end{minipage}
\begin{minipage}{0.5\linewidth}
\begin{center}
\includegraphics[scale=1.1]{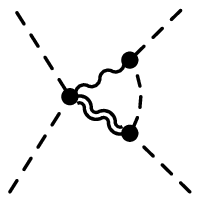}
\end{center}
{\begin{center}3(d)\end{center}}
\end{minipage}
\caption{The set of triangular diagrams contributing to the
non-analytical terms of the potential.}\label{trig1}
\end{figure}

The following triangular diagrams contribute with non-analytic
contributions to the potential. See figure \ref{trig1}.
As for the box diagrams we will only consider one of the diagrams
--- here again the first namely 3(a). The formal expression for
this particular diagram is, we just apply the vertex rules:
\begin{equation}\begin{split}
{i \cal M}_{\rm 3(a)} & = \int\frac{d^4l}{(2\pi)^4}\frac{i}{(l+k_1)^2-m_1^2}\\&
\times \tau_1^{\gamma}(k_1,k_1+l,e_1)\bigg[ \frac{-i\eta_{\gamma\delta}}{l^2}\bigg] \tau_2^{\mu\nu}(l+k_1,k_2,m_1)\\ &
\times \bigg[ \frac{i{\cal P}_{\mu\nu\sigma\rho}}{(l+q)^2}\bigg] \tau_5^{(\delta)\sigma\rho}(k_3,k_4,e_2).
\end{split}\end{equation}
Again all the needed integrals are of the type discussed in the
appendix~\ref{integrals}. Applying our contraction program and doing the
integrations leave us with a result, which Fourier transformed
yields the following contribution to the potential:
\begin{equation}
V_{\rm 3(a)+3(b)+3(c)+3(d)}(r) = \frac{Ge_1e_2(m_1+m_2)}{\pi r^2} - \frac{4e_1e_2G}{\pi^2 r^3}.
\end{equation}
As seen these diagrams yield both a classical, the $\sim
\frac{1}{r^2}$ contribution, as well as a quantum correction
$\sim \frac{1}{r^3}$.

\subsubsection{The circular diagram}
\begin{figure}[h]
\begin{center}
\includegraphics[scale=1.1]{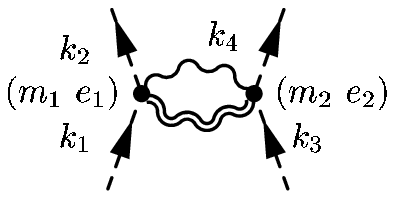}
\end{center}
{\begin{center}4(a)\end{center}} \caption{The circular diagram
with non-analytic contributions.}\label{circ1}
\end{figure}
The circular diagram, see figure \ref{circ1}, has the following formal expression:
\begin{equation}\begin{split}
{i \cal M}_{\rm 4(a)} & = \int\frac{d^4l}{(2\pi)^4} \tau_5^{\mu\nu(\gamma)}(k_1,k_2,e_1)\bigg[ \frac{-i\eta_{\gamma\delta}}{l^2}\bigg] \\&
\times \bigg[ \frac{i{\cal P}_{\mu\nu\sigma\rho}}{(l+q)^2}\bigg] \tau_5^{\sigma\rho(\delta)}(k_3,k_4,e_2).
\end{split}\end{equation}
Doing the contractions and integrations give the following contribution
to the potential:
\begin{equation}
V_{\rm 4(a)}(r) = \frac{2Ge_1e_2}{\pi^2 r^3}.
\end{equation}

\subsubsection{1PR-diagrams}
\begin{figure}[h]
\begin{minipage}{0.5\linewidth}
\begin{center}
\includegraphics[scale=1.1]{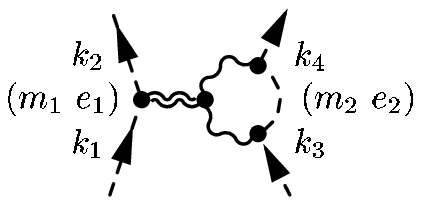}
\end{center}
{\begin{center}5(a)\end{center}}
\end{minipage}
\begin{minipage}{0.5\linewidth}
\begin{center}
\includegraphics[scale=1.1]{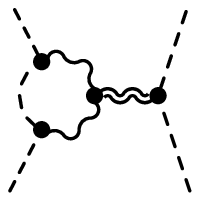}
\end{center}
{\begin{center}5(b)\end{center}}
\end{minipage}\vspace{0.3cm}
\begin{minipage}{0.5\linewidth}
\begin{center}
\includegraphics[scale=1.1]{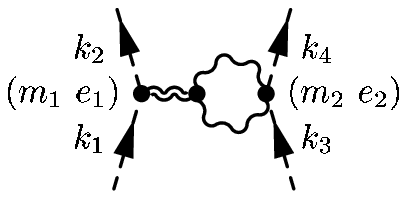}
\end{center}
{\begin{center}5(c)\end{center}}
\end{minipage}
\begin{minipage}[scale=1.1]{0.5\linewidth}
\begin{center}
\includegraphics[scale=1.1]{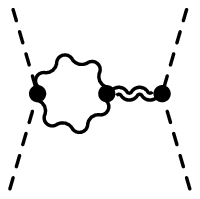}
\end{center}
{\begin{center}5(d)\end{center}}
\end{minipage}
\caption{The class of the graviton 1PR vertex corrections which
yield non-analytical corrections to the potential.}\label{gravi2n1}
\end{figure}

The following class of the set of 1PR-diagrams corresponding to the gravitational
vertex correction will contribute to the potential, see figure \ref{gravi2n1}.

Again we will not treat all diagrams separately. Instead we will
consider two of the diagrams in details --- namely the diagrams
5(a) and 5(c). First we will present the formal expressions for
the diagrams using the Feynman vertex rules. Next we will briefly
consider the calculations and finally we will present the results.

The formal expression for 5(a) is:
\begin{equation}\begin{split}
{i \cal M}_{\rm 5(a)} & =
\int\frac{d^4l}{(2\pi)^4}\frac{i}{(l-k_3)^2-m_2^2}
\tau_2^{\mu\nu}(k_1,k_2,m_1)
\\& \times \bigg[ \frac{i{\cal P}_{\mu\nu\rho\sigma}}{q^2}\bigg] \tau_3^{\rho\sigma(\gamma\delta)}(l,l+q)
\tau_1^{\alpha}(k_3,k_3-l,e_2)\\ & \times\bigg[ \frac{-i\eta_{\alpha\gamma}}{l^2}\bigg]\bigg[ \frac{-i\eta_{\beta\delta}}{(l+q)^2}\bigg] \tau_1^{\beta}(k_3-l,k_4,e_2),
\end{split}\end{equation}
while the expression for 5(c) reads:
\begin{equation}\begin{split}
{i \cal M}_{\rm 5(c)} & = \int\frac{d^4l}{(2\pi)^4}
\tau_2^{\mu\nu}(k_1,k_2,m_1) \bigg[ \frac{i{\cal
P}_{\mu\nu\rho\sigma}}{q^2}\bigg]\\ & \times
\tau_3^{\rho\sigma(\gamma\delta)}(l,l+q)
\tau_4^{\alpha\beta}(k_3,k_4,e_2)\bigg[
\frac{-i{\eta}_{\gamma\alpha}}{l^2}\bigg]\bigg[
\frac{-i{\eta}_{\delta\beta}}{(l+q)^2}\bigg].
\end{split}\end{equation}

Again the calculations of these diagrams diagrams yield no real
complications using our algebraic program.

The result for the diagrams $5(a-d)$ are in terms of the
corrections to the potential:
\begin{equation}\begin{split}
V_{\rm 5(a)+5(b)+5(c)+5(d)}(r) & = \frac{G(e_2^2m_1+e_1^2m_2)}{8\pi r^2}\\ &  - \frac{G\big(\frac{m_1}{m_2}e_2^2+\frac{m_2}{m_1}e_1^2\big)}{3\pi^2 r^3},
\end{split}\end{equation}
where we have associated a factor of one-half due to the symmetry
of the diagrams 5(c-d).

We have checked explicitly, that the above result for correction
to the potential is in complete agreement, with the result for the
gravitational vertex correction calculated
in ref.~\cite{Donoghue:2001qc}.

\begin{figure}[h]
\begin{minipage}{0.5\linewidth}
\begin{center}
\includegraphics[scale=1.1]{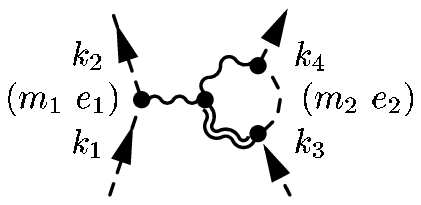}
\end{center}
{\begin{center}6(a)\end{center}}
\end{minipage}
\begin{minipage}{0.5\linewidth}
\begin{center}
\includegraphics[scale=1.1]{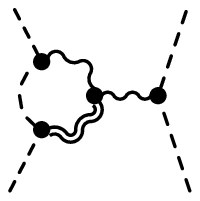}
\end{center}
{\begin{center}6(b)\end{center}}
\end{minipage}\vspace{0.3cm}
\begin{minipage}{0.5\linewidth}
\begin{center}
\includegraphics[scale=1.1]{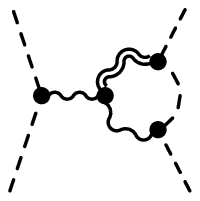}
\end{center}{\begin{center}6(c)\end{center}}
\end{minipage}
\begin{minipage}{0.5\linewidth}
\begin{center}
\includegraphics[scale=1.1]{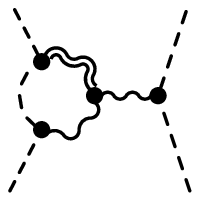}
\end{center}
{\begin{center}6(d)\end{center}}
\end{minipage}
\caption{The first class of the photon vertex 1PR corrections
which yield non-analytical corrections to the potential.}
\label{pho2n1}
\end{figure}

For the photonic vertex correction we consider the following
diagrams. See figure \ref{pho2n1}.

\begin{figure}[h]
\begin{minipage}{0.5\linewidth}
\begin{center}
\includegraphics[scale=1.1]{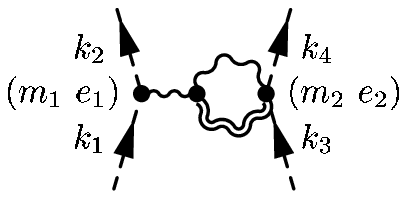}
\end{center}
{\begin{center} 7(a)\end{center}}
\end{minipage}
\begin{minipage}{0.5\linewidth}
\begin{center}
\includegraphics[scale=1.1]{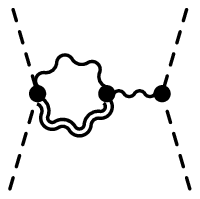}
\end{center}{\begin{center} 7(b)\end{center}}
\end{minipage}
\caption{The remaining photonic vertex 1PR diagrams which yield
non-analytical corrections to the potential.} \label{pho2n2}
\end{figure}

Together with the diagrams. See figure \ref{pho2n2}.

Again we look upon the formal expression for only two of the
diagrams --- namely 6(a) and 7(a):
\begin{equation}\begin{split}
{i \cal M}_{\rm 6(a)} & =
\int\frac{d^4l}{(2\pi)^4}\frac{i}{(l-k_3)^2-m_2^2}
\tau_1^{\gamma}(k_1,k_2,e_1)\\&\times\bigg[
\frac{-i\eta_{\gamma\delta}}{q^2}\bigg]
\tau_3^{\sigma\rho(\delta\alpha)}(q,l+q)
\tau_2^{\mu\nu}(k_3,k_3-l,m_2)\\&\times\bigg[ \frac{i{\cal
P}_{\mu\nu\sigma\rho}}{l^2}\bigg]\bigg[
\frac{-i{\eta}_{\beta\alpha}}{(l+q)^2}\bigg]
\tau_1^{\beta}(k_3-l,k_4,e_2).
\end{split}\end{equation}
\begin{equation}\begin{split}
{i \cal M}_{\rm 7(a)} & = \int\frac{d^4l}{(2\pi)^4}
\tau_1^{\gamma}(k_1,k_2,e_1)\bigg[
\frac{-i\eta_{\gamma\delta}}{q^2}\bigg]
\tau_3^{\mu\nu(\delta\alpha)}(q,l+q)\\& \times \bigg[ \frac{i{\cal
P}_{\mu\nu\sigma\rho}}{l^2}\bigg]\bigg[
\frac{-i{\eta}_{\alpha\beta}}{(l+q)^2}\bigg]
\tau_5^{\sigma\rho(\beta)}(k_3,k_4,e_2).
\end{split}\end{equation}

The result for the diagrams $6(a-d)+7(a-b)$ are in terms of the
corrections to the potential:
\begin{equation}
V_{\rm 6(a)+6(b)+6(c)+6(d)+7(a)+7(b)}(r) =
-\frac{Ge_1e_2(m_1+m_2)}{4\pi r^2}.
\end{equation}

\subsubsection{The vacuum polarization diagram}
\begin{figure}[ht]
\begin{center}
\includegraphics[scale=1.1]{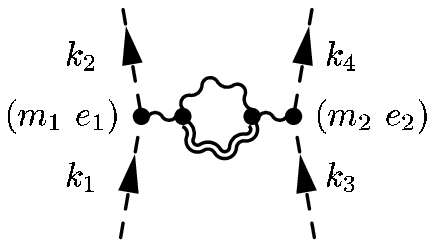}
\end{center}
{\begin{center}8(a)\end{center}} \caption{The only mixed vacuum
polarization diagram to contribute to the potential. There is no
mixed corresponding ghost diagram associated with this
diagram.}\label{vacuum}
\end{figure}

The vacuum diagram, see figure \ref{vacuum}, has the following
formal expression:
\begin{equation}\begin{split}
{i \cal M}_{\rm 8(a)} & = \int\frac{d^4l}{(2\pi)^4}
\tau_1^{\gamma}(k_1,k_2,e_1)\bigg[
\frac{-i\eta_{\gamma\delta}}{q^2}\bigg]\tau_3^{\sigma\rho(\delta\alpha)}(q,-l)\\&\times
\tau_3^{\mu\nu(\beta\epsilon)}(-l,q)\bigg[ \frac{i{\cal
P}_{\mu\nu\sigma\rho}}{(l+q)^2}\bigg] \bigg[
\frac{-i{\eta}_{\beta\alpha}}{l^2}\bigg]\bigg[
\frac{-i{\eta}_{\epsilon\phi}}{q^2}\bigg]\\&\times
\tau_1^{\phi}(k_3,k_4,e_2).
\end{split}\end{equation}

It gives the following contribution to the potential:
\begin{equation}
V_{\rm 8(a)}(r) = \frac{Ge_1e_2}{6 \pi^2 r^3}.
\end{equation}

The exact photon contributions for the 1-loop divergences of the
minimal theory can be found in ref.~\cite{Deser:cz}. Using that the
pole singularity $\left(\frac{1}\epsilon\right)$ always is followed by a
$\left(\ln(-q^2)\right)$-contribution, one can read off the non-analytic result
for the loop diagram using the coefficient of the singular pole
term. We have explicitly checked our result for this diagram with
the result derived in this fashion.

\begin{figure}
\begin{minipage}{0.32\linewidth}
\begin{center}
\includegraphics[scale=1]{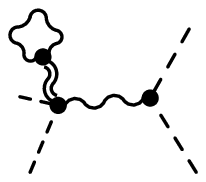}
\end{center}\begin{center}A\end{center}
\end{minipage}
\begin{minipage}{0.32\linewidth}\begin{center}
\includegraphics[scale=1.1]{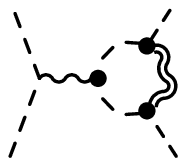}
\end{center}\begin{center}B\end{center}
\end{minipage}
\begin{minipage}{0.32\linewidth}\begin{center}
\includegraphics[scale=1.02]{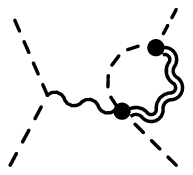}
\end{center}\begin{center}C\end{center}
\end{minipage}
\caption{Diagrams which will only give contributions to the
analytical parts of the potential.}\label{other1}
\end{figure}

The above diagrams generate all the non-analytical contributions
to the potential. There are other diagrams contributing to the
1-loop scattering matrix, but those diagrams will only give
analytical contributions, so we will not discuss them here in much
detail. Examples of such diagrams are shown, see figure
\ref{other1}.

The (diagram A) is a tadpole. Tadpole diagrams will never depend on
the transverse momentum of the diagrams, and will thus never
contribute with a non-analytical term. In fact, massless tadpoles
will be zero in dimensional regularization. The (diagram B) is
interesting --- as it is of the same type as the diagrams 6(a-d),
however with two massive propagators and one massless instead of
two massless and one massive propagator. One can show that this
diagram will not contribute with non-analytic terms, because such
an integral with two massive denominators and one massless
only will give analytical contributions. In the case of (diagram C), the
loop is on one of the external legs. Hence the loop integrations
will not depend on the interchanged momentum of the diagram. Thus
it cannot give any non-analytical contributions to the potential.

\subsection{The results for the diagrams in general relativity}
The results in the pure general relativity case are more
complicated to calculate than in the combined results of general
relativity and scalar QED. This is basically because the vertex
rules are more complicated. The integrals and the algebraic
contractions follow exactly the same principles. The 1-loop
scattering diagrams for the pure gravitational theory will be
presented here. Calculations have been done both by hand and by
the computer algorithm for Maple. As
most calculations follow the same principles as in the preceding
section we will be more brief here and proceed rather directly to
the results.

The discussion for the tree diagram is identical to the discussion
in the previous section. So we will continue straight ahead with
the results for the 1-loop diagrams.

\subsubsection{The box and crossed box diagrams}
\begin{figure}[h]\vspace{0.5cm}
\begin{minipage}[t]{0.5\linewidth}
\begin{center}
\includegraphics[scale=1]{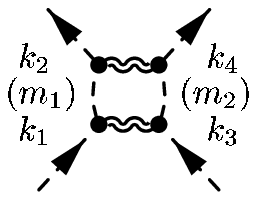}
\end{center}
{\begin{center}{2(a)}\end{center}}
\end{minipage}
\begin{minipage}[t]{0.5\linewidth}
\begin{center}
\includegraphics[scale=1]{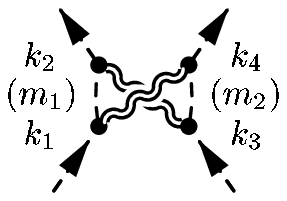}
\end{center}
{\begin{center}2(b)\end{center}}
\end{minipage}
\caption{The box and crossed box diagrams which go in to
a calculation of the 1-loop scattering matrix potential.\label{fig214}}
\end{figure}

In this subsection we will present the results for the box and crossed box
diagrams. Graphically these diagrams can be depicted as, see figure~\ref{fig214}.
Formally we can write the contributions from these two diagrams
in the following manner:
\begin{equation}\begin{aligned}
M_{2a}&=\int\frac{d^4l}{(2\pi)^4}\tau_6^{\mu\nu}(k_1,k_1+l,m_1)
\tau_6^{\rho\sigma}(k_1+l,k_2,m_1)\\ & \times
\tau_6^{\alpha\beta}(k_3,k_3-l,m_2)
\tau_6^{\gamma\delta}(k_3-l,k_4,m_2)\\
& \times
\Big[\frac{i}{(k_1+l)^2-m_1^2}\Big]\Big[\frac{i}{(k_3-l)^2-m_2^2}\Big]\Big[\frac{i{\cal
P}_{\mu\nu\alpha\beta}}{l^2}\Big]\Big[\frac{i{\cal
P}_{\rho\sigma\gamma\delta}}{(l+q)^2}\Big],
\end{aligned}\end{equation}
for the box and:
\begin{equation}\begin{aligned}
M_{2b}&=\int\frac{d^4l}{(2\pi)^4}\tau_6^{\mu\nu}(k_1,k_1+l,m_1)
\tau_6^{\rho\sigma}(k_1+l,k_2,m_1)\\ & \times
\tau_6^{\gamma\delta}(k_3,l+k_4,m_2)
\tau_6^{\alpha\beta}(l+k_4,k_4,m_2)\\
& \times \Big[\frac{i}{(k_1+l)^2-m_1^2}\Big]\cdot
\Big[\frac{i}{(l+k_4)^2-m_2^2}\Big]\cdot \Big[\frac{i{\cal
P}_{\mu\nu\alpha\beta}}{l^2}\Big]\Big[\frac{i{\cal
P}_{\rho\sigma\gamma\delta}}{(l+q)^2}\Big].
\end{aligned}\end{equation}
Preforming all integrals by the same principles as in the
the previous section, and taking the nonrelativistic limit, we end up with:
\begin{equation}
V_{2(a)+2(b)}^{\rm tot}(r) = -\frac{47}{3}\frac{m_1m_2 G^2}{\pi r^3}.
\end{equation}
as the total contribution to the potential from the box and cross
box diagrams. It is seen that we have
agreement with the result of ref.~\cite{ibk}.

\subsubsection{The triangle diagrams}
\begin{figure}[h]
\begin{minipage}[t]{0.5\linewidth}
\begin{center}
\includegraphics[scale=1]{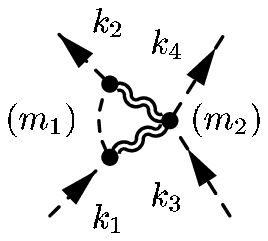}
\end{center}
{\begin{center}3(a)\end{center}}
\end{minipage}
\begin{minipage}[t]{0.5\linewidth}
\begin{center}
\includegraphics[scale=1]{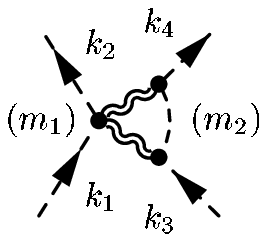}
\end{center}
{\begin{center}3(b)\end{center}}
\end{minipage}
\caption{The triangle diagrams which are contributing
to the scattering potential.}\label{trig2}
\end{figure}

The triangle diagram we graphically express as, see figure~\ref{trig2}.
Formally we can write these diagrams in the following way:
\begin{equation}
\begin{aligned}
M_{3(a)}(q)&=\int{d^4l\over
(2\pi)^4}\tau_6^{\mu\nu}(k_1,l+k_1,m_1)
\tau_6^{\alpha\beta}(l+k_1,k_2,m_1)\tau_8^{\sigma\rho\gamma\delta}(k_3,k_4,m_2)\\
&\times\Big[\frac{i{\cal P}_{\alpha\beta\gamma\delta}}
        {(l+q)^2}\Big] \Big[\frac{i{\cal P}_{\mu\nu\sigma\rho}}{l^2}\Big]
\Big[\frac{i}{(l+k_1)^2-m_1^2}\Big],
\end{aligned}
\end{equation}
and
\begin{equation}
\begin{aligned}
M_{3(b)}(q)&=\int{d^4l\over (2\pi)^4}\tau_6^{\sigma\rho}(k_3,k_3-l,m_2)
\tau_6^{\gamma\delta}(k_3-l,k_4,m_2)\tau_8^{\mu\nu\alpha\beta}(k_1,k_2,m_1)\\
        &\times\Big[\frac{i{\cal P}_{\mu\nu\sigma\rho}}{l^2}\Big]
        \Big[\frac{i{\cal P}_{\alpha\beta\gamma\delta}}{(l+q)^2}\Big]
\Big[\frac{i}{(l-k_3)^2-m_2^2}\Big].
\end{aligned}
\end{equation}
Again using the same type of calculations as previously and
employing simplifications such as:
\begin{equation}
{\cal P}_{\gamma\delta\sigma\rho}{\cal
P}_{\alpha\beta\mu\nu}\tau_8^{\sigma\rho\mu\nu}
(k_1,k_2,m_1)=\tau_{8\gamma\delta\alpha\beta}(k_1,k_2,m_1),
\end{equation}
we end up with the result:
\begin{equation}
V_{3(a)+3(b)}(r) = -4\frac{G^2m_1m_2(m_1+m_2)}{r^2}+28\frac{m_1m_2
G^2}{\pi r^3},
\end{equation}
if we take the nonrelativistic limit and Fourier
transform. The result match ref.~\cite{ibk}.

\subsubsection{The double-seagull diagram}
\begin{figure}[h]
\begin{center}
\includegraphics[scale=1]{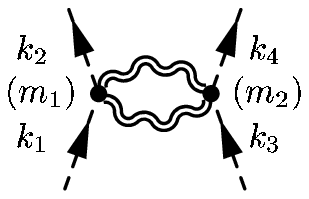}
\end{center}
{\begin{center}4(a)\end{center}} \caption{The double-seagull diagram.}\label{circ2}
\end{figure}

The double-seagull diagram can be depicted graphically as in figure~\ref{circ2}.
Formally we can write this contribution as:
\begin{equation}
\begin{aligned}
M_{4(a)}(q)&={1\over 2!}\int{d^4l\over (2\pi)^4
}\tau_8^{\alpha\beta\gamma\delta}(k_1,k_2,m_1)\tau_8^{\sigma\rho\mu\nu}(k_3,k_4,m_2)
        \times\Big[\frac{i{\cal P}_{\alpha\beta\mu\nu}}{(l+q)^2}\Big]
        \Big[\frac{i{\cal P}_{\gamma\delta\sigma\rho}}{l^2}\Big].
\end{aligned}
\end{equation}
It should be remembered that there is a
symmetry factor of $1\over2!$ in the diagram. The result for the diagram is
found rather straightforwardly and the resulting Fourier transformed amplitude is:
\begin{equation}
V_{4(a)}(r) = -22\frac{m_1m_2 G^2}{\pi r^3}
\end{equation}
Also this result match ref.~\cite{ibk}.

\subsubsection{The vertex correction diagrams}
\begin{figure}[h]
\begin{minipage}[t]{0.5\linewidth}
\begin{center}
\includegraphics[scale=1]{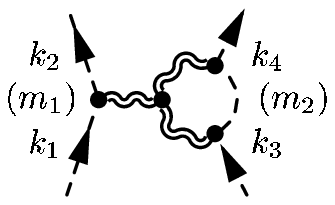}
\end{center}
{\begin{center}5(a)\end{center}}
\end{minipage}
\begin{minipage}[t]{0.5\linewidth}
\begin{center}
\includegraphics[scale=1.1]{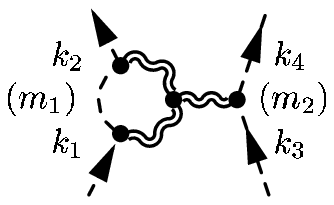}
\end{center}
{\begin{center}5(b)\end{center}}
\end{minipage}
\begin{minipage}[t]{0.5\linewidth}
\begin{center}
\includegraphics[scale=1]{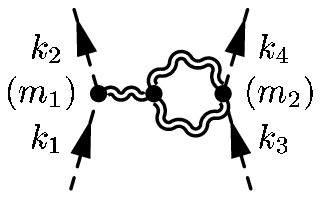}
\end{center}
{\begin{center}5(c)\end{center}}
\end{minipage}
\begin{minipage}[t]{0.5\linewidth}
\begin{center}
\includegraphics[scale=1]{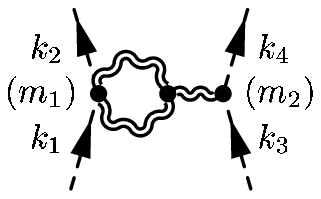}
\end{center}
{\begin{center}5(d)\end{center}}
\end{minipage}
\caption{The vertex correction diagrams which yield non-analytic terms
in the scattering amplitude.}\label{gravi2n2}
\end{figure}

The vertex correction diagrams are graphically represented, see figure~\ref{gravi2n2}.
Two classes of diagrams go into the set of vertex corrections.
There are two massive loop diagrams, which are shown in Figures 5(a) and 5(b).
For these diagrams we can write:
\begin{equation}
\begin{aligned}
M_{5(a)}(q)&={d^4l\over (2\pi)^4
}\tau_6^{\alpha\beta}(k_1,k_2,m_1)\tau_6^{\mu\nu}(k_3,k_3-l,m_2)
\tau_6^{\rho\sigma}(k_3-l,k_4,m_2)\tau_{10}^{\lambda\kappa\phi\epsilon(\gamma\delta)}(l,-q)
\\
 &       \times\Big[\frac{i{\cal P}_{\lambda\kappa\mu\nu}}{l^2}\Big]
 \Big[\frac{i{\cal P}_{\phi\epsilon\rho\sigma}}{(l+q)^2}\Big]
\Big[\frac{i{\cal
P}_{\alpha\beta\gamma\delta}}{q^2}\Big]\Big[\frac{i}{(l-k_3)^2-m_2^2}\Big],
\end{aligned}
\end{equation}
and
\begin{equation}
\begin{aligned}
M_{5(b)}(q)&=\int{d^4l\over (2\pi)^4}\tau_6^{\alpha\beta}(k_1,l+k_1,m_1)
\tau_6^{\mu\nu}(l+k_1,k_2,m_1)\tau_6^{\lambda\kappa}(k_3,k_4,m_2)
\tau_{10}^{\gamma\delta\rho\sigma(\phi\epsilon)}(-l,q)\\
  &      \times\Big[\frac{i{\cal P}_{\alpha\beta\gamma\delta}}{l^2}\Big]
  \Big[\frac{i{\cal P}_{\mu\nu\rho\sigma}}{(l+q)^2}\Big]
\Big[\frac{i{\cal
P}_{\phi\epsilon\lambda\kappa}}{q^2}\Big]\Big[\frac{i}{(l+k_1)^2-m_1^2}\Big].
\end{aligned}
\end{equation}
The other set of diagrams is shown in Figures 5(c) and 5(d), and can be written
as:
\begin{equation}
\begin{aligned}
M_{5(c)}(q)&={1\over 2!}\int{d^4l\over (2\pi)^4
}\tau_8^{\lambda\kappa\epsilon\phi}(k_3,k_4,m_2)
\tau_6^{\alpha\beta}(k_1,k_2,m_1)\tau_{10}^{\mu\nu\rho\sigma(\gamma\delta)}(l,-q)\\
& \times\Big[\frac{i{\cal P}_{\mu\nu\lambda\kappa}}{l^2}\Big]
   \Big[\frac{i{\cal P}_{\rho\sigma\epsilon\phi}}{(l+q)^2}\Big]
\Big[\frac{i{\cal P}_{\alpha\beta\gamma\delta}}{q^2}\Big],
\end{aligned}
\end{equation}
and
\begin{equation}
\begin{aligned}
M_{5(d)}(q)&={1\over 2!}\int{d^4l\over (2\pi)^4
}\tau_8^{\rho\sigma\mu\nu}(k_1,k_2,m_1)\tau_6^{\lambda\kappa}(k_3,k_4,m_2)
\tau_{10}^{\alpha\beta\gamma\delta(\epsilon\phi)}(-l,q)\\
   &     \times\Big[\frac{i{\cal P}_{\mu\nu\gamma\delta}}{l^2}\Big]
   \Big[\frac{i{\cal P}_{\rho\sigma\alpha\beta}}{(l+q)^2}\Big]
\Big[\frac{i{\cal P}_{\lambda\kappa\epsilon\phi}}{q^2}\Big].
\end{aligned}
\end{equation}
These diagrams are among the most complicated to calculate. The
first results for these diagrams date back to the original
calculation of refs.~\cite{Donoghue:1993eb,Donoghue:dn} --- but because
of an algebraic error in the calculation, the original result was
in error and despite various checks of the
calculation refs.~\cite{akh,ibk} the correct result has not been given
until ref.~\cite{B2,Thesis}. Quoting ref.~\cite{B2}, taking the nonrelativistic
limit and Fourier transforming, we get:
\begin{equation}
V_{5(a)+5(b)}(r) = \frac{G^2m_1m_2(m_1+m_2)}{r^2}-\frac53\frac{m_1m_2
G^2}{\pi r^3},
\end{equation}
and
\begin{equation}
V_{5(c)+5(d)}(r) = \frac{26}{3}\frac{m_1m_2 G^2}{\pi r^3}.
\end{equation}
For the diagrams 5(c) and 5(d) a symmetry factor $1\over2!$ have been
included. These results have recently been confirmed in the
independent paper ref.~\cite{Khriplovich:2004cx}.

\section{The result for the potential}
We will first discuss the results for the scattering potentials.
Adding everything up in the scalar QED case, the final result for
the potential reads:
\begin{equation}\begin{split}
V(r) &= -\frac{Gm_1m_2}{r} + \frac{\tilde\alpha \tilde e_1 \tilde
e_2}{r} +\\ & +\frac12 \frac{(m_1\tilde e_2^2+m_2\tilde
e_1^2)G\tilde\alpha}{c^2r^2} +\frac{3\tilde e_1\tilde
e_2(m_1+m_2)G\tilde\alpha}{c^2r^2}\\ & -\frac43
\Big(\frac{m_1^2\tilde e_2^2+m_2^2\tilde
e_1^2}{m_1m_2}\Big)\frac{G\tilde\alpha\hbar}{\pi c^3r^3} -8
\frac{\tilde e_1\tilde e_2G\tilde\alpha\hbar}{\pi c^3r^3}.
\end{split}\end{equation}
In the above expression we have included the appropriate physical
factors of $\left(\hbar\right)$ and $\left(c\right)$, and rescaled the equation in terms of
$\left(\tilde \alpha = \frac{\hbar c}{137}\right)$. The charges $\left(\tilde e_1\right)$
and $\left(\tilde e_2\right)$ are normalized in units of the elementary charge.

It is seen that in the above expression for the potential
there many different terms.
The first two terms are, as noted before, the
well-known Newtonian and Coulomb terms. They represent the lowest
order interactions of the two sources without the inclusion
of radiative corrections. These terms come out of the
formalism as expected, and they will dominate the potential
at sufficiently low energies.

The classical post-Newtonian corrections to the potential are represented
by the two next terms, and they can also be derived from
general relativity with the inclusion of charged particles.

The last two contributions are the most interesting from a quantum
point of view. They represent the leading 1-loop
quantum corrections to the mixed theory of general relativity and
scalar QED, and they were computed in ref.~\cite{B1} for the first
time. It is seen in SI units $\left(\frac{\hbar G}{c^3}\sim 10^{-70}
\ {\rm meters}^2\right)$. Therefore these corrections are thus indeed
very small. Seemingly they will be impossible to detect experimentally.
This is especially due to the large numerical contributions
usually coming from the Coulomb and Newtonian terms.

When looking at the expression for the potential, it is noticed that
the corrections to the potential can be
divided into two classes of terms. There are terms with the
two charges multiplied together, and terms which contain a sum of
squares of each charge. For identical particles the two types of
terms must be exactly the same in form and in this case
one should also include the appropriate diagrams with crossed
particle lines. Our result hold for nonidentical particles only.

In the case of one of the masses or charges being either very large
or very high respectively, some of the contributions in the potential
will dominate over others. The
terms with separated charges will correspond to the dominating
terms, if one of the scattered masses or charges were much larger
than other. In this case the gravitational vertex corrections will
generate the dominating leading contribution to the potential.
For example with a very high charge for one of the particles --- the
probing particle will fell most of the gravitational effect
coming from the electromagnetic field surrounding the heavily
charged particle.

For a very large mass: $\left(M \sim 10^{30}\ {\rm kg}\right)$ but a very low
charged particle: $\left(\tilde e_1 \sim 0\right)$ (the Sun), and a charged:
$\left(\tilde e_2 \sim 1\right)$ but very low mass particle: $\left(m \sim
10^{-31}\ {\rm kg}\right)$ (an electron), one could perhaps test this
effect experimentally, because then the Newton effect is:
$\left(\frac{GmM}r \sim \frac{10^{-10} \ {\rm J \cdot meters}}{r}\right)$, while
the quantum effect is: $\left(G\hbar\tilde\alpha \frac{\tilde e_2^2
M}{mc^3r^3} \sim \frac{10^{-37}\ {\rm J\cdot meters}^3}{r^3}\right)$,
and the classical contribution: $\left(GM\frac{\tilde
e_2^2\tilde\alpha}{c^2r^2}\sim \frac{10^{-25}\ {\rm J \cdot
meters}^2}{r^2}\right)$. The ratio between the post-Newtonian effects and
the quantum correction is for this experimental setup still very
large, but not quite as impossible as often seen in quantum
gravity.

We also have a result for the total potential in
the case of pure gravitational interactions.
If we add up all the corrections to the scattering potential,
we arrive at:
\begin{equation}
V(r) = -{G m_1m_2\over
r}\left[1+3\frac{G(m_1+m_2)}{r}+\frac{41}{10\pi}\frac{ G\hbar}{
r^2}\right].
\end{equation}
It should be noticed that the classical post-Newtonian term in
this expression is in an agreement with (Eq. 2.5) of
Iwasaki ref.~\cite{Iwasaki}. This result was first published in
ref.~\cite{B3}.

The quantum correction is not in agreement with the
results refs.~\cite{Donoghue:1993eb, Donoghue:dn,Muzinich:1995uj,
Hamber:1995cq,akh,ibk} but it is believed that the above
result presents the correct result for the scattering matrix
potential, because all calculations have been carried out extremely
carefully and because all results are done both by hand and by computer.
The recently published paper ref.~\cite{Khriplovich:2004cx} confirms all
calculations.

We can here again easily identify the post-Newtonian correction as
well as the quantum correction and it is seen that a direct experimental
verification of the quantum corrections will indeed be very difficult.

\subsection{A Potential Ambiguity}
A result for the post-Newtonian corrections derived from
classical considerations can be found in ref.~\cite{Barker:bx}. In
the scalar QED case, they read in our notation:
\begin{equation}\begin{split}
V_{\rm post-Newtonian} &= \frac12 \frac{(m_1\tilde e_2^2+m_2\tilde
e_1^2)G\tilde\alpha}{c^2r^2}\\& +(\alpha_p+\alpha_g-1)\frac{\tilde
e_1\tilde e_2(m_1+m_2)G\tilde\alpha}{c^2r^2},
\end{split}\end{equation}
In the pure gravitational case the expected post-Newtonian
terms are:
\begin{equation}\begin{split}
V_{\rm post-Newtonian} &= (\frac12 - \alpha_g)\frac{G^2m_1m_2(m_1+m_2)}{c^2r^2}.
\end{split}\end{equation}
It is seen that the expectations for the classical post-Newtonian
correction terms exactly match the ones derived, and that
the coefficient in front of the first term is equal to ours.
The result for the second term is equivalent in form to the term we
have derived, and the coefficient can be made to match our result
for particular values of $\left(\alpha_p\right)$ and $\left(\alpha_g\right)$. The
physical significance of the arbitrary parameters $\left(\alpha_g\right)$ and
$\left(\alpha_p\right)$, however requires some explanation.

The parameter $\left(\alpha_g\right)$ is related to the
gravitational propagator, while the coefficient
$\left(\alpha_p\right)$ is related to the photonic propagator and
the values of $\left(\alpha_p\right)$ and $\left(\alpha_g\right)$
can be seen as coordinate dependent coefficients for the
potential. They can take on arbitrary values depending on the
coordinate system chosen to represent the potential. This
ambiguity is build into the nonrelativistic potential
refs.~\cite{Iwasaki,Barker:bx,Barker:ae,Okamura}.

Despite the above references the issue concerning the ambiguity of
the classical potential have so far not created much attention.
The references~\cite{Barker:bx, Barker:ae,Okamura}, deal directly
with the post-Newtonian contributions and their arguments for these
terms can be used to look at the ambiguity of the quantum terms as well.
We will here discuss the ambiguity of the the potential and observe that the
quantum corrections at 1-loop order are well defined and unambiguous.

In ref.~\cite{Okamura} the terms present in the
classical post-Newtonian potential are recast under coordinate
transformations such as:
\begin{equation}
r \rightarrow r\Big[1+\alpha\frac{G(m_1+m_2)}{r}\Big]\label{eqn:class} .
\end{equation}
Such a coordinate change can always freely be made, and
the classical terms will be modified in the following manner:
\begin{equation}
\frac{Gm_1m_2}{r} \left[1+c\frac{G(m_1+m_2)}{r}
\right]\rightarrow \frac{Gm_1m_2}{r}
\left[1+(c-\alpha)\frac{G(m_1+m_2)}{r} \right].
\end{equation}
If we follow refs.~\cite{Okamura,Barker:ae} the ambiguity in defining
the coordinates arises from a freedom in the propagator:
\begin{equation}
\frac{1}{q^2} = \frac{1}{q_0^2-{\vec q}^2}.
\end{equation}
Using the energy-conservation, we can namely redefine it for a general $x$:
\begin{equation}
\frac{1}{\frac{1}{2}(1-x)(({p_2}_0-{p_1}_0)^2+({p_4}_0-{p_3}_0)^2))-x({p_2}
_0-{p_1}_0)({p_4}_0-{p_3}_0)-{\vec q}^2},
\end{equation}
and using {\it this} propagator in the derivations, affects the gravitational
potential, so that it will generally depend on $\left(x\right)$. The transformation coefficient
$\left(\alpha\right)$ will depend on $\left(x\right)$, and in our notation, $\left(\alpha=-\frac14(1-x)\right)$.
The terms $\left(\frac{G^2}{r^2}\right)$ as well as $\Big(\frac{G}{r}\Big)\Big(\frac{v}{c}\Big)$,
will therefore generically in the potential have a dependence on $\left(x\right)$.
The post-Newtonian parts in the static gravitational potential are hence
{\it not} well-defined.

Next we can look at the coordinate ambiguity from the point of view of
the quantum corrections. The coordinate redefinition:
\begin{equation}
r \to r\left[1+ \beta {G\hbar \over r^2}\right],\label{eq:quant}
\end{equation}
can be considered. Such a transformation will change the quantum coefficient
in the potential in the following way:
\begin{equation}
\frac{Gm_1m_2}{r} \left[1+d\frac{G\hbar}{r^2}  \right]\rightarrow
\frac{Gm_1m_2}{r} \left[1+(d-\beta)\frac{G\hbar}{r^2} \right].
\end{equation}
It is hence seen that quantum modifications of the potential can
arise, so is the quantum correction of the potential unique? That is what
we will address next.

General relativity is a theory based on the invariance under
coordinate transformations. If we make a coordinate shift of the
potential which generates a modification, a compensating modification of another
term must be
present to leave the physics invariant. In other words in a Hamiltonian
formulation a modification of the post-Newtonian terms in the potential
must be compensated by a change of the kinetic pieces in the Hamiltonian.

We look first at the classical ambiguity.
Two dimensionless variables can be constructed as possible generators
of coordinate transformations:
\begin{equation}
{p^2\over m^2}  ~~,~~ {Gm\over r}.
\end{equation}
At leading order these pieces appear at same order in the Hamiltonian:
\begin{equation}
H_1 \sim {p^2\over m} ,~~{Gm^2 \over r}.
\end{equation}
To leading order these two terms are unambiguous. At next order we
have:
\begin{equation}
H_2 \sim {p^4 \over m^3}, ~~{Gm^2\over r} {Gm\over r}, ~~\left({Gm\over
r}\right)\left({p^2\over m}\right),
\end{equation}
and now interestingly a coordinate change ambiguity between the two
last terms arises. As shown explicitly in ref.~\cite{Okamura} this ambiguity
disappear for physical observables.
In the center of mass frame we can write the Hamiltonian to
post-Newtonian order in the following way:
\begin{eqnarray}
H &=& \left(\frac{{\bf p}^2}{2m_1} +\frac{{\bf p}^2}{2m_2}
\right)-\left(
\frac{{\bf p}^4}{8m_1^3} +\frac{{\bf p}^4}{8m_2^3} \right) \nonumber \\
&-&\frac{Gm_1m_2}{r}\left[ 1+a\frac{\bf p^2}{m_1m_2} +b
\frac{({\bf p\cdot \hat{r}})^2}{m_1m_2} + c{G (m_1+m_2)\over r}
\right].
\end{eqnarray}
The standard Einstein-Infeld-Hoffmann coordinates $\left(a,\ b,\ c\right)$ will
have the values:
\begin{eqnarray}
a &=& \frac12\left[1+3\frac{(m_1+m_2)^2}{m_1m_2}\right],
\nonumber\\
b&=& {1\over 2}, \nonumber \\
c &=& -\frac12,
\end{eqnarray}
and making coordinate transformations such as of the type
in Eq. \ref{eqn:class}, one arrives at:
\begin{eqnarray}
a &\to&\frac12\left[1+(3+2\alpha)\frac{(m_1+m_2)^2}{m_1m_2}\right],
\nonumber\\
b &\to&  \frac12-\alpha \frac{(m_1+m_2)^2}{m_1m_2}, \nonumber \\
c &\to& - \frac12 - \alpha.
\end{eqnarray}
Hence in in the case of the classical corrections to a static
potential, it is necessary to specify the momentum coordinates as well.
In particular if we look at the potential presented by Iwasaki, it will be
appropriate for the choice of Einstein-Infeld-Hoffmann coordinates.

At the next order there will be the terms:
\begin{equation}
H_3 \sim \frac{p^6}{m^5},~~ \frac{p^4}{m^3}\left({Gm\over
r}\right), ~\frac{p^2}{m}\left({Gm\over r}\right)^2,~~{Gm^2\over
r}\left({Gm\over r}\right)^2 ,
\end{equation}
in the Hamiltonian. The term which goes as
$\left(\frac{1}{r^3}\right)$ has the same radial dependence as the
quantum term. However its mass dependence is wrong and it is of
order $\left(G^3\right)$. Without going into details we note that
such terms will also be ambiguous under coordinate
transformations. However cancellations have to appear among the
terms of the same order. Hence a coordinate change affects the
$\left(\frac1{r^3}\right)$ term in the potential, but classical
terms will cancel among themselves and not influence the quantum
terms.

The quantum effects are next dealt with. The generator
of coordinate transformations for the quantum contributions
is:
\begin{equation}
{G\hbar \over r^2}.
\end{equation}
Using the same principles as for the classical terms we can
generate theoretical quantum contributions to the
Hamiltonian, and such terms will be:
\begin{equation}
H_q = {Gm^2\over r}{G\hbar\over r^2},
\end{equation}
together with:
\begin{equation}
H_{qp} \sim {p^2 \over m} {G\hbar \over r^2},
\end{equation}
to first order in $\left(\hbar\right)$.
A quantum change of coordinates, $i.e.$, making a coordinate
transformation as in Eq. \ref{eq:quant} will definitely
generate a term such as $\left(H_q\right)$ in the
Hamiltonian, but in order to cancel this effect
for physical observables a term $\left(H_{qp}\right)$ must be introduced as well.
However a term such as $\left(H_{qp}\right)$ cannot exist in a two-particle potential
for any choice of coordinates, simply because it involves
only a single power of $\left (G\right)$.
A single power of $\left(G\right)$ is possible for one particle quantum observables,
but never for observables involving two particles.

Since no term are of the form $\left(H_{qp}\right)$, the quantum contribution
must be unique and specified in coordinates which make $\left(H_{qp}\right)$ vanish.
Hence at 1-loop order the quantum potential is a well defined quantity.
Without going into details we note that this must too be the case for the
quantum corrections to the scalar QED potential.

\section{Quantum corrections to the metric}
Another way to look at the 1-loop quantum corrections is to
consider the metric instead of a potential.

In ref.~\cite{duff3} quantum corrections to the Schwarzschild solution
are discussed from the view point of including only the vacuum polarization
diagram. This is correct if the source is only classical, meaning
that radiative corrections to the vertex is not included.
However, any matter source will emit gravitons and these fields
will have quantum components.
An effective treatment of the 1-loop scattering diagrams illustrates the
importance of including the radiative quantum corrections
caused by the vertex diagrams.

The diagrams which contribute to the metric corrections are only a subset of
the full scattering matrix.
\begin{figure}
\begin{center}
\includegraphics[height=3cm,]{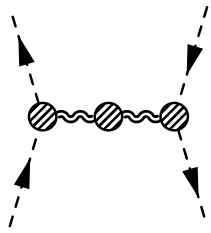} \caption{Diagrams
contributing to the one-particle-reducible potential.}
\end{center}
\end{figure}

We can use this set of the one-particle reducible diagrams to
define a potential. It results in:
\begin{equation}
V({r})=-{Gm_1m_2\over r}\left(1-{G(m_1+m_2)\over {r}}-{167\over
30\pi}{G\hbar\over {r}^2}+\ldots\right).
\end{equation}
The potential obtained this way is not a scattering matrix
potential, but it suggests a way to define a harmonic gauge
running gravitational coupling constant:
\begin{equation}
G(r) = G\left(1-{167\over 30\pi}{G\hbar\over {r}^2}\right).
\end{equation}
The running gravitational coupling is seen to be independent of
the masses of the involved objects. Hence it has the expected
universal character which a running charge should have.
Its independence of spin as can be seen from the fermionic data.

At short distances the gravitational running constant will be
weaker. This can be seen as a sort of gravitational screening. At
large distances the source, when probed, will occur more
point-like, while at small distances the gravitational field
smears out.

Experimental verifications of general relativity as an effective
field will perhaps be a very difficult task. The problem is caused
by the normally very large classical expectations of the theory.
These expectations imply that nearly any quantum effect in
powers of $\left(G\hbar\right)$ will be nearly neglectable compared to $\left(G\right)$.
Therefore the quantum effects will be very hard to extract, using
measurements where classical expectations are involved. The
solution to this obstacle could be to magnify a certain quantum
effect. This could be in cases where the classical effects were
independent of the energy scale, but where the quantum effects
were largely effected by the energy scale. Such effects would
only be observed, when very large interaction energies are
involved.

Another way to observe a quantum effect could be when a certain
classical expectation is zero, but the quantum effect would yield
a contribution. A quantum gravitational ''anomaly''. Such 'null'
experiments maybe used to test a quantum theory for
gravity.

\section{Discussion}
Normally general relativity is viewed as a non-renormalizable
theory, and consequently a quantum theory for general relativity
is believed to be an inconsistent theory. However, treated as an
effective field theory, the renormalization inconsistency of
general relativity is not an issue --- as the theory can be
explicitly renormalized at any given loop order. This fact was
first explored in refs.~\cite{Donoghue:1993eb,Donoghue:dn}. Here we have presented
three recent works which address general relativity from the
view point of treating it as an effective field theory.
We have derived the quantum and post-Newtonian corrections to the
Schwarzschild and Kerr metric, discussed general relativity and the
combined theory of general relativity and scalar QED, and extracted
useful information about the quantum and post-Newtonian long-distance
behavior of the gravitational attraction. Most important we have
directly observed that it is possible in this perspective to extract
information about quantum gravity. The found quantum corrections to the
metric and to the scattering matrix potential are unique and exact
signatures of the quantum nature of gravity.

The effective field theory approach is only valid at sufficiently
low energies, $i.e.$, below the Planck scale $\sim
10^{19}\ {\rm GeV}$, and at long distances. Below these scales the
effective field theory approach present a nice and good way to handle
the challenges of quantum gravity.

At higher energies the effective action will break down and have to
be replaced by a new unknown theory. It will govern the quantum
gravitational effects and might introduce completely new aspects
of physics. Present thoughts circle around a string or $M$-theory, which
is compactified at low energies. However it is important to notice
that the Planck energy scale is much larger than 'traditional' high-energy scales. Optimistic predictions expect a limiting scale for the Standard
Model around $\sim 10^3\ {\rm GeV}$. In this light the effective field
theory approach is seemingly good for all energies presently dealt
with in high energy physics.

Experiments to verify quantum gravity are difficult to propose. This is both
because quantum gravity effects are so small at normal energies and because
classical effects normally are rather large. The nonrelativistic potential
may not be the best offset for an experimental verification of quantum gravity --
however in the quest for an experimental verification of quantum gravity it
is very interesting that quantum effects can be calculated and that such
corrections are unique quantum fingerprints of the nature of gravity.

\chapter{String theory and effective quantum gravity}
\section{Introduction}
In this section we are discussing the thesis work which resulted
in the two papers refs.~\cite{B4,B5}. They survey the implications
of imposing the Kawai, Lewellen and Tye (KLT) string relations
ref.~\cite{KLT} in the effective field theory limit of open and
closed strings. At sufficiently low energy scales the tree
scattering amplitudes derived from string theory are reproduced
exactly by the tree scattering amplitudes, generated from an
effective field theory description with an infinite number of
higher derivative terms ordered in a series in powers of the
string tension $T \sim \frac1{\alpha'}$, see
refs.~\cite{Tsey1,Tsey2,Tsey3,Gross:1986iv,Callan:jb,Gross:1985fr,Hull:yi,Sannan:tz}.
As it probably is well known that closed strings have a natural
interpretation as fundamental theories of gravity -- while open
string theories on the other hand turn up as non-abelian
Yang-Mills vector field theories, the imposed KLT gravity/gauge
correspondence between the tree-amplitudes will create a link
between gravity and Yang-Mills theories. The induced connection
between the two effective Lagrangians is the turning point of the
investigations.

At the tree level the KLT string relations connect on-shell
scattering amplitudes for closed strings with products of
left/right moving amplitudes for open strings. This was first
applied in ref.~\cite{Berends} to gain useful information about
tree gravity scattering amplitudes from the much simpler Yang-Mills
tree amplitudes. The KLT-relations between scattering amplitudes
alone give a rather remarkable non-trivial link between string
theories, but in the field theory regime the existence of such
relationships is almost astonishing, and nonetheless valid.

Yang-Mills theory and general relativity being both non-abelian
gauge theories have some resemblance, but their dynamical and
limiting behaviors at high-energy scales are in fact quite
dissimilar. One of the key results of the paper ref.~\cite{B4} was to
demonstrate that once the KLT-relations were imposed, they uniquely
relate the tree limits of the generic effective Lagrangians of the
two theories. The generic effective Lagrangian for gravity will
have a restricting tree level connection to that of Yang-Mills
theory, through the KLT-relations. In the language of field
theories the KLT-relations present us with a link between the
on-shell field operators in a gravitational theory as a product of
Yang-Mills field theory operators.

Even though the KLT-relations hold only explicitly at the tree level
the factorization of gravity amplitudes into Yang-Mills amplitudes
can be applied with great success in loop calculations. This has been
the focus of various investigations, and much work have been carried
out, $e.g.$, see refs.~\cite{Bern1,Bern2,Bern2a,Bern2b}. Loop diagram cuts
together with properties such as unitarity of the S-matrix are used to
calculate the results of loop diagrams. A good review and a demonstration
of such calculations in QCD is ref.~\cite{Bern11}. See also refs.~\cite{Bern0,Dunbar}.
These methods can be employed in many ways, and as the methods are quite general
they apply in many cases, $e.g.$,  complicated gluon amplitudes are calculated through
cuts of diagrams, and employing the KLT-relations this can be used to derive results for
gravity loop diagrams. An important result derived this way was that N=8 supergravity,
is a less divergent theory in the ultraviolet than has previously been appreciated.
The KLT formalism allows for the presence of matter sources. They can be
introduced in the KLT formalism as done in ref.~\cite{Bern4}. In the papers ref.~\cite{B4,B5}
we did not consider loop extensions, neither the inclusion of matter sources, instead our
focus was on the actual factorization properties of tree amplitudes.

At infinite string tension the KLT-mapping simply factorizes the
amplitudes of the Einstein-Hilbert action:
\begin{equation}
{\cal L}_{\text EH} =\int d^4x\Big[\sqrt{-g}R\Big],
\end{equation}
into a sum of amplitude products of a 'left' and a 'right' moving
Yang-Mills action:
\begin{equation}
{\cal L}_{YM} =\int d^4x\Big[-\frac1{8g^2}{\rm tr}F_{\mu\nu}F_{\mu\nu}\Big],
\end{equation}
where $\left(A_\mu\right)$ is the vector field and $\left(F_{\mu\nu} = \partial_\mu
A_\nu - \partial_\nu A_\mu + g[A_\mu,A_\nu]\right)$.
The factorizations at the amplitude level have been investigated as a method to
explore if gravity vertices might be factorizable into vector vertices at the Lagrangian level,
see ref.~\cite{Bern3}. Remarkable factorizations of gravity vertices have already
been presented in ref.~\cite{Siegel} using a certain vierbein formalism.
The idea that gravity could be factorizable into a product of independent
Yang-Mills theories is indeed very beautiful, the product
of two independent Yang-Mills vectors $\left(A^\mu_L\right)$ and $\left(A^\nu_R\right)$ (without internal
contractions and interchanges of vector indices) should then be related to
the gravitational tensor field $\left(g_{\mu\nu}\right)$ through KLT:
\begin{equation}\boxed{
A_L^{\mu}\times A_R^{\nu} \sim g^{\mu\nu}},
\end{equation}
however such a direct relationship, has so far only been
realized at the tree {\em amplitude} level. At the Lagrangian (vertex)
level gauge issues mix in and complicate the matter. Depending on the
particular gauge choice, $e.g.$, the vertices can take different forms.
Whether this can be resolved or not -- and how -- remains essentially to
be understood. We did not go into the issues about factorizations of
vertices in the papers refs.~\cite{B4,B5}, where the focus was more about gaining
insight into the tree level factorizations of the amplitudes derived from the
effective actions of general relativity and Yang-Mills at order $O(\alpha'^3)$.

As we have seen in the previous chapter, despite common belief,
string theory is not the only possible starting point for the
safe arrival at a successful quantum theory for gravity below
the Planck scale.

An effective field theory appears to be the obvious scene for
quantum gravity at normal energies -- $i.e.$, below $10^{19}$ GeV.
The idea behind the effective field theory approach is so broad and
generic that it appears to be the obvious extension of general relativity
at low-energy scales. At the Planck energy a phase transition is
believed to take place, leaving us in dark about what takes place
at very high energies. String theories can be the answer here but
other theories might be possible to construct. From the effective
field theory point of view there is no need -- however -- to assume
that the high-energy theory {\it a priori} should be a string theory.
The effective field theory viewpoint allows for a broader range of
possibilities, than that imposed by a specific field theory limit
of a string theory. We will see that this also holds true in respect
to imposing the KLT-relations. The effective field approach for the
Yang-Mills theory in four dimensions is not really necessary,
because it is already a
renormalizable theory -- but no principles are forbidding us
to include additional terms in the non-abelian
action and this is anyway needed for a D-dimensional $\left(D > 4\right)$
Yang-Mills Lagrangian. By modern field theory principles the
non-abelian Yang-Mills action can always be regarded as an
effective field theory and treated as such in calculations.

Our aim was to gain additional insight in the mapping
process and to investigate the options of extending the KLT
relations between scattering amplitudes at the effective field
theory level. Profound relations between effective gravity and
Yang-Mills diagrams and between pure effective Yang-Mills diagrams
was found in this process and will be presented in this chapter.
To which level string theory actually is needed in the mapping
process will be another aspect of the investigations.

The KLT-relations also work for mixed closed and
open string amplitudes, the so called heterotic string modes.
In order to explore such possibilities an antisymmetric term
will be augmented to the gravitational
effective action. It is needed to allow for mixed string modes ref.~\cite{Gross:1985fr}.

This chapter is organized as follows. First we review some material
about scattering amplitudes. We will next briefly discuss the
theoretical background for the KLT-relations in string theory and
make a generalization of the mapping. The mapping solution is then
presented, and we give some beautiful diagrammatic relationships
generated by the mapping solution. These relations not only hold
between gravity and Yang-Mills theory, but also in pure Yang-Mills
theory itself. We end the chapter with a discussion about the
implications of such mapping relations between gravity and gauge
theory and try to look ahead. The same conventions as in the
papers refs.~\cite{B4,B5} will be used here, $i.e.$, metric $\left(+ - - -\right)$
and units $\left(c=\hbar=1\right)$.

\section{The effective actions}
In this section we will review the main results of refs.~\cite{B4,B5,Tsey1,
Tsey2,Tsey3}.
The Einstein-Hilbert action is defined as previously and reads:
\begin{equation}
{\cal L} = \sqrt{-g} \bigg[{\frac{2R}{\kappa^2}}\bigg],
\end{equation}
where $\left(g_{\mu\nu}\right)$ denotes the gravitational field,
$\left(g \equiv \det(g_{\mu\nu})\right)$ and $\left(R\right)$ is the scalar curvature tensor.
$\left(\kappa^2 \equiv 32\pi G\right)$, where $\left(G\right)$ is the gravitational
constant. The convention,
$\left(R^\mu_{\ \nu\alpha\beta} = \partial_\alpha \Gamma_{\nu\beta}^\mu - \partial_\beta \Gamma_{\nu\alpha}^\mu + \ldots\right)$ is followed.

For the theory of Yang-Mills non-abelian vector field theory, the pure
Lagrangian for the non-abelian vector field can be written:
\begin{equation}
{\cal L} = -\frac1{8g^2}{\rm tr}{F}^2_{\mu\nu},
\end{equation}
where $\left(A_\mu\right)$ is the vector field and $\left(F_{\mu\nu} = \partial_\mu
A_\nu - \partial_\nu A_\mu + g[A_\mu,A_\nu]\right)$.
The trace in the Lagrangian is over the generators of the non-abelian
Lie-algebra, $e.g.$,
$\left({\rm tr}(T^a T^b T^c)F^a_{\mu\lambda}F^b_{\lambda\nu}F^c_{\nu\mu}\right)$,
where the $\left(T^a\right)$'s are the generators of the
algebra.\footnote{\footnotesize{The notation for the Yang-Mills action,
$i.e.$, the $-\frac18\ldots$ follows the non-standard conventions
of ref.~\cite{Tsey1}. Please note that in this letter the action is not
multiplied with $\left((2\pi\alpha')^2\right)$.}} The convention $\left(g=1\right)$ will also be
employed, as this is convenient when working with the KLT-relations.
The above non-abelian Yang-Mills action is of course renormalizable
at $\left(D=4\right)$, but it still makes sense to treat it as the minimal
Lagrangian in an effective field theory.
In the modern view of field theory, any invariant obeying the
underlying basic field symmetries, such as local gauge invariance, should
in principle always be allowed into the Lagrangian.

In order to generate a scattering amplitude we need
expressions for the generic effective Lagrangians in gravity and
Yang-Mills theory. In principle any higher derivative
gauge or reparametrization invariant term could be included
-- but it turns out that some terms will be ambiguous under
field redefinitions such as:
\begin{equation}\begin{split}
\delta g_{\mu\nu} &= a_1R_{\mu\nu} + a_2g_{\mu\nu}R \ldots\\
\delta A_\mu &= \tilde a_1{\cal D}_\nu F_{\nu\mu} + \ldots,
\end{split}\end{equation}
where $\left({\cal D}\right)$ is the gauge theory covariant
derivative.

Following ref.~\cite{Tsey1} the effective non-abelian Yang-Mills action
for a massless vector field takes the generic form
including all independent invariant field operators:
\begin{equation}\begin{split}
{\cal L} &= -\frac18{\rm tr}\Big[F_{\mu\nu}^2 + \alpha'\big(a_1F_{\mu\lambda}F_{\lambda\nu}F_{\nu\mu}
+a_2{\cal D}_\lambda F_{\lambda\mu}{\cal D}_\rho F_{\rho\mu}\big)\\&
+(\alpha')^2\big(a_3F_{\mu\lambda}F_{\nu\lambda}F_{\mu\rho}F_{\nu\rho}
+a_4F_{\mu\lambda}F_{\nu\lambda}F_{\nu\rho}F_{\mu\rho}\\
&+a_5F_{\mu\nu}F_{\mu\nu}F_{\lambda\rho}F_{\lambda\rho}
+a_6F_{\mu\nu}F_{\lambda\rho}F_{\mu\nu}F_{\lambda\rho}+a_7F_{\mu\nu}{\cal D}_\lambda F_{\mu\nu}{\cal D}_\rho
F_{\rho\lambda}\\& + a_8 {\cal D}_\lambda F_{\lambda \mu}{\cal D}_\rho
F_{\rho\nu}F_{\mu\nu}
+a_9{\cal D}_\rho{\cal D}_\lambda F_{\lambda\mu}{\cal D}_\rho {\cal D}_\sigma F_{\sigma\mu}\big)+\ldots \Big],
\end{split}\end{equation}
where $\left(\alpha'\right)$ orders the derivative expansion. In
string theory $\left(\frac{1}{\alpha'}\right)$ will of course be
the string tension. $\left(F_{\mu\nu} = \partial_\mu A_\nu -
\partial_\nu A_\mu + [A_\mu,A_\nu]\right)$ and $\left({\cal
D}\right)$ is the non-abelian covariant derivative.

In general relativity a similar situation occurs for the effective
action, following ref.~\cite{Tsey2}
the most general Lagrangian one can consider is:
\begin{equation}\begin{split}
{\cal L} &= \frac{2\sqrt{-g}}{\kappa^2}\Big[R + \alpha'\big (a_1
R_{\lambda\mu\nu\rho}^2 + a_2R_{\mu\nu}^2 +a_3R^2\big)\\&
+(\alpha')^2\big(b_1 R^{\mu\nu}_{\ \  \alpha\beta}
R^{\alpha\beta}_{\ \  \lambda\rho} R^{\lambda\rho}_{\ \
\mu\nu}+b_2(R^{\mu\nu}_{\ \  \alpha\beta} R^{\alpha\beta}_{\ \
\lambda\rho} R^{\lambda\rho}_{\ \  \mu\nu}\\&-2R^{\mu\nu\alpha}_{\
\ \ \beta} R_{\ \ \nu\lambda}^{\beta\gamma} R^{\lambda}_{\ \mu
\gamma\alpha}) + b_3 R_{\mu\alpha\beta\gamma}
R^{\alpha\beta\gamma}_{\ \ \ \ \nu}R^{\mu\nu} + b_4
R_{\mu\nu\rho\lambda}R^{\nu\lambda}R^{\mu\rho}\\&+b_5R_{\mu\nu}R^{\nu\lambda}R^\mu_{\
\lambda} +b_6R_{\mu\nu}{\cal
D}^2R^{\mu\nu}+b_7R^2_{\lambda\mu\nu\rho}R
\\&+b_8R_{\mu\nu}^2R+b_9R^3+b_{10}R{\cal D}^2R\big)+\ldots\Big].
\end{split}\end{equation}
As the S-matrix is manifestly invariant under a field redefinition
such ambiguous terms need not to be included in the generic
Lagrangian which give rise to the scattering amplitude,
see also refs.~\cite{Deser,Gross:1986iv}. The Lagrangian obtained from the
reparametrization invariant terms alone will be sufficient, because
we are solely considering scattering amplitudes. Including only
field redefinition invariant contributions in the action one can
write for the on-shell action of the gravitational fields (to
order $\left(O(\alpha'^3)\right)$\footnote{\footnotesize{For completeness we
note; in, $e.g.$, six-dimensions there exist an integral relation
linking the term $\left(R^{\mu\nu}_{\ \  \alpha\beta} R^{\alpha\beta}_{\
\  \lambda\rho} R^{\lambda\rho}_{\ \ \mu\nu}\right)$ with the term
$\left(R^{\mu\nu\alpha}_{\ \ \ \beta} R_{\ \ \nu\lambda}^{\beta\gamma}
R^{\lambda}_{\ \mu\gamma\alpha}\right)$, in four-dimensions there exist
an algebraic relation also linking these terms,
see ref.~\cite{Nieuwenhuizen}. For simplicity we have included all the
reparametrization invariant terms in our approach and not looked
into the possibility that some of these terms actually might be
further related through an algebraic or integral relation.}}:
\begin{equation}\begin{split}\label{eq9}
{\cal L} &= \frac{2\sqrt{-g}}{\kappa^2}\Big[R + \alpha'\big(a_1
R_{\lambda\mu\nu\rho}^2\big) +(\alpha')^2\big(b_1 R^{\mu\nu}_{\ \
\alpha\beta} R^{\alpha\beta}_{\ \  \lambda\rho} R^{\lambda\rho}_{\
\  \mu\nu}\\& +b_2 (R^{\mu\nu}_{\ \  \alpha\beta}
R^{\alpha\beta}_{\ \  \lambda\rho} R^{\lambda\rho}_{\ \
\mu\nu}-2R^{\mu\nu\alpha}_{\ \ \ \beta} R_{\ \
\nu\lambda}^{\beta\gamma} R^{\lambda}_{\
\mu\gamma\alpha})\big)+\ldots\Big],
\end{split}\end{equation}
while the on-shell effective action in the Yang-Mills case (to
order $\left(O(\alpha'^3)\right)$ has the form:
\begin{equation}\begin{split}\label{eq7}
{\cal L}_{YM}^L &= -\frac18{\rm tr}\Big[F^L_{\mu\nu}F^L_{\mu\nu} +
\alpha'\big(a^L_1
F^L_{\mu\lambda}F^L_{\lambda\nu}F^L_{\nu\mu}\big)
+(\alpha')^2\big(a^L_3F^L_{\mu\lambda}F^L_{\nu\lambda}F^L_{\mu\rho}
F^L_{\nu\rho}+a^L_4F^L_{\mu\lambda}F^L_{\nu\lambda}F^L_{\nu\rho}F^L_{\mu\rho}
\\&+a^L_5F^L_{\mu\nu}F^L_{\mu\nu}F^L_{\lambda\rho}F^L_{\lambda\rho}
+a^L_6F^L_{\mu\nu}F^L_{\lambda\rho}F^L_{\mu\nu}F^L_{\lambda\rho}\big)
+\ldots \Big].\end{split}\end{equation}
The $\left(L\right)$ in the above
equation states that this is the effective Lagrangian for a 'left'
moving vector field. To the 'left' field action is associated an
independent 'right' moving action, where 'left' and 'right'
coefficients are treated as generally dissimilar. The 'left' and
the 'right' theories are completely disconnected theories, and have
no interactions. The Yang-Mills coupling constant in the above
equation is left out for simplicity. Four additional double trace
terms at order $\left(O(\alpha'^2)\right)$ have been neglected, see ref.~\cite{Dunbar2}.
These terms have a different Chan-Paton structure than the single
trace terms. We need to augment the gravitational effective
Lagrangian in order to allow for the heterotic
string ref.~\cite{Gross:1985fr}. As shown in ref.~\cite{Tsey2}, one should
in that case add an antisymmetric tensor field coupling:
\begin{equation}
{\cal L}_{\rm Anti} =
\frac{2\sqrt{-g}}{\kappa^2}\Big[-\frac34(\partial_{[\mu}B_{\nu\rho]}+\alpha'c
\omega^{ab}_{[\mu}R^{ab}_{\nu\rho]}+\ldots)(\partial_{[\mu}B_{\nu\rho]}+\alpha'
c \omega^{cd}_{[\mu}R^{cd}_{\nu\rho]}+\ldots) \Big],\end{equation}
where $\left(\omega_{\mu}^{ab}\right)$ represents the spin connection.

This term will contribute to the 4-graviton amplitude.
We will observe how such a term affects the
mapping solution, and forces the 'left' and 'right' coefficients
for the Yang-Mills Lagrangians to be different.\footnote{{See, $e.g.$,
ref.~\cite{Bern:1987bq}, for a discussion of field redefinitions
and the low-energy effective action for Heterotic strings.}}

From the effective Lagrangians above one can expand the field
invariants and thereby derive the scattering amplitudes. The results are:
\begin{equation}
R=-\frac{\kappa^3}4h_{\alpha\beta}(h_{\lambda\rho}\partial_\alpha\partial_\beta h_{\lambda\rho}+
2\partial_\alpha h_{\lambda\rho}\partial_\rho h_{\beta\lambda})+\ldots,
\end{equation}
\begin{equation}
R^2_{\lambda\mu\nu\rho}-4R^2_{\mu\nu}+R^2 = -\kappa^3h_{\mu\nu}\partial_\mu\partial_\rho h_{\alpha\beta}\partial_\alpha\partial_\beta h_{\nu\rho}+\ldots,
\end{equation}
\begin{equation}
R^{\mu\nu}_{\ \ \alpha\beta}R^{\alpha\beta}_{\ \ \lambda\rho}R^{\lambda\rho}_{\ \ \mu\nu} =
\kappa^3\partial_\rho \partial_\alpha h_{\beta\nu}\partial_\beta\partial_\lambda h_{\mu\alpha}\partial_\mu\partial_\nu h_{\lambda\rho}+\ldots,
\end{equation}
where $\left(g_{\mu\nu}\equiv\eta_{\mu\nu}+\kappa h_{\mu\nu}\right)$, and $\left(\eta_{\mu\nu}\right)$ is the flat metric. Because
of the flat background metric no distinctions are made of up and down indices on the right hand
side of the equations, so indices can be put where convenient.
The results are identical to these of ref.~\cite{Tsey2}.
This gives the following generic 3-point scattering amplitude for the effective
action of general relativity:
\begin{equation}\begin{split}
{M_3}&= \kappa \Big[ \zeta_2^{\mu\sigma}\zeta_3^{\mu\rho}
\big(\zeta_1^{\alpha\beta}k_2^{\alpha}k_2^{\beta}\delta^{\sigma\rho}
+\zeta_1^{\sigma\alpha}k_2^{\alpha}k_1^{\rho}
+\zeta_1^{\sigma\alpha}k_3^{\alpha}k_1^{\rho}\big)
+\zeta_1^{\mu\sigma}\zeta_3^{\mu\rho}
\big(\zeta_2^{\alpha\beta}k_3^{\alpha}k_3^{\beta}\delta^{\sigma\rho}
\\&+\zeta_2^{\sigma\alpha}k_1^{\alpha}k_2^{\rho}
+\zeta_2^{\sigma\alpha}k_3^{\alpha}k_2^{\rho}\big)
+\zeta_1^{\mu\sigma}\zeta_2^{\mu\rho}
\big(\zeta_3^{\alpha\beta}k_1^{\alpha}k_1^{\beta}\delta^{\sigma\rho}
+\zeta_3^{\sigma\alpha}k_1^{\alpha}k_3^{\rho}
+\zeta_3^{\sigma\alpha}k_2^{\alpha}k_3^{\rho}\big)
\\&+\alpha'[4a_1\zeta_2^{\mu\sigma}\zeta_3^{\mu\rho}\zeta_1^{\alpha\beta}
k_2^\alpha k_2^\beta k_3^\sigma k_1^\rho
+4a_1\zeta_1^{\mu\sigma}\zeta_3^{\mu\rho}\zeta_2^{\alpha\beta}
k_3^\alpha k_3^\beta k_2^\sigma k_1^\rho\\&
+4a_1\zeta_1^{\mu\sigma}\zeta_2^{\mu\rho}\zeta_3^{\alpha\beta}
k_1^\alpha k_1^\beta k_2^\sigma k_3^\rho]
+(\alpha')^2[12b_1\zeta_1^{\alpha\beta}
\zeta_2^{\gamma\delta}\zeta_3^{\tau\rho}k_1^\tau k_1^\rho
k_2^\alpha k_2^\beta k_3^\gamma k_3^\delta] \Big],
\end{split}\end{equation}
where $\left(\zeta_i^{\mu\nu}\right)$ and $\left(k_i,\ i=1,..,3\right)$ denote the
polarization tensors and momenta for the external graviton legs.

For the gauge theory 3-point amplitude one finds:
\begin{equation}\begin{split}
A_{3L} &= -\Big[(\zeta_3\cdot k_1 \zeta_1 \cdot \zeta_2 + \zeta_2\cdot k_3
\zeta_3 \cdot \zeta_1+\zeta_1\cdot k_2 \zeta_2 \cdot \zeta_3)
+ \frac34\alpha'a_1^L\zeta_1\cdot k_2 \zeta_2\cdot k_3 \zeta_3 \cdot k_1\Big].
\end{split}\end{equation}

For the general 4-point amplitude matters are somewhat more
complicated. The scattering amplitude will consist both of direct
contact terms as well as 3-point contributions combined with a
propagator -- the interaction parts. Furthermore imposing on-shell
constrains will still leave many non-vanishing terms in the
scattering amplitude. A general polynomial expression for the
4-point scattering amplitude has the form, see refs.~\cite{Deser,shimada}:
\begin{equation}
A_4 \sim \sum_{0\leq n+m+k\leq 2} b_{nmk}s^n t^m u^k,
\end{equation}
where $\left(s\right)$, $\left(t\right)$ and $\left(u\right)$ are normal Mandelstam variables and the
factor $\left(b_{nmk}\right)$ consist of scalar contractions of momenta and
polarizations for the external lines. The case $\left(n+m+k=2\right)$
corresponds to a particular simple case, where the coefficients
$\left(b_{nmk}\right)$ consist only of momentum factors contracted with
momentum factors, and polarization indices contracted with other
polarization indices. In the process of comparing amplitudes, we
need only to match coefficients for a sufficient part of the
amplitude; $i.e.$, for a specific choice of factors $\left(b_{nmk}\right)$, --
gauge symmetry will then do the rest and dictate that once
adequate parts of the amplitudes match, we have achieved matching
of the full amplitude. We will choose the case where $\left(b_{nmk}\right)$ has
this simplest form.
Expanding the Lagrangian, one finds the following contributions
\begin{equation}
R=-\frac{\kappa^3}4h_{\mu\nu}h_{\nu\rho}\partial^2h_{\rho\mu}+\ldots,
\end{equation}
\begin{equation}
R^2_{\lambda\mu\nu\rho}-4R^2_{\mu\nu}+R^2 = -\frac{\kappa^3}2\partial_\alpha h_{\mu\nu}\partial_\alpha h_{\nu\rho}\partial^2 h_{\rho\mu}+\ldots,
\end{equation}
\begin{equation}\begin{split}
R^{\mu\nu}_{\ \ \alpha\beta}R^{\alpha\beta}_{\ \ \lambda\rho}R^{\lambda\rho}_{\ \ \mu\nu} &=
-\frac{3\kappa^3}2\partial_\alpha\partial_\beta h_{\mu\nu} \partial_\alpha\partial_\beta h_{\nu\rho}\partial^2 h_{\rho\mu}
\\&+3\kappa^4(h_{\mu\nu}\partial_\alpha\partial_\beta h_{\nu\rho} \partial_\alpha\partial_\gamma h_{\rho\sigma} \partial_\beta\partial_\gamma h_{\sigma\mu}
\\&+\frac{1}2\partial_\alpha h_{\mu\nu}\partial_\beta h_{\nu\rho} \partial_\alpha\partial_\gamma h_{\rho\sigma}
\partial_\beta\partial_\gamma h_{\sigma\mu})+\ldots,
\end{split}\end{equation}
\begin{equation}\begin{split}
R^{\mu\nu\alpha}_{\ \ \ \beta}R_{\ \ \nu\lambda}^{\beta\gamma}R^{\lambda}_{\ \mu\gamma\alpha}&=
-\frac{3\kappa^4}8\partial_\alpha h_{\mu\nu}\partial_\beta h_{\nu\rho} \partial_\beta \partial_\gamma h_{\rho\sigma} \partial_\alpha\partial_\gamma h_{\sigma\mu}+\ldots.
\end{split}\end{equation}
The same conventions as for 3-point terms are used in these equations.
The above results simply verify ref.~\cite{Tsey2} and they result
the following expression for 4-point amplitude:
\begin{equation}\begin{split}
{M_4} &= \frac12
\frac{\kappa^2}{\alpha'}\Big[\zeta_1\zeta_2\zeta_3\zeta_4(z + a_1
z^2 -3b_1(z^3-4xyz) -3b_2(z^3-\frac72xyz)\\& +(a_1^2+3b_1 +
3b_2)(z^3-3xyz)+\frac14c^2(z^3-xyz))\\& +
\zeta_1\zeta_2\zeta_4\zeta_3(y + a_1 y^2
-3b_1(y^3-4xyz)-3b_2(y^3-\frac72xyz)\\& +(a_1^2+3b_1 +
3b_2)(y^3-3xyz)+\frac14c^2(y^3-xyz))\\& +
\zeta_1\zeta_3\zeta_2\zeta_4(x + a_1 x^2
-3b_1(x^3-4xyz)-3b_2(x^3-\frac72xyz)\\& +(a_1^2+3b_1 +
3b_2)(x^3-3xyz)+\frac14c^2(x^3-xyz))\Big]
\end{split}\end{equation}
where we use the definitions: $\left(\zeta_1\zeta_2\zeta_3\zeta_4 =
\zeta_1^{\alpha\beta}
\zeta_2^{\beta\gamma}\zeta_3^{\gamma\delta}\zeta_4^{\delta\alpha}\right)$
as well as $\left(x = -2\alpha'(k_1\cdot k_2)\right)$, $\left(y = -2\alpha'(k_1\cdot
k_4)\right)$ and $\left(z=-2\alpha'(k_1\cdot k_3)\right)$. In the above expression the
antisymmetric term (with generic coefficient $\left(c\right)$) needed in the
heterotic string scattering amplitude has been
included refs.~\cite{Tsey2,Gross:1985fr}.

The corresponding 4-point gauge amplitude can be written:
\begin{equation}\begin{split}
A_{4L} &= \bigg[
\Big[\zeta_{1324}+\frac{z}{x}\zeta_{1234}+\frac{z}{y}\zeta_{1423}\Big]
+\Big[-\frac{3a^L_1}{8}z(\zeta_{1324}+\zeta_{1234}+\zeta_{1423})\Big]
\\&+\Big[\frac{9(a^L_1)^2}{128}(x(z-y)\zeta_{1234}+y(z-x)\zeta_{1423})\Big]
\\&-\frac14\Big[(\frac12a^L_3)xy\zeta_{1324}-(\frac14a_3^L+2a_6^L)z^2\zeta_{1324}
+(\frac14a^L_3 +\frac12a^L_4)yz\zeta_{1234} + (\frac14a^L_3 +
a^L_5)zx\zeta_{1234}\\&+(\frac12a_4^L+a_5^L)yx\zeta_{1234}
+(\frac14a^L_3 +a_5^L)yz\zeta_{1423}+(\frac14a_3^L+
\frac12a_4^L)xz\zeta_{1423}+(\frac12a_4^L+a_5^L)xy\zeta_{1423}\Big]\bigg],
\end{split}\end{equation}
where $\left(x\right)$, $\left(y\right)$ and $\left(z\right)$ are defined as above, and, $e.g.$,
$\left(\zeta_{1234} = (\zeta_1 \cdot \zeta_2)(\zeta_3 \cdot \zeta_4)\right)$
and etc. The contact terms in the above expression were calculated explicitly
and the results agree with those of ref.~\cite{Tsey2}.
The non-contact terms were adapted from ref.~\cite{Tsey2}.

The expressions for the scattering amplitudes are quite general.
They relate the general generic effective Lagrangians in the
gravity and Yang-Mills case to their 3- and 4-point scattering
amplitudes respectively in any dimension. The next section will be
dedicated to showing how the 3-point amplitudes and 4-point
amplitudes in gravity and Yang-Mills have to be related through
the KLT-relations.

\section{The open-closed string relations}
The KLT-relations between closed and open string are discussed in literature on
string theory, but let us recapitulate the essentials here. Following conventional string
theory ref.~\cite{GSW} the general $M$-point scattering amplitude for a
closed string is related to that of an open string in the
following manner:
\begin{equation}\begin{split}
{A}^M_{\rm closed} \sim \sum_{\Pi,\tilde \Pi}
e^{i\pi\Phi(\Pi,\tilde\Pi)}{A}_M^\text{left open}(\Pi)
{A}_M^\text{right open}(\tilde \Pi),
\end{split}\end{equation}
in this expression $\left(\Pi\right)$ and $\left(\tilde\Pi\right)$ corresponds to particular
cyclic orderings of the external lines of the open string. While
the $\left(\Pi\right)$ ordering corresponds to a left-moving open string, the
$\left(\tilde \Pi\right)$ ordering corresponds to the right-moving string. The
factor $\left(\Phi(\Pi,\tilde \Pi)\right)$ in the exponential is a phase factor
chosen appropriately with the cyclic permutations $\left(\Pi\right)$ and
$\left(\tilde \Pi\right)$. In the cases of 3- and 4-point amplitudes, the
following specific KLT-relations can be adapted from the $M$-point
amplitude\footnote{\footnotesize{The specific
forms of the KLT-relations do not follow any specific convention, see e.g., ref.~\cite{Bern4}.
In order the keep the conventions of ref.~\cite{Tsey1,Tsey2}, the employed
relations are normalized
to be consistent for the $\left (O(1)\right)$ terms of the 3- and 4-point scattering
amplitudes.}}:
\begin{equation}\begin{split}\label{eq22}
{M}_{\rm 3 \ gravity}^{\mu\tilde\mu\nu\tilde\nu\rho\tilde\rho}&(1,2,3) =
\kappa{A}^{\mu\nu\rho}_\text{3 L-gauge}(1,2,3)\times
{A}^{\tilde\mu\tilde\nu\tilde\rho}_\text{3 R-gauge}(1,2,3)\\
{M}_{\rm 4 \
gravity}^{\mu\tilde\mu\nu\tilde\nu\rho\tilde\rho\sigma\tilde\sigma}&(1,2,3,4)
= \frac{\kappa^2}{4\pi\alpha'}\sin(\pi x)\times
{A}^{\mu\nu\rho\sigma}_\text{4
L-gauge}(1,2,3,4)\times{A}^{\tilde\mu\tilde\nu\tilde\rho\tilde\sigma}_\text{4
R-gauge}(1,2,4,3),
\end{split}\end{equation}
where $\left(M\right)$ is a tree amplitude in gravity, and we have the color
ordered amplitude $\left(A\right)$ for the gauge theory with coupling constant $\left(g=1\right)$.
As previously the definition $\left(x = -2\alpha'(k_1\cdot k_2), \ldots\right)$
is employed,
where $\left(k_1\right)$ and $\left(k_2\right)$ are particular momenta of the external lines.

In string theory the specific mapping relations originate through
the comparison of open and closed string amplitudes. However
considering effective field theories there is no need to assume
anything {\it a priori} about the tree amplitudes. It makes sense
to investigate the mapping of scattering amplitudes in the
broadest possible setting. We will thus try to generalize the
above mapping relations. In the case of the 4-amplitude mapping it
seems that a possibility is to replace the specific sine function
with a general Taylor series, $e.g.$:
\begin{equation}
\frac{\sin(\pi x)}{\pi} \rightarrow xf(x) = x(1+P_1x+P_2x^2+\ldots)
\end{equation}
If this is feasible and what it means for the mapping, will be explored below.

\section{Generalized mapping relations}
Insisting on the ordinary mapping relation for the 3-point
amplitude and replacing in the 4-point mapping relation the sine
function with a general polynomial the following relations between
the coefficients in the scattering amplitudes are found to be
necessary in order for the generalized KLT-relations to hold. From
the 3-amplitude at order $\left(\alpha'\right)$ we have:
\begin{equation}\begin{split}
3a_1^L+3a_1^R&=16a_1,
\end{split}\end{equation}
while from the 3-amplitude at order $\left(\alpha'^2\right)$ one gets:
\begin{equation}\begin{split}
3a_1^L\, a_1^R &= 64b_1
\end{split}\end{equation}
From the 4-amplitude at order $\left(\alpha'\right)$ we have:
\begin{equation}\begin{split}
16a_1 = 3a_1^L+3a_1^R, \ \ P_1=0
\end{split}\end{equation}
while the 4-amplitude at order $\left(\alpha'^2\right)$ states:
\begin{equation}\begin{split}
6a_5^L+3a_4^L+\frac{27(a_1^L)^2}{16} &= 0,\ \
6a_5^R+3a_4^R+\frac{27(a_1^R)^2}{16} = 0,
\end{split}\end{equation}
together with the equations:
\begin{equation}\begin{split}
24c^2+96a_1^2&= 6a_3^L+3a_3^R+18a_4^L+12a_5^L+24a_6^R+\frac{81(a_1^L)^2}{8}+96P_2+18(a_1^R+a_1^L)P_1,\\
24c^2+96a_1^2&= 3a_3^L+6a_3^R+18a_4^R+12a_5^R+24a_6^L+\frac{81(a_1^R)^2}{8}+96P_2+18(a_1^R+a_1^L)P_1,\\
24c^2+96a_1^2&= 6a_3^L-3a_3^R-6a_4^L-36a_5^L-12a_5^R-\frac{27}{8}(a_1^L)^2+\frac{27}{8}(a_1^R)^2+96P_2\\&-18(a_1^R+a_1^L)P_1,\\
24c^2+96a_1^2&=-3a_3^L+6a_3^R-6a_4^R-12a_5^L-36a_5^R+\frac{27}{8}(a_1^L)^2-\frac{27}{8}(a_1^R)^2+96P_2\\&-18(a_1^R+a_1^L)P_1,
\end{split}\end{equation}
and furthermore:
\begin{equation}\begin{split}
96b_1+48b_2+16c^2&=4a_3^L+5a_3^R+2a_4^L+6a_4^R+8a_6^L
+\frac{9(a_1^R)^2}{4}+\frac{9a_1^L\, a_1^R}{2}+96P_2\\&+18(a_1^R+a_1^L)P_1,\\
96b_1+48b_2+16c^2&=4a_3^R+5a_3^L+2a_4^R+6a_4^L+8a_6^R
+\frac{9(a_1^L)^2}{4}+\frac{9a_1^L\,a_1^R}{2}+96P_2\\&+18(a_1^R+a_1^L)P_1,
\end{split}\end{equation}
as well as:
\begin{equation}\begin{split}
96a_1^2-96b_1-48b_2+8c^2& = a_3^L+2a_4^R-12a_5^L-12a_5^R-\frac{9a_1^L\, a_1^R}{2}+\frac{9((a_1^L)^2+(a_1^R)^2)}{8}\\&+64P_2-6(a_1^R+a_1^L)P_1,\\
96a_1^2-96b_1-48b_2+8c^2& = 2a_4^R-4a_5^L-12a_5^R+8a_6^L-\frac{9a_1^L\, a_1^R}{2}+\frac{9((a_1^L)^2+(a_1^R)^2)}{8}\\&+32P_2,\\
96a_1^2-96b_1-48b_2+8c^2& = 2a_4^L-12a_5^L-4a_5^R+8a_6^R-\frac{9a_1^L\, a_1^R}{2}+\frac{9((a_1^L)^2+(a_1^R)^2)}{8}\\&+32P_2,\\
96a_1^2-96b_1-48b_2+8c^2& =
a_3^R+2a_4^L-12a_5^L-12a_5^R-\frac{9a_1^L\,a_1^R}{2}+\frac{9((a_1^L)^2+(a_1^R)^2)}{8}\\&+64P_2-6(a_1^R+a_1^L)P_1.
\end{split}\end{equation}
These equations are found by relating similar scattering
components, $e.g.$, the product resulting from the generalized KLT
relations: $\left(\zeta_{1234}\zeta_{1324}y^2x\right)$ on the gauge side, with
$\left(\zeta_1\zeta_2\zeta_4\zeta_3 y^2x\right)$ on the gravity side. The
relations can be observed to be a generalization of the mapping
equations we found in ref.~\cite{B4}. Allowing
for a more general mapping and including the antisymmetric term
needed in the case of an heterotic string generates additional
freedom in the mapping equations. The generalized equations can
still be solved and the solution is unique. One one ends up with
the following solution\footnote{{We have employed an algebraic
equation solver {\em Maple}, to solve the equations, (Maple and
Maple V are registered trademarks of Maple Waterloo Inc.)}}:
\begin{eqnarray}
a_1^L=\frac8{3}a_1\pm \frac43c,&&
a_1^R=\frac8{3}a_1\mp \frac43c,\\
a_3^L=-8P_2,&&
a_3^R=-8P_2,\\
a_4^L=-4P_2,&&
a_4^R=-4P_2,\\
a_5^L=\mp 2a_1c-2a_1^2-\frac12c^2+2P_2,&&
a_5^R=\pm 2a_1c-2a_1^2-\frac12c^2+2P_2,\\
a_6^L=\pm 2a_1c+2a_1^2+\frac12c^2+P_2,&&
a_6^R=\mp 2a_1c+2a_1^2+\frac12c^2+P_2,\\
b_1=\frac{1}{3}a_1^2-\frac{1}{12}c^2,&&
b_2=\frac{2}{3}a_1^2-\frac16c^2.
\end{eqnarray}
This is the unique solution to the generalized mapping relations.
As seen, the original KLT solution ref.~\cite{B4} is still
contained but the generalized mapping solution is not as
constraining as the original KLT solution was. In fact now one can
freely choose $\left(c\right)$, and $\left(P_2\right)$ as well as
$\left(a_1\right)$. Given a certain gravitational action with or
without the possibility of terms needed for heterotic strings, one
can choose between different mappings from the gravitational
Lagrangian to the given Yang-Mills action. Traditional string
solutions are contained in this and are possible to reproduce --
but the solution space for the generalized solution is broader and
allows seemingly for a wider range of possible effective actions
on the gravity and the Yang-Mills side. It is important to note
that this does not imply that the coefficients in the effective
actions can be chosen freely -- the generalized KLT-relations
still present rather restricting constraints on the effective
Lagrangians. To which extent one may be able to reproduce the full
solution space by string theory is not definitely answered.
Clearly superstrings cannot reproduce the full solution space
because of space-time supersymmetry which does not allow for terms
in the effective action like $\left({\rm
tr}(F_{\mu\nu}F_{\nu\alpha}F_{\alpha\mu})\right)$ on the open
string side and
$\left(R^{\mu\nu}_{\alpha\beta}R^{\alpha\beta}_{\gamma\lambda}R^{\gamma\lambda}_{\mu\nu}\right)$
on the gravity side ref.~\cite{grisaru1,grisaru2,grisaru3}. For
non-supersymmetric string theories constraints on the effective
actions such as the above do not exist, and it is therefore
possible that in this case some parts or all of the solution space
{\it might} be reproduced by the variety of non-supersymmetric
string theories presently known. It is, $e.g.$, observed that the
bosonic non-supersymmetric string solution in fact covers parts of
the solution space not covered by the supersymmetric string
solution.

The possibility of heterotic strings on the gravity side, $i.e.$, a
non-vanishing $\left(c\right)$, will as observed always generate dissimilar
'left' and 'right' Yang-Mills coefficients. That is for nonzero $\left(c\right)$,
$e.g.$, $\left(a_1^L \neq a_1^R\right)$, $\left(a_5^L \neq a_5^R\right)$ and $\left(a_6^L \neq
a_6^R\right)$. It is also seen that for $\left(c=0\right)$ that 'left' is equivalent
to 'right'. The coefficients $\left(a_3^L\right)$, $\left(a_3^R\right)$ and $\left(a_4^L\right)$, $\left(a_4^R\right)$ are
completely determined by the coefficient $\left(P_2\right)$. In the generalized
mapping relation the only solution for the coefficient $\left(P_1\right)$ is
zero.

The KLT or generalized KLT-mapping equations can be seen as
coefficient constraints linking the generic terms in the
gauge/gravity Lagrangians. To summarize the gravitational
Lagrangian has to take the following form, dictated by the
generalized KLT-relations:
\begin{equation}\begin{split}
{\cal L} &= \frac{2\sqrt{-g}}{\kappa^2}\Big[R + \alpha'\Big(a_1
R_{\lambda\mu\nu\rho}^2\Big)
+(\alpha')^2\Big(\big(\frac{1}{3}a_1^2-\frac{1}{12}c^2\big)
R^{\mu\nu}_{\ \ \alpha\beta} R^{\alpha\beta}_{\ \  \lambda\rho}
R^{\lambda\rho}_{\ \  \mu\nu}
\\&+\big(\frac{2}{3}a_1^2-\frac16c^2\big) \big(R^{\mu\nu}_{\ \
\alpha\beta} R^{\alpha\beta}_{\ \  \lambda\rho} R^{\lambda\rho}_{\
\ \mu\nu}-2R^{\mu\nu\alpha}_{\ \ \ \beta} R_{\ \
\nu\lambda}^{\beta\gamma} R^{\lambda}_{\
\mu\gamma\alpha}\big)\Big)\\&-\frac34\Big(\partial_{[\mu}B_{\nu\rho]}+\alpha'
c \omega^{ab}_{[\mu}R^{ab}_{\nu\rho]}+\ldots\Big)^2+\ldots\Big],
\end{split}\end{equation}
and the corresponding 'left' or 'right' Yang-Mills action are then
forced to be:
\begin{equation}\begin{split}
{\cal L}_{YM} &= -\frac18{\rm tr}\Big[F^L_{\mu\nu}F^L_{\mu\nu} +
\alpha'\Big(\big(\frac8{3}a_1\pm \frac43 c \big)
F^L_{\mu\lambda}F^L_{\lambda\nu}F^L_{\nu\mu}\Big)
\\&+(\alpha')^2\Big(-8P_2F^L_{\mu\lambda}F^L_{\nu\lambda}F^L_{\mu\rho}
F^L_{\nu\rho}-4P_2F^L_{\mu\lambda}F^L_{\nu\lambda}F^L_{\nu\rho}F^L_{\mu\rho}
\\&+\big(\mp 2a_1 c-2a_1^2+2P_2-\frac12c^2\big)F^L_{\mu\nu}F^L_{\mu\nu}F^L_{\lambda\rho}F^L_{\lambda\rho}
\\&+\big(\pm 2a_1
c+2a_1^2+\frac12c^2+P_2\big)F^L_{\mu\nu}F^L_{\lambda\rho}F^L_{\mu\nu}F^L_{\lambda\rho}\Big)
+\ldots \Big],\end{split}\end{equation} where 'left' and 'right,'
reflect opposite choices of signs, in the above equation. One
sees that the Yang-Mills Lagrangian is fixed once a gravitational
action is chosen; the only remaining freedom in the Yang-Mills
action is then the choice of $\left(P_2\right)$, corresponding to different
mappings.

\begin{figure}[h]
\begin{center}\includegraphics{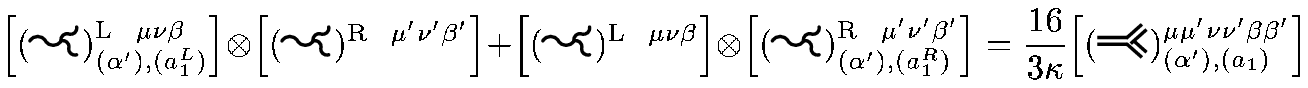}\end{center} \caption{A diagrammatical
expression for the generalized mapping of the 3-point gravity
amplitude into the product of Yang-Mills amplitudes at order
$\alpha'$.\label{fig1a_p}}
\end{figure}

\begin{figure}[h]
\begin{center}\includegraphics{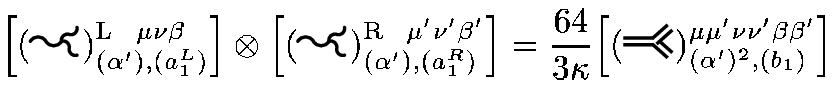}\end{center} \caption{A diagrammatical
expression for the generalized mapping of the 3-point gravity
amplitude into the product of Yang-Mills amplitudes at order
$\alpha'^2$.\label{fig1b_p}}
\end{figure}

\begin{figure}[h]
\begin{center}\includegraphics{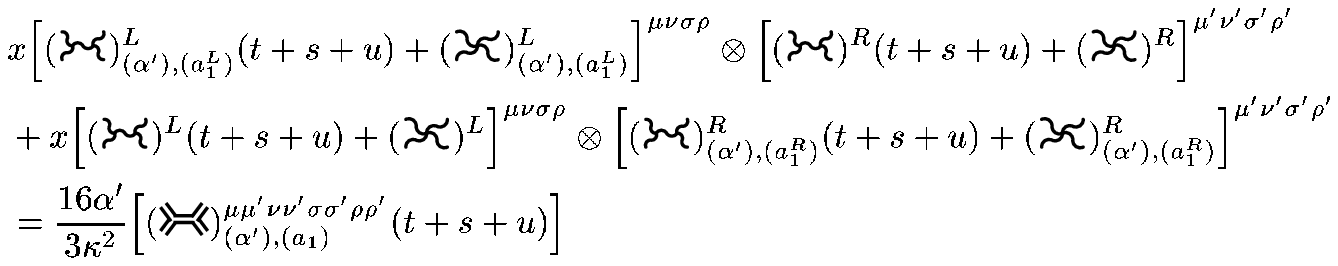}\end{center} \caption{The diagrammatical
expression for the generalized mapping of the 4-point gravity
amplitude as a product of Yang-Mills amplitudes at order
$\alpha'$.\label{fig1c_p}}
\end{figure}

\begin{figure}[h]
\begin{center}\includegraphics{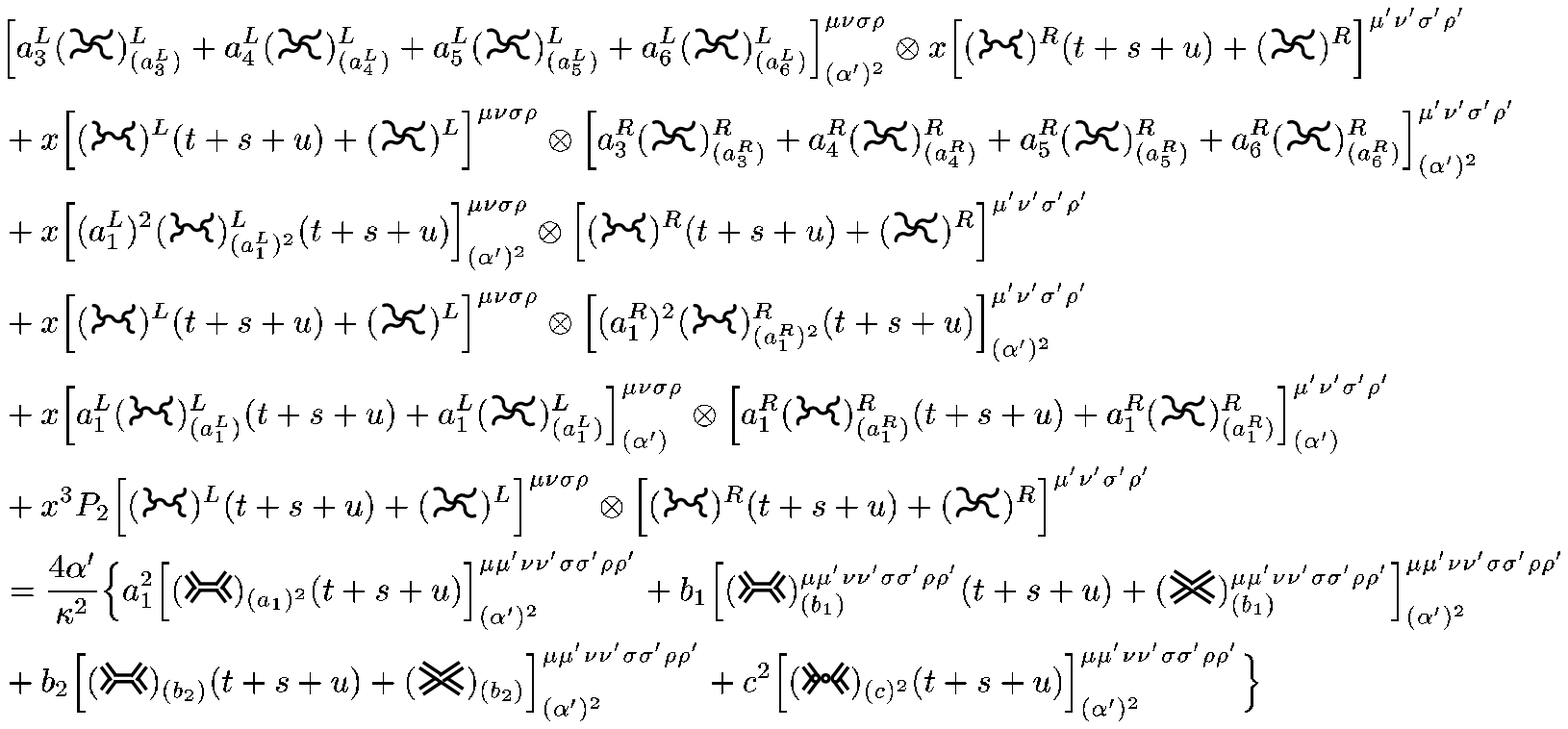}\end{center} \caption{The diagrammatical
expression for the generalized mapping of the 4-point gravity
amplitude into the product of Yang-Mills amplitudes at order
$\alpha'^2$.\label{fig1d_p}}
\end{figure}

\begin{figure}[h]
\begin{center}\includegraphics{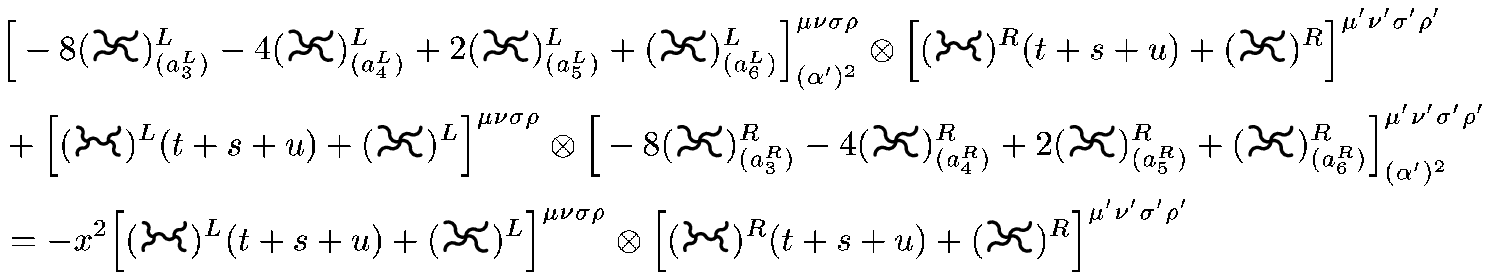}\end{center} \caption{A diagrammatic
relationship on the Yang-Mill side for the tree operators in the
effective action between operators of unit order and order
$\alpha'^2$.\label{fig1e_p}}
\end{figure}

\begin{figure}[h]
\begin{center}\includegraphics{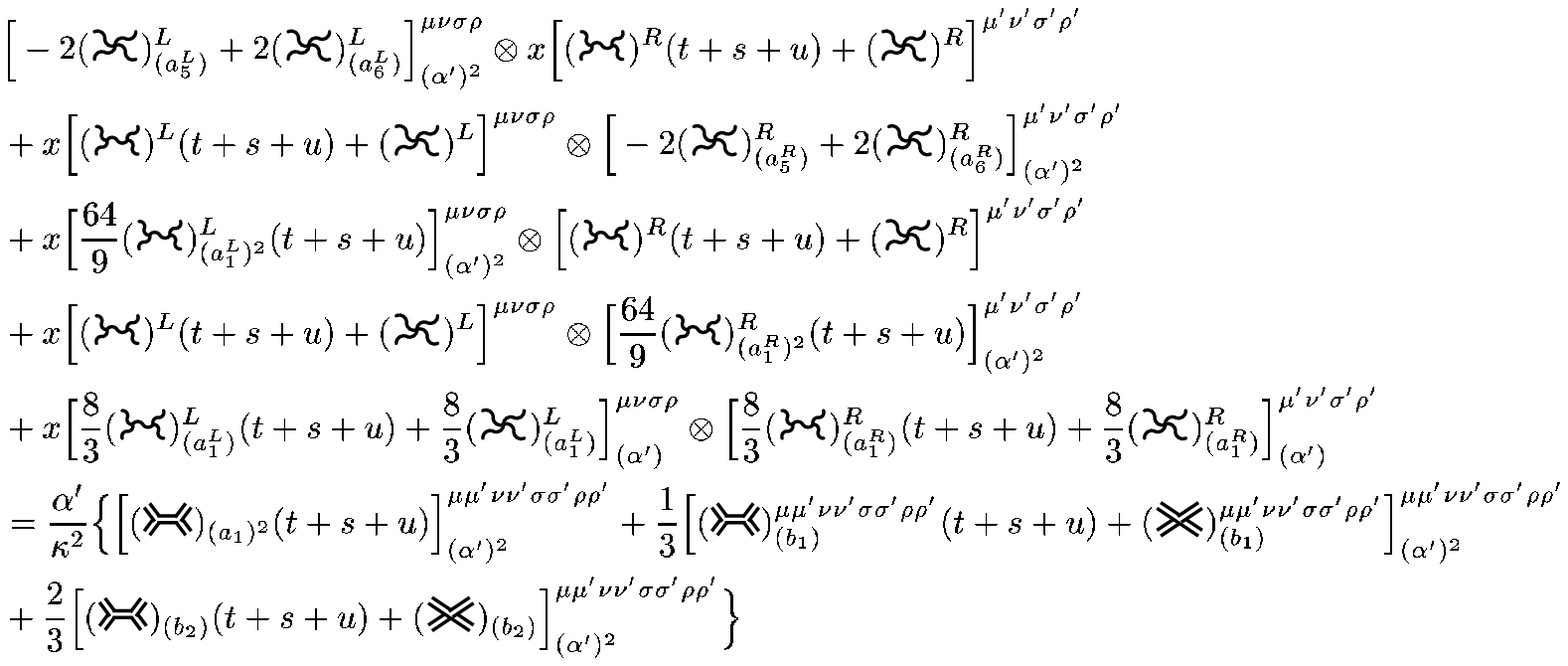}\end{center} \caption{The essential
diagrammatic relationship between gauge and gravity diagrams.\label{fig1f_p}}
\end{figure}

\begin{figure}[h]
\begin{center}\includegraphics{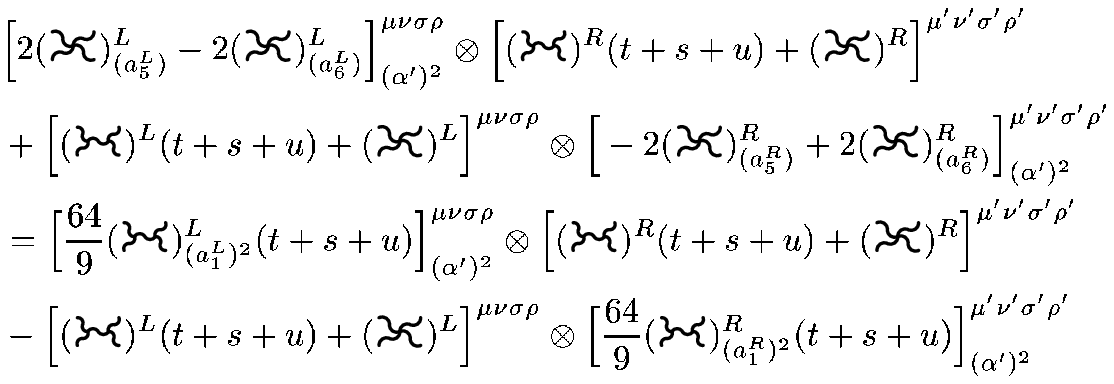}\end{center} \caption{Another diagrammatic
relationship on the Yang-Mills side between order $\alpha'^2$ and
$\alpha'$ operators.\label{fig1g_p}}
\end{figure}

It is directly seen that the graviton 3-amplitude to order
$\left(\alpha'\right)$ is re-expressible in terms of 3-point Yang-Mills
amplitudes. This can be expressed diagrammatically in figure~\ref{fig1a_p}.
At order $\left(\alpha'^2\right)$ we have the expression as shown in figure~\ref{fig1b_p}.
At the amplitude level this factorization is no surprise; this is
just what the KLT-relations tell us. At the Lagrangian level
things usually get more complicated and no factorizations of
gravity vertex rules are readily available. At order $\left(O(\alpha')\right)$
the factorization of gravity vertex rules was investigated in
ref.~\cite{Bern3}. An interesting task would be to continue this
analysis to order $\left(O(\alpha'^3)\right)$, and investigate the KLT
relations for effective actions directly at the Lagrangian vertex
level. Everything is more complicated in the 4-point case as
contact and non-contact terms mix in the mapping relations. This
originates from the fact that we are actually relating S-matrix
elements. Diagrammatically the 4-point generalized KLT-relation at
order $\left(\alpha'\right)$ is depicted in figure~\ref{fig1c_p}.
This relation is essentially equivalent to the 3-vertex
relation at order $\left(\alpha'\right)$. The 4-point relation at order
$\left(\alpha'\right)$ rules out the possibility of a $\left(P_1\right)$ term in the
generalized mapping. The full KLT-relation at order $\left(\alpha'^2\right)$
is shown in figure~\ref{fig1d_p}.
The coefficients in the above expression need to be
taken in agreement with the generalized KLT solution in order for
this identity to hold. A profound consequence of the KLT-relations
is that they link the sum of a certain class of diagrams in
Yang-Mills theory with a corresponding sum of diagrams in gravity,
for very specific values of the constants in the Lagrangian. At
glance we do not observe anything manifestly about the
decomposition of, $e.g.$, vertex rules, -- however this should be
investigated more carefully before any conclusions can be drawn.
The coefficient equations cannot directly be transformed into
relations between diagrams, -- this can be seen by calculation.
However reinstating the solution for the coefficients it is
possible to turn the above equation for the 4-amplitude at order
$\left(\alpha'^2\right)$ into interesting statements about diagrams. Relating
all $\left(P_2\right)$ terms give, $e.g.$, the following remarkable diagrammatic
statement as shown in figure~\ref{fig1e_p}.
The surprising fact is that from the relations which
apparently link gravity and Yang-Mills theory only, one can
eliminate the gravity part to obtain relations entirely in pure
Yang-Mills theory.
Similarly, relating all contributions with $a_1^2$
give the result shown in figure~\ref{fig1f_p}.
Furthermore we have a relationship generated by the
$ca_1$ parts as shown in figure~\ref{fig1g_p}.

These diagrammatical relationships can be readily checked by
explicit calculations. It is quite remarkable that the KLT
relations provide such detailed statements about pure Yang-Mills
effective field theories, without any reference to gravity at all.
To explain the notation in the above diagrammatic statements. An
uppercase $\left(L\right)$ or $\left(R\right)$ states that the scattering amplitude
originates from a 'left' or a 'right' mover respectively. The lower
indices, $e.g.$, $\left(\alpha'\right)$ and the coupling constants, $e.g.$, $\left(a_3^L\right)$
or $\left((a_1^L)^2\right)$ denote respectively the order of $\left(\alpha'\right)$ in the
particular amplitude and the coupling constant prefactor obtained
when this particular part of the amplitude is generated from the
generic Lagrangian. The parentheses with $\left[(t+s+u)\right]$ denote that we
are supposed to sum over all $\left(t\right)$, $\left(s\right)$ and $\left(u\right)$ channels for this
particular amplitude.

\section{Discussion}
The solution space of the above equations suggests the following
interpretation.
For a given coefficient $\left(a_1\right)$ in the gravitational effective field theory action
through order $\left(\alpha'^2\right)$ and the string value of $\left(P=-\frac{\pi^2}6\right)$,
there is a set of unique ('left'-'right') effective Yang-Mills Lagrangians,
up to a field redefinition, that satisfies
the constraint of the KLT-relations. The set of solutions
spans out a space of operator solutions, where the different
types of strings correspond to certain points along the path.
For string theories with similar 'left' and 'right' Lagrangians,
one can find a bosonic string corresponding to the
solution: $\left(a_1^L = a_1^R = \frac83\right)$,
$\left(a_3^L=a_3^R=\frac{4\pi^2}3\right)$, $\left(a_4^L=a_4^R=\frac{3\pi^2}2\right)$,
$\left(a_5^L=a_5^R=-\frac{\pi^2}{3}-2\right)$,
and $\left(a_6^L=a_6^R=-\frac{\pi^2}{6}+2\right)$, on the open string
side, and $\left(a_1=1\right)$, $\left(b_1=\frac13\right)$ and $\left(b_2 = \frac23\right)$, on the closed
string side, while the superstring correspond to:
$\left(a_1^L = a_1^R =0\right)$, $\left(a_3^L=a_3^R=\frac{4\pi^2}3\right)$,
$\left(a_4^L=a_4^R=\frac{2\pi^2}3\right)$, $\left(a_5^L=a_5^R=-\frac{\pi^2}{3}\right)$
and $\left(a_6^L=a_6^R=-\frac{\pi^2}{6}\right)$, for the open string,
and $\left(a_1=b_1=b_2=0\right)$ for the closed string. So the string
solutions are contained in the above solution space.

However it has been observed that it is possible to generalize
the KLT open/closed string relations in the effective field
theory framework and that the solution space is somewhat bigger
than the space justified by string theory.
The KLT-relations are seen to serve as mapping
relations between the {\em tree} effective field theories for
Yang-Mills and gravity. The belief, that general relativity is in
fact an effective field theory at loop orders, makes
investigations of its tree level manifestations and connections to
other effective actions interesting. Links such as KLT, which are
applicable and very simplifying in actual calculations should be
exploited and investigated at the effective Lagrangian level. An
important result of our investigations is the generalization of
the KLT-relations. It is found that one cannot completely replace
the sine function in the KLT-relations by an arbitrary function.
To order $\left(O(\alpha'^3)\right)$ it is possible to replace the sine with an
odd third order polynomial in $\left(x\right)$. However, the degree of freedom
represented by $\left(P_2\right)$ can be completely absorbed into a rescaling
of $\left(\alpha'\right)$ at this order, so additional investigations are
required before any conclusions can be drawn. The mapping
relations between the effective theories are found to be broader
than those given completely from the KLT-relations as the
coefficient $\left(P_2\right)$ in the generalized framework can be chosen
freely. Such a rescaling of $\left(P_2\right)$ represents an additional freedom
in the mapping. It has been shown that despite this
generalization, the effective extension of the KLT-relations is
still rather restrictive.

The possibility of an antisymmetric coupling of gravitons needed
in the effective action of a heterotic string has also been
allowed for and is seen to be consistent with the original and
generalized KLT-relations. We have learned that detailed
diagrammatic statements can be deduced from the KLT-relations.
This presents very interesting aspects which perhaps can be used
to gain additional insight in issues concerning effective
Lagrangian operators. Furthermore we expect this process to
continue, -- at order $\left(O(\alpha'^4)\right)$ we assume that the
KLT-relations will tell us about new profound diagrammatic
relationships between effective field theory on-shell operators in
gauge theory and gravity.

The generalized mapping relations represent an effective field
theory version of the well known KLT-relations. We have used what
we knew already from string theory about the KLT-relations to
produce a more general description of mapping relations between
gravity and Yang-Mills theory. String theory is not really needed
in the effective field theory setting. All that is used here is
the tree scattering amplitudes. Exploring if or if not a mapping
from the general relativity side to the gauge theory side is
possible produces the generalized KLT-relations. The KLT-relations
could also be considered in the case of external matter. In this
case, operators as the Ricci tensor and the scalar curvature will
not vanish on-shell. Such an approach is presently the subject for an
ongoing research project, see ref.~\cite{Kasper}.

KLT-relations involving loops are not yet resolved. One can
perform some calculations by making cuts of the diagrams using the
unitarity of the S-matrix but no direct factorizations of loop
scattering amplitudes have been seen to support a true loop
extension of the KLT-relations. Progress in this direction still
waits to be seen, and perhaps such loop extensions are not
possible. Loops in string theory and in field theory are not
directly comparable, and complicated issues with additional string
theory modes in the loops seem to be unavoidable in extensions of
the KLT-relations beyond tree level. Perhaps since 4-point
scattering amplitudes are much less complicated compared to
5-point amplitudes, it would be actually more important first to
check the mapping solutions at the 5-point level before
considering the issue of loop amplitudes.

\chapter{Quantum gravity at large numbers of dimensions}
\section{Introduction}
In this chapter we will discuss the last part of the thesis
work which is described in the paper, ref.~\cite{B6}.

The idea is to look upon quantum gravity at a large number of
dimensions. An intriguing paper by Strominger ref.~\cite{Strominger:1981jg}
deals with the perspectives of using the basic Einstein-Hilbert action
to define a theory of quantum gravity but to let the theory have
an infinite number of dimensions. The fundamental idea in this approach
is to let the spatial dimension be a parameter in which one is allowed
to expand the theory. Each graph of the theory is then associated with a
dimensional factor. Formally one can expand every Greens function
of the theory as a series in $\left(\frac1D\right)$ and the
gravitational coupling constant ($\kappa$):
\begin{equation}
G = \sum_{i,j}(\kappa)^i\left(\frac1D\right)^j {\cal G}_{i,j}.
\end{equation}
The contributions with highest dimensional dependence will be the
leading ones in the large-$D$ limit. Concentrating on these graphs
only simplifies the theory, and makes explicit calculations easier.
In the effective field theory we expect an expansion of the theory
of the type:
\begin{equation}
G = \sum_{i,j,k}(\kappa)^i\left(\frac1D\right)^j ({\cal
E})^{2k}{\cal G}_{i,j,k},
\end{equation}
where (${\cal E}$) represents a parameter for the effective
expansion of the theory in terms of the energy, {\it i.e.}, each
derivative acting on a massless field will correspond to a factor
of (${\cal E}$). The basic Einstein-Hilbert scale will correspond
to the (${\cal E}^2\sim
\partial^2 g$) contribution, while higher order effective
contributions will be of order (${\cal E}^4\sim\partial^4 g$),
(${\cal E}^6\sim \partial^6 g$), $\ldots$ etc.

The idea of making a large-$D$ expansion in gravity is somewhat
similar to the large-$N$ Yang-Mills planar diagram limit
considered by 't Hooft ref.~\cite{tHooft}. In large-$N$ gauge theories,
one expands in the internal symmetry index ($N$) of the gauge
group, {\it e.g.}, ($SU(N)$) or ($SO(N)$). The physical
interpretations of the two expansions are of course completely
dissimilar, {\it e.g.}, the number of dimensions in a theory
cannot really be compared to the internal symmetry index of a
gauge theory. A comparison of the Yang Mills large-$N$ limit (the
planar diagram limit), and the large-$D$ expansion in gravity is
however still interesting to perform, and as an example of a
similarity between the two expansions; the leading graphs in the
large-$D$ limit in gravity consist of a subset of planar
diagrams.

Higher dimensional models for gravity are well known from string
and supergravity theories. The mysterious $M$-theory has to exist
in an 11-dimensional space-time. So there are many good reasons to
believe that on fundamental scales, we might experience additional
space-time dimensions. Additional dimensions could be treated as
free or as compactified below the Planck scale, as in the case of
a Kaluza-Klein mechanism. One application of the large-$D$
expansion in gravity could be to approximate Greens functions at
finite dimension, {\it e.g.}, ($D=4$). The successful and various
uses of the planar diagram limit in Yang-Mills theory might
suggest other possible scenarios for the applicability of the
large-$D$ expansion in gravity. For example, is effective quantum
gravity renormalizable in its leading large-$D$ limit, {\it i.e.},
is it renormalizable in the same way as some non-renormalizable
theories are renormalizable in their planar diagram limit? It
could also be, that quantum gravity at large-$D$ is a completely
different theory, than Einstein gravity. A planar diagram limit is
essentially a string theory at large distances, {\it i.e.}, could
gravity at large-$D$ be interpreted as a large distance truncated
string limit?

We will here discuss the results of ref.~\cite{B6}, $i.e.$, we
will combine the successful effective field theory
approach which holds in any dimension, with the expansion of
gravity in the large-$D$ limit. The treatment will be mostly
conceptual, but we will also address some of the phenomenological
issues of this theory. The structure of this chapter will be as
follows. First we will briefly review the basic quantization of the
Einstein-Hilbert action, and then we will go on with the large-$D$
behavior of gravity, {\it i.e.}, we will show how to derive a
consistent limit for the leading graphs. The effective extension
of the theory will next be taken up in the large-$D$ framework,
and we will especially focus on the implications of the effective
extension of the theory in the large-$D$ limit. The
$D$-dimensional space-time integrals will also be looked
upon, and we will make a conceptual comparison of the large-$D$ in
gravity and the large-$N$ limit in gauge theory. Here we will
point out some similarities and some dissimilarities of the two
theories. We will also discuss some issues in the original paper
by Strominger, ref.~\cite{Strominger:1981jg}.

We will work in units $(\hbar=c=1)$, and metric (${\rm
diag}(\eta_{\mu\nu})=(1,-1,-1,-1,\dots,-1)$), {\it i.e.}, with
$(D-1)$ minus signs.

\section{Review of the large-$D$ limit of Einstein gravity}
In this section we will review the main idea of the large-$D$ expansion
of quantum gravity. We have already discussed how to quantize a gravitational
theory at ($D=4$) using the background field method. Here we will apply a slightly
different expansion of the metric field and we will keep an arbitrary number
of dimensions ($D$) at every step in the quantization. For completeness we quickly
summarize the essentials of the quantization here.

As is well known that the $D$-dimensional Einstein-Hilbert Lagrangian
has the form:
\begin{equation}
{\cal L} = \int d^Dx \sqrt{-g}\Big(\frac{2R}{\kappa^2}\Big).
\end{equation}
The notation in the above equation is identical to what previously have
been used. If we neglect the renormalization problems of this action, it is
possible to make a formal quantization of the theory using the path integral.
To make a diagram expansion of this theory we have to introduce a gauge breaking
term in the action to fix the propagator, and because gravity is a non-abelian
theory we have to introduce a proper Faddeev-Popov ref.~\cite{Faddeev} ghost action
as well.

The action for the quantized theory will consequently be:
\begin{equation}
{\cal L} = \int d^Dx \sqrt{-g}\Big(\frac{2R}{\kappa^2}+{\cal
L}_{\rm gauge\ fixing}+ {\cal L}_{\rm ghosts}\Big),
\end{equation}
where (${\cal L}_{\rm gauge\ fixing}$) is the gauge fixing term,
and (${\cal L}_{\rm ghosts}$) is the ghost contribution. In order
to generate vertex rules for this theory an expansion of the
action has to be carried out, and vertex rules have to be
extracted from the gauge fixed action. The vertices will depend on
the gauge choice and on how we define the gravitational field.

It is conventional to define:
\begin{equation}
g_{\mu\nu} \equiv \eta_{\mu\nu}+\kappa h_{\mu\nu},
\end{equation}
and work in harmonic gauge ($\partial^\lambda h_{\mu\lambda} =
\frac12\partial_\mu h^\lambda_\lambda$). For this gauge choice the
vertex rules for the 3- and 4-point Einstein vertices can be found
in refs.~\cite{Dewitt,Sannan:tz}.

Yet another possibility is to use the background field method as we
did previously:
\begin{equation}
g_{\mu\nu} \equiv \tilde g_{\mu\nu}+\kappa h_{\mu\nu},
\end{equation}
and expand the quantum fluctuation ($h_{\mu\nu}$) around an
external source, the background field ($\tilde g_{\mu\nu} \equiv
\eta_{\mu\nu} + \kappa H_{\mu\nu}$). In the background field
approach we have to differ between vertices with internal lines
(quantum lines: $\sim h_{\mu\nu}$) and external lines (external
sources: $\sim H_{\mu\nu}$). There are no real problems in using
either method; it is mostly a matter of notation and conventions.
Here we will concentrate our efforts on the conventional approach.

Another possibility is to define:
\begin{equation}
\tilde g^{\mu\nu} \equiv \sqrt{-g}g^{\mu\nu} = \eta^{\mu\nu} +
\kappa h^{\mu\nu},
\end{equation}
this field definition makes a transformation of the
Einstein-Hilbert action into the following form:
\begin{equation}\begin{split}
{\cal L} &= \int d^D x \frac{1}{2\kappa^2}(\tilde
g^{\rho\sigma}\tilde g_{\lambda\alpha}\tilde g_{\kappa\tau}\tilde
g^{\alpha\kappa}_{\ \ \ , \rho}\tilde g^{\lambda\tau}_{\ \ \ ,
\sigma} -\frac1{(D-2)}\tilde g^{\rho\sigma}\tilde
g_{\alpha\kappa}\tilde g^{\lambda\tau}\tilde g^{\lambda\tau}_{\ \
\ , \sigma} -2\tilde g_{\alpha\tau}\tilde g^{\alpha\kappa}_{\ \ \
, \rho}\tilde g^{\rho\tau}_{\ \ \ , \kappa}),
\end{split}\end{equation}
in, ($D=4$), this form of the Einstein-Hilbert action is known as
the Goldberg action ref.~\cite{Capper:pv}. In this description of the
Einstein-Hilbert action its dimensional dependence is more
obvious. It is however more cumbersome to work with in the course
of practical large-$D$ considerations.

In the standard expansion of the action the propagator for
gravitons take the form:
\begin{equation}
D_{\alpha\beta,\gamma\delta}(k) =
-\frac{i}{2}\frac{\Big[\eta_{\alpha\gamma}\eta_{\beta\delta}
+\eta_{\alpha\delta}\eta_{\beta\gamma}-\frac{2}{D-2}
\eta_{\alpha\beta}\eta_{\gamma\delta}\Big]}{k^2-i\epsilon}.
\end{equation}
The 3- and 4-point vertices for the standard expansion of the
Einstein-Hilbert action can be found in ref.~\cite{Dewitt,Sannan:tz} and
has the following form:
\begin{equation}\begin{split}
V^{(3)}_{\mu\alpha,\nu\beta,\sigma\gamma}(k_1,k_2,k_3) &= \kappa\,{\rm sym} \Big[
-\frac12P_3(k_1\cdot k_2\,\eta_{\mu\alpha}\eta_{\nu\beta}\eta_{\sigma\gamma})
 -\frac12P_6(k_{1\nu}k_{1\beta}\eta_{\mu\alpha}\eta_{\sigma\gamma})
\\&+ \frac12P_3(k_1\cdot
k_2\,\eta_{\mu\nu}\eta_{\alpha\beta}\eta_{\sigma\gamma})
+P_6(k_1\cdot
k_2\,\eta_{\mu\alpha}\eta_{\nu\sigma}\eta_{\beta\gamma})
+2P_3(k_{1\nu}k_{1\gamma}\eta_{\mu\alpha}\eta_{\beta\sigma})
\\&-P_3(k_{1\beta}k_{2\mu}\eta_{\alpha\nu}\eta_{\sigma\gamma})
+P_3(k_{1\sigma}k_{2\gamma}\eta_{\mu\nu}\eta_{\alpha\beta})
+P_6(k_{1\sigma}k_{1\gamma}\eta_{\mu\nu}\eta_{\alpha\beta})
\\&+2P_6(k_{1\nu}k_{2\gamma}\eta_{\beta\mu}\eta_{\alpha\sigma})
+2P_3(k_{1\nu}k_{2\mu}\eta_{\beta\sigma}\eta_{\gamma\alpha})
-2P_3(k_1\cdot
k_2\,\eta_{\alpha\nu}\eta_{\beta\sigma}\eta_{\gamma\mu})\Big],
\end{split}\end{equation}
and
\begin{equation}\begin{split}
V^{(4)}_{\mu\alpha,\nu\beta,\sigma\gamma,\rho\lambda}(k_1,k_2,k_3,k_4)&=\kappa^2\,{\rm
sym}\Big[ -\frac14P_6(k_1\cdot
k_2\,\eta_{\mu\alpha}\eta_{\nu\beta}\eta_{\sigma\gamma}\eta_{\rho\lambda})
-\frac14P_{12}(k_{1\nu}k_{1\beta}\eta_{\mu\alpha}\eta_{\sigma\gamma}\eta_{\rho\lambda})
\\&-\frac12P_6(k_{1\nu}k_{2\mu}\eta_{\alpha\beta}\eta_{\sigma\gamma}\eta_{\rho\lambda})
+\frac14P_6(k_1\cdot k_2\,
\eta_{\mu\nu}\eta_{\alpha\beta}\eta_{\sigma\gamma}\eta_{\rho\lambda})\\&
+\frac12P_6(k_1\cdot k_2\,
\eta_{\mu\alpha}\eta_{\nu\beta}\eta_{\sigma\rho}\eta_{\gamma\lambda})
+\frac12P_{12}(k_{1\nu}k_{1\beta}\eta_{\mu\alpha}\eta_{\sigma\rho}\eta_{\gamma\lambda})\\&
+P_6(k_{1\nu}k_{2\mu}\eta_{\alpha\beta}\eta_{\sigma\rho}\eta_{\gamma\lambda})
-\frac12P_6(k_1\cdot
k_2\,\eta_{\mu\nu}\eta_{\alpha\beta}\eta_{\sigma\rho}\eta_{\gamma\lambda})\\&
+\frac12P_{24}(k_1\cdot
k_2\,\eta_{\mu\alpha}\eta_{\nu\sigma}\eta_{\beta\gamma}\eta_{\rho\lambda})
+\frac12P_{24}(k_{1\nu}k_{1\beta}\eta_{\mu\sigma}\eta_{\alpha\gamma}\eta_{\rho\lambda}\\&
+\frac12P_{12}(k_{1\sigma}k_{2\gamma}\eta_{\mu\nu}\eta_{\alpha\beta}\eta_{\rho\lambda})
+P_{24}(k_{1\nu}k_{2\sigma}\eta_{\beta\mu}\eta_{\alpha\gamma}\eta_{\rho\lambda})\\&
-P_{12}(k_1\cdot
k_2\,\eta_{\alpha\nu}\eta_{\beta\sigma}\eta_{\gamma\mu}\eta_{\rho\lambda})
+P_{12}(k_{1\nu}k_{2\mu}\eta_{\beta\sigma}\eta_{\gamma\alpha}\eta_{\rho\lambda})\\&
+P_{12}(k_{1\nu}k_{1\sigma}\eta_{\beta\gamma}\eta_{\mu\alpha}\eta_{\rho\lambda})
-P_{24}(k_1\cdot
k_2\,\eta_{\mu\alpha}\eta_{\beta\sigma}\eta_{\gamma\rho}\eta_{\lambda\nu})\\&
-2P_{12}(k_{1\nu}k_{1\beta}\eta_{\alpha\sigma}\eta_{\gamma\rho}\eta_{\lambda\mu})
-2P_{12}(k_{1\sigma}k_{2\gamma}\eta_{\alpha\rho}\eta_{\lambda\nu}\eta_{\beta\mu})\\&
-2P_{24}(k_{1\nu}k_{2\sigma}\eta_{\beta\rho}\eta_{\lambda\mu}\eta_{\alpha\gamma})
-2P_{12}(k_{1\sigma}k_{2\rho}\eta_{\gamma\nu}\eta_{\beta\mu}\eta_{\alpha\lambda})\\&
+2P_{6}(k_1\cdot
k_2\,\eta_{\alpha\sigma}\eta_{\gamma\nu}\eta_{\beta\rho}\eta_{\lambda\mu})
-2P_{12}(k_{1\nu}k_{1\sigma}\eta_{\mu\alpha}\eta_{\beta\rho}\eta_{\lambda\gamma})\\&
-P_{12}(k_1\cdot
k_2\,\eta_{\mu\sigma}\eta_{\alpha\gamma}\eta_{\nu\rho}\eta_{\beta\lambda})
-2P_{12}(k_{1\nu}k_{1\sigma}\eta_{\beta\gamma}\eta_{\mu\rho}\eta_{\alpha\lambda}\\&
-P_{12}(k_{1\sigma}k_{2\rho}\eta_{\gamma\lambda}\eta_{\mu\nu}\eta_{\alpha\beta})
-2P_{24}(k_{1\nu}k_{2\sigma}\eta_{\beta\mu}\eta_{\alpha\rho}\eta_{\lambda\gamma})\\&
-2P_{12}(k_{1\nu}k_{2\mu}\eta_{\beta\sigma}\eta_{\gamma\rho}\eta_{\lambda\alpha})
+4P_{6}(k_1\cdot
k_2\,\eta_{\alpha\nu}\eta_{\beta\sigma}\eta_{\gamma\rho}\eta_{\lambda\mu})
\Big],
\end{split}\end{equation}
in the above two expressions, '${\rm sym}$', means that each pair
of indices: ($\mu\alpha$), ($\nu\beta$), $\ldots$ will have to be
symmetrized. The momenta factors: ($k_1$, $k_2$, $\ldots$) are
associated with the index pairs: ($\mu\alpha$, $\nu\beta$,
$\ldots$) correspondingly. The symbol: ($P_{\#}$) means that a
$\#$-permutation of indices and corresponding momenta has to
carried out for this particular term. As seen the algebraic
structures of the 3- and 4-point vertices are already rather
involved and complicated. 5-point vertices will not be considered
explicitly here.

The explicit prefactors of the terms in the vertices will not be
essential in this treatment. The various algebraic structures
which constitute the vertices will be more important. Different
algebraic terms in the vertex factors will generate dissimilar
traces in the final diagrams. It is useful to adapt an index line
notation for the vertex structure similar to that used in
large-$N$ gauge theory.

\begin{figure}[h]
\begin{tabular}{llll}\vspace{0.05cm}
$\left(\parbox{1cm}{\includegraphics[height=0.8cm]{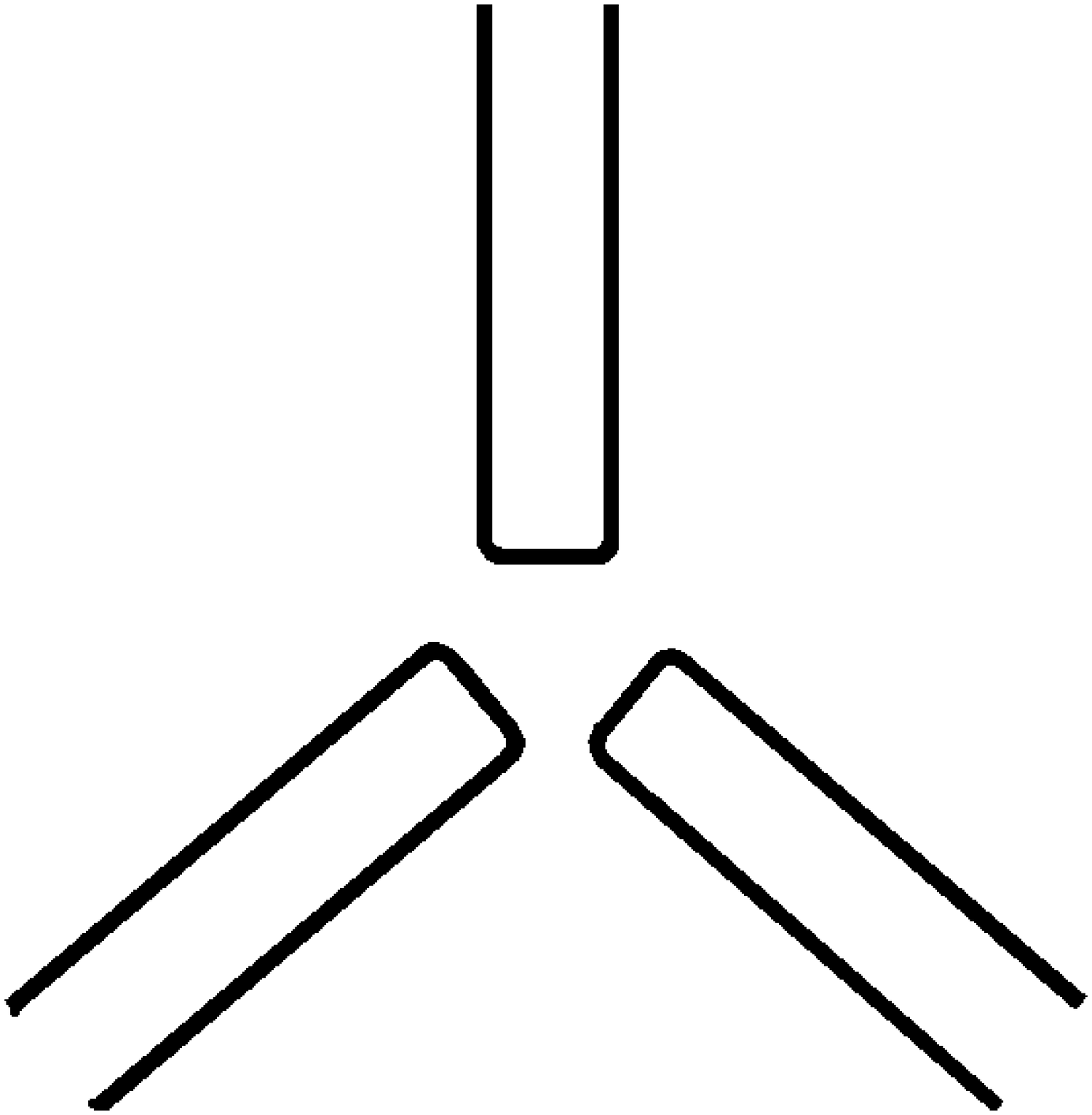}}\right)_{\rm
3A}$&{$\sim {\rm sym}[-\frac12P_3(k_1\cdot
k_2\eta_{\mu\alpha}\eta_{\nu\beta}\eta_{\sigma\gamma})],$}
&\\
$\left(\parbox{1cm}{\includegraphics[height=0.8cm]{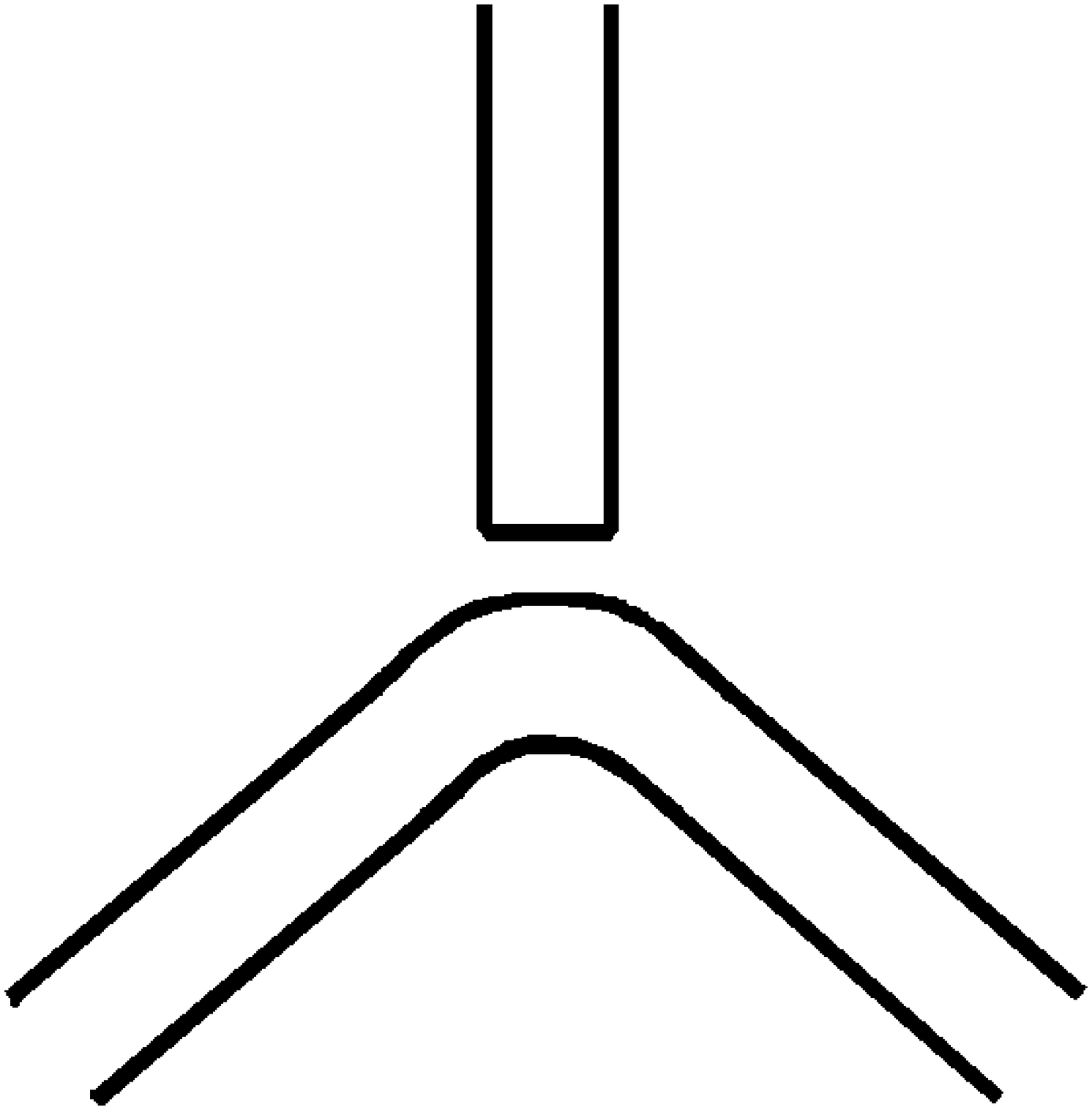}}\right)_{\rm
3B}$&$\sim {\rm sym}[ \frac12P_3(k_1\cdot k_2
\eta_{\mu\nu}\eta_{\alpha\beta}\eta_{\sigma\gamma})+P_6(k_1\cdot
k_2 \eta_{\mu\alpha}\eta_{\nu\sigma}\eta_{\beta\gamma})],$
\\ \vspace{0.05cm}
$\left(\parbox{1cm}{\includegraphics[height=0.8cm]{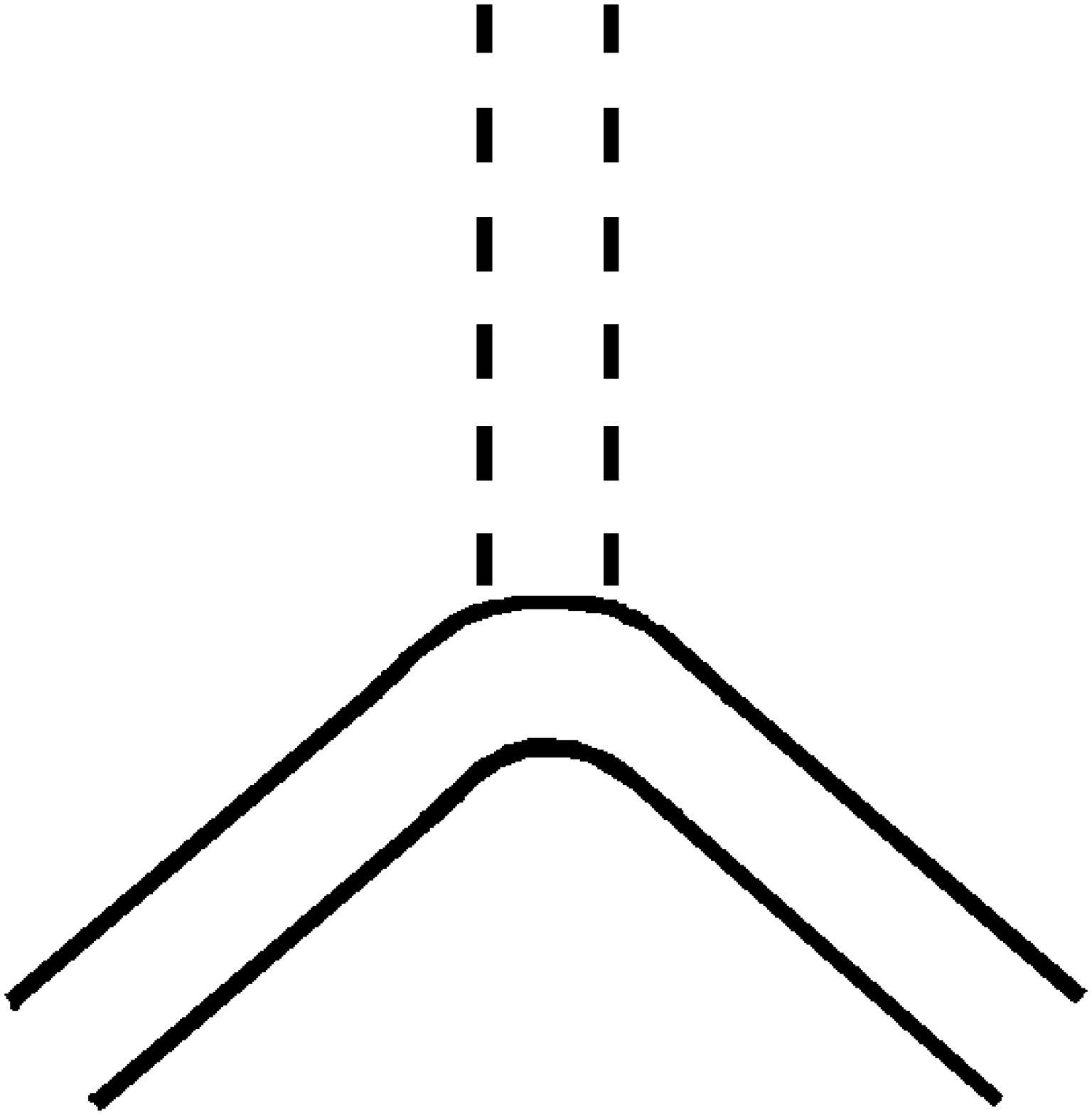}}\right)_{\rm
3C}$&$\sim {\rm sym}[
P_3(k_{1\sigma}k_{2\gamma}\eta_{\mu\nu}\eta_{\alpha\beta})],$
&\\
$\left(\parbox{1cm}{\includegraphics[height=0.8cm]{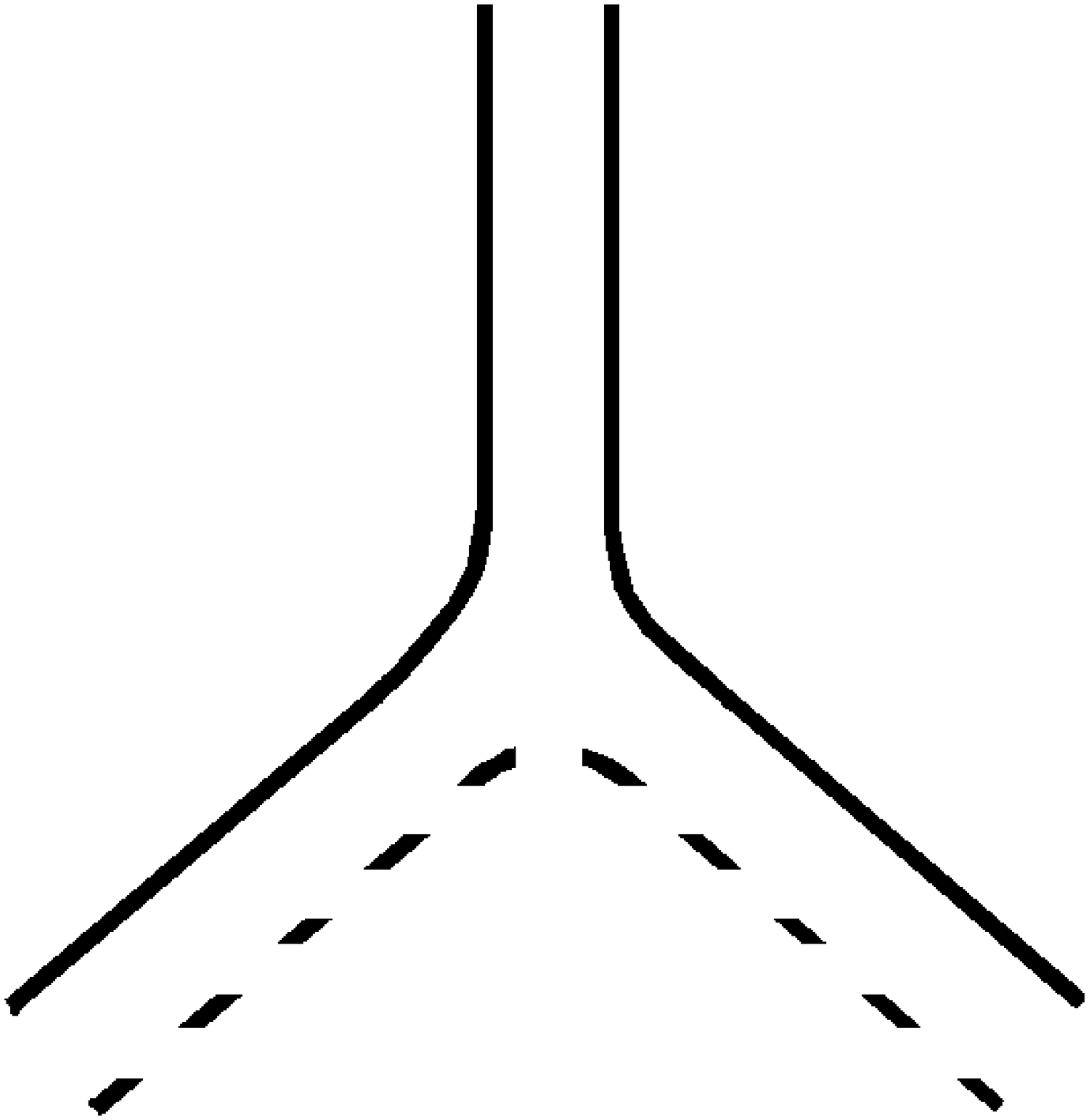}}\right)_{\rm
3D}$&$\sim {\rm sym}[
2P_6(k_{1\nu}k_{2\gamma}\eta_{\beta\mu}\eta_{\alpha\sigma})+2P_3(k_{1\nu}k_{2\mu}\eta_{\beta\sigma}\eta_{\gamma\alpha})],$\\
\vspace{0.1cm}
$\left(\parbox{1cm}{\includegraphics[height=0.8cm]{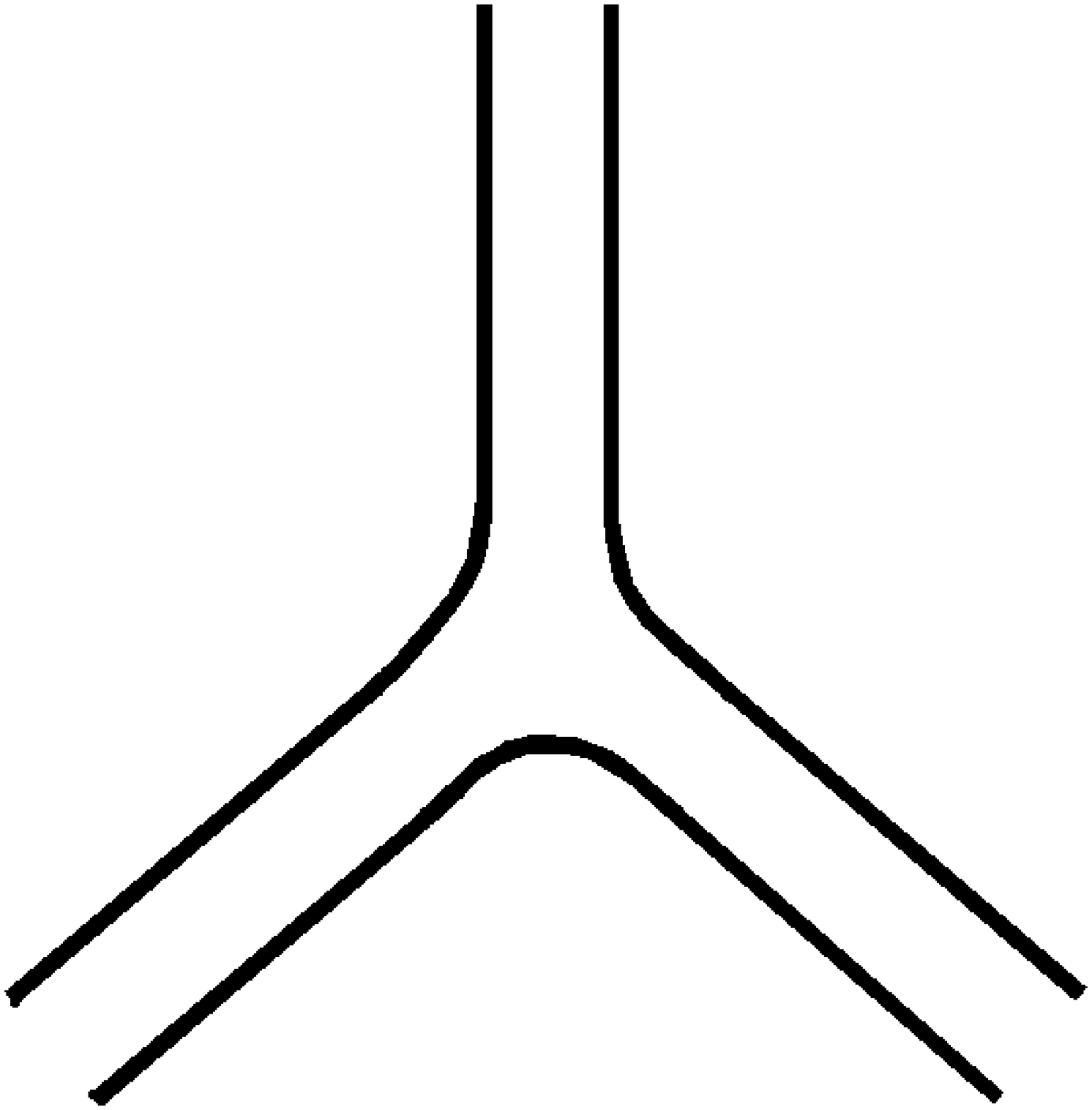}}\right)_{\rm
3E}$&$\sim{\rm sym}[ -2P_3(k_1\cdot
k_2\eta_{\alpha\nu}\eta_{\beta\sigma}\eta_{\gamma\mu})],$
&\\
$\left(\parbox{1cm}{\includegraphics[height=0.8cm]{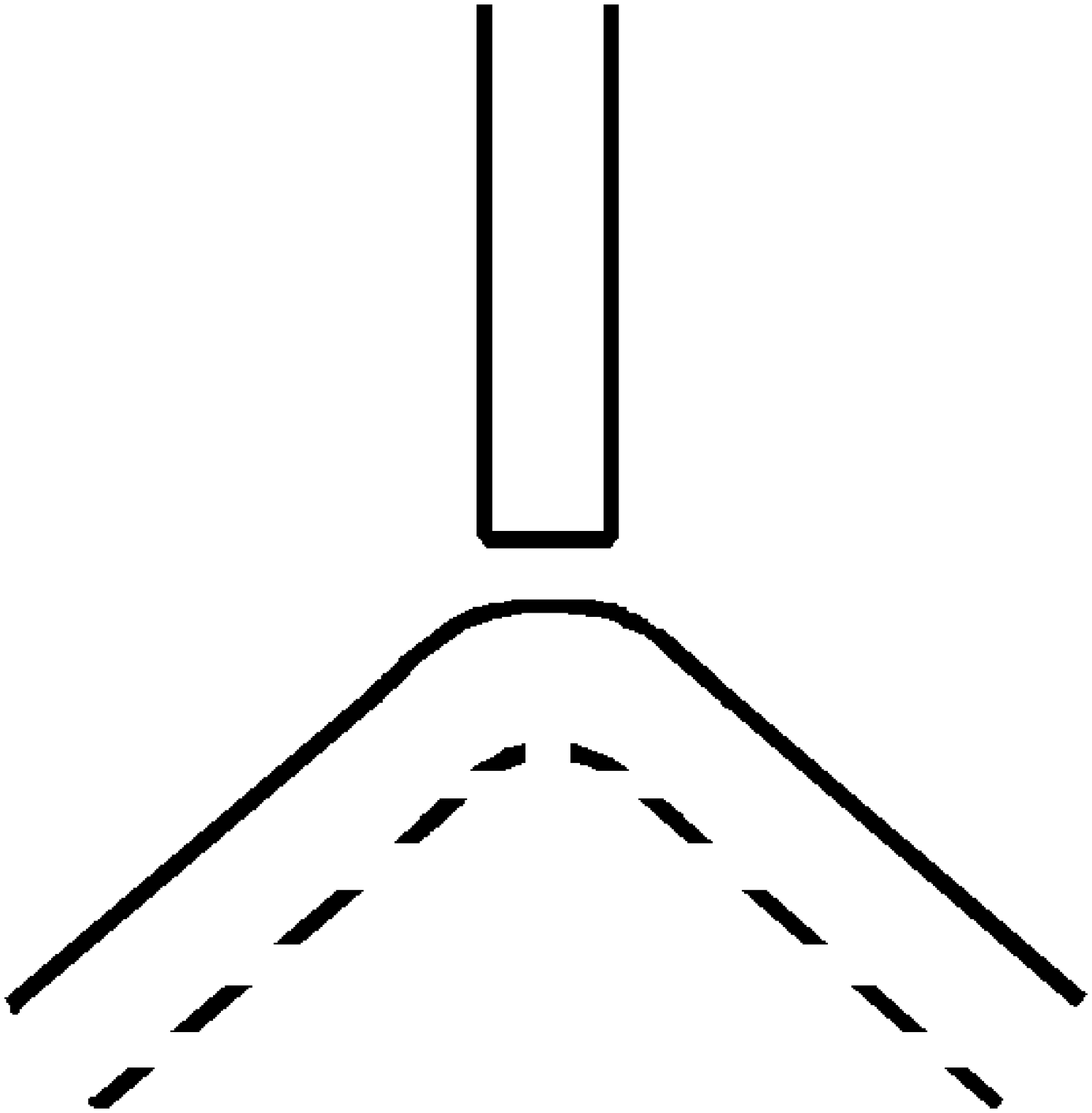}}\right)_{\rm
3F}$&$\sim {\rm
sym}[2P_3(k_{1\nu}k_{1\gamma}\eta_{\mu\alpha}\eta_{\beta\sigma})
-P_3(k_{1\beta}k_{2\mu}\eta_{\alpha\nu}\eta_{\sigma\gamma}) ],$\\
\vspace{0.05cm}
$\left(\parbox{1cm}{\includegraphics[height=0.8cm]{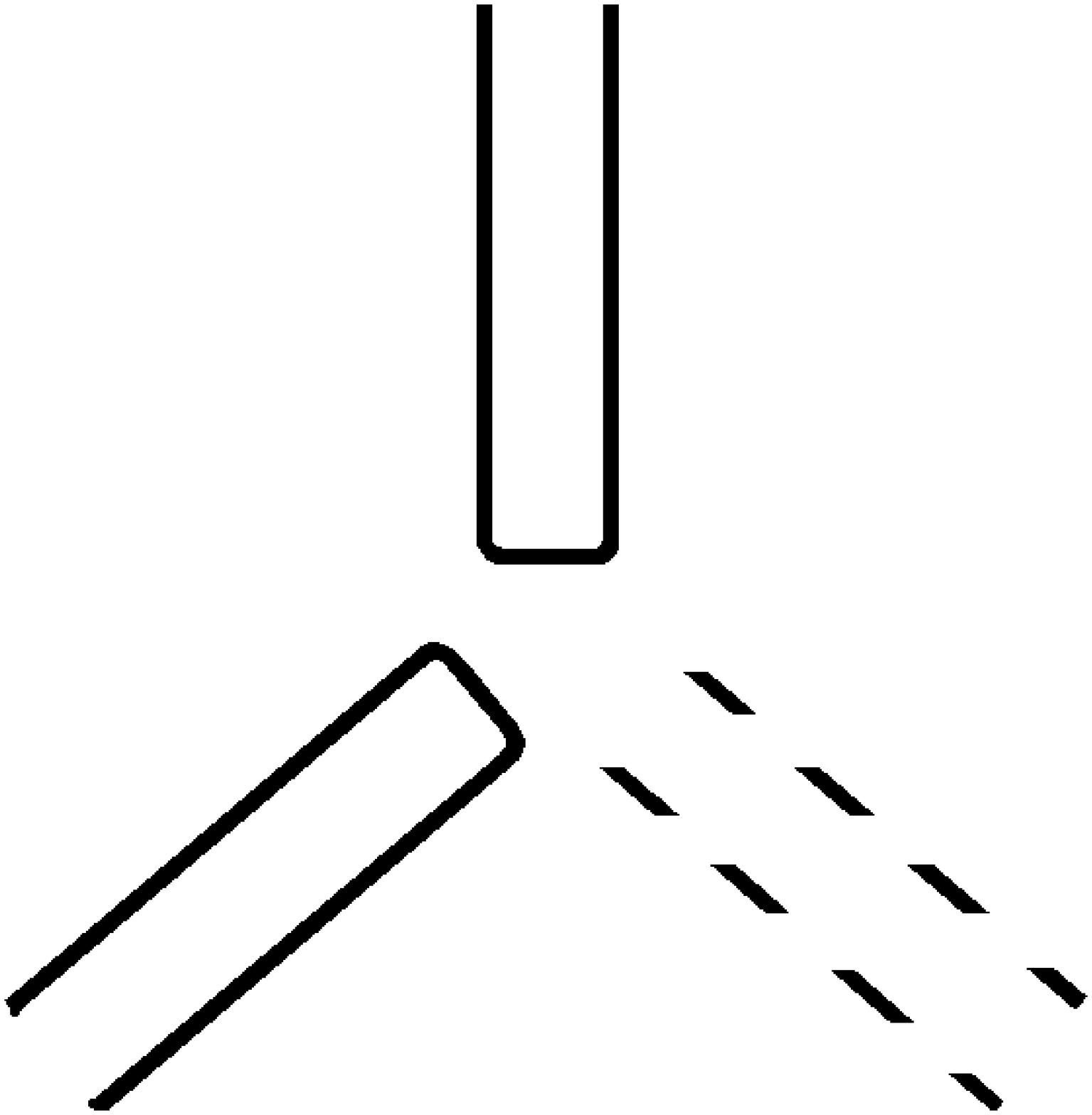}}\right)_{\rm
3G}$&$\sim {\rm
sym}[-\frac12P_6(k_{1\nu}k_{1\beta}\eta_{\mu\alpha}\eta_{\sigma\gamma})].$&
\end{tabular}
\caption{A graphical representation of the various terms in the
3-point vertex factor. A dashed line represents a contraction of an
index with a momentum line. A full line means a contraction of two
index lines. The above vertex notation for the indices and momenta
also apply here.\label{3vert}}
\end{figure}

In this notation we can represent the different algebraic terms of
the 3-point vertex (see figure \ref{3vert}).

\begin{figure}[h]
\begin{tabular}{ll}\vspace{0.05cm}
$\left(\parbox{1cm}{\includegraphics[height=0.8cm]{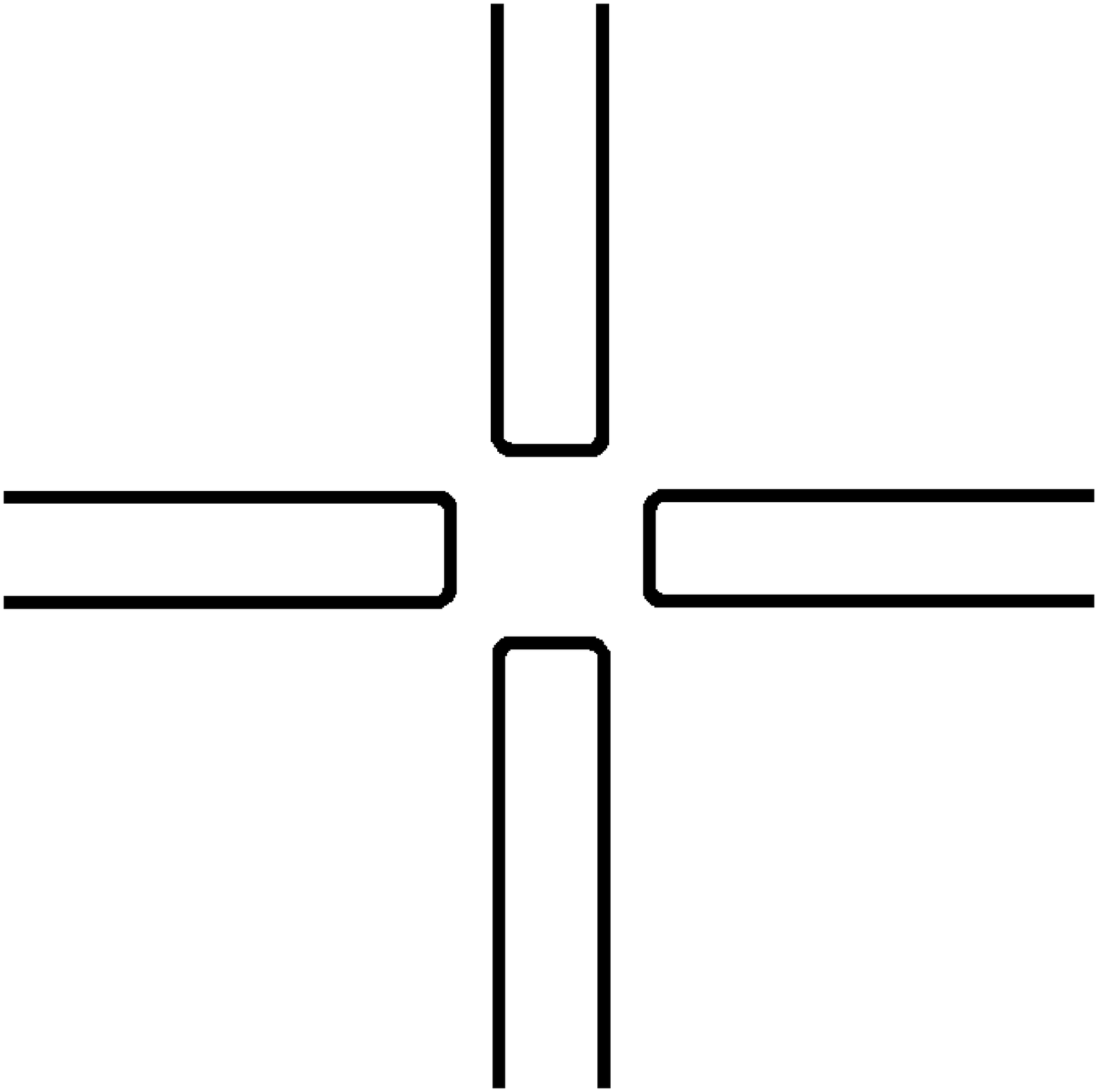}}\right)_{\rm
4A}$&$\sim {\rm sym} [-\frac14P_6(k_1\cdot k_2
\eta_{\mu\alpha}\eta_{\nu\beta}\eta_{\sigma\gamma}\eta_{\rho\lambda})],$\\
\vspace{0.05cm}
$\left(\parbox{1cm}{\includegraphics[height=0.8cm]{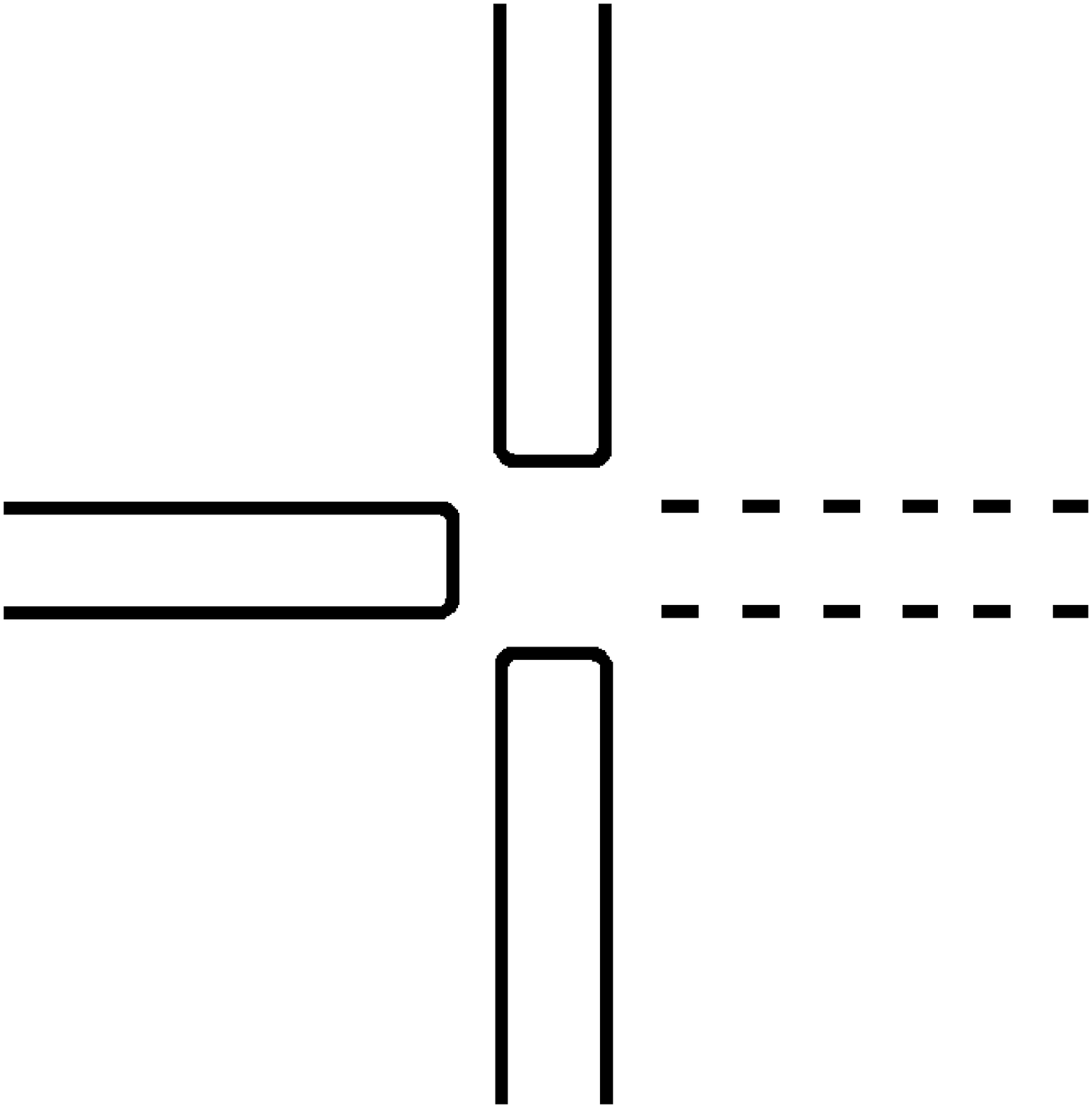}}\right)_{\rm
4B}$&$\sim {\rm sym}
[-\frac14P_{12}(k_{1\nu}k_{1\beta}\eta_{\mu\alpha}\eta_{\sigma\gamma}\eta_{\rho\lambda})],$\\
\vspace{0.05cm}
$\left(\parbox{1cm}{\includegraphics[height=0.8cm]{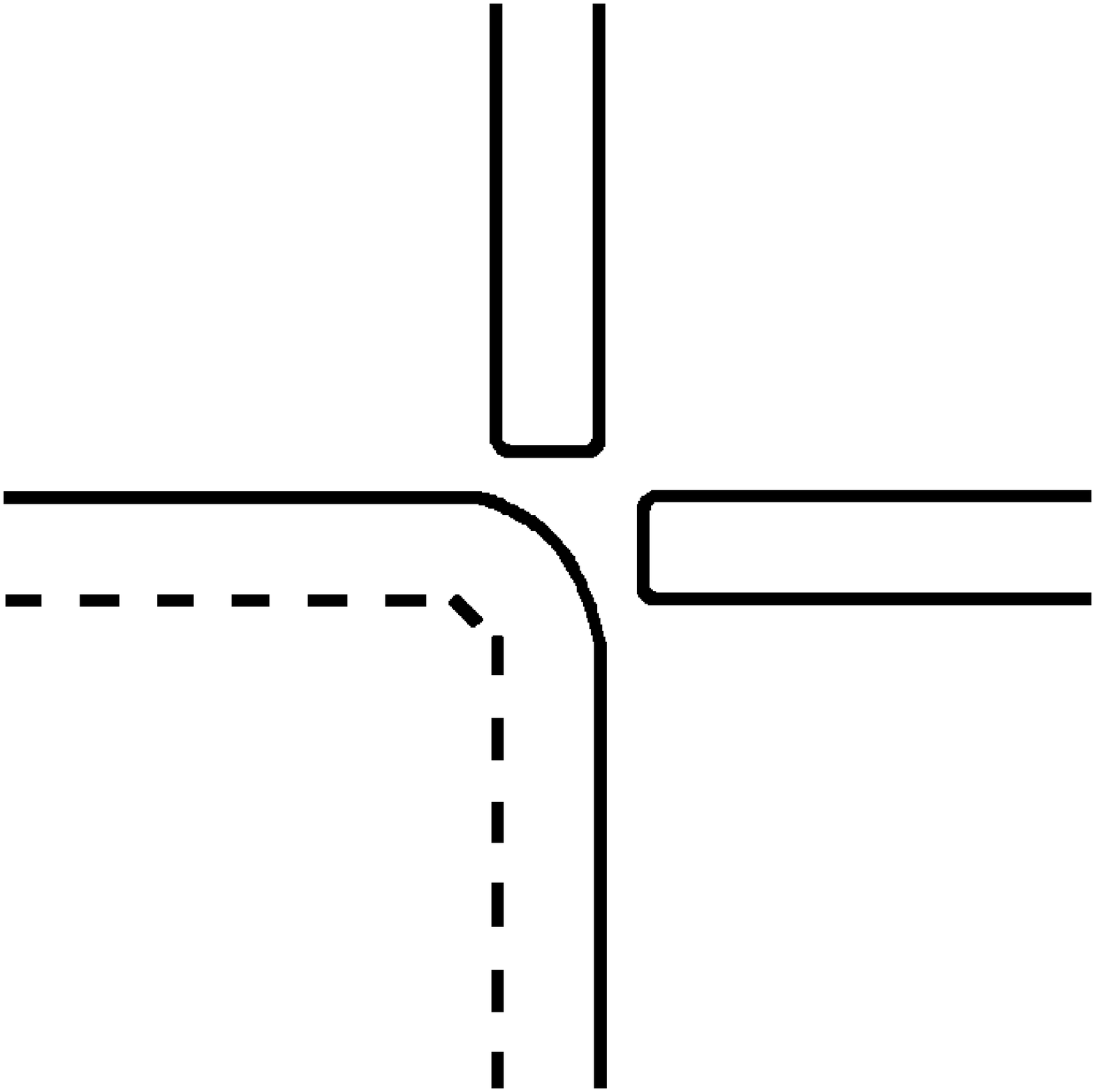}}\right)_{\rm
4C}$&$\sim {\rm sym}
[-\frac12P_6(k_{1\nu}k_{2\mu}\eta_{\alpha\beta}\eta_{\sigma\gamma}\eta_{\rho\lambda})],$
\\ \vspace{0.05cm}
$\left(\parbox{1cm}{\includegraphics[height=0.8cm]{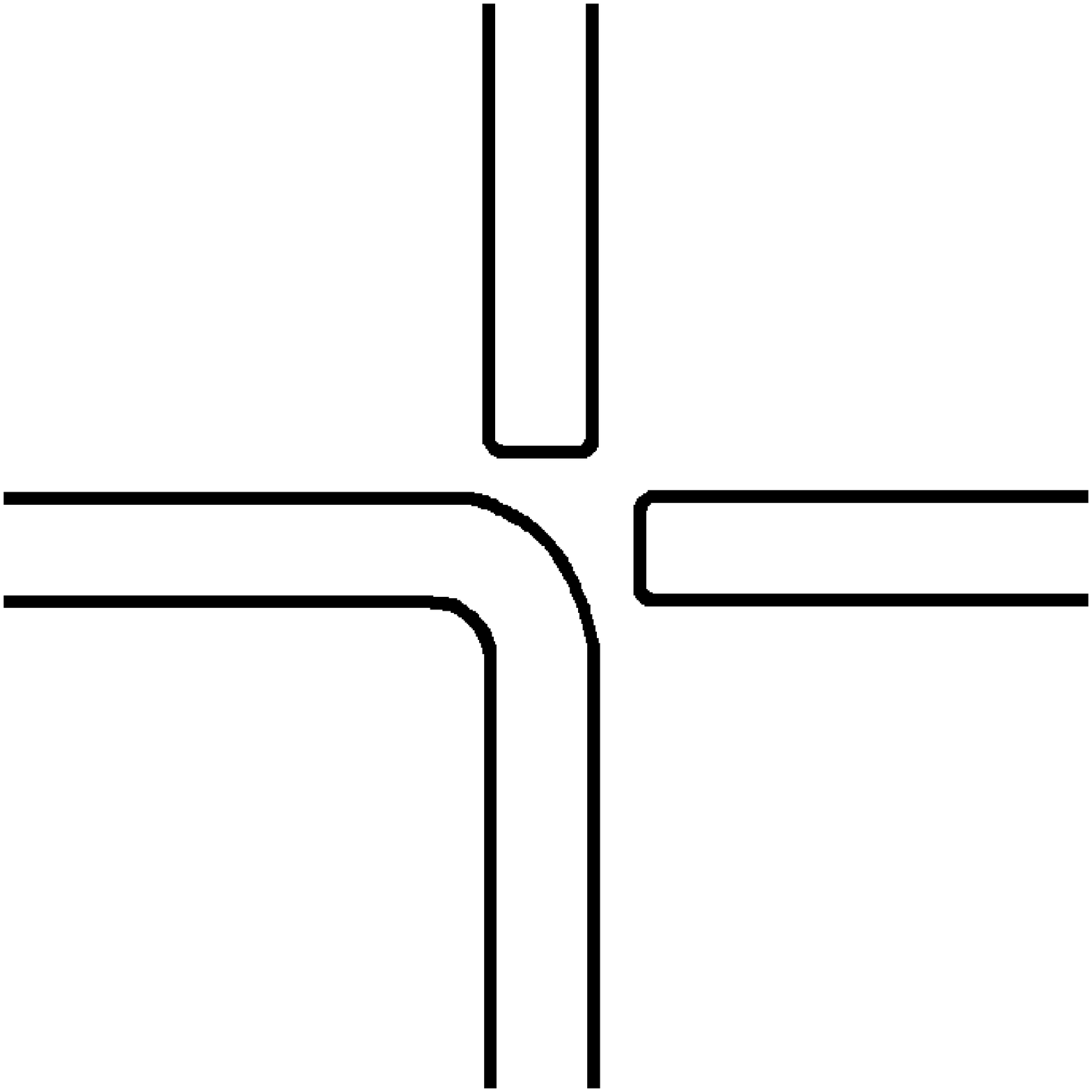}}\right)_{\rm
4D}$&$\sim {\rm sym} [\frac14P_6(k_1\cdot k_2
\eta_{\mu\nu}\eta_{\alpha\beta}\eta_{\sigma\gamma}\eta_{\rho\lambda})
+\frac12P_6(k_1\cdot k_2
\eta_{\mu\alpha}\eta_{\nu\beta}\eta_{\sigma\rho}\eta_{\gamma\lambda})
$\\ &$+\frac12P_{24}(k_1\cdot k_2
\eta_{\mu\alpha}\eta_{\nu\sigma}\eta_{\beta\gamma}\eta_{\rho\lambda})],$\\
\vspace{0.05cm}
$\left(\parbox{1cm}{\includegraphics[height=0.8cm]{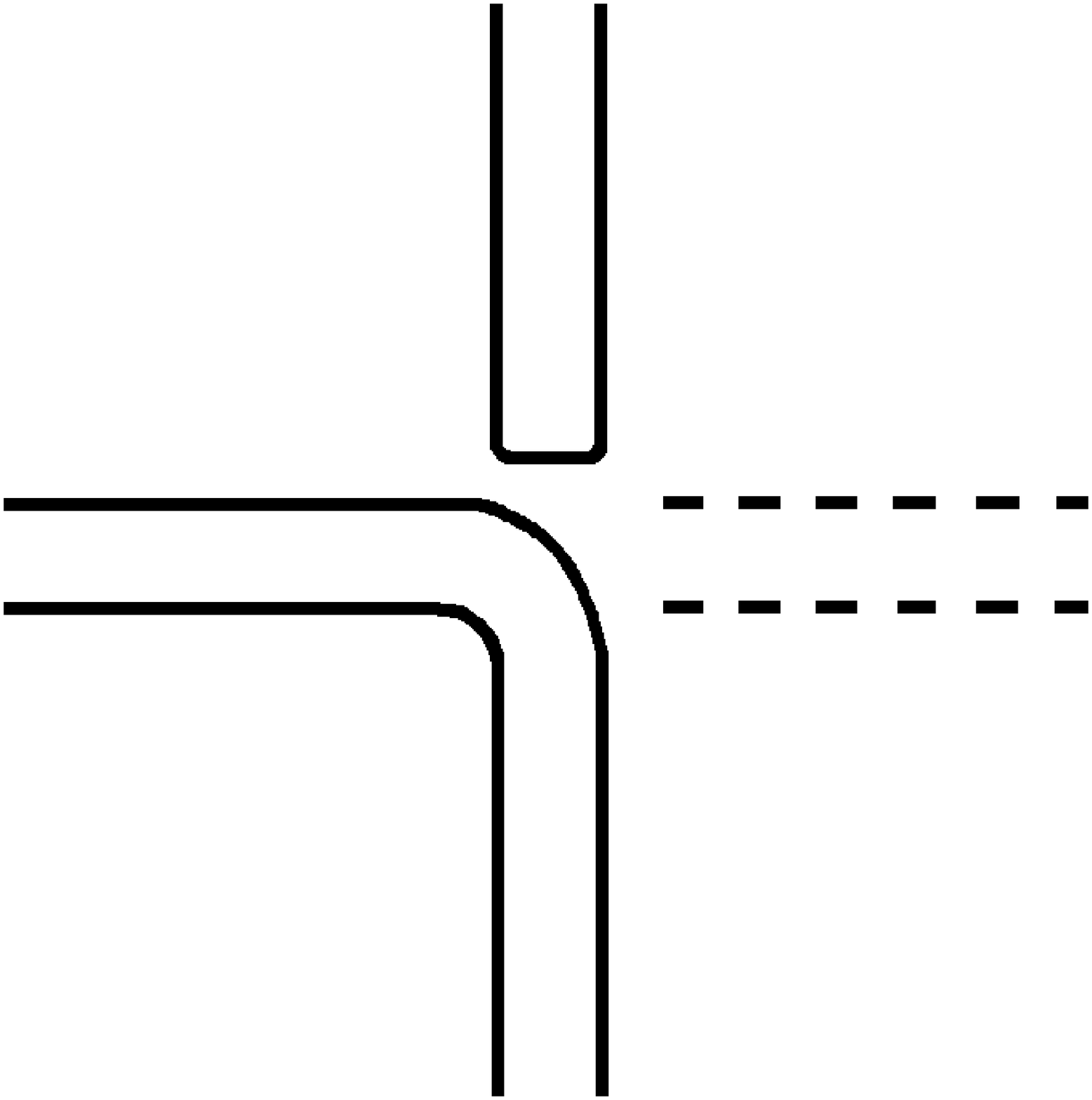}}\right)_{\rm
4E}$&$\sim {\rm sym}
[\frac12P_{12}(k_{1\nu}k_{1\beta}\eta_{\mu\alpha}\eta_{\sigma\rho}\eta_{\gamma\lambda})
+\frac12P_{24}(k_{1\nu}k_{1\beta}\eta_{\mu\sigma}\eta_{\alpha\gamma}\eta_{\rho\lambda})
$\\&$+\frac12P_{12}(k_{1\sigma}k_{2\gamma}\eta_{\mu\nu}\eta_{\alpha\beta}\eta_{\rho\lambda})],$
\\ \vspace{0.05cm}
$\left(\parbox{1cm}{\includegraphics[height=0.8cm]{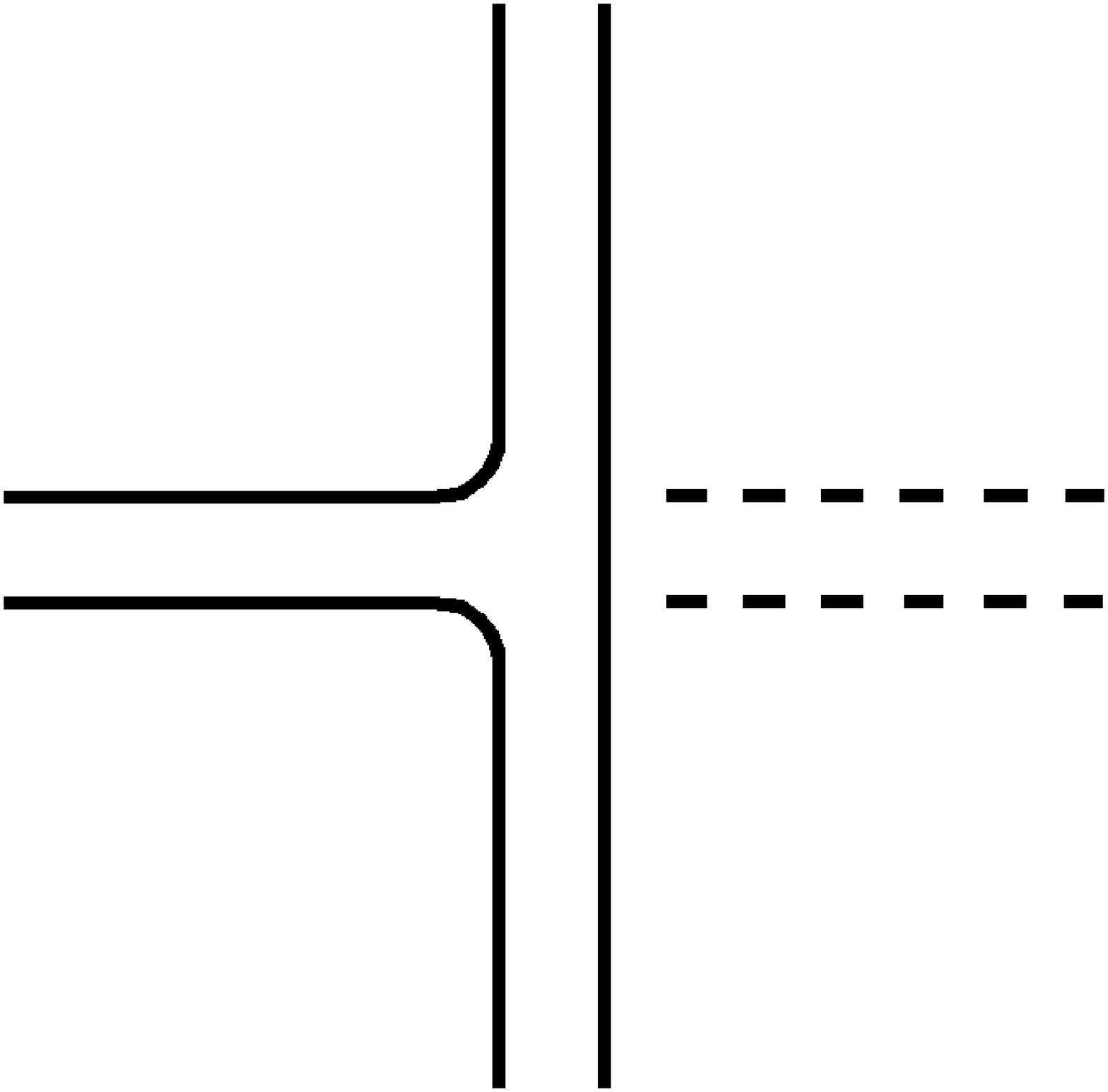}}\right)_{\rm
4F}$&$\sim {\rm sym}
[-2P_{12}(k_{1\nu}k_{1\beta}\eta_{\alpha\sigma}\eta_{\gamma\rho}\eta_{\lambda\mu})
-2P_{12}(k_{1\sigma}k_{2\gamma}\eta_{\alpha\rho}\eta_{\lambda\nu}\eta_{\beta\mu})],$\\
\vspace{0.05cm}
$\left(\parbox{1cm}{\includegraphics[height=0.8cm]{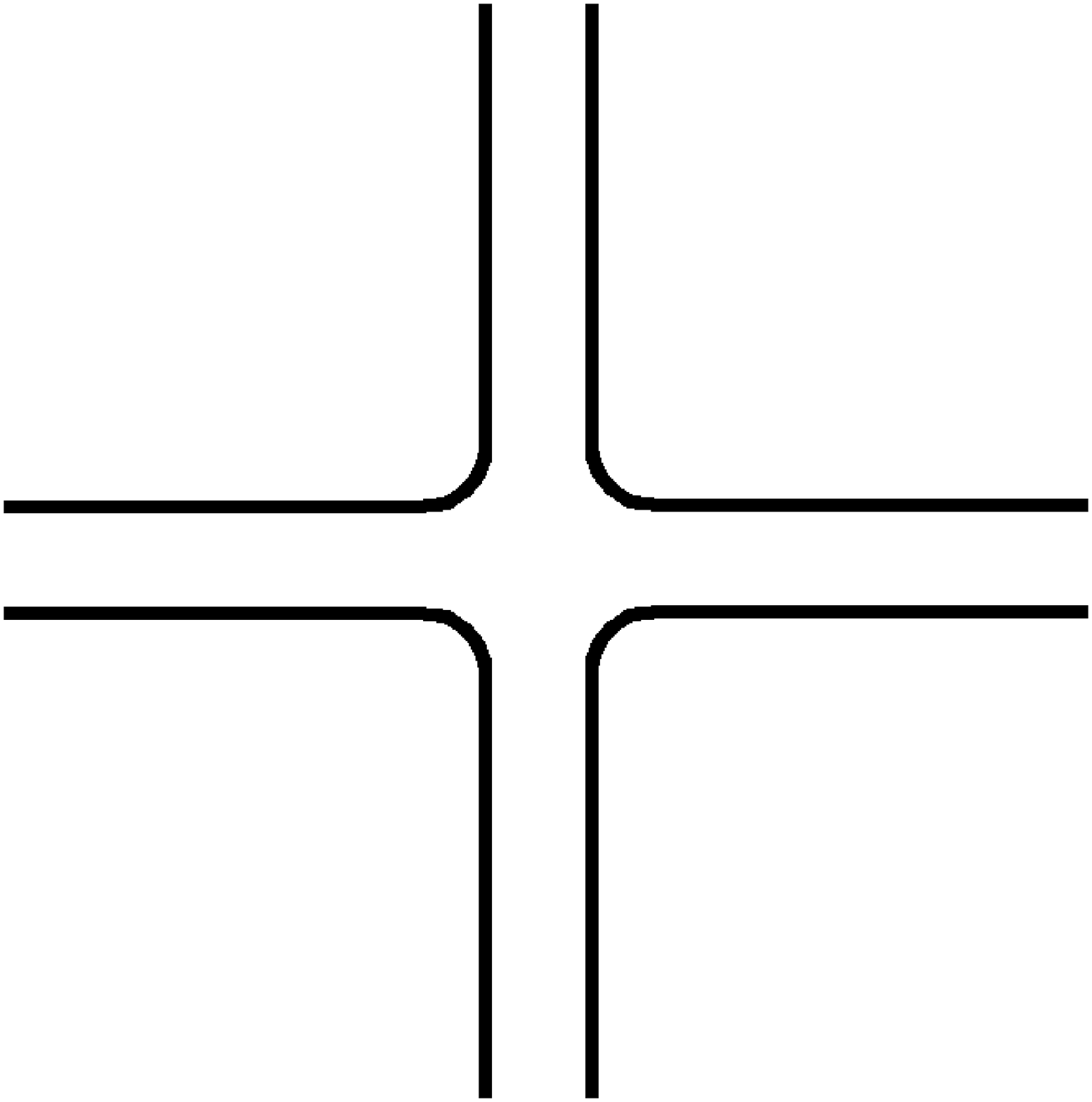}}\right)_{\rm
4G}$&$\sim {\rm sym} [2P_6(k_1\cdot k_2
\eta_{\alpha\sigma}\eta_{\gamma\nu}\eta_{\beta\rho}\eta_{\lambda\mu})
+4P_6(k_1\cdot k_2
\eta_{\alpha\nu}\eta_{\beta\sigma}\eta_{\gamma\rho}\eta_{\lambda\mu})],$
\\ \vspace{0.05cm}
$\left(\parbox{1cm}{\includegraphics[height=0.8cm]{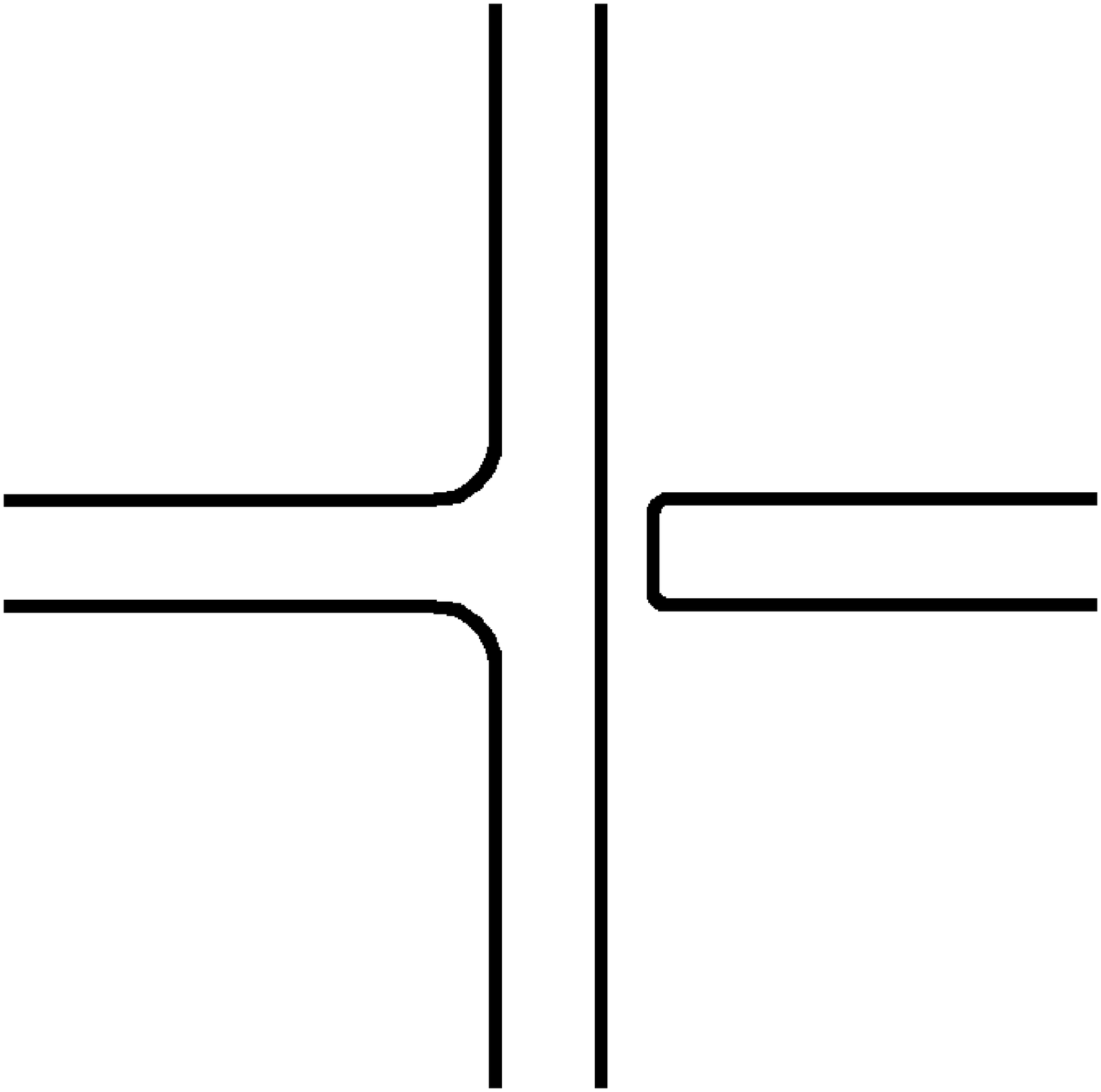}}\right)_{\rm
4H}$&$\sim {\rm sym} [-P_{12}(k_1\cdot k_2
\eta_{\alpha\nu}\eta_{\beta\sigma}\eta_{\gamma\mu}\eta_{\rho\lambda})
-P_{24}(k_1\cdot k_2
\eta_{\mu\alpha}\eta_{\beta\sigma}\eta_{\gamma\rho}\eta_{\lambda\nu})],$\\
\vspace{0.05cm}
$\left(\parbox{1cm}{\includegraphics[height=0.8cm]{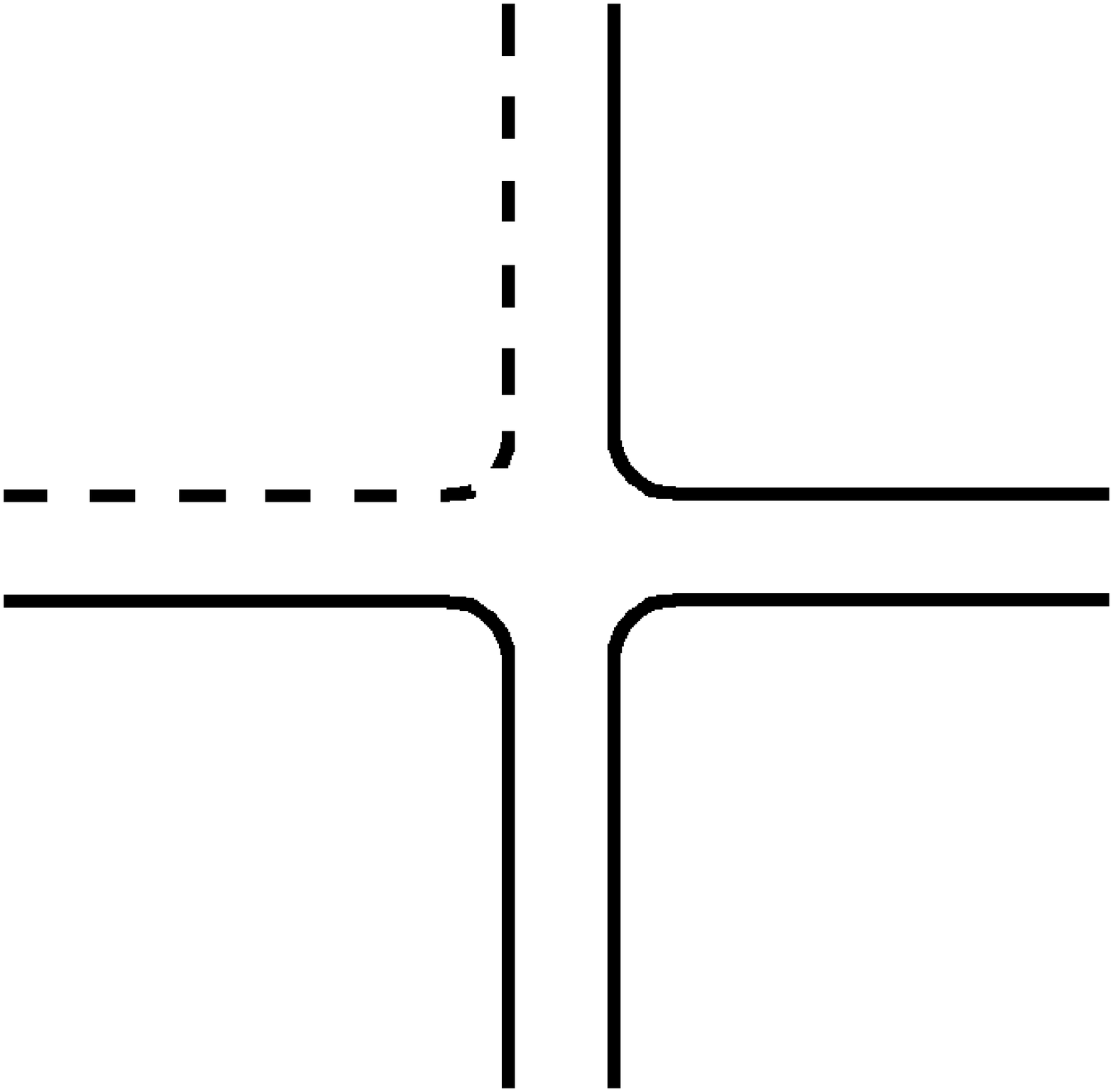}}\right)_{\rm
4I}$&$\sim {\rm sym}
[-2P_{24}(k_{1\nu}k_{2\sigma}\eta_{\beta\rho}\eta_{\lambda\mu}\eta_{\alpha\gamma})
-2P_{12}(k_{1\sigma}k_{2\rho}\eta_{\gamma\nu}\eta_{\beta\mu}\eta_{\alpha\lambda})
$\\&$-2P_{12}(k_{1\sigma}k_{2\rho}\eta_{\gamma\lambda}\eta_{\mu\nu}\eta_{\alpha\beta})
-2P_{24}(k_{1\nu}k_{2\sigma}\eta_{\beta\mu}\eta_{\alpha\rho}\eta_{\lambda\gamma})
-2P_{12}(k_{1\nu}k_{2\mu}\eta_{\beta\sigma}\eta_{\gamma\rho}\eta_{\lambda\alpha})],$\\
\vspace{0.05cm}
$\left(\parbox{1cm}{\includegraphics[height=0.8cm]{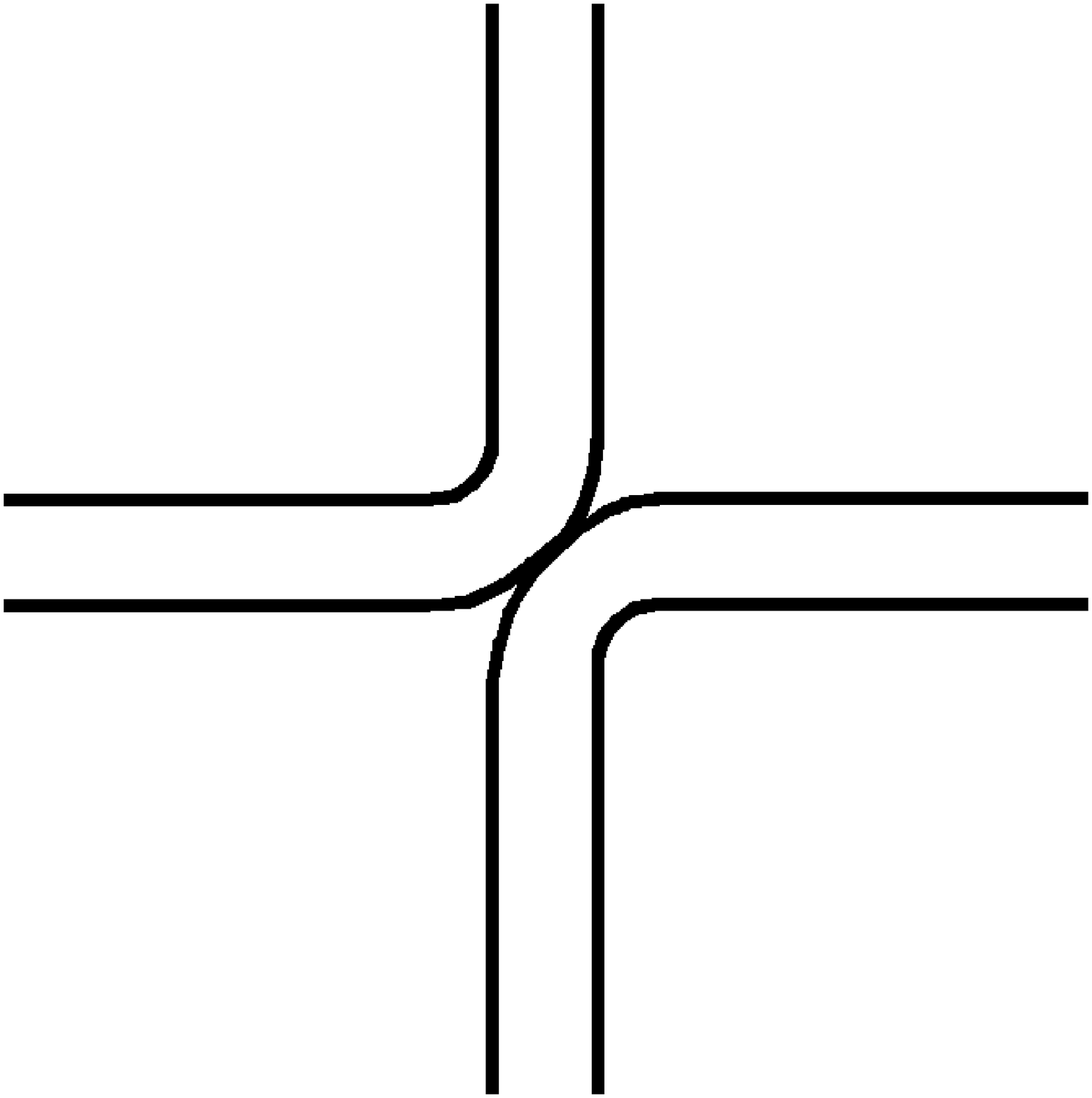}}\right)_{\rm
4J}$&$\sim {\rm sym} [-\frac12P_6(k_1\cdot k_2
\eta_{\mu\nu}\eta_{\alpha\beta}\eta_{\sigma\rho}\eta_{\gamma\lambda})
-P_{12}(k_1\cdot k_2
\eta_{\mu\sigma}\eta_{\alpha\gamma}\eta_{\nu\rho}\eta_{\beta\lambda})],$\\
\vspace{0.05cm}
$\left(\parbox{1cm}{\includegraphics[height=0.8cm]{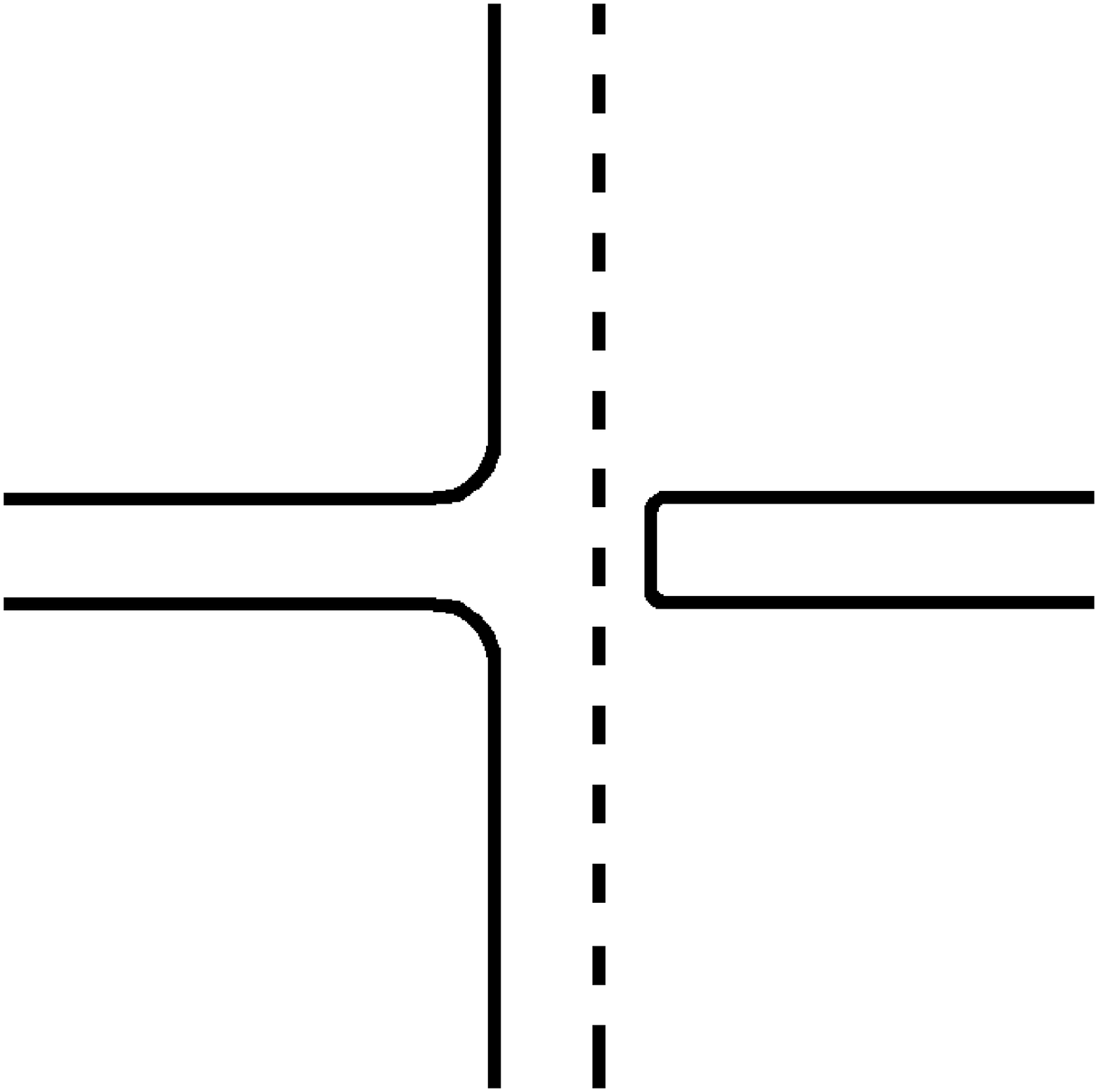}}\right)_{\rm
4K}$&$\sim {\rm sym}
[P_{24}(k_{1\nu}k_{2\sigma}\eta_{\beta\mu}\eta_{\alpha\gamma}\eta_{\rho\lambda})
+P_{12}(k_{1\nu}k_{2\mu}\eta_{\beta\sigma}\eta_{\gamma\alpha}\eta_{\rho\lambda})
$\\&$-2P_{12}(k_{1\nu}k_{1\sigma}\eta_{\mu\alpha}\eta_{\beta\rho}\eta_{\lambda\gamma})],$\\
\vspace{0.05cm}
$\left(\parbox{1cm}{\includegraphics[height=0.8cm]{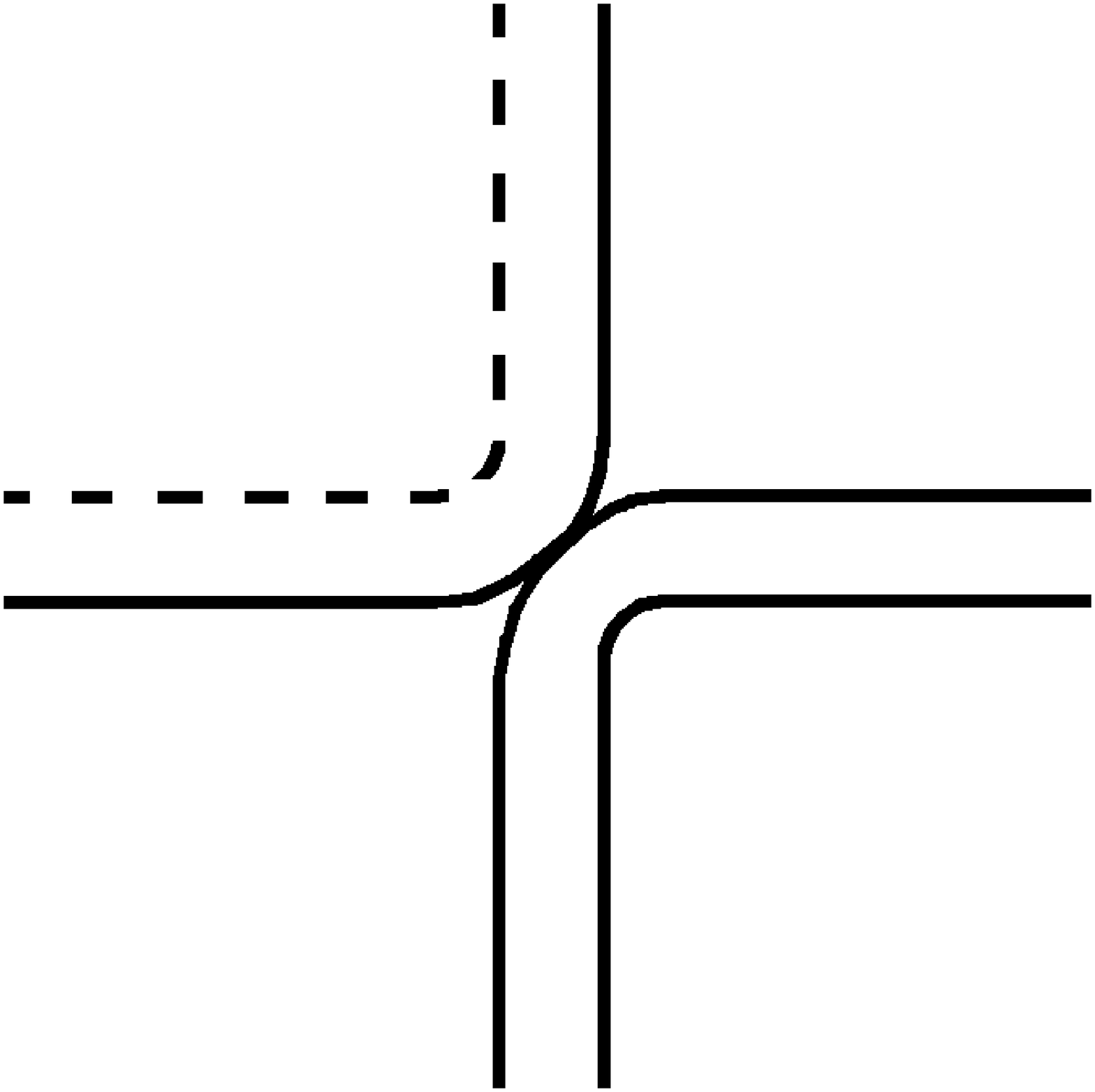}}\right)_{\rm
4L}$&$\sim {\rm sym}
[P_6(k_{1\nu}k_{2\mu}\eta_{\alpha\beta}\eta_{\sigma\rho}\eta_{\gamma\lambda}
-2P_{12}(k_{1\nu}k_{1\sigma}\eta_{\beta\gamma}\eta_{\mu\rho}\eta_{\alpha\lambda})
-P_{12}(k_{1\sigma}k_{2\rho}\eta_{\gamma\lambda}\eta_{\mu\nu}\eta_{\alpha\beta})].$
\end{tabular}
\caption{A graphical representation of the terms in the 4-point
vertex factor. The notation here is the same as the above for the
3-point vertex factor.\label{4-point}}
\end{figure}

For the 4-point vertex we can use a similar diagrammatic notation
(see figure \ref{4-point}).

There are only two sources of factors of dimension in a diagram. A
dimension factor can arise either from the algebraic contractions
in that particular diagram or from the space-time integrals. No
other possibilities exists. In order to arrive at the large-$D$
limit in gravity both sources of dimension factors have to be
considered and the leading contributions will have to be derived
in a systematic way.

Comparing to the case of the large-$N$ planar diagram expansion,
only the algebraic structure of the diagrams has to be considered.
The symmetry index ($N$) of the gauge group does not go into the
evaluation of the integrals. Concerns with the $D$-dimensional
integrals only arise in gravity. This is a great difference
between the two expansions.

A dimension factor will arise whenever there is a trace over a
tensor index in a diagram, {\it e.g.},
($\eta_{\mu\alpha}\eta^{\alpha\mu},\
\eta_{\mu\alpha}\eta^{\alpha\beta}
\eta_{\beta\gamma}\eta^{\gamma\mu},\ \ldots \sim D$). The diagrams
with the most traces will dominate in the large-$D$ limit in gravity.
Traces of momentum lines will never generate a factor of ($D$).

It is easy to be assured that some diagrams naturally will
generate more traces than other diagrams. The key result of
ref.~\cite{Strominger:1981jg} is that only a particular class of
diagrams will constitute the large-$D$ limit, and that only
certain contributions from these diagrams will be important there.
The large-$D$ limit of gravity will correspond to a truncation of
the Einstein theory of gravity, where in fact only a subpart,
namely the leading contributions of the graphs will dominate. We
will use the conventional expansion of the gravitational field,
and we will not separate conformal and traceless parts of the
metric tensor.

\begin{figure}[h]
\begin{center}
\includegraphics[height=0.5cm]{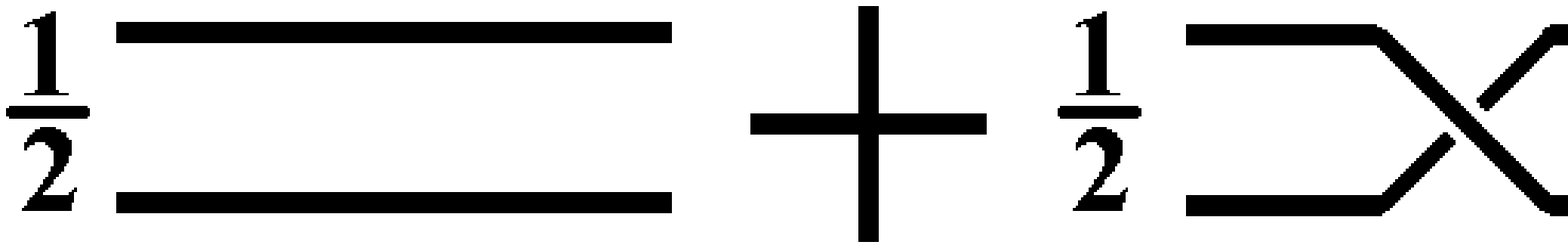}
\end{center}
 \caption{A graphical representation of the graviton
propagator, where a full line corresponds to a contraction of two
indices.\label{prop}}
\end{figure}

To justify this we begin by looking on how traces occur from contractions of the propagator.\\
We can graphically write the propagator in the way shown in figure
\ref{prop}.

\begin{figure}[h]
\begin{center}\parbox{7cm}{\includegraphics[height=2.2cm]{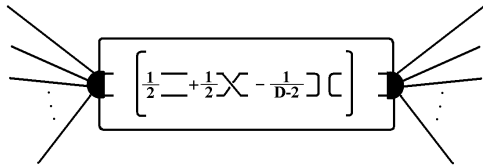}}\end{center}
\caption{A graphical representation of a propagator contraction in
an amplitude. The black half dots represent an arbitrary internal
structure for the amplitude, the full lines between the two half
dots are internal contractions of indices, the outgoing lines
represents index contractions with external sources.
\label{figprop1}}
\end{figure}

The only way we can generate index loops is through a propagator.
We can graphically represent a propagator contraction of two
indices in a particular amplitude in the following way (see figure
\ref{figprop1}).

\begin{figure}[h]
\parbox{8.7cm}{\includegraphics[height=2.2cm]{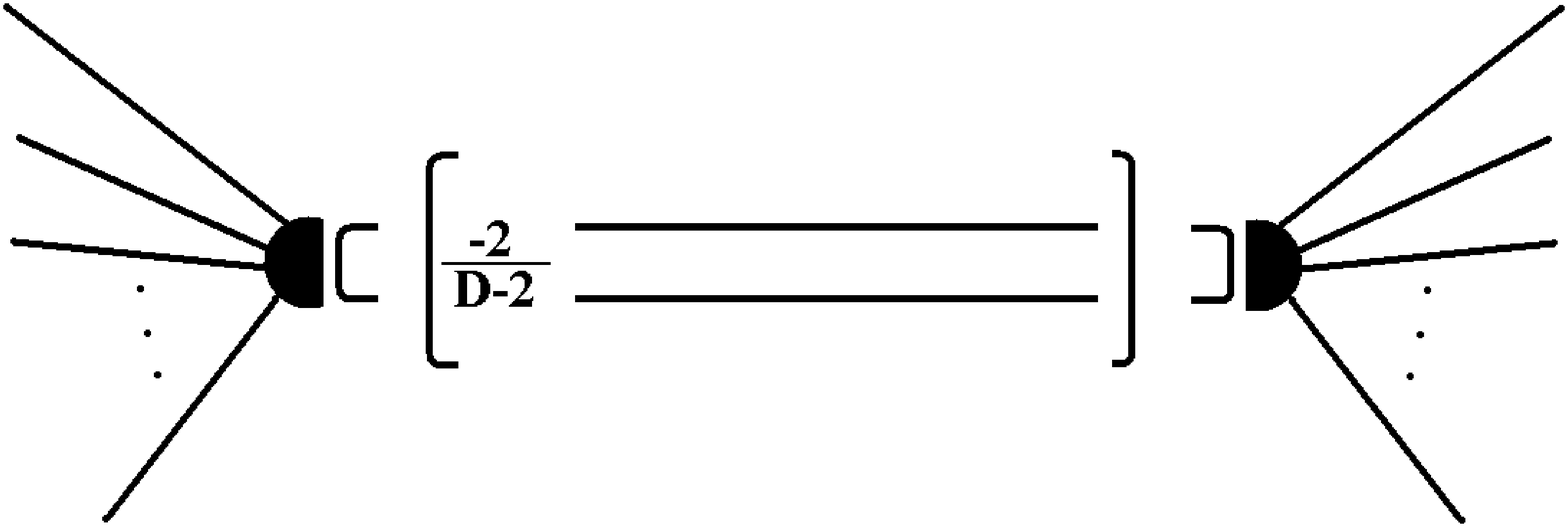}}
\parbox{8.7cm}{\includegraphics[height=2.2cm]{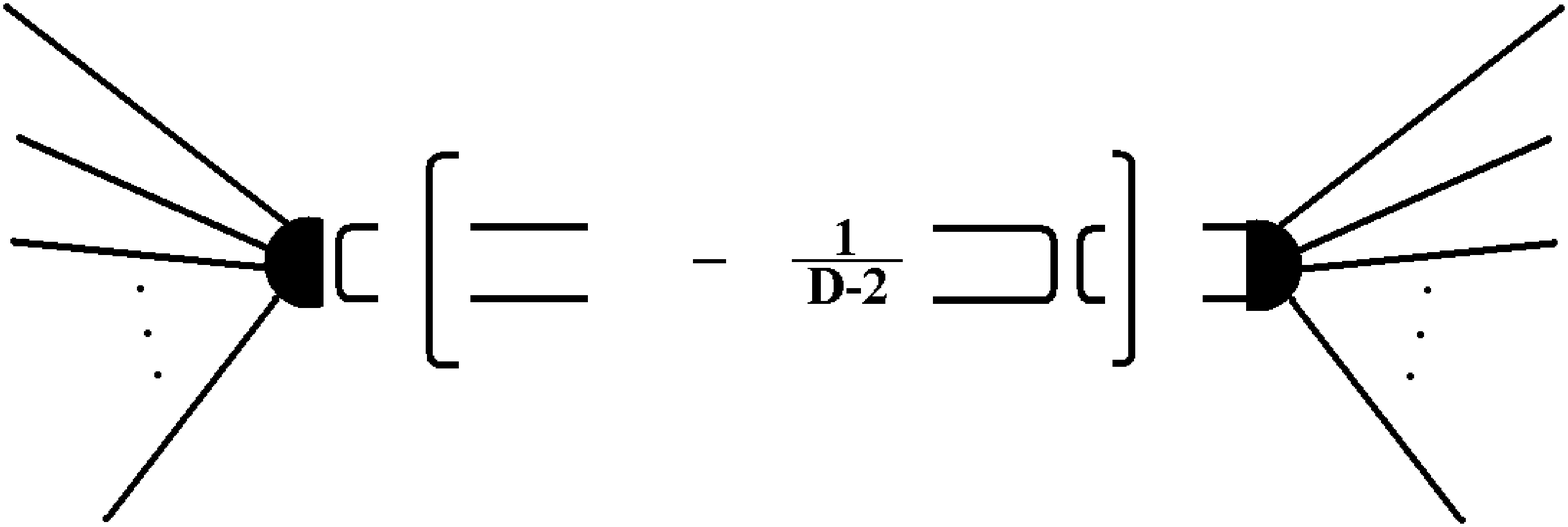}}
\caption{Two possible types of contractions for the propagator.
Whenever we have an index loop we have a contraction of indices.
It is seen that none of the above contractions will generate
something with a positive power of ($D$).\label{figprop2}}
\end{figure}

\begin{figure}[h]
\parbox{8.7cm}{\includegraphics[height=2.2cm]{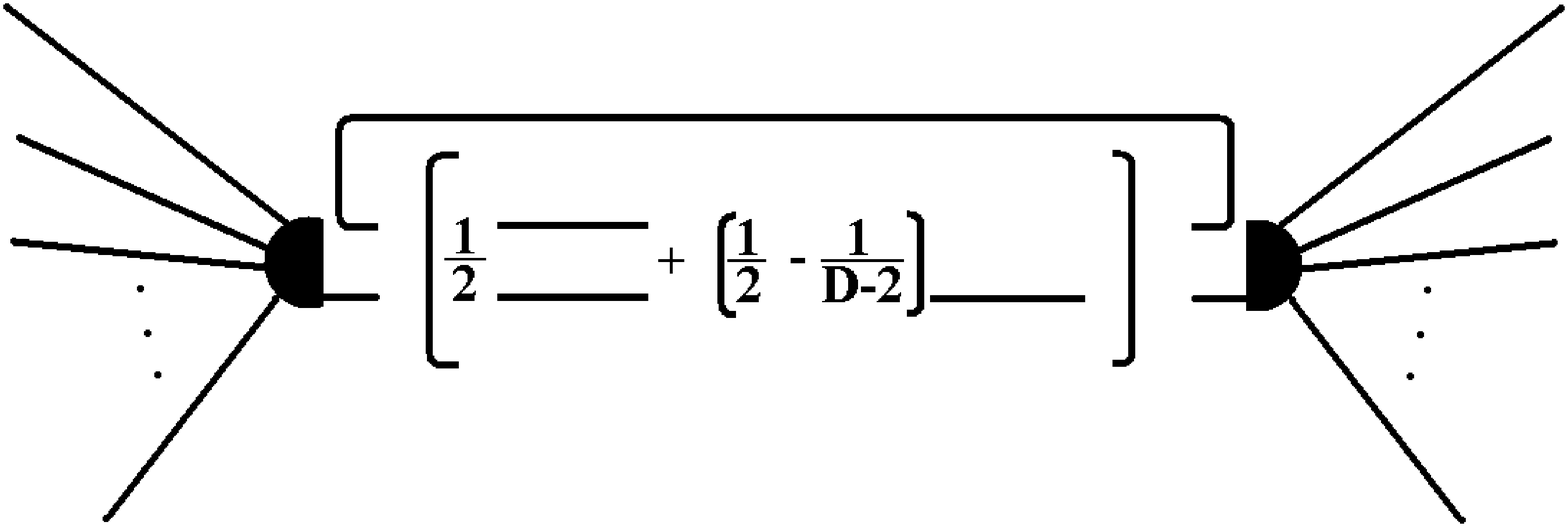}}
\parbox{8.7cm}{\includegraphics[height=2.2cm]{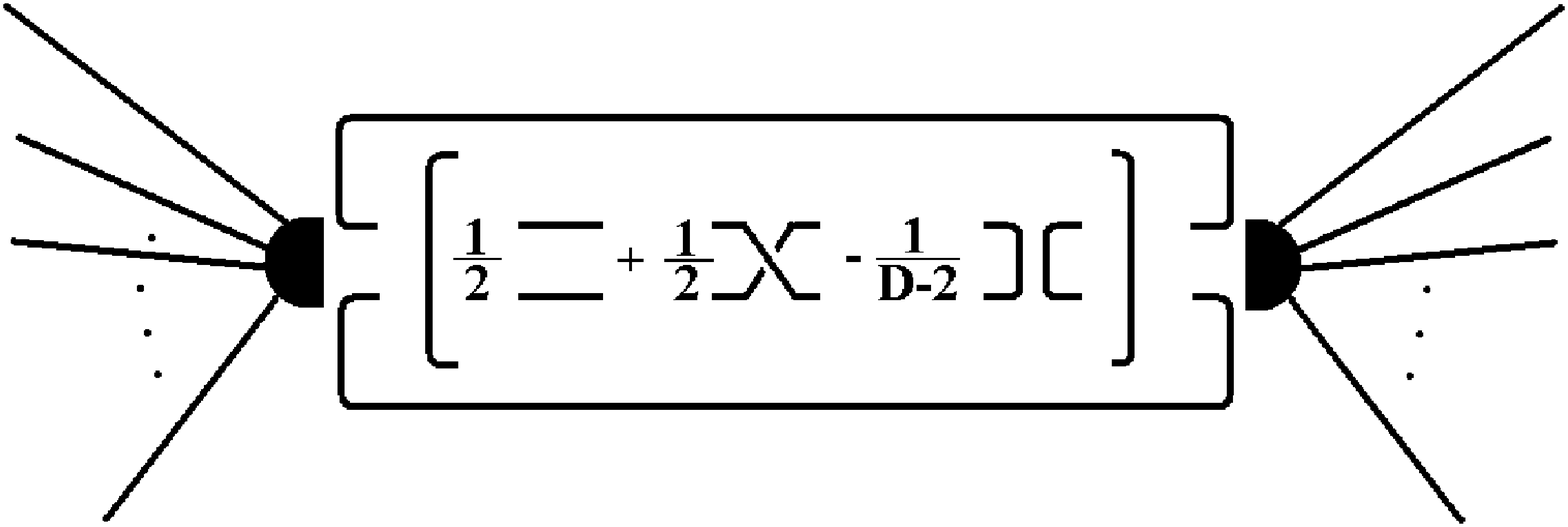}}
\caption{Two other types of index contractions for the propagator.
Again it is seen that only the
$(\eta_{\alpha\gamma}\eta_{\beta\delta}
+\eta_{\alpha\delta}\eta_{\beta\gamma})$-part of the propagator is
important in large-$D$ considerations.\label{figprop3}}
\end{figure}

Essentially (disregarding symmetrizations of indices etc) only the
following contractions of indices in the conventional graviton
propagator can occur in an given amplitude, we can contract, {\it
e.g.}, ($\alpha$ and $\beta$) or both ($\alpha$ and $\beta$) and
($\gamma$ and $\delta$). The results of such contractions are
shown below and also graphically depicted in figure
\ref{figprop2}.
\begin{equation}\begin{split}
\frac12[\eta_{\alpha\gamma}\eta_{\beta\delta}+\eta_{\alpha\delta}\eta_{\beta\gamma}-\frac{2}{D-2}\eta_{\alpha\beta}\eta_{\gamma\delta}\Big]\eta^{\alpha\beta}
= \eta_{\gamma\delta}
-\frac{D}{D-2}\eta_{\gamma\delta} \\
\frac12[\eta_{\alpha\gamma}\eta_{\beta\delta}+\eta_{\alpha\delta}\eta_{\beta\gamma}-\frac{2}{D-2}\eta_{\alpha\beta}\eta_{\gamma\delta}\Big]\eta^{\alpha\beta}\eta^{\gamma\delta}
=-\frac{2D}{D-2}
\end{split}\end{equation}
We can can also contract, {\it e.g.}, ($\alpha$ and $\gamma$) or
both ($\alpha$ and $\gamma$) and ($\beta$ and $\delta$) again the
results are shown below and depicted graphically in figure
\ref{figprop3}.
\begin{equation}\begin{split}
\frac12[\eta_{\alpha\gamma}\eta_{\beta\delta}+\eta_{\alpha\delta}\eta_{\beta\gamma}-\frac{2}{D-2}\eta_{\alpha\beta}\eta_{\gamma\delta}\Big]\eta^{\alpha\gamma} =
\frac12\eta_{\beta\delta}D+\eta_{\beta\delta}\left(\frac12+\frac{1}{D-2}\right)\\
\frac12[\eta_{\alpha\gamma}\eta_{\beta\delta}+\eta_{\alpha\delta}\eta_{\beta\gamma}-\frac{2}{D-2}\eta_{\alpha\beta}\eta_{\gamma\delta}\Big]\eta^{\alpha\gamma}\eta^{\beta\delta}
= D^2 - \frac{D}{D-2}\end{split}\end{equation}

As it is seen, only the
$(\eta_{\alpha\gamma}\eta_{\beta\delta}+\eta_{\alpha\delta}\eta_{\beta\gamma})$
part of the propagator will have the possibility to contribute
with a positive power of ($D$). Whenever the
$\left(\frac{\eta_{\mu\nu}\eta_{\alpha\beta}}{D-2}\right)$-term in
the propagator is in play, {\it e.g.}, what remains is something
that goes as $\left(\sim \frac{D}{D-2}\right)$ and which
consequently will have no support to the large-$D$ leading loop
contributions.

\begin{figure}[h]\centering
\parbox{3.3cm}{\includegraphics[height=0.7cm]{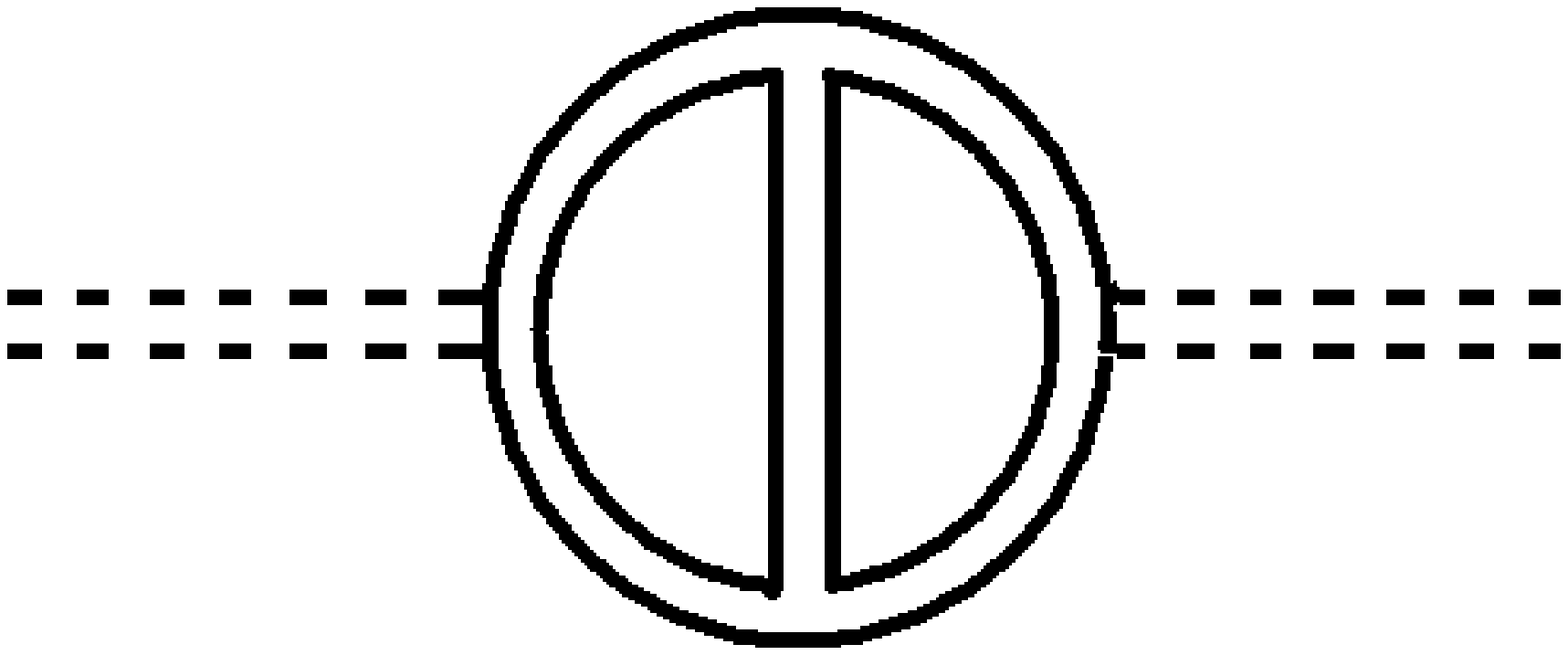}}
\parbox{3.3cm}{\includegraphics[height=0.7cm]{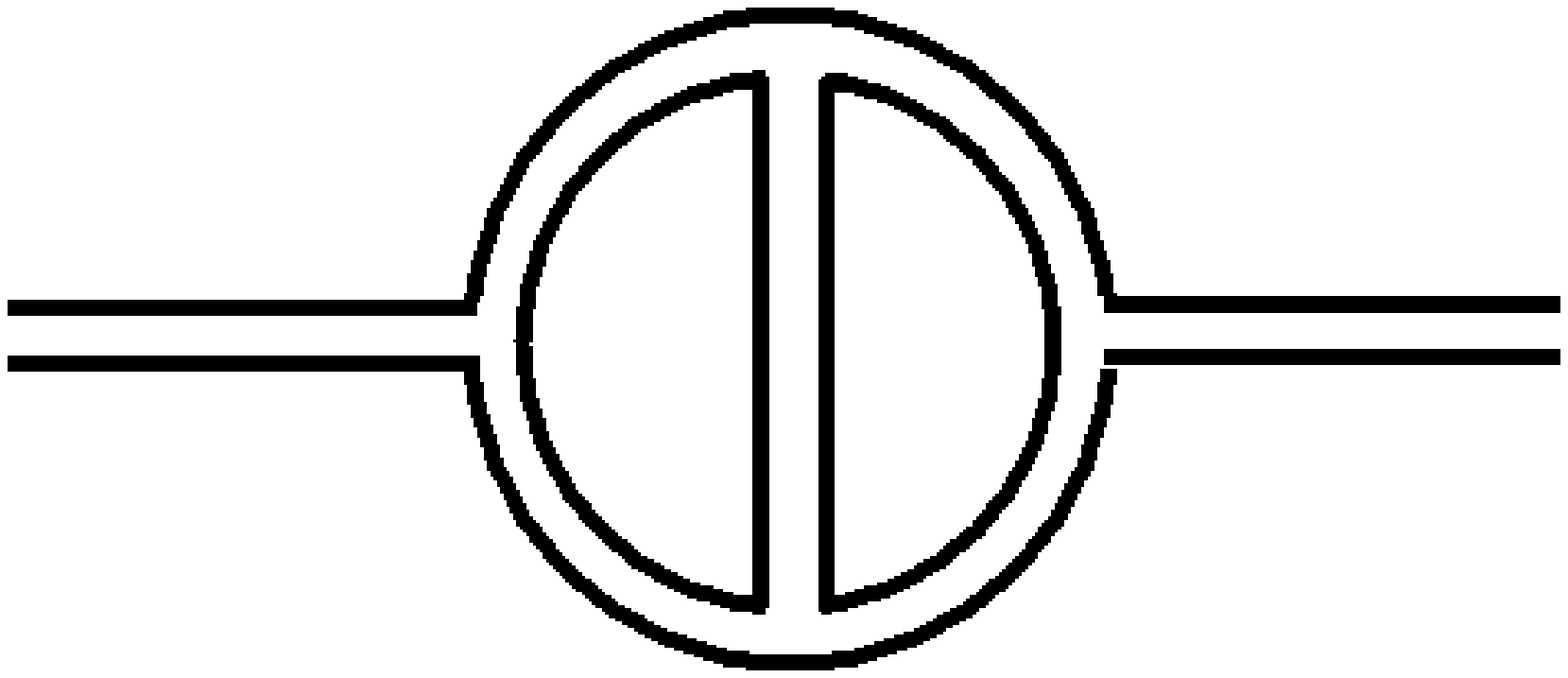}}
\caption{Different trace structures in the same diagram, generated
by different index structures in the vertex ($i.e.$, (3C) and (3E)
respectively), the first diagram have three traces, $\sim D^3$,
the other diagram have only two traces, $\sim D^2$.\label{dif3}}
\end{figure}

We have seen that different index structures go into the same
vertex factor. Below (see figure \ref{dif3}), we depict the same
2-loop diagram, but for different index structures of the external
3-point vertex factor ($i.e.$, (3C) and (3E)) for the external
lines.
Different types of trace structures will hence occur in the same
loop diagram. Every diagram will consist of a sum of contributions
with a varying number of traces. It is seen that the graphical
depiction of the index structure in the vertex factor is a very
useful tool in determining the trace structure of various
diagrams.

\begin{figure}[h]
\begin{minipage}{\linewidth}\centering
\parbox{3.2cm}{\includegraphics[height=0.7cm]{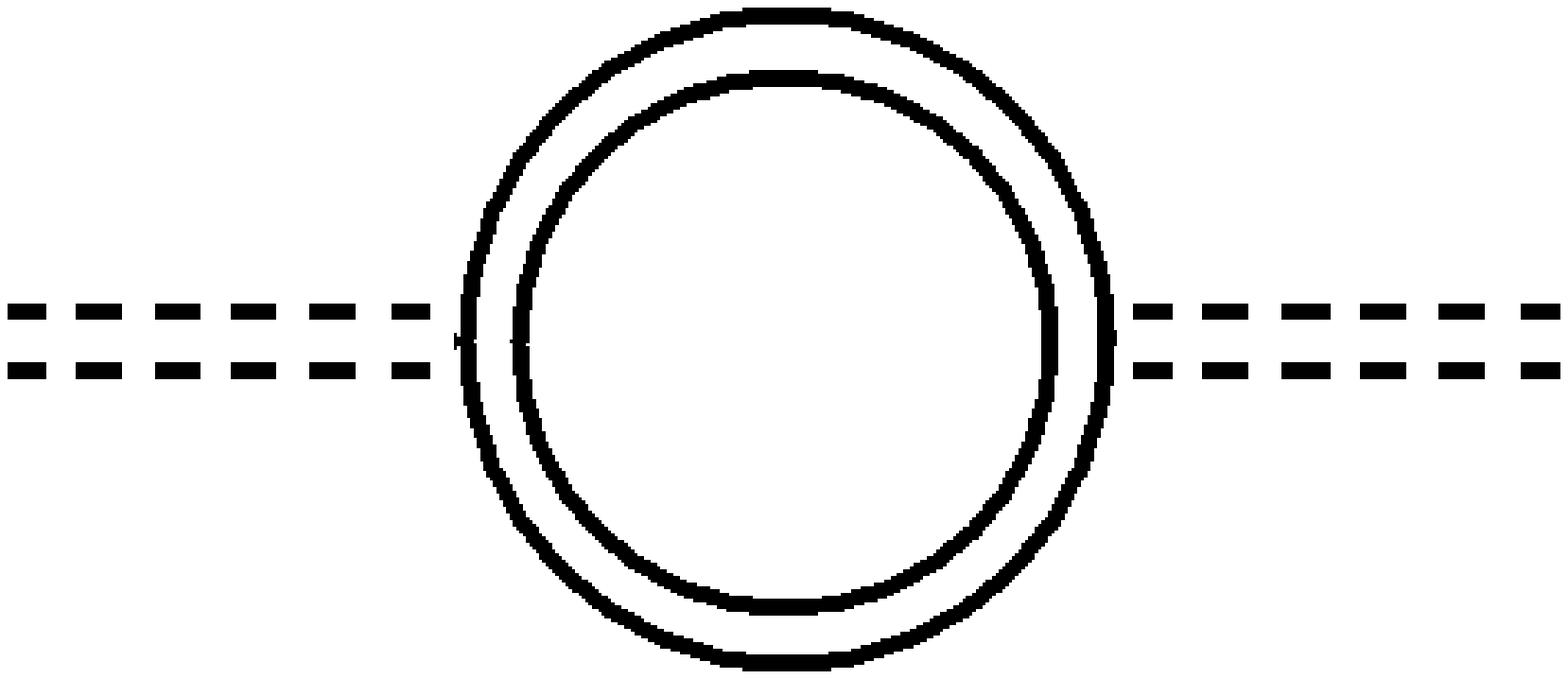}}\parbox{3.2cm}
{\includegraphics[height=0.7cm]{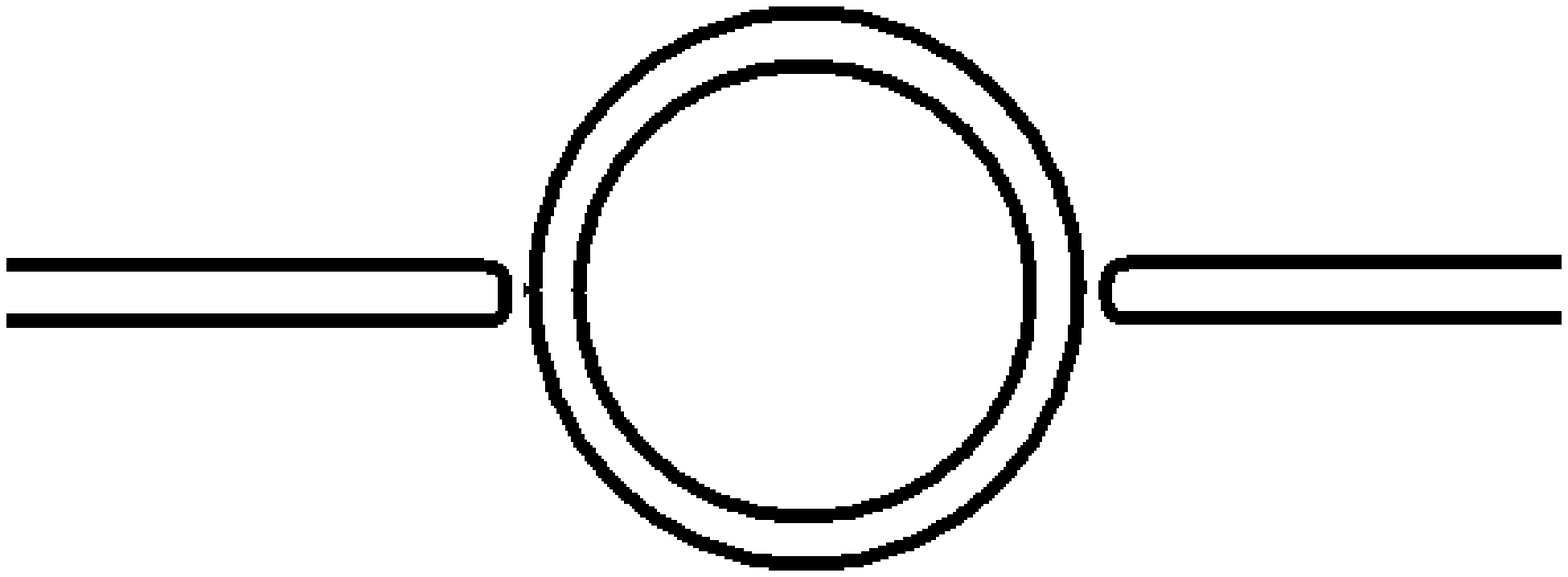}} \caption{A
diagrammatic representation of some leading contributions to the
2-point 1-loop graph in the large-$D$ limit.\label{1-loop}}
\end{minipage}\vspace{0.5cm}
\begin{minipage}{\linewidth}\centering
\parbox{3.2cm}{\includegraphics[height=0.7cm]{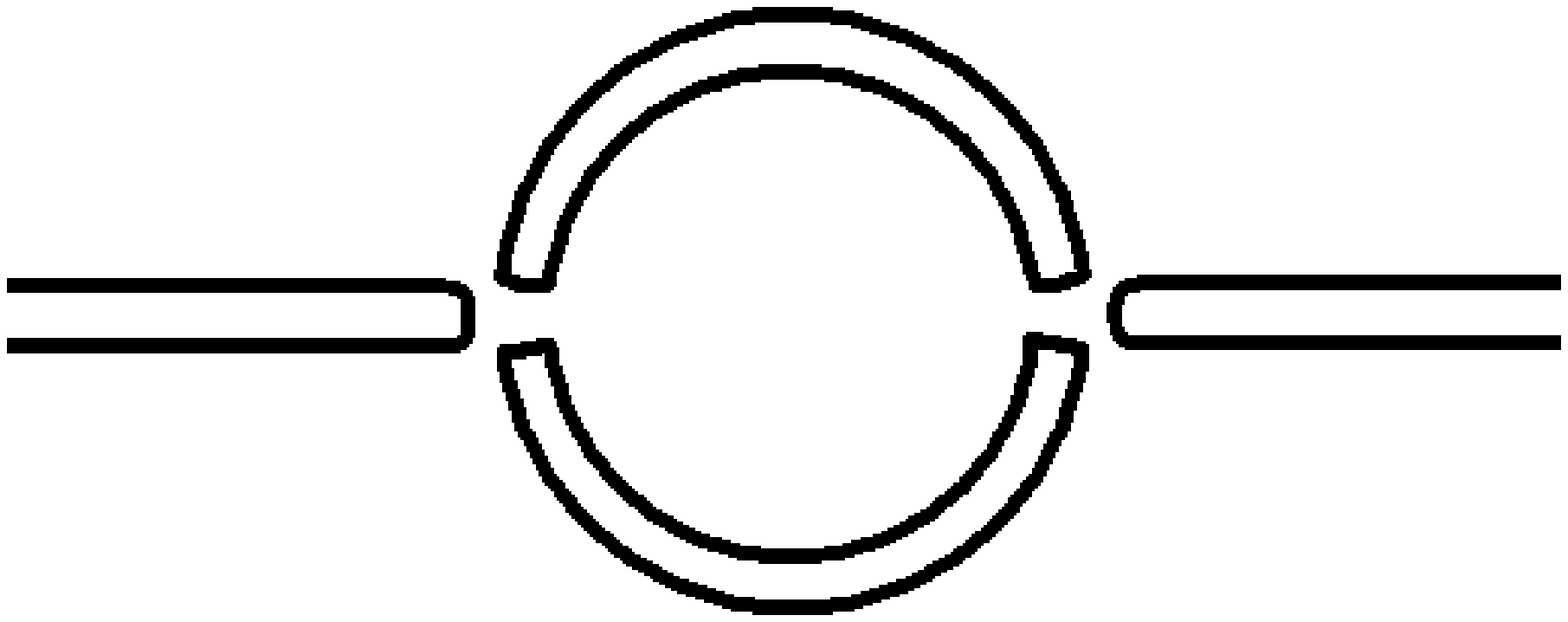}}
\parbox{3.2cm}{\includegraphics[height=0.7cm]{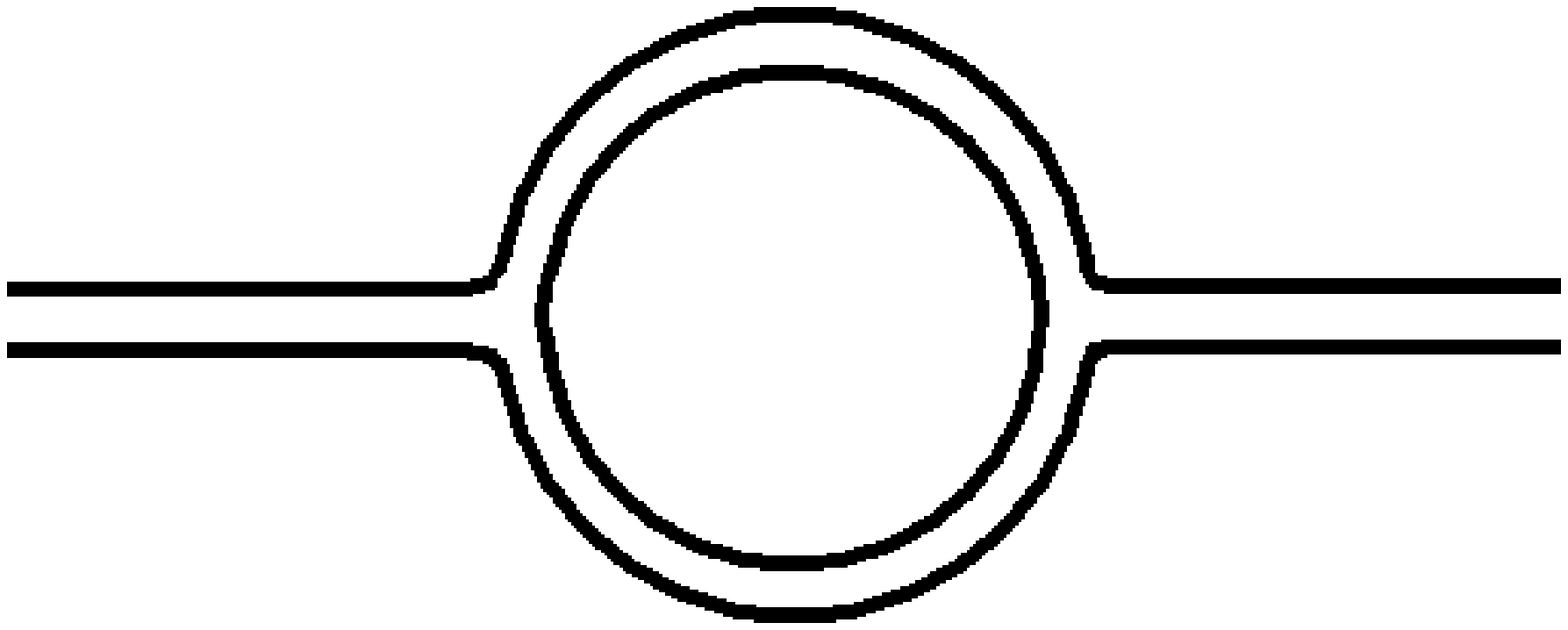}}
\caption{Some non-leading contributions to the 2-point 1-loop
graph in the large-$D$ limit.\label{1-loop2}}
\end{minipage}\end{figure}

The following type of two-point diagram (see figure \ref{1-loop})
build from the (3B) or (3C) parts of the 3-point vertex will
generate a contribution which carries a maximal number of traces
compared to the number of vertices in the diagram.
The other parts of the 3-point vertex factor, {\it e.g.}, the (3A)
or (3E) parts respectively will not generate a leading
contribution in the large-$D$ limit, however such contributions
will go into the non-leading contributions of the theory (see
figure \ref{1-loop2}). The diagram built from the (3A) index
structure will, {\it e.g.}, not carry a leading contribution due
to our previous considerations regarding contractions of the
indices in the propagator (see figure \ref{figprop2}).

It is crucial that we obtain a consistent limit for the graphs,
when we take ($D \rightarrow \infty$). In order to cancel the
factor of ($D^2$) from the above type of diagrams, {\it i.e.}, to
put it on equal footing in powers of ($D$) as the simple
propagator diagram, it is seen that we need to rescale
$\left(\kappa \rightarrow \frac{\kappa}{D}\right)$. In the
large-$D$ limit $\left(\frac{\kappa}{D}\right)$ will define the
'new' redefined gravitational coupling. A finite limit for the
2-point function will be obtained by this choice, $i.e.$, the leading
$D$-contributions will not diverge, when the limit ($D\rightarrow
\infty$) is imposed. By this choice all leading $n$-point diagrams
will be well defined and go as: $\left(\sim
\left(\frac{\kappa}{D}\right)^{(n-2)}\right)$ in a
$\left(\frac1D\right)$ expansion. Each external line will carry a
gravitational coupling\footnote{\footnotesize{A tadpole diagram
will thus not be well defined by this choice but instead go as
$(\sim D)$ however because we can dimensionally regularize any
tadpole contribution away, any idea to make the tadpoles well
defined in the large-$D$ limit would occur to be strange.}}.
However, it is important to note that our rescaling of $(\kappa)$
is solely determined by the requirement of creating a finite
consistent large-$D$ limit, where all $n$-point tree diagrams and
their loop corrections are on an equal footing. The above
rescaling seems to be the obvious to do from the viewpoint of
creating a physical consistent theory. However, we note that other
limits for the graphs are in fact possible by other redefinitions
of $(\kappa)$ but such rescalings will, {\it e.g.}, scale away the
loop corrections to the tree diagrams, thus such rescalings must
be seen as rather unphysical from the viewpoint of traditional
quantum field theory.

\begin{figure}[h]
\begin{minipage}{\linewidth}\centering
\parbox{3.2cm}{\includegraphics[height=0.7cm]{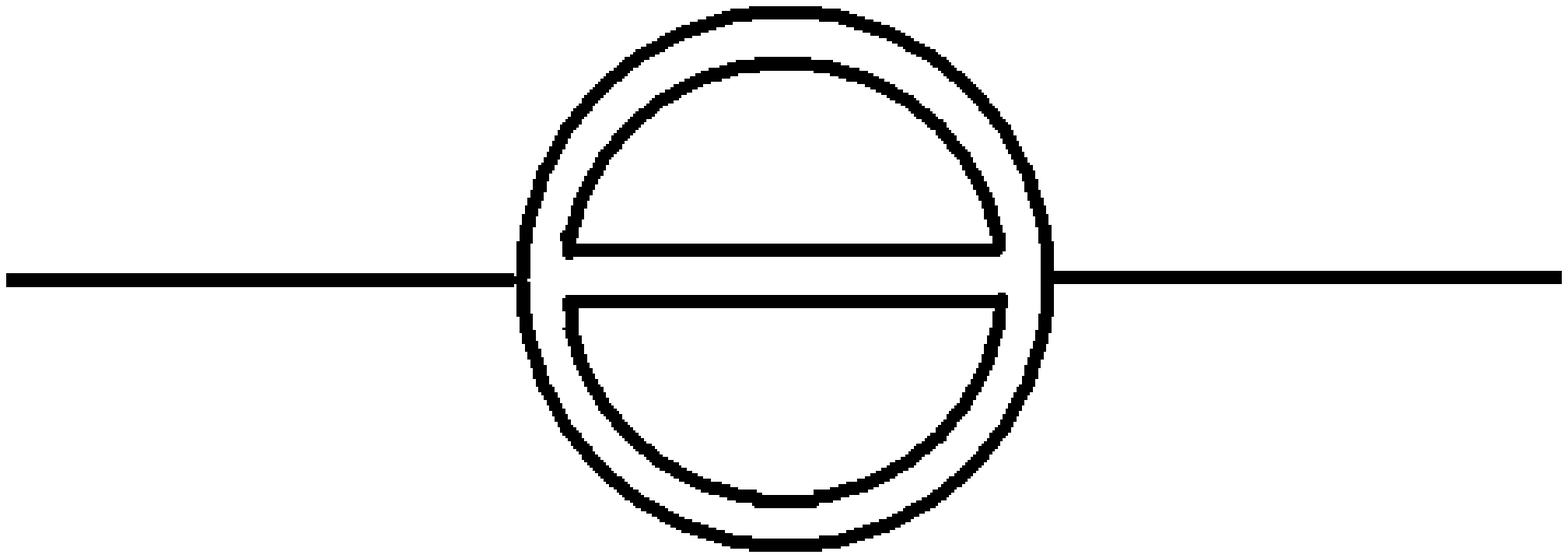}}
\parbox{3.2cm}{\includegraphics[height=0.7cm]{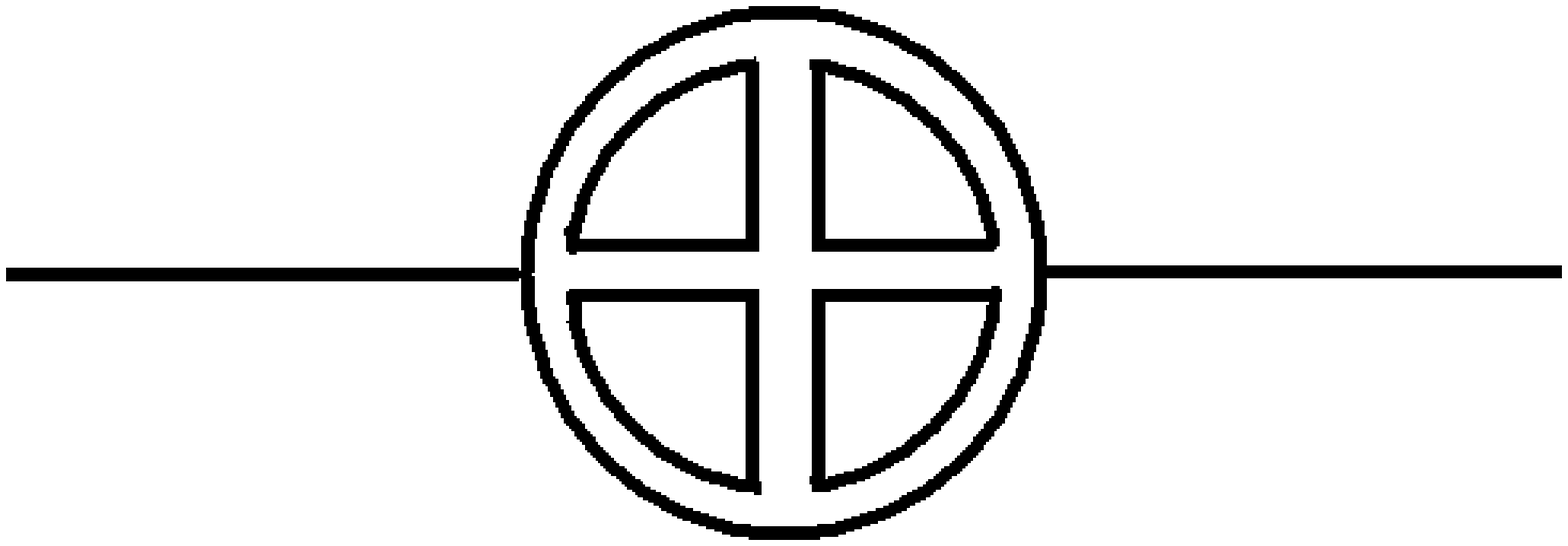}}
\caption{Two multi-loop diagrams with many traces. The rescaling
of $(\kappa)$ will however render both diagrams as non-leading
contributions compared to diagrams with separate loops. Here and
in the next coming figures we will employ the following notation
for the index structure of the lines originating from loop
contributions. Whenever a line originates from a loop
contribution it should be understood that there are only two
possibilities for its index structure, {\it i.e.}, the lines have
to be of the same type as in the contracted index lines in the
index structure (3B) or as the dashed lines in (3C). No
assumptions on the index structures for tree external lines are
made, here any viable index structures are present, but we still
represent such lines with a single full line for simplicity.
\label{other2}}
\end{minipage}
\end{figure}

For any loop with two propagators the maximal number of traces are
obviously two. However, we do not know which diagrams will
generate the maximal number of powers of $D$ in total. Some
diagrams will have more traces than others, however because of the
rescaling of $(\kappa)$ they might not carry a leading
contribution after all. For example,  if we look at the two graphs
depicted in figure \ref{other2}, they will in fact go as: $\left(D^3
\times \frac{\kappa^4}{D^4}\times D^3 \sim \frac{1}{D}\right)$ and
$\left(D^5 \times \frac{\kappa^8}{D^8}\times D^5 \sim
\frac{1}{D^3}\right)$ respectively and thus not be leading
contributions.

In order to arrive at the result of ref.~\cite{Strominger:1981jg},
we have to consider the rescaling of $(\kappa)$ which will
generate negative powers of $(D)$ as well. Let us now discuss the
main result of ref.~\cite{Strominger:1981jg}.

Let ($P_{\rm \max}$) be the maximal power of ($D$) associated with
a graph. Next we can look at the function ($\Pi$):
\begin{equation}
\Pi = 2L - \sum_{m = 3}^\infty(m-2)V_m.
\end{equation}
Here ($L$) is the number of loops in a given diagram, and ($V_m$)
counts the number of $m$-point vertices. It is obviously true that
($P_{\rm max}$) must be less than or equal to ($\Pi$).

The above function counts the maximal number of positive factors
of ($D$) occurring for a given graph. The first part of the
expression simply counts the maximal number of traces ($\sim D^2$)
from a loop, the part which is subtracted counts the powers of
($\kappa$ $\sim \frac1D$) arising from the rescaling of
($\kappa$).

Next, we employ the two well-known formulas for the topology of
diagrams:
\begin{equation}
L = I - V + 1,
\end{equation}
\begin{equation}
\sum_{m=3}^\infty m V_m = 2I + E,
\end{equation}
where ($I$) is the number of internal lines, $\left(V =
\sum_{m=3}^\infty V_m\right)$ is total number of vertices and
($E$) is the number of external lines.

Putting these three formulas together one arrives at:
\begin{equation}
P_{\rm max} \leq 2 - E - 2\sum_{m=3}^\infty V_m.
\end{equation}
Hence, the maximal power of ($D$) is obtained in the case of
graphs with a minimal number of external lines, and two traces per
diagram loop. In order to generate the maximum of two traces per
loop, separated loops are necessary, because whenever a propagator
index line is shared in a diagram we will not generate a maximal
number of traces. Considering only separated bubble graphs, thus
the following type of $n$-point (see figure \ref{bubble}) diagrams
will be preferred in the large-$D$ limit:
\begin{figure}[h]\centering
\vspace{0cm}\parbox{5cm}{\includegraphics[height=3.4cm]{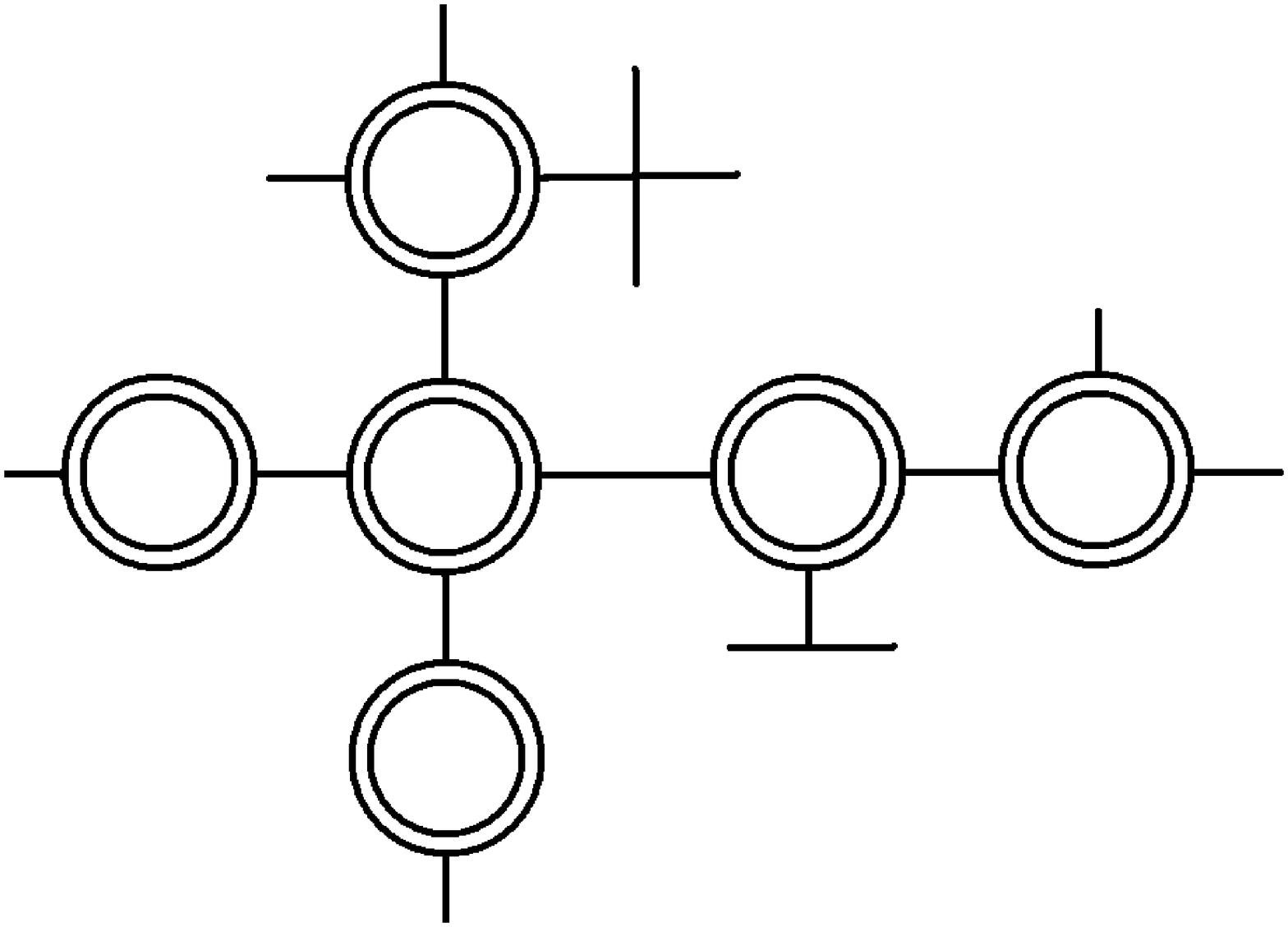}}
\vspace{1.2cm}\parbox{5cm}{\includegraphics[height=3.4cm]{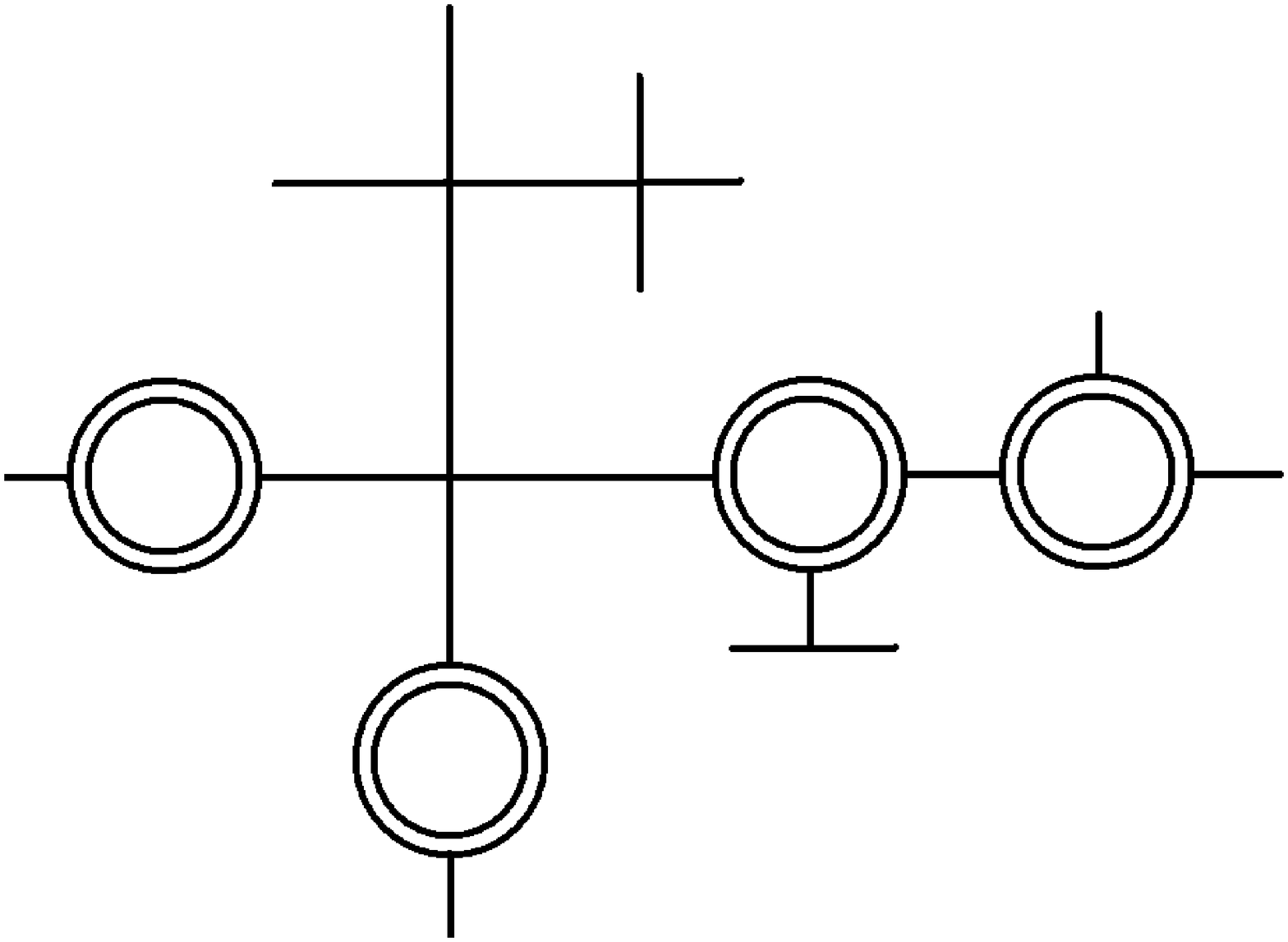}}
\parbox{5cm}{\includegraphics[height=3.4cm]{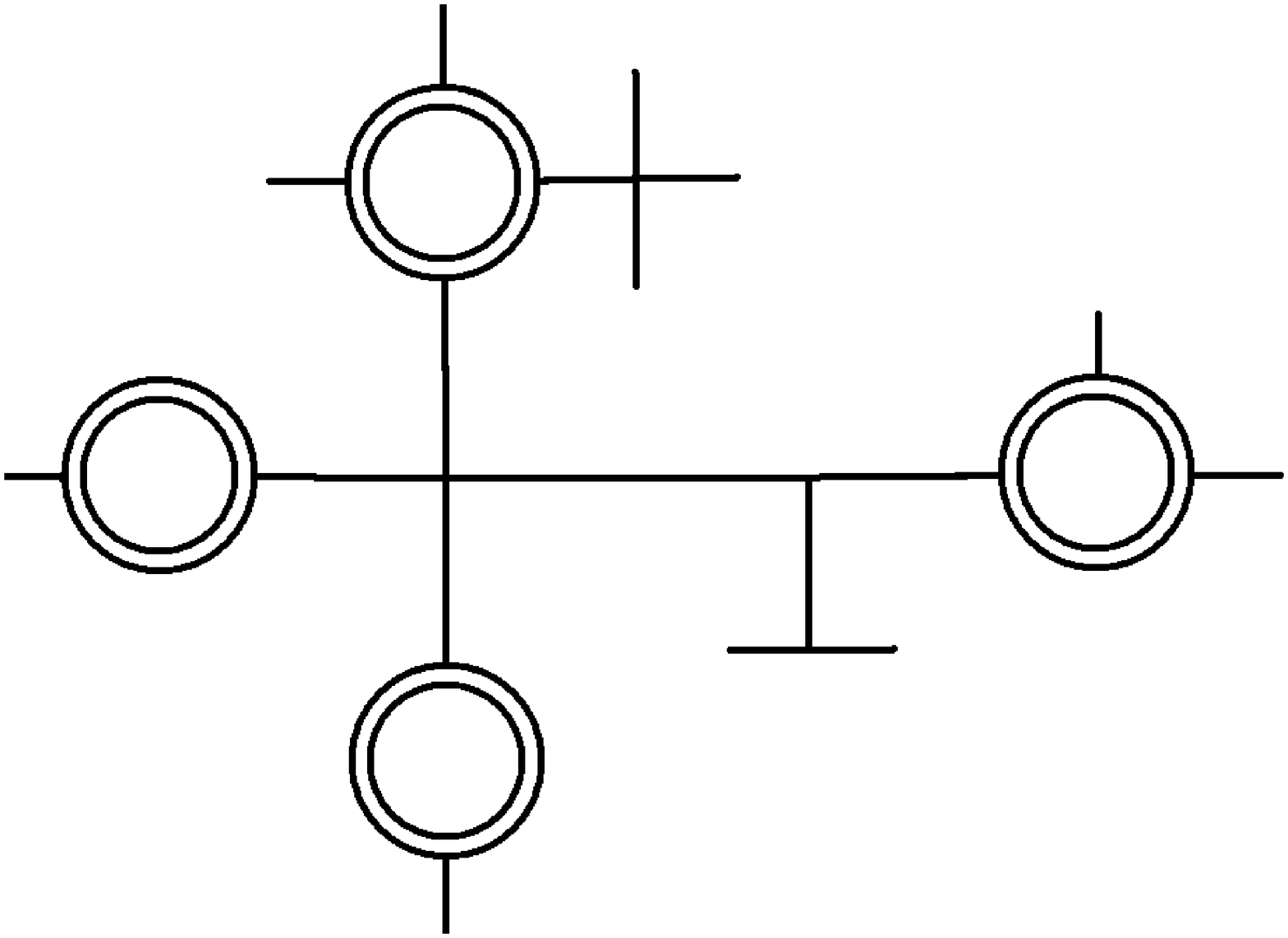}}\vspace{-1.2cm}
\caption{Examples of multi-loop diagrams with leading
contributions in the large-$D$ limit.\label{bubble}}
\end{figure}

\begin{figure}[h]\centering
\parbox{7cm}{\includegraphics[height=0.7cm]{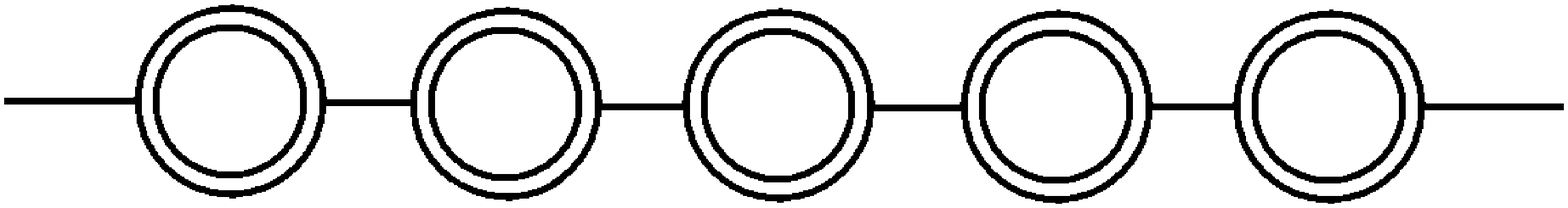}}
\parbox{7cm}{\includegraphics[height=1.6cm]{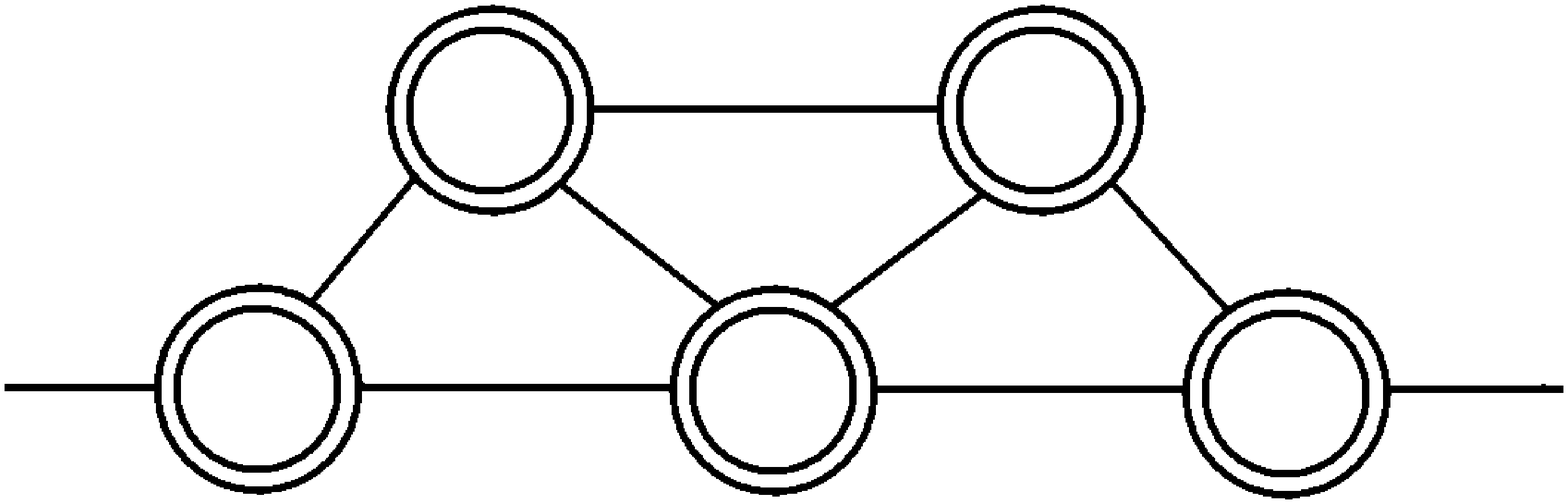}}
\caption{Of the following two diagrams with double trace loops the
first one will be leading.\label{multi}}
\end{figure}

Vertices, which generate unnecessary additional powers of
$\left(\frac1D\right)$, are suppressed in the large-$D$ limit;
multi-loop contributions with a minimal number of vertex lines
will dominate over diagrams with more vertex lines. That is, in
figure \ref{multi} the first diagram will dominate over the second
one.

\begin{figure}[h]\centering
\parbox{2.5cm}{\includegraphics[height=0.7cm]{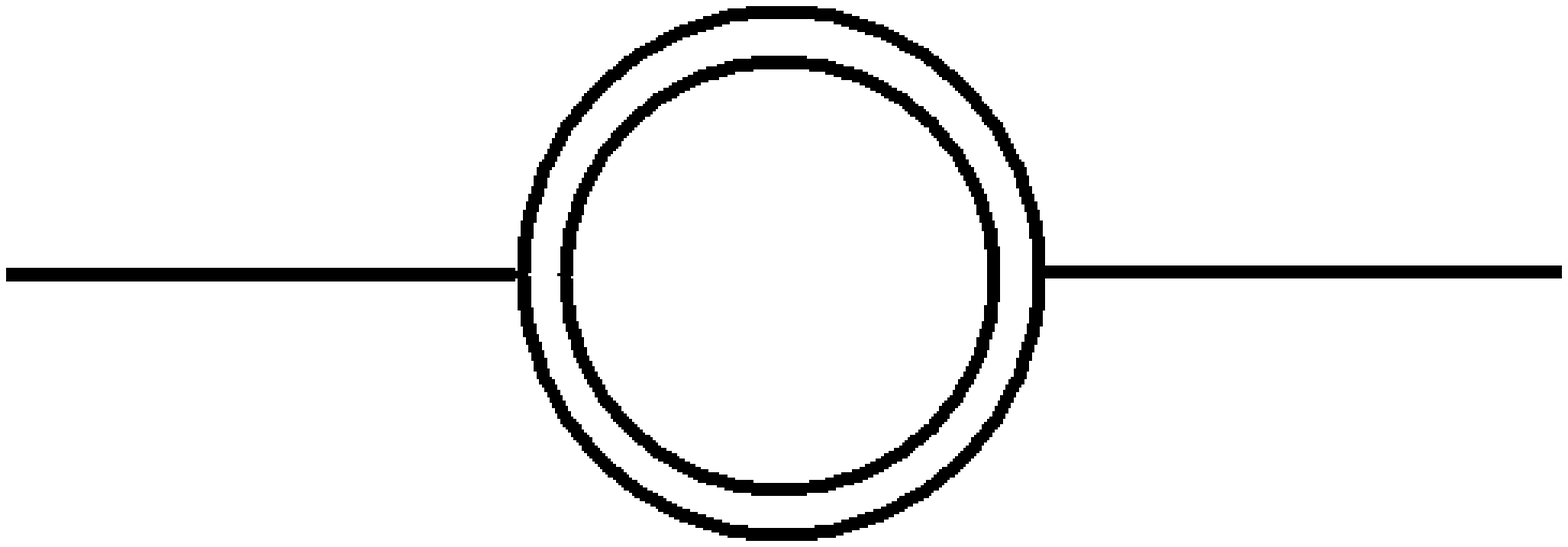}}
\parbox{3.5cm}{\includegraphics[height=0.7cm]{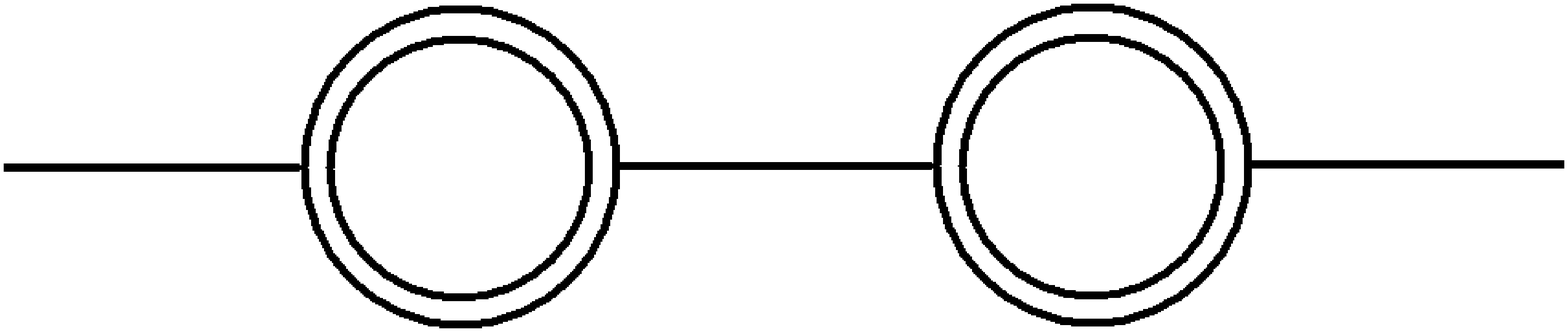}}
\parbox{3.9cm}{\includegraphics[height=0.7cm]{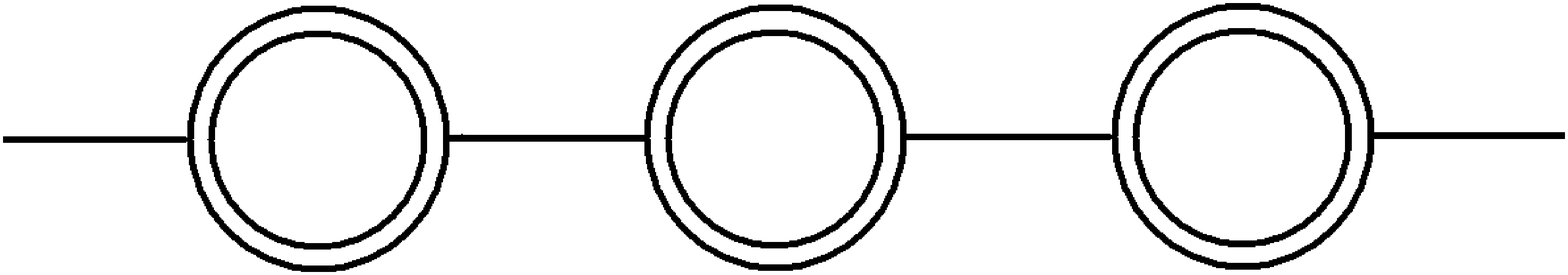}}
\parbox{4.4cm}{\includegraphics[height=0.7cm]{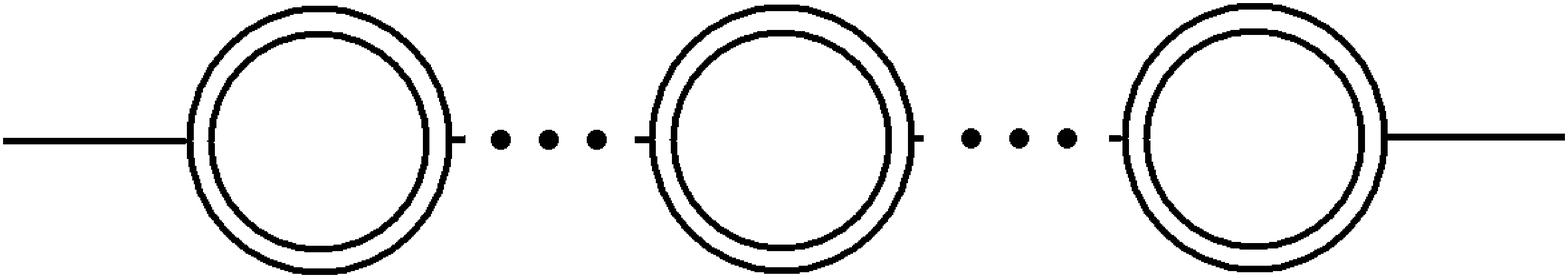}}
\caption{Leading bubble 2-point diagrams.\label{bubble2}}
\end{figure}

Thus, separated bubble diagrams will contribute to the large-$D$
limit in quantum gravity. The 2-point bubble diagram limit is
shown in figure \ref{bubble2}.
\begin{figure}[h]\centering
\parbox{4cm}{\includegraphics[height=0.7cm]{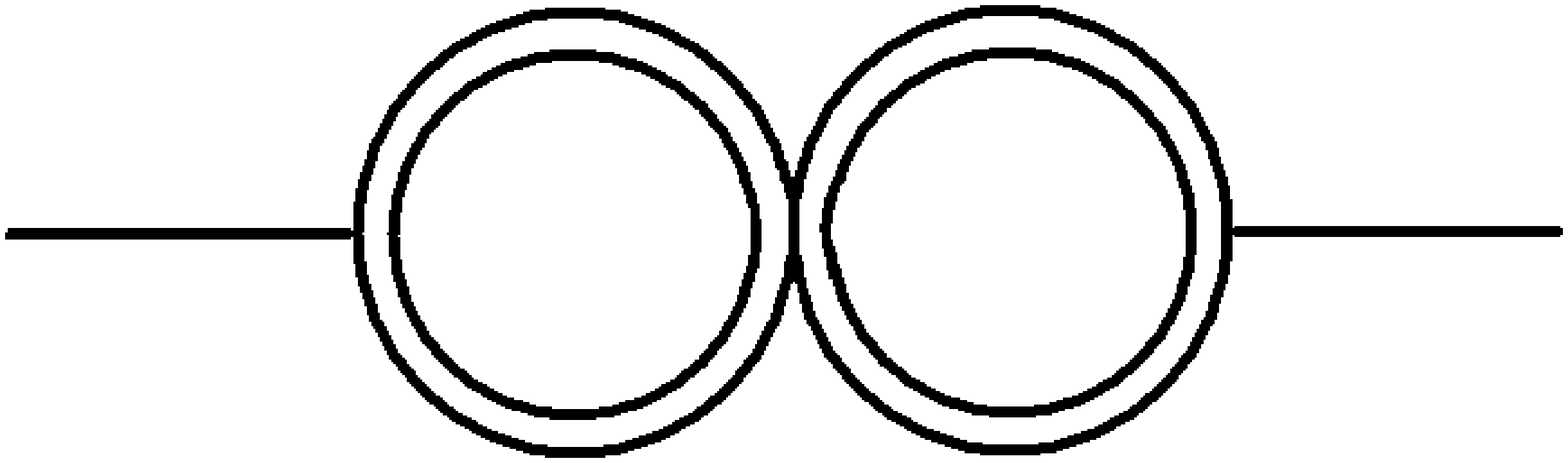}}
\parbox{4cm}{\includegraphics[height=0.7cm]{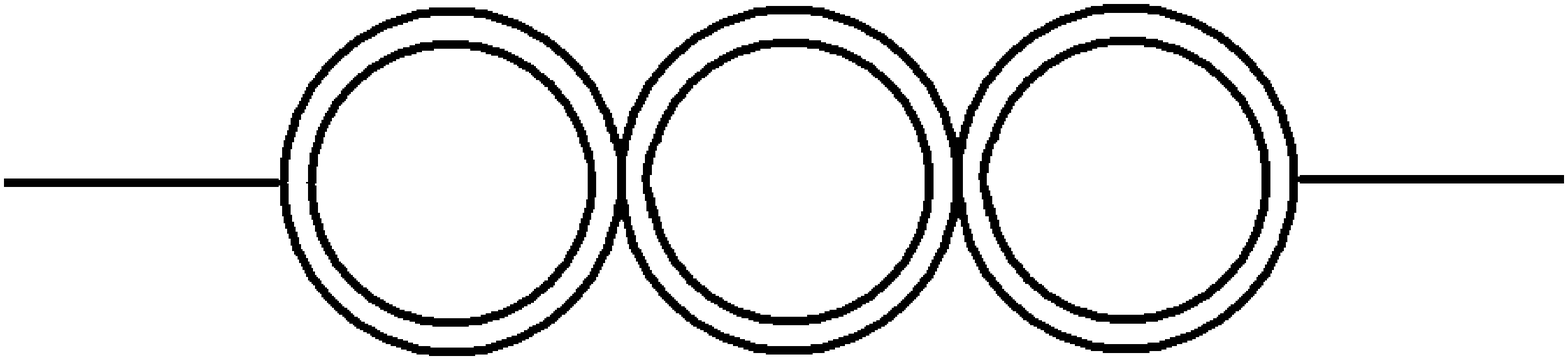}}
\parbox{4cm}{\includegraphics[height=0.7cm]{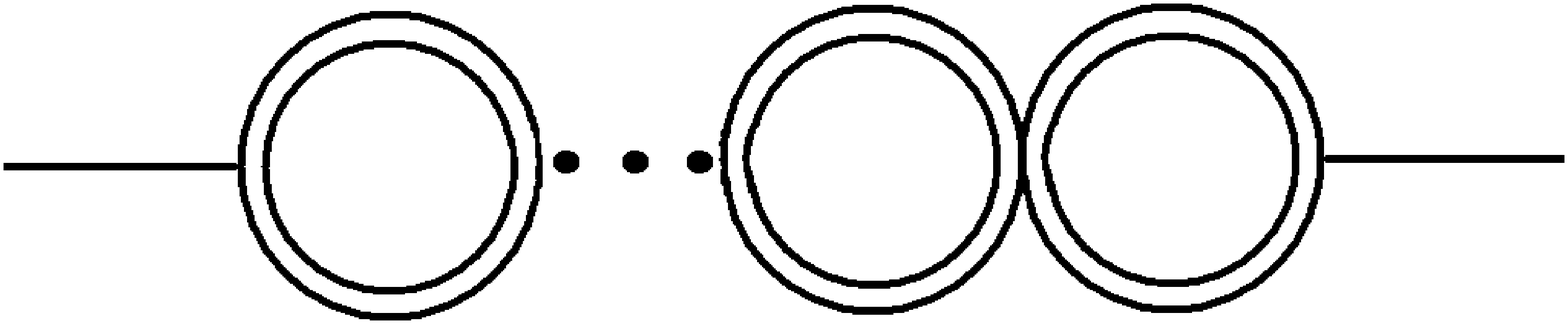}}
\caption{Another type of 2-point diagram build from the index
structure (4J), which will be leading in the large-$D$
limit.\label{4-vertex}}
\end{figure}

Another type of diagrams, which can make leading contributions, are
the diagrams depicted in figure \ref{4-vertex}. These
contributions are of a type where a propagator directly combines
different legs in a vertex factor with a double trace loop.
Because of the index structure (4J) in the 4-point vertex such
contributions are possible. These types of diagrams will have the
same dimensional dependence: $\left(\frac{\kappa^4}{D^4}\times
D^4\sim 1\right)$, as the 2-loop separated bubble diagrams. It is
claimed in ref.~\cite{Strominger:1981jg}, that such diagrams will be
non-leading. We have however found no evidence of this in our
investigations of the large-$D$ limit.

\begin{figure}[h]\centering
\parbox{3cm}{\includegraphics[height=1.3cm]{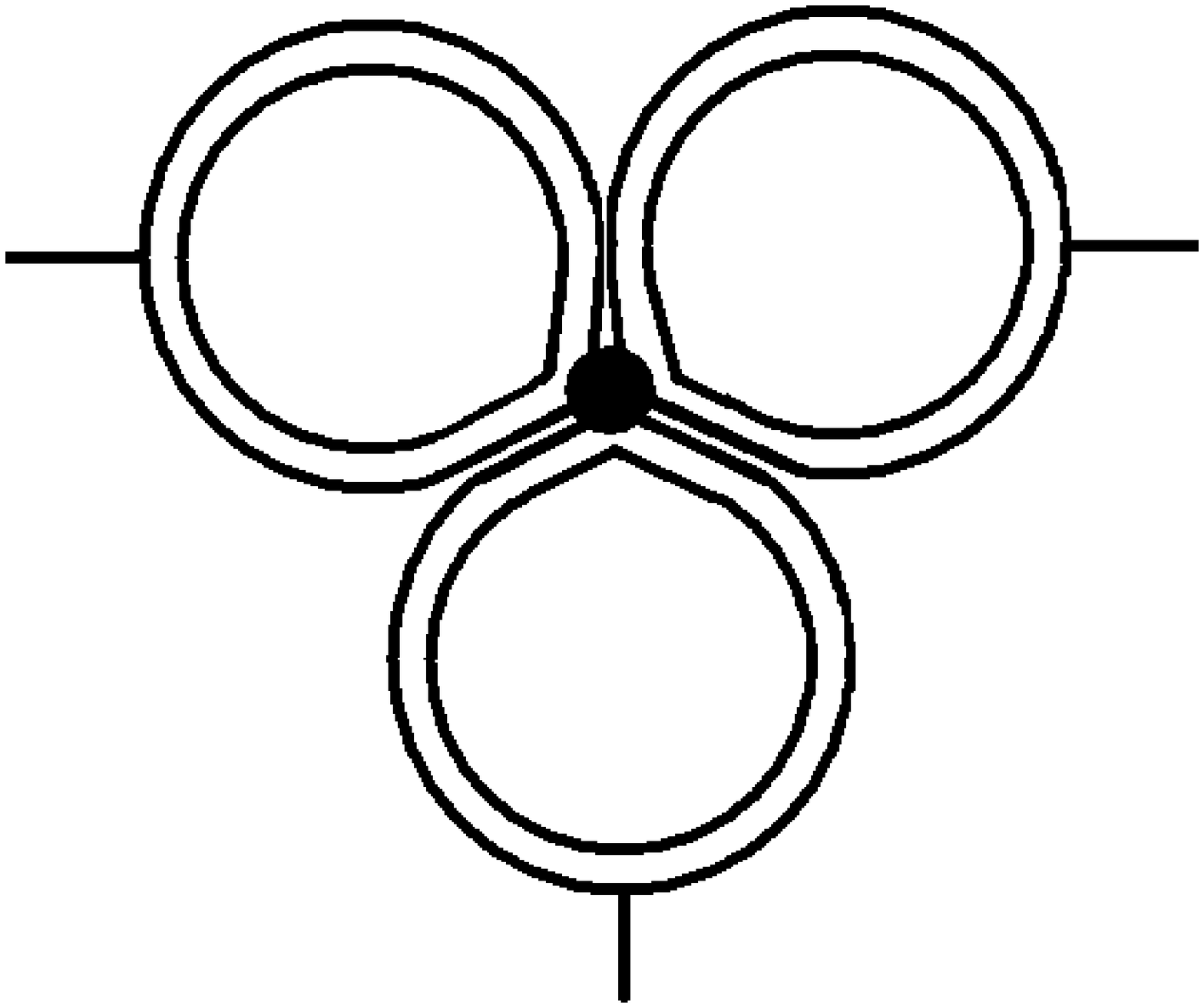}}
\parbox{3cm}{\includegraphics[height=1.3cm]{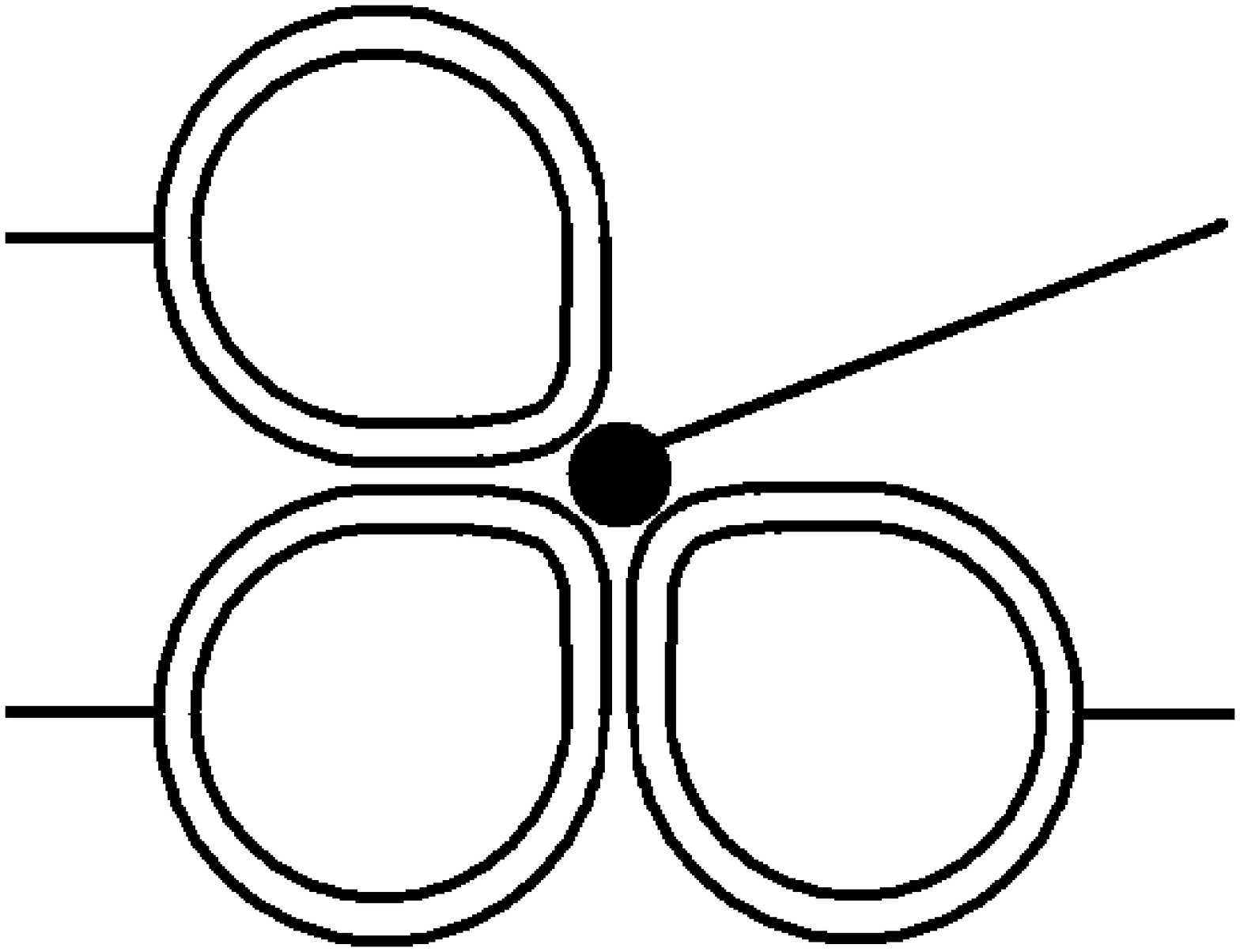}}
\parbox{3cm}{\includegraphics[height=1.3cm]{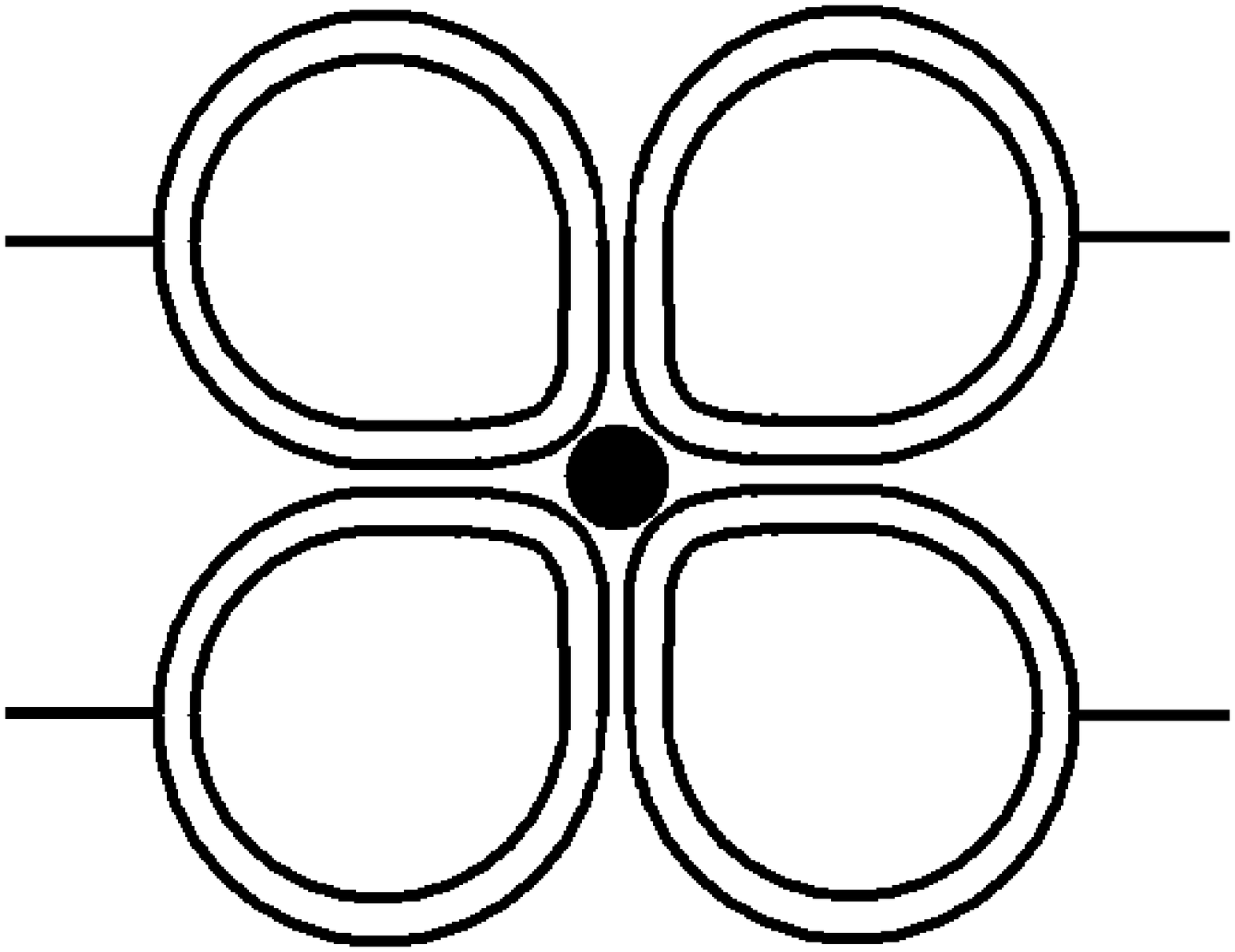}}
\caption{Particular examples of leading vertex-loop diagrams build
from 6-point and 8-point vertex factors
respectively.\label{6-point}}
\end{figure}

Higher order vertices with more external legs can generate
multi-loop contributions of the vertex-loop type, {\it e.g.}, a
6-point vertex can in principle generate a 3-loop contribution to
the 3-point function, a 8-point vertex a 4-loop contribution the
the 4-point function etc (see figure \ref{6-point}).

The 2-point large-$D$ limit will hence consist of the sum of the
propagator and all types of leading 2-point diagrams which one
considered above, {\it i.e.}, bubble diagrams and vertex-loops
diagrams and combinations of the two types of diagrams. Ghost
insertions will be suppressed in the large-$D$ limit, because such
diagrams will not carry two traces, however they will go into the
theory at non-leading order. The general $n$-point function will
consist of diagrams built from separated bubbles or vertex-loop
diagrams. We have devoted appendix~\ref{app2} to a study of the
generic $n$-point functions.

\section{The large-$D$ limit and the effective extension of gravity}
So far we have avoided the renormalization
difficulties of Einstein gravity. To obtain an effective
renormalizable theory up to the Planck scale we will have to introduce
the effective field theory description.

The most general effective Lagrangian of a $D$-dimensional
gravitational theory has the form:
\begin{equation}
{\cal L} = \int d^Dx \sqrt{-g}\bigg(\Big(\frac{2}{\kappa^2}R+
c_1R^2+c_2R_{\mu\nu}R^{\mu\nu}+\ldots\Big)+{\cal L}_{\rm eff.\
matter}\bigg),
\end{equation}
where ($R^\alpha_{\ \mu\nu\beta}$) is the curvature tensor ($g =
-{\rm det}(g_{\mu\nu})$) and the gravitational coupling is defined
as before, {\it i.e.}, ($\kappa^2 = 32\pi G_D$). The matter
Lagrangian includes in principle everything which couple to a
gravitational field, {\it i.e.}, any effective or
higher derivative couplings of gravity to bosonic and or fermionic
matter as has been discussed previously. Here we will
however look solely at pure gravitational interactions.
The effective Lagrangian for the theory is then reduced to
invariants built from the Riemann tensors:
\begin{equation}
{\cal L} = \int d^Dx \sqrt{-g}\Big(\frac{2R}{\kappa^2}+c_1R^2+
c_2R_{\mu\nu}R^{\mu\nu}+\ldots\Big).
\end{equation}
In an effective field theory higher derivative couplings of the
fields are allowed for, while the underlying physical symmetries
of the theory are kept manifestly intact. In gravity the general
action of the theory has to be covariant with respect to the
external gravitational fields.

The renormalization problems of traditional Einstein gravity are
trivially solved by the effective field theory approach. The
effective treatment of gravity includes all possible invariants
and hence all divergences occurring in the loop diagrams can be
absorbed in a renormalized effective action.

The effective expansion of the theory will be an expansion of the
Lagrangian in powers of momentum, and the minimal powers of
momentum will dominate the effective field theory at low-energy
scales. In gravity this means that the ($R \sim ({\rm
momentum})^2$) term will be dominant at normal energies, {\it
i.e.}, gravity as an effective field theory at normal energies
will essentially still be general relativity. At higher energy
scales the higher derivative terms ($R^2 \sim ({\rm momentum})^4)
,\ \ldots, (\ R^3\sim ({\rm momentum})^6) \ldots$, corresponding
to higher powers of momentum will mix in and become increasingly
important.

The large-$D$ limit in Einstein gravity is arrived at by expanding
the theory in ($D$) and subsequently taking ($D\rightarrow
\infty$). This can be done in its effective extension too.
Treating gravity as an effective theory and deriving the large-$D$
limit results in a double expansion of the theory, {\it i.e.}, we
expand the theory both in powers of momentum and in powers of
$\left(\frac1D\right)$. In order to understand the large-$D$ limit
in the case of gravity as an effective field theory, we have to
examine the vertex structure for the additional effective field
theory terms. Clearly any vertex contribution from, {\it e.g.},
the effective field theory terms, ($R^2$), and, ($R_{\mu\nu}^2$),
will have four momentum factors and hence new index structures are
possible for the effective vertices.

\begin{figure}[h]
\begin{tabular}{cccccc}\vspace{0.1cm}
$V_3^{\rm eff}=$& $\kappa^3\left(\parbox{1cm}
{\includegraphics[height=0.8cm]{fig1.1.ps}}\right)^{\rm eff}_{\rm
3A}$+&
$\kappa^3\left(\parbox{1cm}{\includegraphics[height=0.8cm]{fig1.2.ps}}\right)^{\rm
eff}_{\rm 3B}$+&
$\kappa^3\left(\parbox{1cm}{\includegraphics[height=0.8cm]{fig1.3.ps}}\right)^{\rm
eff}_{\rm 3C}$+&
$\kappa^3\left(\parbox{1cm}{\includegraphics[height=0.8cm]{fig1.4.ps}}\right)^{\rm
eff}_{\rm 3D}$&\\&+
$\kappa^3\left(\parbox{1cm}{\includegraphics[height=0.8cm]{fig1.5.ps}}\right)^{\rm
eff}_{\rm 3E}$+
&$\kappa^3\left(\parbox{1cm}{\includegraphics[height=0.8cm]{fig1.6.ps}}\right)^{\rm
eff}_{\rm 3F}$+&
$\kappa^3\left(\parbox{1cm}{\includegraphics[height=0.8cm]{fig1.7.ps}}\right)^{\rm
eff}_{\rm 3G}$+&
$\kappa^3\left(\parbox{1cm}{\includegraphics[height=0.8cm]{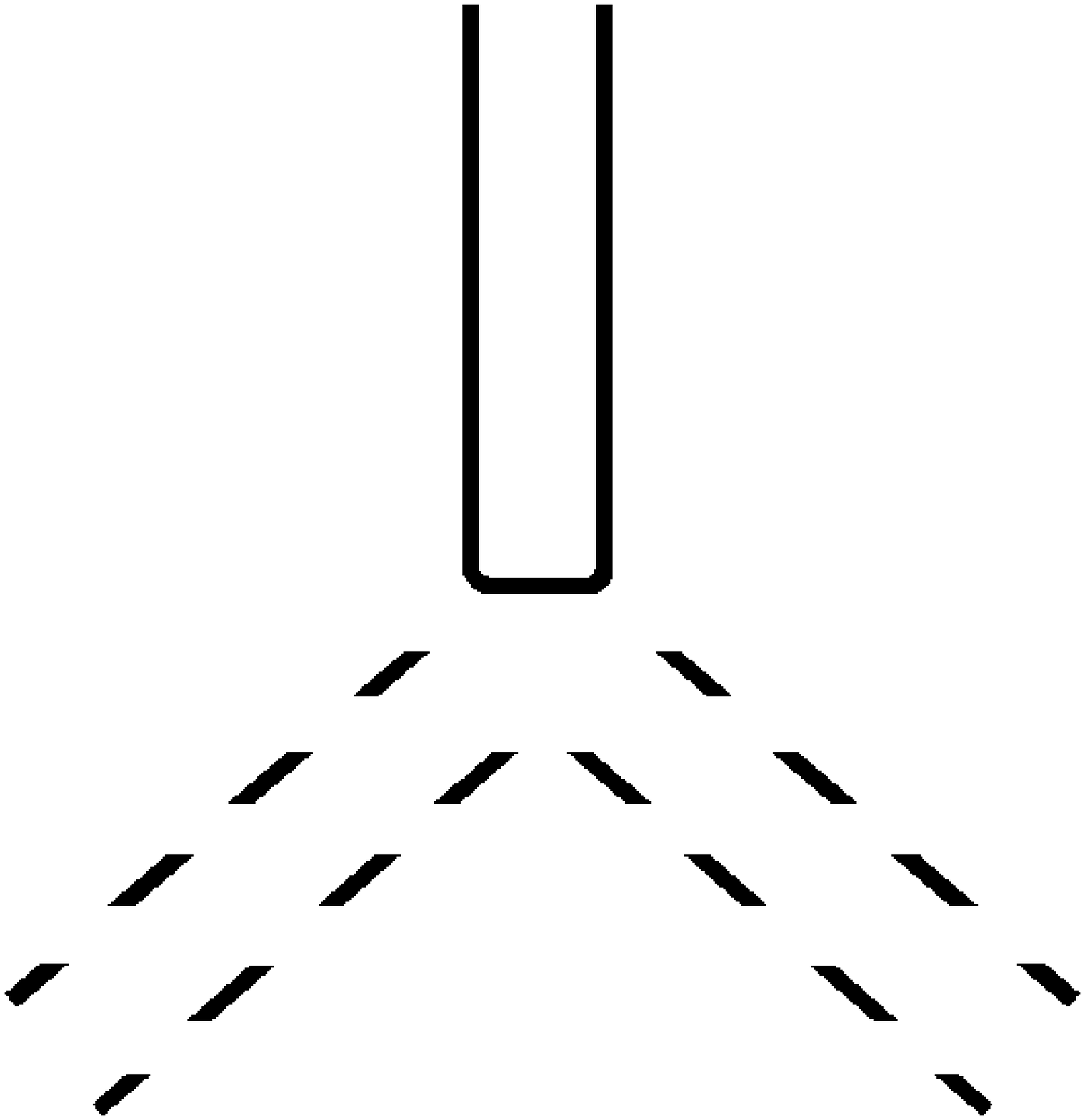}}\right)^{\rm
eff}_{\rm 3H}$\\&+
$\kappa^3\left(\parbox{1cm}{\includegraphics[height=0.8cm]{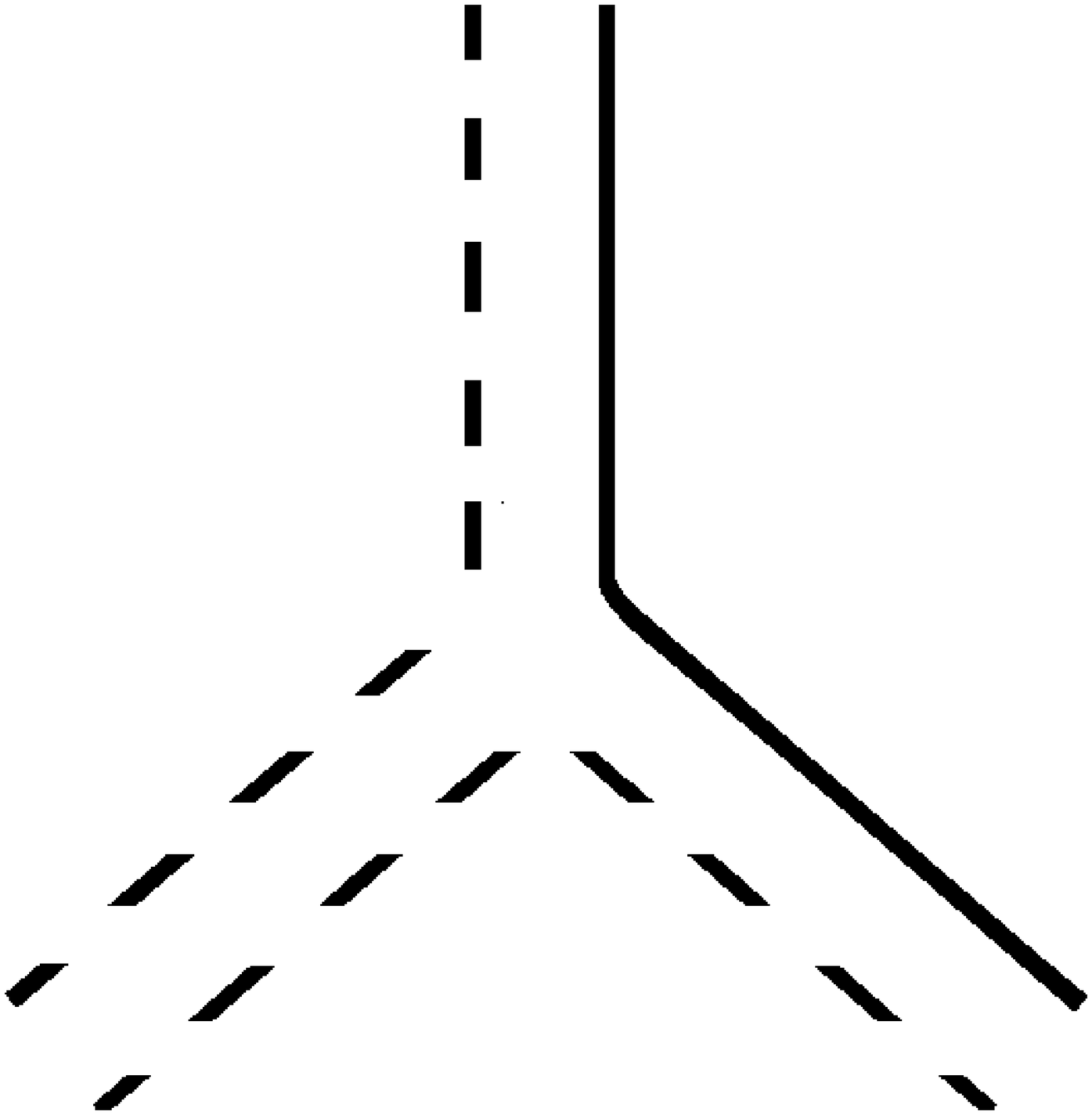}}\right)^{\rm
eff}_{\rm 3I}$.&&
\end{tabular}
\caption{A graphical representation of the ($R^2$) and the
($R^2_{\mu\nu}$) index structures in the effective 3-point vertex
factor. A dashed line represents a contraction of an index with a
momentum line. A full line symbolizes a contraction of two index
lines.\label{efft}}
\end{figure}

We find the following vertex structure for the ($R^2$) and the
($R_{\mu\nu}^2$) terms of the effective action (see figure
\ref{efft} and appendix~\ref{app1}).

\begin{figure}[h]
\begin{tabular}{ll}\vspace{0.1cm}
$\left(\parbox{1cm}{\includegraphics[height=0.8cm]{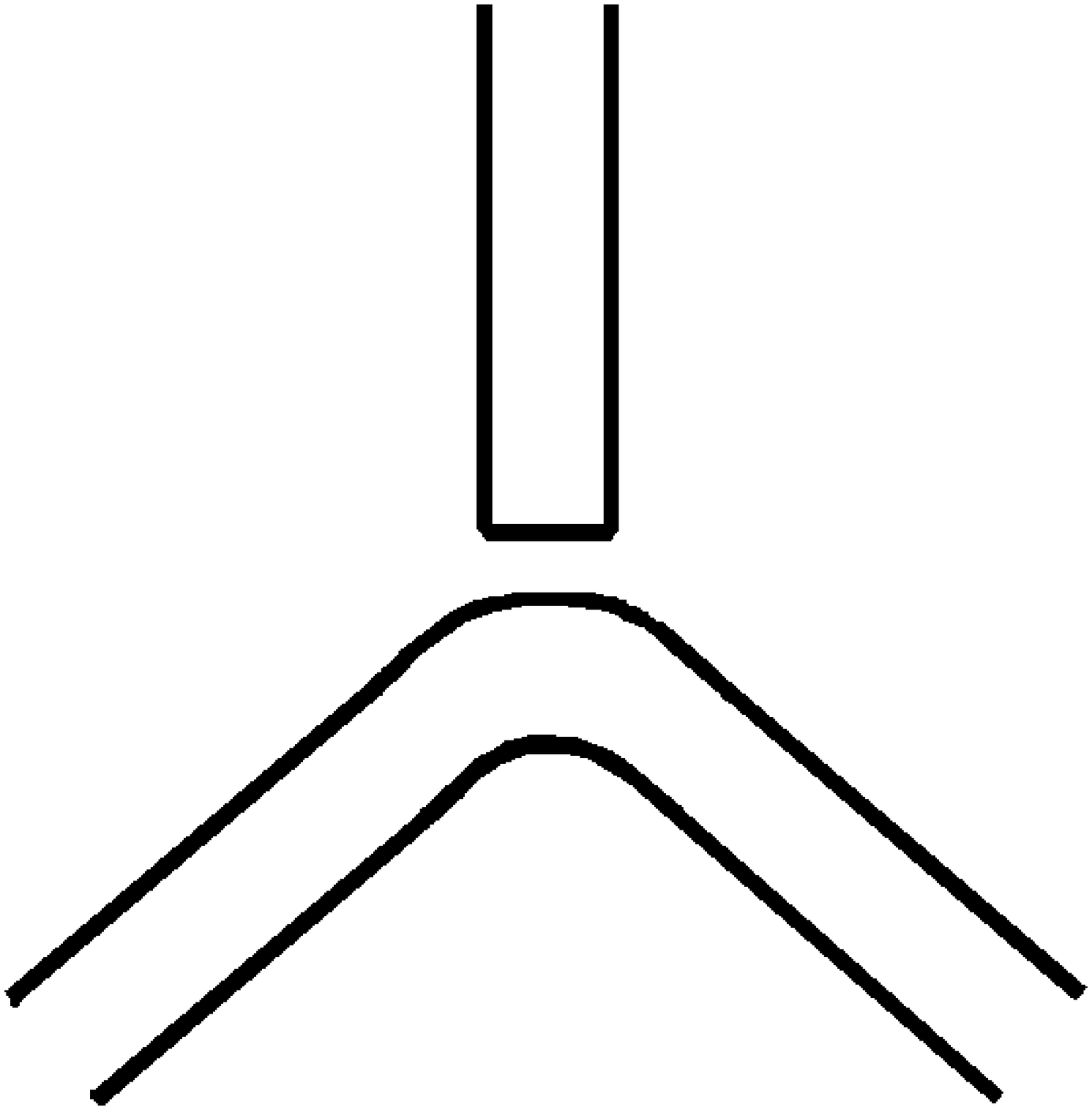}}\right)^{\rm
eff}_{\rm 3B}$ &$\sim {\rm
sym}[-P_3\left(\eta_{\mu\alpha}\eta_{\nu\sigma}\eta_{\beta\gamma}[3c_1k_1^2(k_2\cdot
k_3)+c_2(\frac12(k_1\cdot k_2)(k_1\cdot
k_3)-\frac14k_2^2k_3^2)]\right)$\\ &
$-P_6\left(\eta_{\mu\alpha}\eta_{\nu\sigma}\eta_{\beta\gamma}[2c_1k_1^2k_3^2+c_2(\frac12(k_1\cdot
k_3)^2-\frac14k_2^2(k_1\cdot k_3))]\right)],$
\\ \vspace{0.05cm}
$\left(\parbox{1cm}{\includegraphics[height=0.8cm]{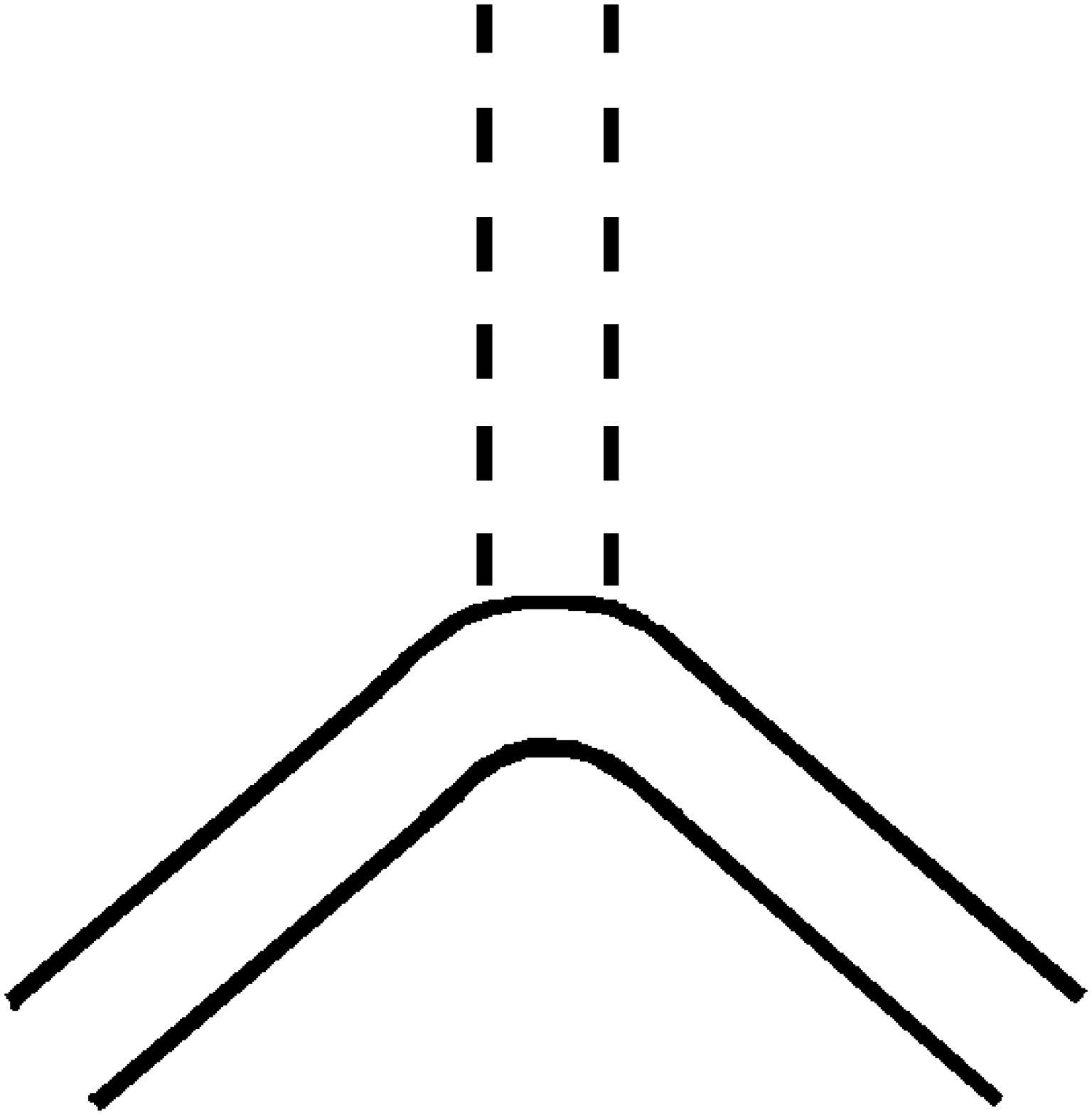}}\right)^{\rm
eff}_{\rm 3C}$ &$\sim {\rm
sym}[-P_3\left(k_{1\mu}k_{1\alpha}\eta_{\nu\sigma}\eta_{\beta\gamma}[3c_1k_2\cdot
k_3]\right)
-2P_6\left(k_{1\mu}k_{1\alpha}\eta_{\nu\sigma}\eta_{\beta\gamma}[c_1
k_3^2]\right)
$\\&$+\frac12P_6\left(k_{1\mu}k_{2\alpha}\eta_{\nu\sigma}\eta_{\beta\gamma}[c_2k_1\cdot
k_3]\right)
$+$P_6\left(k_{1\mu}k_{3\alpha}\eta_{\nu\sigma}\eta_{\beta\gamma}[c_2k_1\cdot
k_3]\right)
$\\&$-\frac12P_3\left(k_{2\mu}k_{3\alpha}\eta_{\nu\sigma}\eta_{\beta\gamma}[c_2k_1^2]\right)
$
$-\frac12P_6\left(k_{3\mu}k_{3\alpha}\eta_{\nu\sigma}\eta_{\beta\gamma}[c_2k_1^2]\right)$
\\& $
-\frac12P_6\left(k_{1\beta}k_{3\nu}\eta_{\mu\sigma}\eta_{\alpha\gamma}[c_2k_2^2]\right)
-\frac12P_6\left(k_{3\beta}k_{3\nu}\eta_{\mu\sigma}\eta_{\alpha\gamma}[c_2k_2^2]\right)].$
\end{tabular}
\caption{Explicit index structure terms in the effective vertex
3-point factor, which can contribute with leading contributions in
the large-$D$ limit.\label{3B3C}}
\end{figure}

We are particularly interested in the terms which will give
leading trace contributions. For the (3B)$^{\rm eff}$ and
(3C)$^{\rm eff}$ index structures of we have shown some explicit
results (see figure \ref{3B3C}).

It is seen that also in the effective theory there are terms which
will generate double trace structures. Hence an effective 1-loop
bubble diagram with two traces is possible. Such diagrams will go
as $\left(\frac{\kappa^4}{D^4}c_1\times D^2 \sim
\frac{c_1}{D^2}\right)$. In order to arrive at the same limit for
the effective tree diagrams as the tree diagrams in Einstein
gravity we need to rescale ($c_1$) and ($c_2$) by ($D^2$), $i.e.$,
($c_1 \rightarrow c_1D^2$) and ($c_2 \rightarrow c_2D^2$). Thereby
every $n$-point function involving the effective vertices will go
into the $n$-point functions of Einstein gravity. In fact this
will hold for any ($c_i$) provided that the index structures
giving double traces also exist for the higher order effective
terms. Other rescalings of the ($c_i$)'s are possible. But such
other rescalings will give a different tree limit for the
effective theory at large-$D$. The effective tree-level graphs
will be thus be scaled away from the Einstein tree graph limit.

The above rescaling with ($c_i \rightarrow c_iD^2$) corresponds to
the maximal rescaling possible. Any rescaling with, {\it e.g.},
($c_i \rightarrow c_iD^n$, $n > 2$) will not be possible, if we
still want to obtain a finite limit for the loop graphs at ($D
\rightarrow \infty$). A rescaling with $(D^2)$ will give maximal
support to the effective terms at large-$D$.

The effective terms in the action are needed in order to
renormalize infinities from the loop diagrams away. The
renormalization of the effective field theory can be carried out
at any particular ($D$). Of course the exact cancellation of pole
terms from the loops will depend on the integrals, and the algebra
will change with the dimension, but one can take this into account
for any particular order of $(D=D_p)$ by an explicit calculation
of the counter-terms at dimension ($D_p$) followed by an
adjustment of the coefficients in the effective action. The exact
renormalization will in this way take place order by order in
$\left(\frac1D\right)$. For any rescaling of the $(c_i)$'s it is
thus always possible carry out a renormalization of the theory.
Depending on how the rescaling of the coefficients $(c_i)$ are
done, there will be different large-$D$ limits for the effective
field theory extension of the theory.

The effective field theory treatment does in fact not change the
large-$D$ of gravity as much as one could expect. The effective
enlargement of gravity does not introduce new exciting leading
diagrams. Most parts of our analysis of the large-$D$ limit
depend only on the index structure for the various vertices, and
the new index structures provided by the effective field theory
terms only add new momentum lines, which do not give additional
traces over index loops. Therefore the effective extension of
Einstein gravity is not a very radical change of its large-$D$
limit. In the effective extension of gravity the action is
trivially renormalizable up to a cut-off scale at ($M_{\rm
Planck}$) and the large-$D$ expansion just as well defined as an
large-$N$ expansion of a renormalizable planar expansion of a
Yang-Mills action.

As a subject for further investigations, it might be useful to
employ more general considerations about the index structures.
Seemingly all possible types of index-structures are present in
the effective field theory extension of gravity in the
conventional approach. As only certain index structures go into
the leading large-$D$ limit, the {\it interesting} index structures
can be classified at any given loop order. In fact one could
consider loop amplitudes with only {\it interesting} index structures
present. Such applications might useful in very complicated
quantum gravity calculations, where only the leading large-$D$
behavior is interesting.

\begin{figure}[h]
\parbox{1.8cm}{\includegraphics[height=0.5cm]{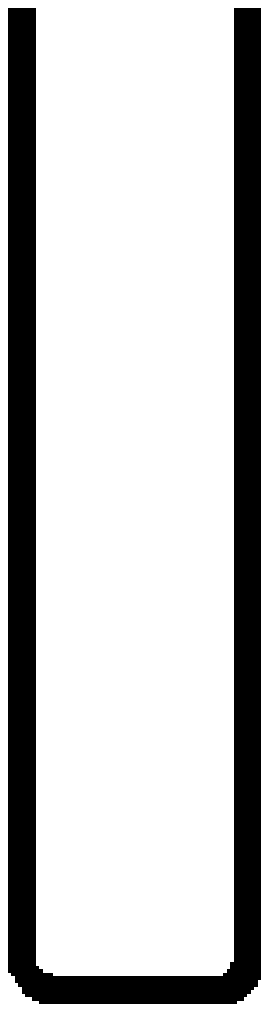}}
\parbox{1.8cm}{\includegraphics[height=0.5cm]{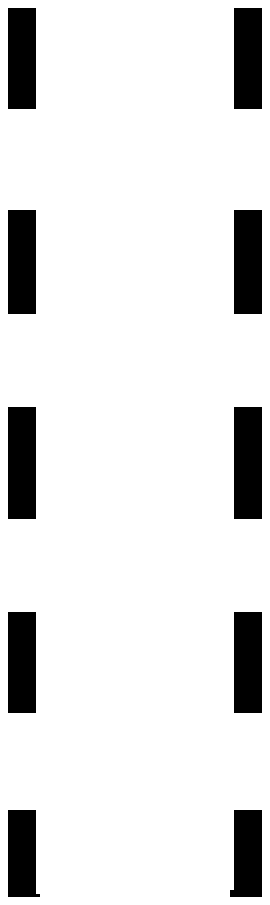}}
\parbox{1.8cm}{\includegraphics[height=0.5cm]{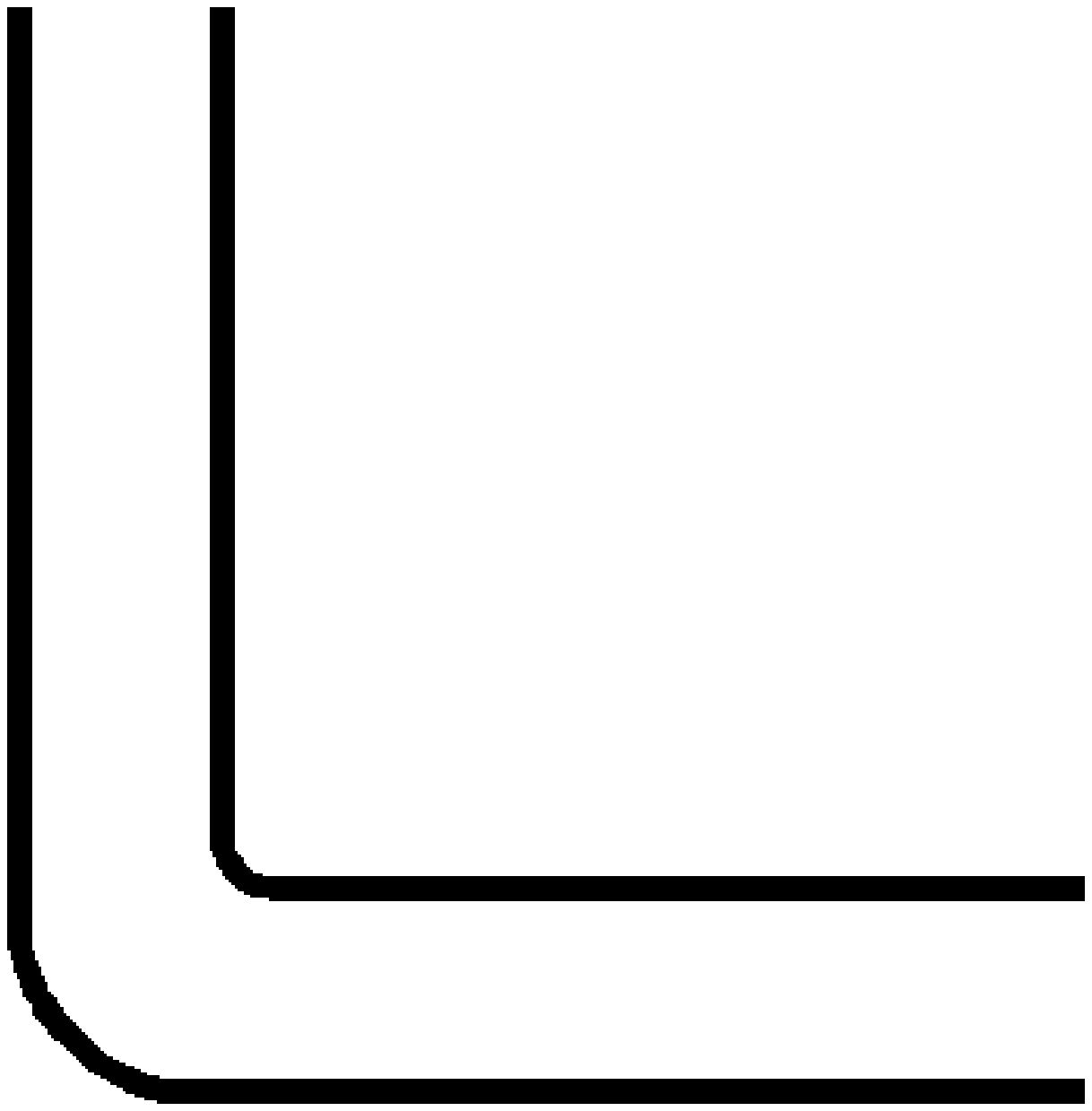}}
\parbox{1.8cm}{\includegraphics[height=0.5cm]{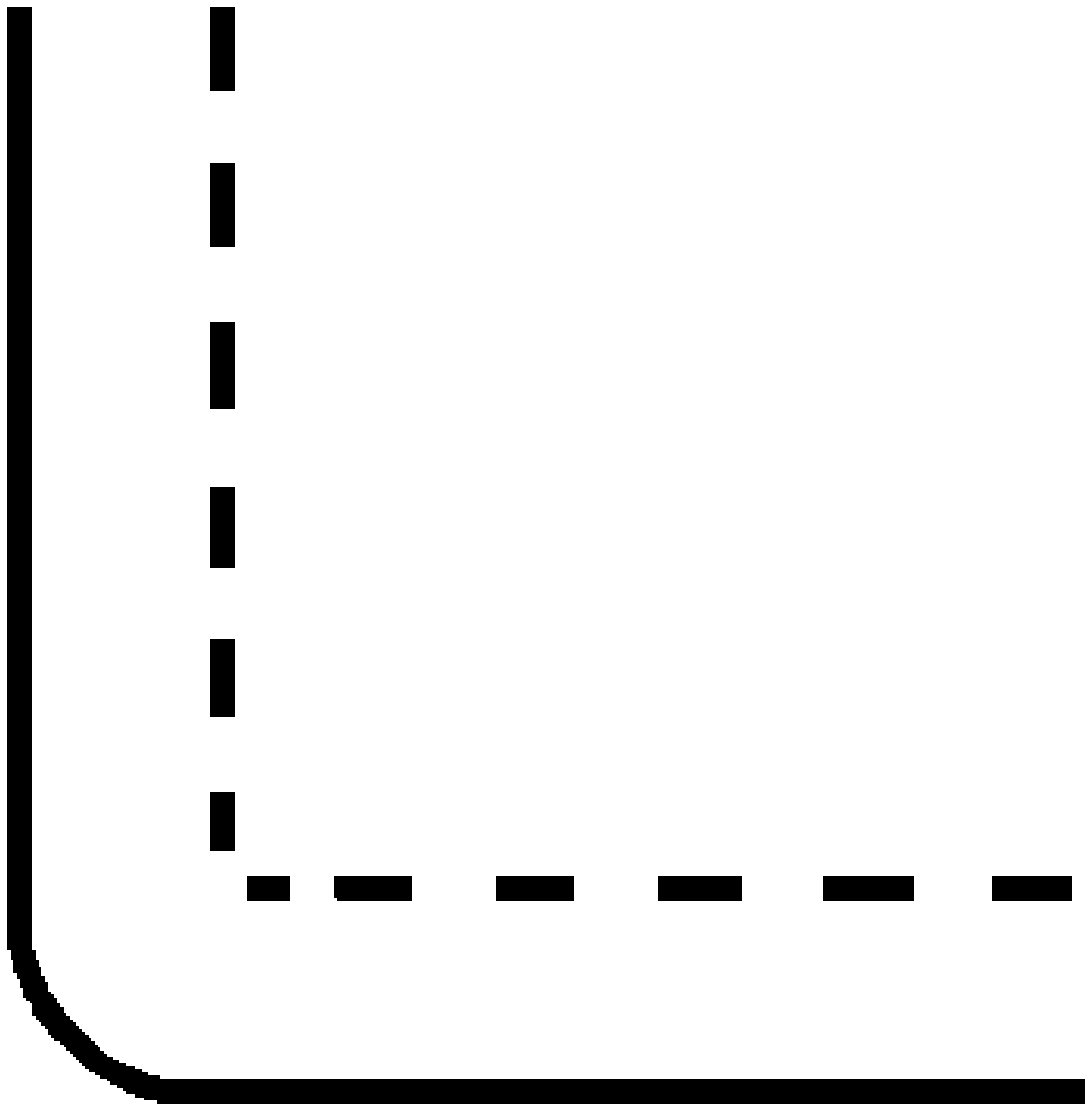}}
\parbox{1.8cm}{\includegraphics[height=1cm]{fig1.5.ps}}
\parbox{1.8cm}{\includegraphics[height=1cm]{fig1.6.ps}}
\parbox{1.8cm}{\includegraphics[height=1cm]{fig2.7.ps}}
\parbox{1.8cm}{\includegraphics[height=1cm]{fig2.9.ps}}
\caption{The first 8 basic index structures which builds up any
conventional vertex factor.\label{indexs}}
\end{figure}

A comparison of the index structures present in other field
expansions and the conventional expansion might also be a working
area for further investigations. Every index structure in the
conventional expansion of the field can be build up from the
following types of index structures, see figure \ref{indexs}.

Thus, it is not possible to consider any index structures in a 3-
or 4-point vertex, which cannot be decomposed into some product of
the above types of index contractions. It might be possible to
extend this to something useful in the analysis of the large-$D$
limit, {\it e.g.}, a simpler description of the large-$D$
diagrammatic truncation of the theory. This is another interesting
working area for further research.

\section{Considerations about the space-time integrals}
The space-time integrals pose certain fundamental restrictions to
our analysis of the large-$D$ limit in gravity. So far we have
considered only the algebraic trace structures of the $n$-point
diagrams. This analysis lead us to a consistent large-$D$ limit of
gravity, where we saw that only bubble and vertex-loop diagrams
will carry the leading contributions in $\left(\frac1D\right)$. In
a large-$N$ limit of a gauge theory this would have been adequate,
however the space-time dimension is much more that just a symmetry
index for the gauge group, and has a deeper significance for the
physical theory than just that of a symmetry index. Space-time
integrations in a $D$-dimensional world will contribute extra
dimensional dependencies to the graphs, and it thus has to be
investigated if the $D$-dimensional integrals could upset the
algebraic large-$D$ limit. The issue about the $D$-dimensional
integrals is not resolved in ref.~\cite{Strominger:1981jg}, only
explicit examples are discussed and some conjectures about the
contributions from certain integrals in the graphs at large-$D$
are stated. It is clear that in the case of the 1-loop separated
integrals the dimensional dependence will go as $\left(\sim
\Gamma(\frac{D-1}{2})\right)$. Such a dimensional dependence can
always be rescaled into an additional redefinition of $(\kappa)$,
{\it i.e.}, we can redefine $\left(\kappa\rightarrow
\frac{\kappa}{\sqrt{C}}\right)$ where $\left(C \sim
\Gamma(\frac{D-1}{2})\right)$. The dimensional dependence from
both bubble graphs and vertex-loop graphs will be possible to
scale away in this manner. Graphs which have a nested structure,
{\it i.e.}, which have shared propagator lines, will be however be
suppressed in $(D)$ compared to the separated loop diagrams. Thus
the rescaling of $(\kappa)$ will make the dimensional dependence
of such diagrams even worse.

In the lack of rigorously proven mathematical statements, it is
hard to be completely certain. But it is seen, that in the
large-$D$ limit the dimensional dependencies from the integrations
in $D$-dimensions will favor the same graphs as the algebraic
graphs limit.

The effective extension of the gravitational action does not pose
any problems in these considerations, concerning the dimensional
dependencies of the integrals and the extra rescalings of
$(\kappa)$. The conjectures made for the Einstein-Hilbert types of
leading loop diagrams hold equally well in the effective field
theory case. A rescaling of the effective coupling constants have
of course to be considered in light of the additional rescaling of
($\kappa$). We will have ($c_i\rightarrow Cc_i$) in this case.
Nested loop diagrams, which algebraically will be down in
$\left(\frac1D\right)$ compared to the leading graphs, will also
be down in $\left(\frac1D\right)$ in the case of an effective
field theory.

To directly resolve the complications posed by the space-time
integrals, in principle all classes of integrals going into the
$n$-point functions of gravity have to be investigated. Such an
investigation would indeed be a very ambitious task and it is
outside the scope of this study. However, as a point for
further investigations, this would be an interesting place to
begin a more rigorous mathematical justification of the large-$D$
limit.

To avoid any complications with extra factors of ($D$) arising
from the evaluation of the integrals, one possibility is to treat
the extra dimensions in the theory as compactified Kaluza-Klein
space-time dimensions. The extra dimensions are then only excited
above the Planck scale. The space-time integration will then only
have to be performed over a finite number of space-time
dimensions, {\it e.g.}, the traditional first four. In such a
theory the large-$D$ behavior is solely determined by the
algebraic traces in the graphs and the integrals cannot not
contribute with any additional factors of ($D$) to the graphs.
Having a cutoff of the theory at the Planck scale is fully
consistent with an effective treatment of the gravitational
action. The effective field theory description is anyway only
valid up until the Planck scale, so an effective action having a
Planckian Kaluza-Klein cutoff of the momentum integrations, is a
possibility.

\section{Comparison of the large-$N$ limit and the large-$D$ limit}
The large-$N$ limit of a gauge theory and the large-$D$ limit in
gravity has some similarities and some dissimilarities which we
will look upon here.

The large-$N$ limit in a gauge theory is the limit where the
symmetry indices of the gauge theory can go to infinity. The idea
behind the gravity large-$D$ limit is to do the same, where in
this case ($D$) will play the role of ($N$), we thus treat the
spatial dimension ($D$) as if it is only a symmetry index for the
theory, {\it i.e.}, a physical parameter, in which we are allowed
to make an expansion.

\begin{figure}[h]\centering
\includegraphics[height=4cm]{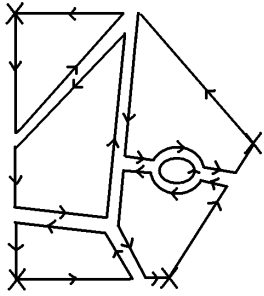}
\caption{A particular planar diagram in the large-$N$ limit. A
$(\times)$ symbolizes an external source, the full lines are index
lines, a full internal index loop gives a factor of
($N$).\label{planar}}
\end{figure}

As it was shown by 't Hooft in ref.~\cite{tHooft} the large-$N$
limit in a gauge theory will be a planar diagram limit, consisting
of all diagrams which can be pictured in a plane. The leading
diagrams will be of the type depicted in figure \ref{planar}.

The arguments for this diagrammatic limit follow similar arguments
to those we applied in the large-$D$ limit of gravity. However the
index loops in the gauge theory are index loops of the internal
symmetry group, and for each type of vertex there will only be one
index structure. Below we show the particular index structure for
a 3-point vertex in a gauge theory.
\begin{figure}[h]\centering
\includegraphics[height=2.5cm]{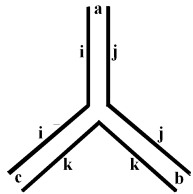}
\caption{The index structure for the 3-point in vertex in a gauge theory.}
\end{figure}

This index structure is also present in the gravity case, {\it
i.e.}, in the (3E) index structure, but various other index
structure are present too, {\it e.g.}, (3A), (3B), $\ldots$, and
each index structure will generically give dissimilar traces, {\it
i.e.}, dissimilar factors of ($D$) for the loops.

In gravity only certain diagrams in the large-$D$ limit carry
leading contributions, but only particular parts of the graph's
amplitudes will survive when ($D\rightarrow\infty$) because of the
different index structures in the vertex factor. The gravity
large-$D$ limit is hence more like a truncated graph limit, where
only some parts of the graph amplitude will be important, {\it
i.e.}, every leading graph will also carry contributions which are
non-leading.

In the Yang-Mills large-$N$ limit the planar graph's amplitudes
will be equally important as a whole. No truncation of the graph's
amplitudes occur. Thus, in a Yang-Mills theory the large-$N$
amplitude is found by summing the full set of planar graphs. This
is a major dissimilarity between the two expansions

The leading diagram limits in the two expansions are not identical
either. The leading graphs in gravity at large-$D$ are given by
the diagrams which have a possibility for having a closed double
trace structure, hence the bubble and vertex-loop diagrams are
favored in the large-$D$ limit. The diagrams contributing to the
gravity large-$D$ limit will thus only be a subset of the full
planar diagram limit.

\section{Other expansions of the fields}
In gravity it is a well-known fact that different definitions of
the fields will lead to different results for otherwise similar
diagrams. Only the full amplitudes will be gauge invariant and
identical for different expansions of the gravitational field.
Therefore comparing the large-$D$ diagrammatical expansion for
dissimilar definitions of the gravitational field, diagram for
diagram, will have no meaning. The analysis of the large-$D$ limit
carried out in this chapter is based on the conventional definition
of the gravitational field, and the large-$D$ diagrammatic
expansion should hence be discussed in this light.

Other field choices may be considered too, and their diagrammatic
limits should be investigated as well. It may be worth some
efforts and additional investigations to see, if the diagrammatic
large-$D$ limit considered here is an unique limit for every
definition of the field, {\it i.e.}, to see if the same
diagrammatic limit always will occur for any definition of the
gravitational field. A scenario for further investigations of
this, could, $e.g.$, be to look upon if the field expansions can be
dressed in such a way that we reach a less complicated
diagrammatic expansion at large-$D$, at the cost of having a more
complicated expansion of the Lagrangian. For example, it could be
that vertex-loop corrections are suppressed for some definitions
of the gravitational field, but that more complicated trace
structures and cancellations between diagrams occurs in the
calculations. In the case of the Goldberg definition of the
gravitational field the vertex index structure is less
complicated ref.~\cite{Strominger:1981jg}, however the trace
contractions of diagrams appears to much more complicated than in
the conventional method. It should be investigated more carefully
if the Goldberg definition of the field may be easier to employ in
calculations. Indeed the diagrammatical limit considered
by ref.~\cite{Strominger:1981jg} is easier, but it is not clear if
results from two different expansions of the gravitational field
are used to derive this. Further investigations should sheet more
light on the large-$D$ limit, and on how a field redefinition may
or may not change its large-$D$ diagrammatic expansion.

Furthermore one can consider the background field method in the
context of the large-$D$ limit as well. This is another working
area for further investigations. The background field method is
very efficient in complicated calculations and it may be useful to
employ it in the large-$D$ analysis of gravity as well. No
problems in using background field methods together with a
large-$D$ quantum gravity expansion seem to be present.

\section{Discussion}
In this chapter we have discussed the large-$D$ limit of effective
quantum gravity. In the large-$D$ limit, a particular subset of
planar diagrams will carry all leading $\left(\frac1D\right)$
contributions to the $n$-point functions. The large-$D$ limit for
any given $n$-point function will consist of the full tree
$n$-point amplitude, together with a set of loop corrections which
will consist of the leading bubble and vertex-loop graphs we have
considered. The effective treatment of the gravitational action
insures that a renormalization of the action is possible and that
none of the $n$-point renormalized amplitudes will carry
uncancelled divergent pole terms. An effective renormalization of
the theory can be performed at any particular dimensionality.

The leading $\left(\frac1D\right)$ contributions to the theory
will be algebraically less complex than the graphs in the full
amplitude. Calculating only the $\left(\frac1D\right)$ leading
contributions simplify explicit calculations of graphs in the
large-$D$ limit. The large-$D$ limit we have found is completely
well defined as long as we do not extend the space-time
integrations to the full $D$-dimensional space-time. That is, we
will always have a consistent large-$D$ theory as long as the
integrations only gain support in a finite dimensional space-time
and the remaining extra dimensions in the space-time are left as
compactified, {\it e.g.}, below the Planck scale.

We hence have a renormalizable, definite and consistent large-$D$
limit for effective quantum gravity -- good below the Planck
scale!

No solution to the problem of extending the space-time integrals
to a full $D$-dimensional space-time has been presented in this
paper, however it is clear that the effective treatment of gravity
do not impose any additional problems in such investigations. The
considerations about the $D$-dimensional integrals discussed in
ref.~\cite{Strominger:1981jg} hold equally well in an effective
field theory, however as well as we will have to make an
additional rescaling of ($\kappa$) in order to account for the
extra dimensional dependence coming from the $D$-dimensional
integrals, we will have make an extra rescaling of the
coefficients ($c_i$) in the effective Lagrangian too.

Our large-$D$ graph limit is not in agreement with the large-$D$
limit of ref.~\cite{Strominger:1981jg}, the bubble graph are
present in both limits, but vertex-loops are not allowed and
claimed to be down in $\left(\frac1D\right)$ the latter. We do not
agree on this point, and believe that this might be a consequence
of comparing similar diagrams with different field definitions. It
is a well-known fact in gravity that dissimilar field definitions
lead to different results for the individual diagrams. Full
amplitudes are of course unaffected by any particular definition
of the field, but results for similar diagrams with different
field choices cannot be immediately compared.

Possible extensions of the large-$D$ considerations discussed here
would be to allow for external matter in the Lagrangian. This
might present some new interesting aspects in the analysis, and
would relate the theory more directly to external physical
observables, {\it e.g.}, scattering amplitudes, and corrections to
the Newton potential at large-$D$ or to geometrical objects such
as a space-time metric. The investigations carried out
in refs.~\cite{B1,B2,B3} could be discussed from a large-$D$ point of view.

Investigations in quantum gravitational cosmology may also present
an interesting work area for applications of the large-$D$ limit,
$e.g.$, quantum gravity large-$D$ big bang models etc.

The analysis of the large-$D$ limit in quantum gravity have
prevailed that using the physical dimension as an expansion
parameter for the theory, is not as uncomplicated as expanding a
Yang-Mills theory at large-$N$. The physical dimension goes into
so many aspects of the theory, {\it e.g.}, the integrals, the
physical interpretations etc, that it is far more complicated to
understand a large-$D$ expansion in gravity than a large-$N$
expansion in a Yang-Mills theory. In some sense it is hard to tell
what the large-$D$ limit of gravity really physically relates to,
{\it i.e.}, what describes a gravitational theory with infinitely
many spatial dimensions? One way to look at the large-$D$ limit in
gravity, might be to see the large-$D$ as a physical phase
transition of the theory at the Planck scale. This suggests that
the large-$D$ expansion of gravity should be seen as a very high
energy scale limit for a gravitational theory.

Large-$N$ expansions of gauge theories ref.~\cite{tHooft} have many
interesting calculational and fundamental applications in, {\it
e.g.}, string theory, high energy, nuclear and condensed matter
physics. A planar diagram limit for a gauge theory will be a
string theory at large distance, this can be implemented in
relating different physical limits of gauge theory and fundamental
strings. In explicit calculations, large-$N$ considerations have
many important applications, {\it e.g.}, in the approximation of
QCD amplitudes from planar diagram considerations or in the theory
of phase-transitions in condensed matter physics, {\it e.g.}, in
theories of superconductivity, where the number of particles will
play the role of ($N$).

High energy physics and string theory are not the only places
where large symmetry investigations can be employed to obtain
useful information. As a calculational tool large-$D$
considerations can be imposed in quantum mechanics to obtain
useful information about eigenvalues and wave-functions of
arbitrary quantum mechanical potentials with a extremely high
precision ref.~\cite{B7}.

Practical applications of the large-$D$ limit in explicit
calculations of $n$-point functions could be the scope for further
investigations as well as certain limits of string theory might be
related to the large-$D$ limit in quantum gravity. The planar
large-$D$ limit found in our investigation might resemble that of
a truncated string limit at large distances? Investigations of the
(Kawai-Lewellen-Tye ref.~\cite{KLT}) open/closed string, gauge
theory/gravity relations, could also be interesting to perform in
the context of a large-$D$ expansion. Such questions are outside
the scope of this investigation, however knowing the huge
importance of large-$N$ considerations in modern physics,
large-$D$ considerations of gravity might pose very interesting
tasks for further research.

\chapter{Summary and outlook}
In this thesis we have have looked at different aspects of quantum
gravity from the view point of general relativity being an effective
field theory. We have discussed in detail how to calculate quantum
gravitational corrections, and presented results of such calculations
of the quantum corrections to both the Schwarzschild and
Kerr metrics as well as to the scattering matrix potentials. The
uniqueness of the quantum corrections has also been dealt with.

Low-energy string limits have also been discussed from the natural
point of view of being specific effective actions at low
energies. Using the Kawai-Lewellen-Tye string relations
as a proposal for a connection between gauge theories and gravity we
have derived a mysterious diagrammatic relationship between higher
derivative operators.
Quantum gravity with an infinitely number of spatial dimensions
have also been looked at. In this case it has been shown that
it is possible to treat general relativity as an effective field
theory even in a large-$D$ limit. As as spin-off of this investigation
the less-complicated diagrammatic limit at ($D=\infty$) might be
used to approximate $n$-point gravity loop amplitudes at ($D=4$).

It has been illustrated quite clearly that general relativity in
its effective extension propose a rather good way to deal with the
challenges of quantum gravity. The theory make perfect sense below
the energies at the Planck scale, meaning that we have excellent theory for
quantum gravity at all energies that can be reached by present
experiments.

The effective approach is in some sense the most general and
natural extension of general relativity one can think of.
It is hard to think of any principles which should prevent
the gravitational action from containing terms which
are only of first order in the curvature tensor. The
effective approach clearly is a very obvious way to deal
with non-renormalizable theories.
Any high-energy theory should in principle correspond to a specific
effective theory and perhaps that is the most problematic issue in the effective
field theory approach. Because effective coupling constants are experimental
quantities they lack predictability, where the energies scales are
to high to carry out experiments.

However, as it hopefully has been shown in this thesis, taking on the effective
approach and investigating various issues concerning quantum gravity lead
in fact to important results, which are unique and interesting from the
point of view of understanding any theory of quantum gravity.

At very high energy scales the quantum picture of the effective field theory
breaks down. About this theory we still do not know much. Further investigations
are needed before we truly understand quantum gravity at such energy scales.
However until a complete theory for quantum gravity is proposed we should
use every aspect of quantum gravity as an effective field theory to uncover
Nature's secrets.

\appendix
\setcounter{equation}{0}
\setcounter{figure}{0}
\renewcommand{\thefigure}{\thesection.\arabic{figure}}
\renewcommand{\theequation}{\thesection.\arabic{equation}}

\chapter{Effective field theory treatment of gravity}\label{Appendix A}
\section{Summary of the vertex rules}\label{vertex}
\subsubsection{Scalar propagator}\noindent
The massive scalar propagator is well known:\\ \\

\begin{minipage}[h]{0.4\linewidth}\vspace{0.4cm}
\centering\includegraphics[scale=1.1]{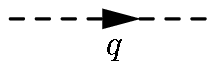}
\end{minipage}
\begin{minipage}[h]{0.65\linewidth}
\centering$\displaystyle = \frac i{q^2-m^2+i\epsilon}$
\end{minipage}

\subsubsection{Photon propagator}\noindent
The photon propagator is also known from the literature.
We have applied Feynman gauge which gives the least
complicated propagator:\\ \\

\begin{minipage}[h]{0.4\linewidth}\vspace{0.4cm}
\centering\includegraphics[scale=1.1]{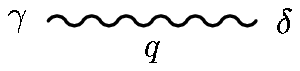}
\end{minipage}
\begin{minipage}[h]{0.65\linewidth}
\centering $\displaystyle = \frac {-i\eta^{\gamma\delta}}{q^2+i\epsilon}$
\end{minipage}

\subsubsection{Graviton propagator}\noindent
The graviton propagator in harmonic gauge is discussed in the
literature refs.~\cite{Veltman1,Donoghue:dn},
but can be derived quite easily explicitly refs.~\cite{Thesis}. We shall write it in the form:\\ \\

\begin{minipage}[h]{0.4\linewidth}\vspace{0.4cm}
\centering\includegraphics[scale=1.1]{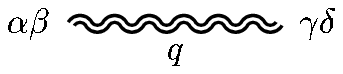}
\end{minipage}
\begin{minipage}[h]{0.65\linewidth}
\centering $\displaystyle = \frac {i{\cal P}^{\alpha\beta\gamma\delta}}{q^2+i\epsilon}$
\end{minipage}\\
where:
\begin{equation}
{\cal P}^{\alpha\beta\gamma\delta} = \frac12\left[\eta^{\alpha\gamma}\eta^{\beta\delta} +
\eta^{\beta\gamma}\eta^{\alpha\delta}
-\eta^{\alpha\beta}\eta^{\gamma\delta}\right]
\end{equation}

\subsubsection{2-scalar-1-photon vertex}\noindent
The 2-scalar-1-photon vertex is well known in the literature. We will
write this vertex as:\\

\begin{minipage}[h]{0.4\linewidth}\vspace{0.15cm}
\centering\includegraphics[scale=1.3]{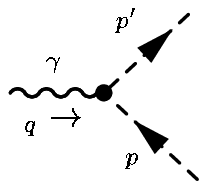}
\end{minipage}
\begin{minipage}[h]{0.65\linewidth}
\centering$\displaystyle = \tau_1^{\gamma}(p,p',e)$
\end{minipage}\vspace{0.3cm}
where:
\begin{equation}
\tau_1^{\gamma}(p,p',e) = -ie\left(p+p'\right)^\gamma.
\end{equation}

\subsubsection{2-photon-1-graviton vertex}\noindent
For the 2-photon-1-graviton vertex we have derived:\\

\begin{minipage}[h]{0.4\linewidth}\vspace{0.15cm}
\centering\includegraphics[scale=1.3]{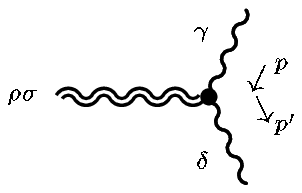}
\end{minipage}
\begin{minipage}[h]{0.65\linewidth}
\centering$\displaystyle = \tau_3^{\rho\sigma(\gamma \delta) }(p,p')$
\end{minipage}\vspace{0.3cm}
where:
\begin{equation}\begin{aligned}
\tau_3^{\rho\sigma(\gamma\delta)}(p,p') &= i\kappa\Big[{\cal
P}^{\rho\sigma(\gamma\delta)}(p\cdot p')
+\frac12\Big(\eta^{\rho\sigma} p^\delta p^{\prime\gamma} +
\eta^{\gamma\delta} (p^\rho p^{\prime \sigma} + p^\sigma p^{\prime
\rho})\\& - (p^{\prime\gamma}p^\sigma \eta^{\rho\delta} +
p^{\prime\gamma}p^\rho \eta^{\sigma\delta} +
p^{\prime\rho}p^\delta \eta^{\sigma\gamma} +
p^{\prime\sigma}p^\delta \eta^{\rho\gamma})\Big)\Big].
\end{aligned}\end{equation}

\subsubsection{2-scalar-2-photon vertex}\noindent
The 2-scalar-2-photon vertex is also well known from scalar QED. We will
write it as:\\

\begin{minipage}[h]{0.4\linewidth}\vspace{0.15cm}
\centering\includegraphics[scale=1.3]{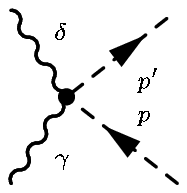}
\end{minipage}
\begin{minipage}[h]{0.65\linewidth}
\centering $\displaystyle = \tau_4^{\gamma\delta}(p,p',e)$
\end{minipage}\vspace{0.3cm}
where:
\begin{equation}
\tau_4^{\gamma\delta}(p,p',e) = 2ie^2\eta^{\gamma\delta}.
\end{equation}

\subsubsection{2-scalar-1-photon-1-graviton vertex}\noindent
For the 2-scalar-1-photon-1-graviton vertex we have derived:\\

\begin{minipage}[h]{0.4\linewidth}\vspace{0.15cm}
\centering\includegraphics[scale=1.3]{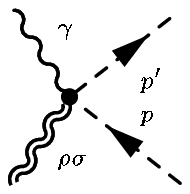}
\end{minipage}
\begin{minipage}[h]{0.65\linewidth}
\centering $\displaystyle = \tau_5^{\rho\sigma(\gamma)}(p,p',e)$
\end{minipage}\vspace{0.3cm}
where:
\begin{equation}
\tau_5^{\rho\sigma(\gamma)}(p,p',e) = ie\kappa\left[{\cal P}^{\rho\sigma\alpha\gamma}(p+p')_\alpha\right],
\end{equation}
and ${\cal P}^{\rho\sigma\alpha\gamma}$ is defined as above.

For all vertices the rules of momentum conservation have been applied.
For the external scalar lines we associate a factor of 1. At each loop we will
integrate over the undetermined loop momentum.

For a certain diagram we will divide with the appropriate symmetry factor
of the Feynman diagram.

\subsubsection{2-scalar-1-graviton vertex}\noindent The
2-scalar-1-graviton vertex is well known. We will
write it in the following way:\\

\begin{minipage}[h]{0.4\linewidth}\vspace{0.15cm}
\centering\includegraphics[scale=1.3]{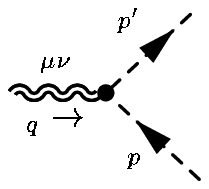}
\end{minipage}
\begin{minipage}[h]{0.65\linewidth}
\centering$\displaystyle = \tau_6^{\mu \nu}(p,p',m)$
\end{minipage}\vspace{0.3cm}
where:
\begin{equation}
\tau_6^{\mu\nu}(p,p',m) = -\frac{i\kappa}2\left[p^\mu p^{\prime \nu}
+p^\nu p^{\prime \mu} - \eta^{\mu\nu}\left((p\cdot
p^\prime)-m^2\right)\right].
\end{equation}

\subsubsection{2-fermion-1-graviton vertex}\noindent The
2-fermion-1-graviton vertex has been derived and is known from the
literature. We will write it in the following way:\\

\begin{minipage}[h]{0.4\linewidth}\vspace{0.15cm}
\centering\includegraphics[scale=1.3]{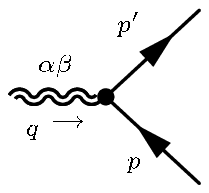}
\end{minipage}
\begin{minipage}[h]{0.65\linewidth}
\centering$\displaystyle = \tau_{7}^{\alpha\beta}(p,p',m)$
\end{minipage}\vspace{0.3cm}
where:
\begin{equation}
\tau_{7}^{\alpha\beta}(p,p',m)={-i\kappa\over 2}\left[{1\over
4}(\gamma^\alpha(p+p')^\beta+\gamma^\beta(p+p')^\alpha)-{1\over
2}\eta^{\alpha\beta}({1\over
2}(\not\!\!{p}+\not\!\!{p}')-m)\right].
\end{equation}

\subsubsection{2-scalar-2-graviton vertex} A discussion of the
2-scalar-2-graviton vertex can be found in the literature. In the
expression below it will be written with the full
symmetry of the two gravitons:\\

\begin{minipage}[h]{0.4\linewidth}\vspace{0.15cm}
\centering\includegraphics[scale=1.3]{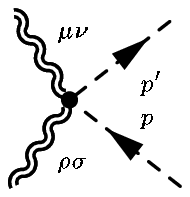}
\end{minipage}
\begin{minipage}[h]{0.65\linewidth}
\centering$\displaystyle = \tau_8^{\eta\lambda\rho\sigma}(p,p',m)$
\end{minipage}\vspace{0.3cm}
where:
\begin{equation}{\begin{aligned}
\tau_8^{\eta \lambda \rho \sigma}(p,p')&= {i\kappa^2} \bigg [
\left \{I^{\eta \lambda\alpha \delta} {I}^{\rho \sigma\beta}_{\ \
\ \ \delta} - \frac14\left\{\eta^{\eta \lambda} I^{\rho
\sigma\alpha \beta} +
 \eta^{\rho \sigma} I^{\eta \lambda\alpha \beta} \right \}
\right \} \left (p_\alpha p^\prime_{\beta} + p^\prime_{\alpha}
p_\beta
\right ) \\
&-\frac12 \left \{ I^{\eta \lambda\rho \sigma} - \frac12\eta^{\eta
\lambda}\eta^{\rho \sigma} \right \} \left [ (p\cdot p') - m^2
\right]\bigg].\end{aligned}}
\end{equation}

\subsubsection{2-fermion-2-graviton vertex}\noindent The
2-fermion-2-graviton vertex has been derived.
We will write it in the following way:\\

\begin{minipage}[h]{0.4\linewidth}\vspace{0.15cm}
\centering\includegraphics[scale=1.3]{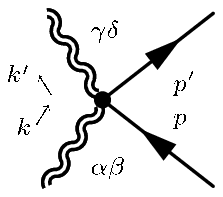}
\end{minipage}
\begin{minipage}[h]{0.65\linewidth}
\centering$\displaystyle = \tau_9^{\alpha\beta,\gamma\delta}(p,p',m)$
\end{minipage}\vspace{0.3cm}
where:
\begin{equation}\begin{aligned}
\tau_9^{\alpha\beta,\gamma\delta}(p,p',m)&=i\kappa^2\left\{-{1\over
2}({1\over
2}(\not\!\!{p}+\not\!\!{p}')-m){\cal P}^{\alpha\beta,\gamma\delta}\right.\nonumber
-\left.{1\over
16}[\eta^{\alpha\beta}(\gamma^\gamma(p+p')^\delta+\gamma^\delta(p+p')^\gamma)
\right.\nonumber\\
&+\left.\eta^{\gamma\delta}(\gamma^\alpha
(p+p')^\beta+\gamma^\beta(p+p')^\alpha)]\right.\nonumber+\left.{3\over
16}(p+p')_{\epsilon}\gamma^{\xi}(I^{\xi\phi,\alpha\beta}{I_{\phi}}^{\epsilon,\gamma\delta}
+I^{\xi\phi,\gamma\delta}{I_{\phi}}^{\epsilon,\alpha\beta})\right.\nonumber\\
&+\left.{i\over 16}\epsilon^{\rho\sigma\eta\lambda}\gamma_\lambda
\gamma_5({I^{\alpha\beta}_{\ \ \ ,\eta\nu}}
I^{\gamma\delta}_ {\ \ \ ,\sigma\nu}{k'}_\rho-{I^{\gamma\delta}_{\ \ \ ,\eta\nu}}
I^{\alpha\beta}_{\ \ \ , \sigma\nu}k_\rho)\right\}.
\end{aligned}\end{equation}

\subsubsection{3-graviton vertex} The 3-graviton vertex in
the background field method has the
form refs.~\cite{Donoghue:1993eb,Donoghue:dn}

\begin{minipage}[h]{0.4\linewidth}\vspace{0.15cm}
\centering\includegraphics[scale=1.3]{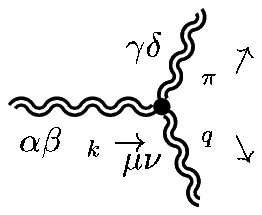}
\end{minipage}
\begin{minipage}[h]{0.65\linewidth}
\centering$\displaystyle =
{\tau_{10}}_{\alpha\beta\gamma\delta}^{\mu\nu}(k,q)$
\end{minipage}\vspace{0.3cm}
where:
\begin{equation}{
\begin{aligned}
{\tau_{10}}_{\alpha  \beta \gamma \delta }^{\mu
\nu}(k,q)&=-\frac{i\kappa}2\times \bigg({\cal P}_{\alpha \beta
\gamma \delta }\bigg[k^\mu k^\nu+ (k-q)^\mu (k-q)^\nu +q^\mu
q^\nu-
\frac32\eta^{\mu \nu}q^2\bigg]\\[0.00cm]&
+2q_\lambda q_\sigma\bigg[ I_{\alpha \beta }^{\ \ \
\sigma\lambda}I_{\gamma \delta }^{\ \ \ \mu \nu} + I_{\gamma
\delta }^{\ \ \ \sigma\lambda}I_{\alpha \beta }^{\ \ \ \mu \nu}
-I_{\alpha \beta }^{\ \ \ \mu  \sigma} I_{\gamma \delta }^{\ \ \
\nu \lambda} - I_{\gamma \delta }^{\ \ \ \mu \sigma} I_{\alpha
\beta }^{\ \ \ \nu \lambda}
\bigg]\\[0cm]&
+\bigg[q_\lambda q^\mu \bigg(\eta_{\alpha \beta }I_{\gamma \delta
}^{\ \ \ \nu \lambda}+\eta_{\gamma \delta }I_{\alpha \beta }^{\ \
\ \nu \lambda}\bigg) +q_\lambda q^\nu \left(\eta_{\alpha \beta
}I_{\gamma \delta }^{\ \ \ \mu \lambda}+\eta_{\gamma \delta
}I_{\alpha \beta }^{\ \ \ \mu  \lambda}\right)\\&
-q^2\left(\eta_{\alpha \beta }I_{\gamma \delta }^{\ \ \ \mu
\nu}-\eta_{\gamma \delta }I_{\alpha \beta }^{\ \ \ \mu \nu}\right)
-\eta^{\mu \nu}q_\sigma q_\lambda\left(\eta_{\alpha \beta
}I_{\gamma \delta }^{\ \ \ \sigma\lambda} +\eta_{\gamma \delta
}I_{\alpha \beta }^{\ \ \
\sigma\lambda}\right)\bigg]\\[0cm]&
+\bigg[2q_\lambda\big(I_{\alpha \beta }^{\ \ \
\lambda\sigma}I_{\gamma \delta \sigma}^{\ \ \ \ \nu}(k-q)^\mu
+I_{\alpha \beta }^{\ \ \ \lambda\sigma}I_{\gamma \delta
\sigma}^{\ \ \ \ \mu }(k-q)^\nu -I_{\gamma \delta }^{\ \ \
\lambda\sigma}I_{\alpha \beta \sigma}^{\ \ \ \ \nu}k^\mu
-I_{\gamma \delta }^{\ \ \ \lambda\sigma}I_{\alpha \beta
\sigma}^{\ \ \ \ \mu }k^\nu \big)\\& +q^2\left(I_{\alpha \beta
\sigma}^{\ \ \ \ \mu }I_{\gamma \delta }^{\ \ \ \nu \sigma} +
I_{\alpha \beta }^{\ \ \ \nu \sigma}I_{\gamma \delta \sigma}^{\ \
\ \ \mu }\right) +\eta^{\mu \nu}q_\sigma q_\lambda\left(I_{\alpha
\beta }^{\ \ \ \lambda\rho}I_{\gamma \delta  \rho}^{\ \ \ \
\sigma} +I_{\gamma \delta }^{\ \ \ \lambda\rho}I_{\alpha \beta
\rho}^{\ \ \ \
\sigma}\right)\bigg]\\[0cm]&
+\bigg\{(k^2+(k-q)^2)\big[I_{\alpha \beta }^{\ \ \ \mu
\sigma}I_{\gamma \delta \sigma}^{\ \ \ \ \nu} +I_{\gamma \delta
}^{\ \ \ \mu  \sigma}I_{\alpha \beta \sigma}^{\ \ \ \ \nu}
-\frac12\eta^{\mu \nu}{\cal P}_{\alpha \beta \gamma \delta
}\big]\\&-\left(I_{\gamma \delta }^{\ \ \ \mu \nu}\eta_{\alpha
\beta }k^2+I_{\alpha \beta }^{\ \ \ \mu \nu}\eta_{\gamma \delta
}(k-q)^2\right)\bigg\}\bigg).
\end{aligned}}
\end{equation}

\section{Needed integrals in the calculation of diagrams}\label{integrals}
To calculate the diagrams the following integrals are needed
\begin{eqnarray}\displaystyle
J=&\displaystyle \int\frac{d^4l}{(2\pi)^4} \frac{1}{l^2(l+q)^2} & = \frac{i}{32\pi^2}\big[-2L\big] + \ldots,\\
J_\mu=&\displaystyle\int\frac{d^4l}{(2\pi)^4} \frac{l_\mu}{l^2(l+q)^2} & =   \frac{i}{32\pi^2}\Big[q_\mu L\Big]+\ldots,\\
J_{\mu\nu}=&\displaystyle\int\frac{d^4l}{(2\pi)^4} \frac{l_\mu l_\nu}{l^2(l+q)^2}
& = \frac{i}{32\pi^2}\bigg[q_\mu q_\nu \Big(-\frac23L\Big) - q^2\eta_{\mu\nu}\Big(-\frac16 L\Big)\bigg]+\ldots,
\end{eqnarray}
together with
\begin{equation}\begin{split}
I=\int\frac{d^4l}{(2\pi)^4} \frac{1}{l^2(l+q)^2((l+k)^2-m^2)}  = \frac{i}{32\pi^2m^2}\big[-L-S\big]+\ldots,
\end{split}\end{equation}
\begin{equation}\begin{split}
I_\mu&=\int\frac{d^4l}{(2\pi)^4} \frac{l_\mu}{l^2(l+q)^2((l+k)^2-m^2)}
\\ &=  \frac{i}{32\pi^2m^2}\bigg[k_\mu\bigg(\Big(-1-\frac12 \frac{q^2}{m^2}\Big)L-
\frac14\frac{q^2}{m^2}S\bigg)+q_\mu\bigg(L+\frac12S\bigg)\bigg]+\ldots,
\end{split}\end{equation}
\begin{equation}\begin{split}
I_{\mu\nu}&=\int\frac{d^4l}{(2\pi)^4} \frac{l_\mu l_\nu}{l^2(l+q)^2((l+k)^2-m^2)}\\
 & =  \frac{i}{32\pi^2m^2}\bigg[q_\mu q_\nu\bigg(-L-\frac38 S\bigg)
+k_\mu k_\nu \bigg(-\frac12\frac{q^2}{m^2}L-\frac18\frac{q^2}{m^2}S\bigg)\\
&+\big(q_\mu k_\nu + q_\nu k_\mu \big)\bigg(\Big(\frac12 + \frac12\frac{q^2}{m^2}\Big)L + \frac{3}{16}\frac{q^2}{m^2}S\bigg)+
q^2\eta_{\mu\nu}\Big(\frac14L+\frac18S\Big)
\bigg]+\ldots,
\end{split}\end{equation}
\begin{equation}\begin{split}
I_{\mu\nu\alpha}&=\int\frac{d^4l}{(2\pi)^4} \frac{l_\mu l_\nu l_\alpha}{l^2(l+q)^2((l+k)^2-m^2)}\\ & =  \frac{i}{32\pi^2m^2}\bigg[
q_\mu q_\nu q_\alpha\bigg(L+\frac5{16}S\bigg)+k_\mu k_\nu k_\alpha\bigg(-\frac16 \frac{q^2}{m^2}\bigg)
\\ &+\big(q_\mu k_\nu k_\alpha + q_\nu k_\mu k_\alpha + q_\alpha k_\mu k_\nu\big)\bigg(\frac13\frac{q^2}{m^2}L+
\frac1{16}\frac{q^2}{m^2}S\bigg)
\\&+\big(q_\mu q_\nu k_\alpha + q_\mu q_\alpha k_\nu + q_\nu q_\alpha k_\mu \big)\bigg(\Big(-\frac13 - \frac12\frac{q^2}{m^2}\Big)L
-\frac{5}{32}\frac{q^2}{m^2}S\bigg)
\\ &+\big(\eta_{\mu\nu}k_\alpha + \eta_{\mu\alpha}k_\nu + \eta_{\nu\alpha}k_\mu\big)\Big(\frac1{12}q^2L\Big)
\\ &+\big(\eta_{\mu\nu}q_\alpha + \eta_{\mu\alpha}q_\nu + \eta_{\nu\alpha}q_\mu\big)\Big(-\frac16q^2L -\frac1{16}q^2S\Big)
\bigg]+\ldots,
\end{split}\end{equation}
where $\left(L=\ln(-q^2)\right)$ and
$\left(S=\frac{\pi^2m}{\sqrt{-q^2}}\right)$. In the above
integrals only the lowest order non-analytical terms are
presented. The ellipses denote higher order non-analytical
contributions as well as the neglected analytical terms.
Furthermore the following identities hold true for the on shell
momenta, $\left(k\cdot q = \frac{q^2}2\right)$, where $\left(k-k'=q\right)$ and
$\left(k^2=m^2=k^{\prime 2}\right)$. In some cases the integrals are needed
with $\left(k\right)$ replaced by $\left(-k'\right)$, where $\left(k' \cdot q = -\frac{q^2}{2}\right)$,
these results, are obtained by replacing everywhere $\left(k\right)$ with
$\left(-k'\right)$. This can be verified explicitly. The above integrals check
with the results of ref.~\cite{Donoghue:dn}.

The following integrals are needed to do the box diagrams. The ellipses denote higher
order contributions of non-analytical terms as well as neglected analytical
terms:
\begin{equation}\begin{split}
K  &= \int\frac{d^4l}{(2\pi)^4} \frac{1}{l^2(l+q)^2((l+k_1)^2-m_1^2)((l-k_3)^2-m_2^2)}\\ & =   \frac{i}{16\pi^2m_1 m_2 q^2}\bigg[\Big(1-\frac{w}{3m_1m_2}\Big)L - {i\pi m_1 m_2\over
(m_1 + m_2)p}\bigg]+\ldots, \\
K' &= \int\frac{d^4l}{(2\pi)^4} \frac{1}{l^2(l+q)^2((l+k_1)^2-m_1^2)((l+k_4)^2-m_2^2)}\\ & =   \frac{i}{16\pi^2m_1 m_2 q^2}\bigg[\Big(-1+\frac{W}{3m_1m_2}\Big)L\bigg]+\ldots.
\end{split}\end{equation}
where
\begin{equation}
p=\bigg[{(s-(m_1+m_2)^2)(s-(m_1-m_2)^2)\over 4s}\bigg]^{1\over 2}
\end{equation}
equals the center of mass momentum.
Here $\left(k_1\cdot q = \frac{q^2}2\right)$,
$\left(k_2\cdot q = -\frac{q^2}2\right)$,
$\left(k_3\cdot q = -\frac{q^2}2\right)$
and $\left(k_4\cdot q = \frac{q^2}2\right)$, where
$\left(k_1-k_2=k_4-k_3=q\right)$ and $\left(k_1^2=m_1^2=k_2^2\right)$
together with $\left(k_3^2=m_2^2=k_4^2\right)$. Furthermore we have defined
$\left(w = (k_1\cdot k_3)-m_1m_2\right)$
and $\left(W = (k_1\cdot k_4) - m_1m_2\right)$.
The above results for the integrals
check with ref.~\cite{Donoghue:1996mt}.

The algebraic structure of the calculated diagrams is rather involved and complicated, but
yields no complications using our algebraic program. The integrals are rather
complicated to do, but one can make use of various contraction rules for the integrals which
hold true on the mass-shell.

From the choice $\left(q=k_1-k_2=k_4-k_3\right)$ one can easily derive:
\begin{equation}\begin{split}
k_1\cdot q & = k_4 \cdot q = -k_2\cdot q = -k_3 \cdot q = \frac{q^2}{2},\\
k_1\cdot k_2 & = m_1^2 - \frac{q^2}{2},\\
k_3\cdot k_4 & = m_2^2 - \frac{q^2}{2},
\end{split}\end{equation}
where $\left(k_1^2=k_2^2=m_1^2\right)$ and  $\left(k_3^2=k_4^2=m_2^2\right)$ on the mass shell.

On the mass shell we have identities like:
\begin{equation}\begin{split}
l \cdot q & = \frac{(l+q)^2-q^2-l^2}2,\\
l \cdot k_1 & = \frac{(l+k_1)^2-m_1^2-l^2}2,\\
l \cdot k_3 & = -\frac{(l-k_3)^2-m_2^2-l^2}2.
\end{split}\end{equation}

Now clearly, $e.g.$:
\begin{equation}\begin{split}
&\int\frac{d^4l}{(2\pi)^4} \frac{l\cdot q} {l^2(l+q)^2((l+k_1)^2-m_1^2)((l-k_3)^2-m_2^2)}  \\ & =
\frac12\int\frac{d^4l}{(2\pi)^4} \frac{(l+q)^2-q^2-l^2} {l^2(l+q)^2((l+k_1)^2-m_1^2)((l-k_3)^2-m_2^2)},
\end{split}\end{equation}
as only the integral with $\left(q^2\right)$ yield the non-analytical terms we let:
\begin{equation}\begin{split}
&\int\frac{d^4l}{(2\pi)^4} \frac{l\cdot q} {l^2(l+q)^2((l+k_1)^2-m_1^2)((l-k_3)^2-m_2^2)} \\ & \rightarrow
\frac{-q^2}{2}\int\frac{d^4l}{(2\pi)^4} \frac{1} {l^2(l+q)^2((l+k_1)^2-m_1^2)((l-k_3)^2-m_2^2)}.
\end{split}\end{equation}
A perhaps more significant reduction of the integrals is with the contraction of the source's momenta, $e.g.$:
\begin{equation}\begin{split}
&\int\frac{d^4l}{(2\pi)^4} \frac{l\cdot k_1} {l^2(l+q)^2((l+k_1)^2-m_1^2)((l-k_3)^2-m_2^2)} \\ & =
\frac12\int\frac{d^4l}{(2\pi)^4} \frac{(l+k_1)^2-m_1^2-l^2} {l^2(l+q)^2((l+k_1)^2-m_1^2)((l-k_3)^2-m_2^2)} \\ &
\rightarrow
\frac12\int\frac{d^4l}{(2\pi)^4} \frac{1}{l^2(l+q)^2((l-k_3)^2-m_2^2)},
\end{split}\end{equation}
or
\begin{equation}\begin{split}
&\int\frac{d^4l}{(2\pi)^4} \frac{l\cdot k_3} {l^2(l+q)^2((l+k_1)^2-m_1^2)((l-k_3)^2-m_2^2)}\\ & =
\int\frac{d^4l}{(2\pi)^4} \frac{-(l-k_3)^2+m_2^2+l^2} {l^2(l+q)^2((l+k_1)^2-m_1^2)((l-k_3)^2-m_2^2)}
\\ & \rightarrow
-\frac12\int\frac{d^4l}{(2\pi)^4} \frac{1} {l^2(l+q)^2((l+k_1)^2-m_1^2)},
\end{split}\end{equation}
as seen the contraction of a loop momentum factor with a source's momentum factor removes one
of the propagators leaving a much simpler loop integral.
Such reductions in the box diagram integrals help to do the calculations.

For the above integrals the following constraints for the non-analytical terms can be verified directly on
the mass-shell:
\begin{equation}
I_{\mu\nu\alpha}\eta^{\alpha\beta} = I_{\mu\nu}\eta^{\mu\nu} = J_{\mu\nu}\eta^{\mu\nu} = 0,
\end{equation}
\begin{equation}\begin{split}
I_{\mu\nu\alpha}q^\alpha = -\frac{q^2}{2}I_{\mu\nu}, \ \ \ \ I_{\mu\nu}q^\nu  = -\frac{q^2}{2}I_{\mu}, \ \ \ \   I_{\mu}q^\mu  = -\frac{q^2}{2}I \\
J_{\mu\nu}q^\nu = -\frac{q^2}{2}J_{\mu}, \ \ \ \ J_{\mu}q^\mu   = -\frac{q^2}{2}J,
\end{split}\end{equation}
\begin{equation}
I_{\mu\nu\alpha}k^\alpha = \frac{1}{2}J_{\mu\nu}, \ \ \ \ I_{\mu\nu}k^\nu  = \frac{1}{2}J_{\mu}, \ \ \ \   I_{\mu}k^\mu  = \frac{1}{2}J,
\end{equation}
These mass-shell constraints can be used to derive the algebraic expressions
for the non-analytic parts of the integrals.

\section{Fourier transforms}\label{Fourier}
In these section we present the Fourier transforms used to
calculate the long-range corrections to the energy-momentum tensor in
ref.~\cite{B2}. The integrals needed in the calculations
of the classical corrections are:
\begin{eqnarray}
& &\int{d^3q\over (2\pi)^3}e^{i\vec{q}\cdot\vec{r}}|\vec{q}|=-{1\over
\pi^2r^4},\nonumber\\
&
&\int{d^3q\over(2\pi)^3}e^{i\vec{q}\cdot\vec{r}}q_j|\vec{q}|={-4ir_j\over
\pi^2r^6},\nonumber\\
& &\int{d^3q\over (2\pi)^3}e^{i\vec{q}\cdot\vec{r}}{q_iq_j\over
|\vec{q}|} = {1\over
\pi^2r^4}\left(\delta_{ij}-4{r_ir_j\over r^2}\right),\label{eq:r1}
\end{eqnarray}
while we have for the quantum effects:
\begin{eqnarray}
& &\int{d^3q\over (2\pi)^3}e^{i\vec{q}\cdot\vec{r}}\vec{q}^2\log
\vec{q}^2 ={3\over \pi r^5},\nonumber\\
& &  \int{d^3q\over (2\pi)^3}e^{i\vec{q}\cdot\vec{r}}{q_j\vec{q}^2\log
\vec{q}^2}={i15r_j\over \pi r^7},\nonumber\\
& & \int{d^3q\over (2\pi)^3}e^{i\vec{q}\cdot\vec{r}}q_iq_j\log
\vec{q}^2={{-3\over 2\pi r^5}} \left({\delta_{ij}-5{r_i r_j \over r^2}}\right).\label{eq:r2}
\end{eqnarray}
We also need the integrals:
\begin{eqnarray}
& &   \int{d^3q\over
   (2\pi)^3}e^{i\vec{q}\cdot\vec{r}}{1\over |\vec{q}|}
={1\over 2\pi^2r^2}, \nonumber \\
 & &   \int{d^3q\over
   (2\pi)^3}e^{i\vec{q}\cdot\vec{r}}{q_j\over |\vec{q}|}
={ir_j\over \pi^2r^4},\nonumber\\
& &   \int{d^3q\over
   (2\pi)^3}e^{i\vec{q}\cdot\vec{r}}{q_iq_j\over |\vec{q}|^3}={1\over
   2\pi^2r^2}\left(\delta_{ij}-2{r_ir_j\over r^2}\right).\label{eq:r3}
   \end{eqnarray}
and
\begin{eqnarray}
& &\int{d^3q\over (2\pi)^3}e^{i\vec{q}\cdot\vec{r}}
\log\vec{q}^2=-{1\over 2\pi r^3}, \nonumber \\
& &\int{d^3q\over (2\pi)^3}e^{i\vec{q}\cdot\vec{r}}
q_j\log\vec{q}^2={-i3r_j\over 2\pi
r^5},\nonumber\\
& &\int{d^3q\over (2\pi)^3}e^{i\vec{q}\cdot\vec{r}}
\left({q_iq_j\over
\vec{q}^2}\right)\log\vec{q}^2
={1\over 2\pi r^3}\left(\delta_{ij} {(1-\log r)} - {r_ir_j \over r^2}(4-3\log r)\right).
\end{eqnarray}

\chapter{Quantum gravity at large numbers of dimensions}\label{Appendix B}
\section{Effective 3- and 4-point vertices}\label{app1}
The effective Lagrangian takes the form:
\begin{equation}
{\cal L} = \int d^Dx
\sqrt{-g}\Big(\frac{2R}{\kappa^2}+c_1R^2+c_2R_{\mu\nu}R^{\mu\nu}+\ldots\Big).
\end{equation}
In order to find the effective vertex factors we need to expand:
$\left(\sqrt{-g}R^2\right)$ and $\left(\sqrt{-g}R_{\mu\nu}^2\right)$.

We are working in the conventional expansion of the field, so we
define:
\begin{equation}
g_{\mu\nu} \equiv \eta_{\mu\nu}+\kappa h_{\mu\nu}.
\end{equation}
To second order we find the following expansion for ($\sqrt{-g}$):
\begin{equation}
\sqrt{-g} = \exp\left(\frac12\log (\eta_{\mu\nu}+\kappa
h_{\mu\nu})\right) =
\left(1+\frac\kappa2h_\alpha^\alpha-\frac{\kappa^2}4h_\alpha^\beta
h^\alpha_\beta+\frac{\kappa^2}8(h_\alpha^\alpha)^2\ldots\right).
\end{equation}
To first order in ($\kappa$) we have the following expansion for
($R$):
\begin{equation}\begin{split}
R^{(1)}=\kappa\left[\partial_{\alpha}\partial^{\alpha}h_\beta^\beta-\partial^\alpha\partial^\beta
h_{\alpha\beta}\right],
\end{split}\end{equation}
and to second order in ($\kappa^2$) we find:
\begin{equation}\begin{split}
R^{(2)}&=\kappa^2\Big[-\frac12\partial_\alpha\left[h_\mu^\beta\partial^\alpha
h_\beta^\mu\right]+\frac12\partial_\beta\left[h_\nu^\beta(2\partial_\alpha
h^{\nu\alpha} - \partial_\nu
h_\alpha^\alpha)\right]\\&+\frac14\left[\partial_\alpha h_\beta^\nu +
\partial_\beta h_\alpha^\nu - \partial^\nu h_{\beta\alpha}\right]\left[\partial^\alpha h_\nu^\beta + \partial_\nu h^{\beta\alpha}-\partial^\beta
h_\nu^\alpha\right]\\
&-\frac14\left[2\partial_\alpha h^{\nu\alpha}-\partial^\nu
h_\alpha^\alpha\right]\partial_\nu h_\beta^\beta
-\frac12h^{\nu\alpha}\partial_\nu\partial_\alpha h^\beta_\beta +
\frac12h_\alpha^\nu\partial_\beta\left[\partial^\alpha
h_\nu^\beta+\partial_\nu h^{\beta\alpha} - \partial^\beta
h_\nu^\alpha\right]\Big]
\end{split}\end{equation}
In the same way we find for ($R_{\mu\nu}$):
\begin{equation}\begin{split}
R^{(1)}_{\nu\alpha} = \frac{\kappa}2
\left[\partial_\nu\partial_\alpha h_\beta^\beta-
\partial_\beta\partial_\alpha h_\nu^\beta-
\partial_\beta\partial_\nu h_\alpha^\beta + \partial^2
h_{\nu\alpha}\right],
\end{split}\end{equation}
and
\begin{equation}\begin{split}
R^{(2)}_{\nu\alpha} &=\kappa^2\Big[
-\frac12\partial_\alpha\left[h^{\beta\lambda}\partial_\nu
h_{\lambda\beta}\right] +
\frac12\partial_\beta\left[h^{\beta\lambda}(\partial_\nu
h_{\lambda\alpha}+\partial_\alpha h_{\nu\lambda} -
\partial_\lambda h_{\nu\alpha})\right]\\
&+\frac14\left[\partial_\beta h_\nu^\lambda +
\partial_\nu h_\beta^\lambda - \partial^\lambda h_{\nu\beta}\right]\left[\partial^\lambda h^\beta_\alpha + \partial_\alpha h^{\beta}_\lambda-\partial^\beta h_{\nu\alpha}\right]
-\frac14\left[\partial_\alpha h_\nu^\lambda + \partial_\nu
h_\alpha^\lambda -
\partial^\lambda_{\alpha\nu}\right]\partial_\lambda
h^\beta_\beta\Big].
\end{split}\end{equation}
From these equations we can expand to find  ($R^2$) and
($R^2_{\mu\nu}$).

Formally we can write:
\begin{equation}\begin{split}
\sqrt{-g}R^2 &= \frac12h_\alpha^\alpha R^{(1)}R^{(1)} + 2
R^{(1)}R^{(2)}\\ &=\frac{\kappa^3}2 h_\gamma^\gamma
\Big[\partial_{\alpha}\partial^{\alpha}h_\beta^\beta-\partial^\alpha\partial^\beta
h_{\alpha\beta}\Big]
\Big[\partial_{\sigma}\partial^{\sigma}h_\rho^\rho-\partial^\sigma\partial^\rho
h_{\sigma\rho}\Big]\\& +
2\kappa^3\Big[\partial_{\sigma}\partial^{\sigma}h_\rho^\rho-\partial^\sigma\partial^\rho
h_{\sigma\rho}\Big]\bigg[-\frac12\partial_\alpha\left[h_\mu^\beta\partial^\alpha
h_\beta^\mu\right]\\&+\frac12\partial_\beta\left[h_\nu^\beta(2\partial_\alpha
h^{\nu\alpha} - \partial_\nu
h_\alpha^\alpha)\right]+\frac14\left[\partial_\alpha h_\beta^\nu +
\partial_\beta h_\alpha^\nu - \partial^\nu h_{\beta\alpha}\right]\left[\partial^\alpha h_\nu^\beta + \partial_\nu h^{\beta\alpha}-\partial^\beta
h_\nu^\alpha\right]\\
&-\frac14\left[2\partial_\alpha h^{\nu\alpha}-\partial^\nu
h_\alpha^\alpha\right]\partial_\nu h_\beta^\beta
-\frac12h^{\nu\alpha}\partial_\nu\partial_\alpha h^\beta_\beta +
\frac12h_\alpha^\nu\partial_\beta\left[\partial^\alpha
h_\nu^\beta+\partial_\nu h^{\beta\alpha} - \partial^\beta
h_\nu^\alpha\right] \bigg],
\end{split}\end{equation} and
\begin{equation}\begin{split}
\sqrt{-g}R^2_{\mu\nu} &= \frac12h_\gamma^\gamma
R_{\mu\nu}^{(1)}R^{(1) \mu\nu} -
2h^{\alpha\beta}R_{\mu\alpha}^{(1)}R^{(1) \mu}_\beta+ 2
R_{\mu\nu}^{(1)}R^{(2) \mu\nu} \\ &=\frac{\kappa^3}8
h_\gamma^\gamma \Big[\partial_\nu\partial_\alpha h_\beta^\beta-
\partial_\beta\partial_\alpha h_\nu^\beta-
\partial_\beta\partial_\nu h_\alpha^\beta + \partial^2 h_{\nu\alpha}\Big]\\&
\Big[\partial^\nu\partial^\alpha h_\rho^\rho-
\partial_\rho\partial^\alpha h^{\nu\rho}-
\partial_\rho\partial^\nu h^{\alpha\rho} + \partial^2
h^{\nu\alpha}\Big]\\ & -\frac{\kappa^3}2h^{\rho\sigma}
\Big[\partial_\mu\partial_\sigma h_\gamma^\gamma-
\partial_\gamma\partial_\sigma h_\mu^\gamma-
\partial_\gamma\partial_\mu h_\sigma^\gamma + \partial^2 h_{\mu\sigma}\Big]\\&
\Big[\partial^\mu\partial_\rho h_\beta^\beta-
\partial_\beta\partial_\rho h^{\mu\beta}-
\partial_\beta\partial^\mu h_\rho^\beta + \partial^2 h^{\mu}_{\rho}\Big]
\\&+\kappa^3 \Big[\partial_\nu\partial_\alpha h_\gamma^\gamma-
\partial_\gamma\partial_\alpha h_\nu^\gamma-
\partial_\gamma\partial_\nu h_\alpha^\gamma + \partial^2 h_{\nu\alpha}\Big]\\&
\Big[ -\frac12\partial^\alpha\left[h^{\beta\lambda}\partial^\nu
h_{\lambda\beta}\right] +
\frac12\partial_\beta\left[h^{\beta\lambda}(\partial^\nu
h_{\lambda}^{\alpha}+\partial^\alpha h^{\nu}_{\lambda} -
\partial_\lambda h^{\nu\alpha})\right]\\
&+\frac14\Big[\partial_\beta h^{\nu\lambda} +
\partial^\nu h_\beta^\lambda - \partial^\lambda h^{\nu}_{\beta}\Big]\left[\partial_\lambda h^{\beta\alpha} + \partial^\alpha h^{\beta}_\lambda-\partial^\beta h^{\alpha}_\lambda\right]
\\&-\frac14\left[\partial^\alpha h^{\nu\lambda} + \partial^{\nu}
h^{\alpha\lambda} -
\partial^\lambda h^{\alpha\nu}\right]\partial_\lambda
h^\beta_\beta\Big].
\end{split}\end{equation}

Concentrating only on terms which go into the $(3B)^{\rm eff}$ and
$(3C)^{\rm eff}$ index structures, and putting ($\kappa = 1$) for
simplicity, we have found for the ($R^2$) contribution:
\begin{equation}
(R^2)_{(3B)^{\rm eff}} = \partial^2 h^\rho_\rho\Big (-\frac32
\partial_\alpha h^\beta_\mu\partial^\alpha h_\beta^\mu-2h_\beta^\mu
\partial^2 h^\beta_\mu\Big),
\end{equation}
and
\begin{equation}
(R^2)_{(3C)^{\rm eff}} = -\partial^\sigma\partial^\rho
h_{\sigma\rho}\Big (-\frac32
\partial_\alpha h^\beta_\mu\partial^\alpha h_\beta^\mu-2h_\beta^\mu
\partial^2 h^\beta_\mu\Big),
\end{equation}
For the ($R^2_{\mu\nu}$) contribution we have found:
\begin{equation}\begin{split}
(R^2_{\mu\nu})_{(3B)^{\rm eff}} &= \frac18h_\gamma^\gamma
\partial^2 h_{\nu\alpha}\partial^2
h^{\nu\alpha}-\frac14\partial_\lambda\partial_\beta
h_\gamma^\gamma
\partial^\beta h^{\nu\alpha}\partial^\lambda h_{\nu\alpha}-\frac12\partial_\lambda\partial_\beta h_\gamma^\gamma
 h^{\nu\alpha}\partial^\beta\partial^\lambda h_{\nu\alpha} \\&+\frac14 \partial_\lambda h_\gamma^\gamma \partial^2 h_{\nu\alpha} \partial^\lambda
 h^{\nu\alpha},
\end{split}\end{equation}
and
\begin{equation}\begin{split}
(R^2_{\mu\nu})_{(3C)^{\rm eff}} &=
\frac12\partial_\lambda\partial_\beta h^\lambda_\sigma
\partial^\sigma h^{\nu\alpha} \partial^\beta h_{\nu\alpha}
+\partial_\lambda\partial_\beta h_\sigma^\lambda
h^{\nu\alpha}\partial^\sigma\partial^\beta h_{\nu\alpha}
-\frac14\partial^2h_{\lambda\beta}\partial^\lambda h^{\nu\alpha}
\partial^\beta h_{\nu\alpha}\\& -\frac12 \partial^2 h_{\lambda\beta}
h^{\nu\alpha} \partial^\lambda\partial^\beta h_{\nu\alpha}
-\frac12\partial_\beta h^{\beta\lambda}\partial^2 h_{\nu\alpha}
\partial_\lambda h^{\nu\alpha}
-\frac12h^{\beta\lambda}\partial^2 h_{\nu\alpha} \partial_\lambda
\partial_\beta h^{\nu\alpha}.
\end{split}\end{equation}
This leads to the following results for the (3B)
and (3C) terms:
\begin{figure*}[h]
\begin{tabular}{ll}\vspace{0.1cm}
$\left(\parbox{1cm}{\includegraphics[height=0.8cm]{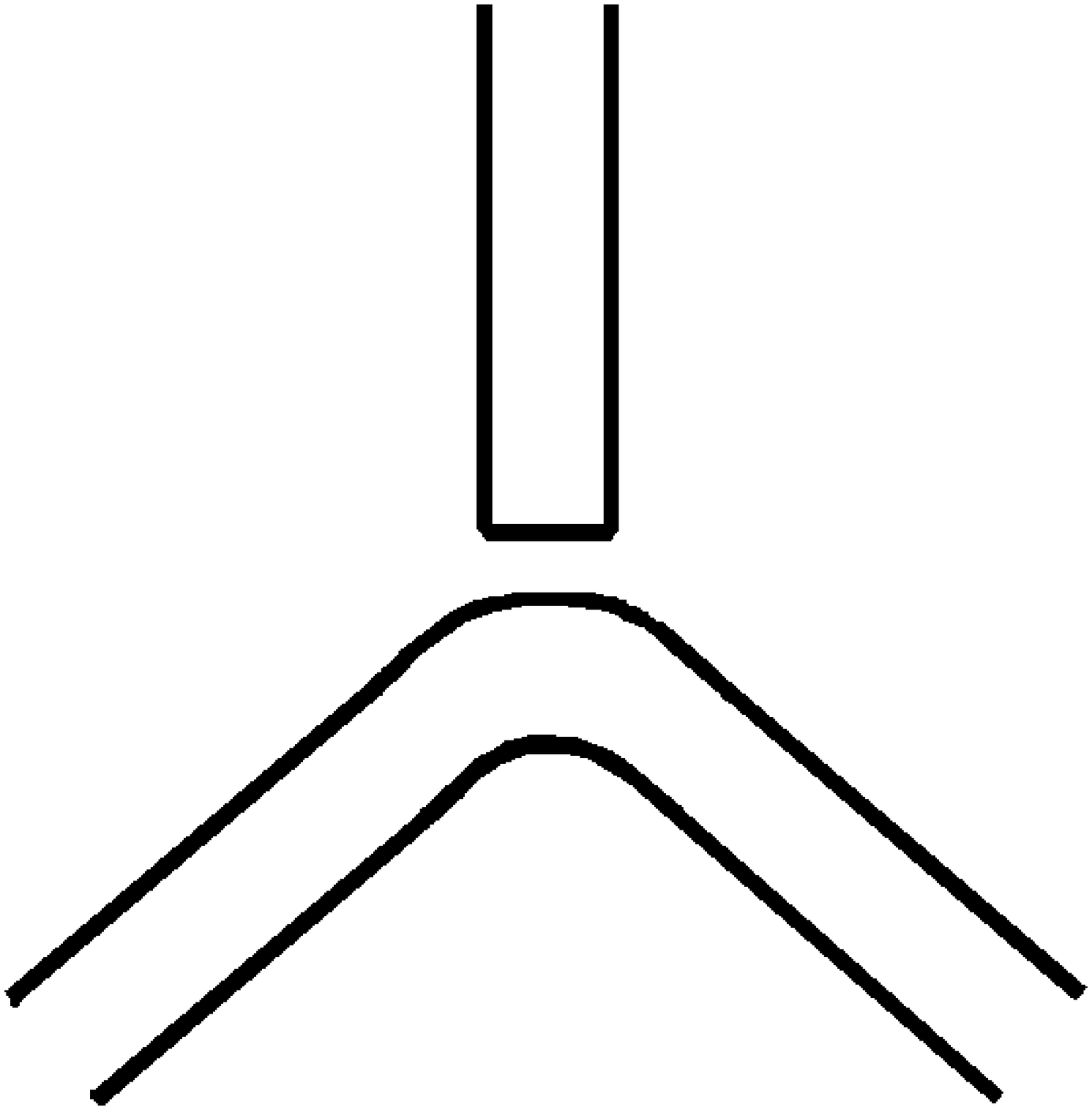}}\right)^{\rm
eff}_{\rm 3B}$ &$\sim {\rm
sym}[-P_3\left(\eta_{\mu\alpha}\eta_{\nu\sigma}\eta_{\beta\gamma}[3c_1k_1^2(k_2\cdot
k_3)+c_2(\frac12(k_1\cdot k_2)(k_1\cdot
k_3)-\frac14k_2^2k_3^2)]\right)$\\ &
$-P_6\left(\eta_{\mu\alpha}\eta_{\nu\sigma}\eta_{\beta\gamma}[2c_1k_1^2k_3^2+c_2(\frac12(k_1\cdot
k_3)^2-\frac14k_2^2(k_1\cdot k_3))]\right)],$
\\ \vspace{0.05cm}
$\left(\parbox{1cm}{\includegraphics[height=0.8cm]{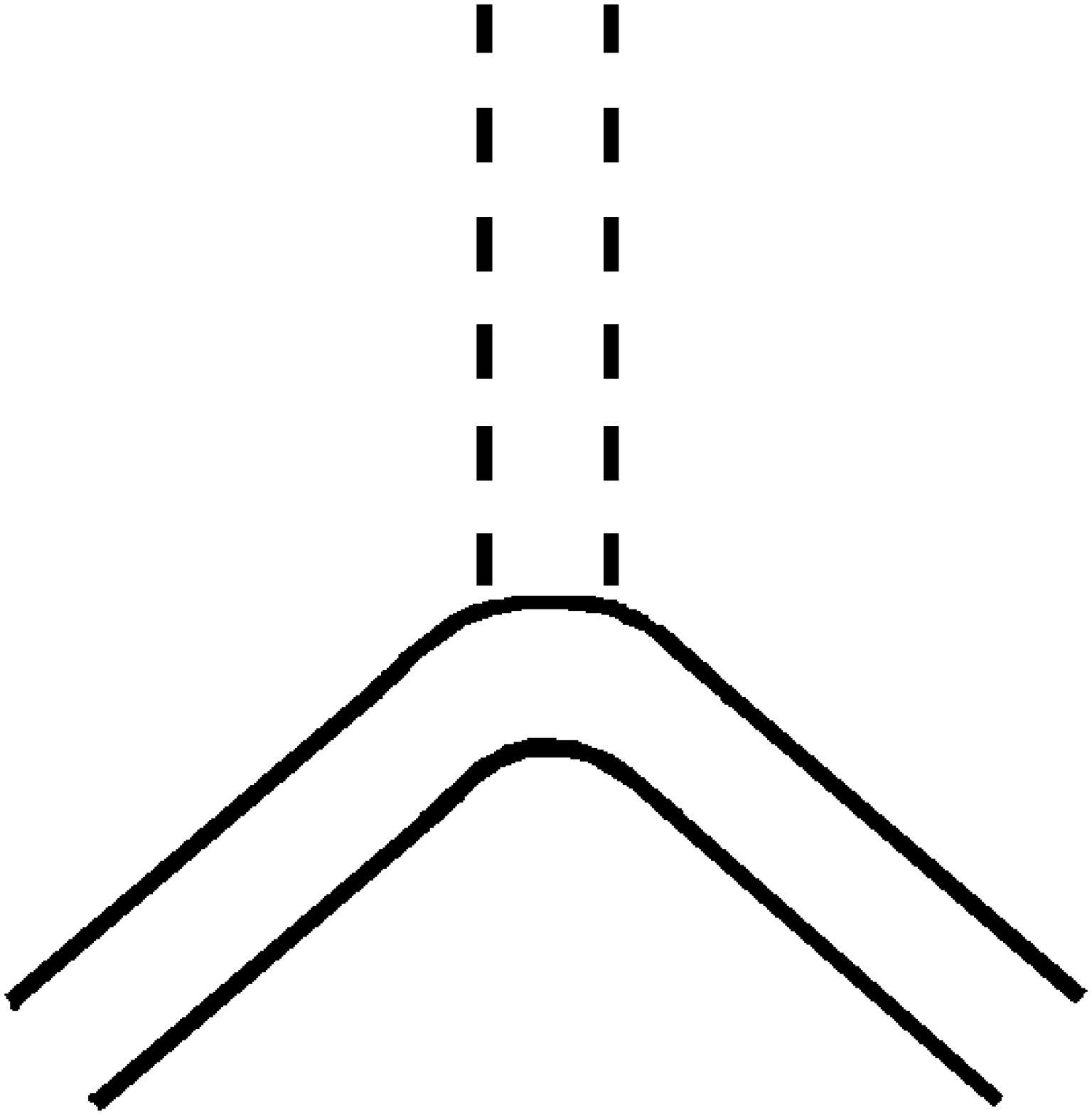}}\right)^{\rm
eff}_{\rm 3C}$ &$\sim {\rm
sym}[-P_3\left(k_{1\mu}k_{1\alpha}\eta_{\nu\sigma}\eta_{\beta\gamma}[3c_1k_2\cdot
k_3]\right)
-2P_6\left(k_{1\mu}k_{1\alpha}\eta_{\nu\sigma}\eta_{\beta\gamma}[c_1
k_3^2]\right)
$\\&$+\frac12P_6\left(k_{1\mu}k_{2\alpha}\eta_{\nu\sigma}\eta_{\beta\gamma}[c_2k_1\cdot
k_3]\right)
$+$P_6\left(k_{1\mu}k_{3\alpha}\eta_{\nu\sigma}\eta_{\beta\gamma}[c_2k_1\cdot
k_3]\right)
$\\&$-\frac12P_3\left(k_{2\mu}k_{3\alpha}\eta_{\nu\sigma}\eta_{\beta\gamma}[c_2k_1^2]\right)
-\frac12P_6\left(k_{3\mu}k_{3\alpha}\eta_{\nu\sigma}\eta_{\beta\gamma}[c_2k_1^2]\right)$\\&
$
-\frac12P_6\left(k_{1\beta}k_{3\nu}\eta_{\mu\sigma}\eta_{\alpha\gamma}[c_2k_2^2]\right)
-\frac12P_6\left(k_{3\beta}k_{3\nu}\eta_{\mu\sigma}\eta_{\alpha\gamma}[c_2k_2^2]\right)].$
\end{tabular}
\end{figure*}

Using the same procedure, we can find expression for non-leading
effective contributions, however this will have no implications
for the loop contributions. We have only calculated 3-point
effective vertex index structures in this appendix, 4-point index
structures are tractable too by the same methods, but the algebra
gets much more complicated in this case.

\section{Comments on general $n$-point functions}\label{app2}
In the large-$D$ limit only certain graphs will give leading
contributions to the $n$-point functions. The diagrams favored in
the large-$D$ will have a simpler structure than the full
$n$-point graphs. Calculations which would be too complicated to
do in a full graph limit, might be tractable in the large-$D$
regime.

The $n$-point contributions in the large-$D$ limit are possible to
employ in approximations of $n$-point functions at any finite
dimension, exactly as the $n$-point planar diagrams in gauge
theories can be used, {\it e.g.}, at ($N=3$), to approximate
$n$-point functions. Knowing the full $n$-point limit is hence
very useful in the course of practical approximations of generic
$n$-point functions at finite ($D$). This appendix will be devoted
to the study of what is needed in order to calculate $n$-point
functions in the large-$D$ limit.

\begin{figure}[h]\centering
\includegraphics[height=1.2cm]{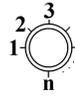}
\caption{The leading 1-loop $n$-point diagram, the index structure
for the external lines is not decided, they can have the momentum
structure, ($i.e.$, as in (3C)), or they can be of the contracted
type, ($i.e.$, as in (3B)). External lines can of course also
originate from 4-point vertex or a higher vertex index structure
such as, $e.g.$, the index structures (4D) or (4E). This possibility
will however not affect the arguments in this appendix, so we will
leave this as a technical issue to be dealt with in explicit
calculations of $n$-point functions.\label{n-point}}
\end{figure}

In the large-$D$ limit the leading 1-loop $n$-point diagrams will
be of the type, as displayed in figure \ref{n-point}.

\begin{figure}[h]\centering
\includegraphics[height=4cm]{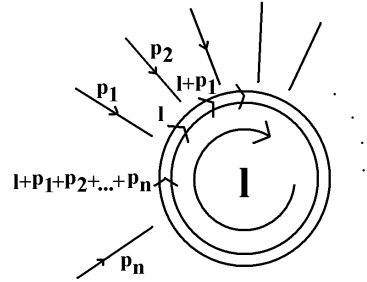}
\caption{A diagrammatic expression for the generic $n$-point
1-loop correction.\label{n-point3}}
\end{figure}

The completely generic $n$-point 1-loop correction will have the
diagrammatic expression, as shown in figure \ref{n-point3}.

\begin{figure}[h]\centering
\parbox{3.5cm}{\includegraphics[height=1.2cm]{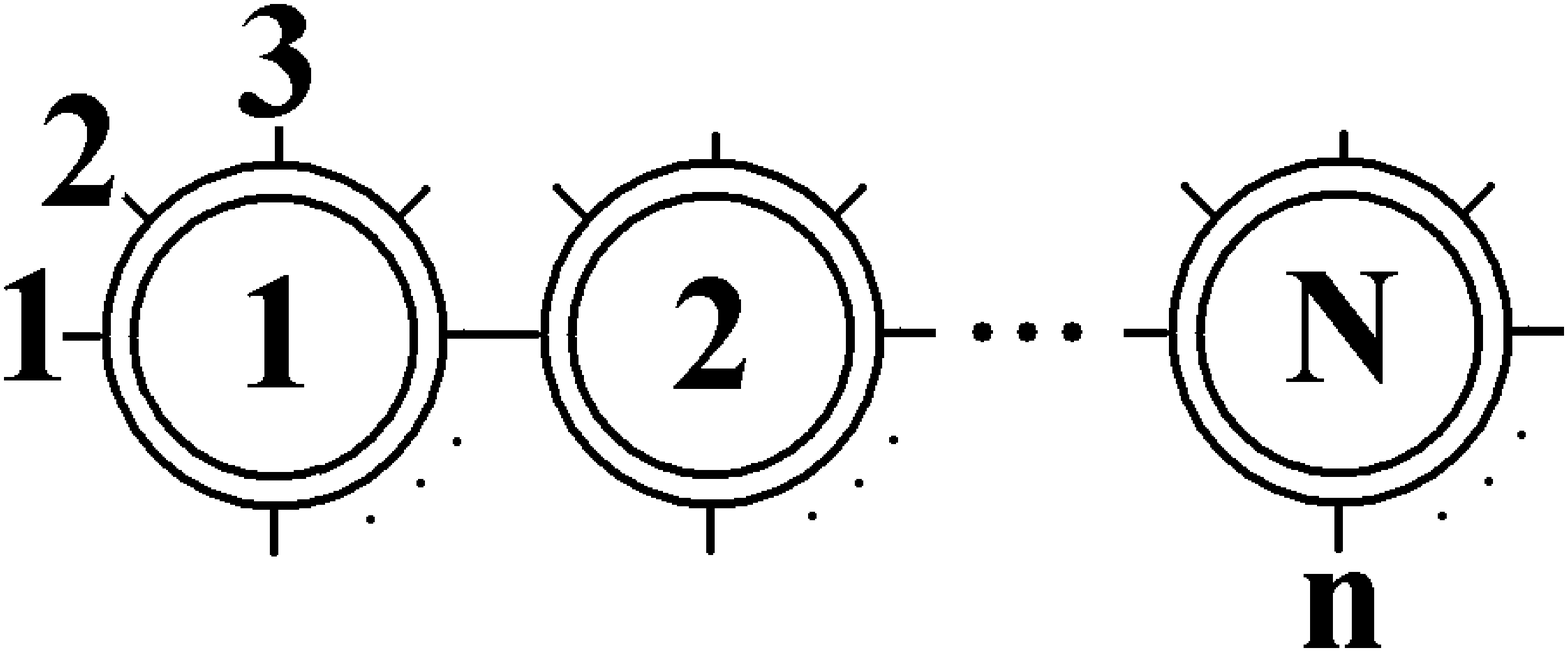}}
\parbox{3.5cm}{\includegraphics[height=1.2cm]{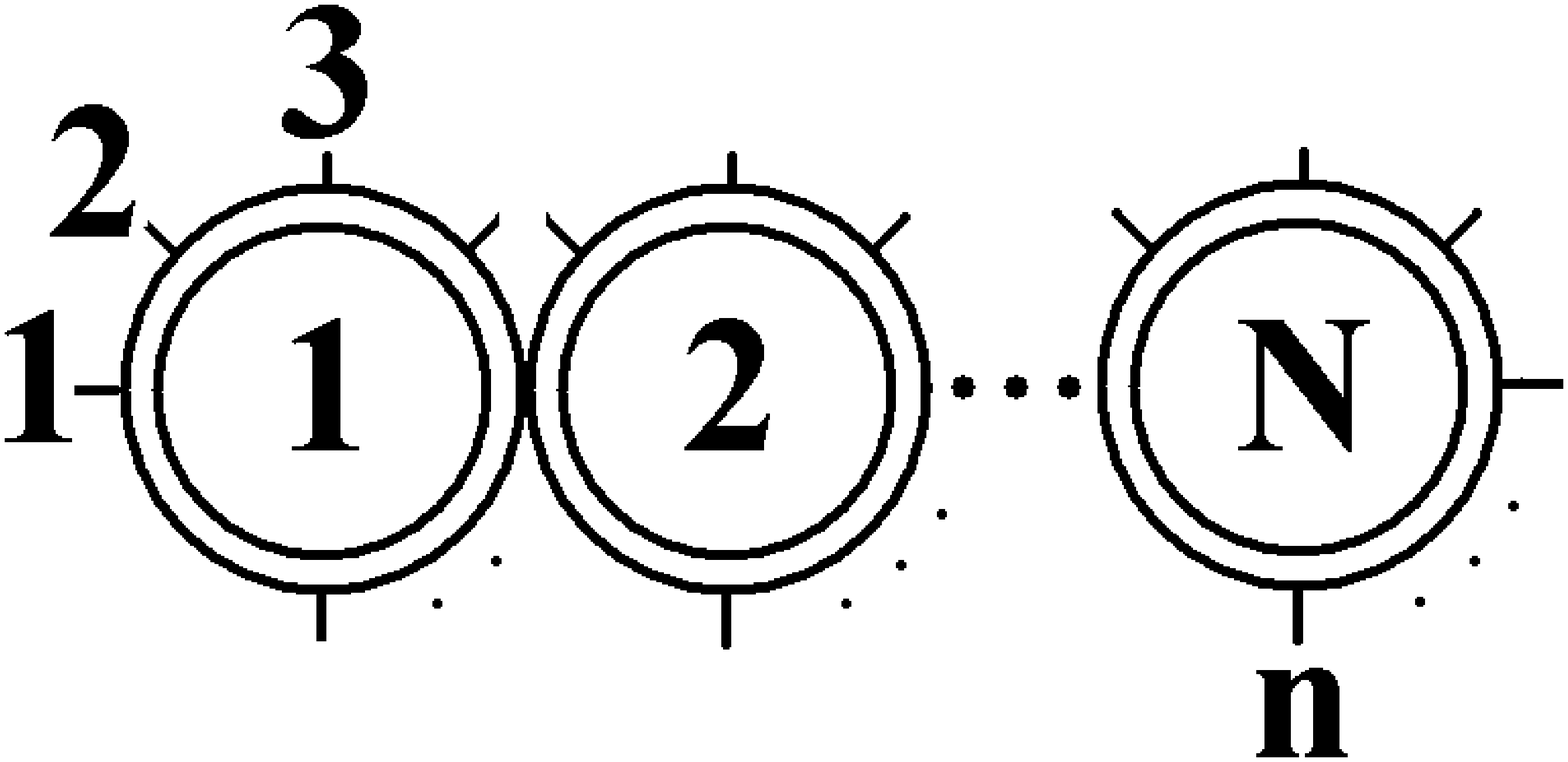}}
\parbox{4.7cm}{\includegraphics[height=1.2cm]{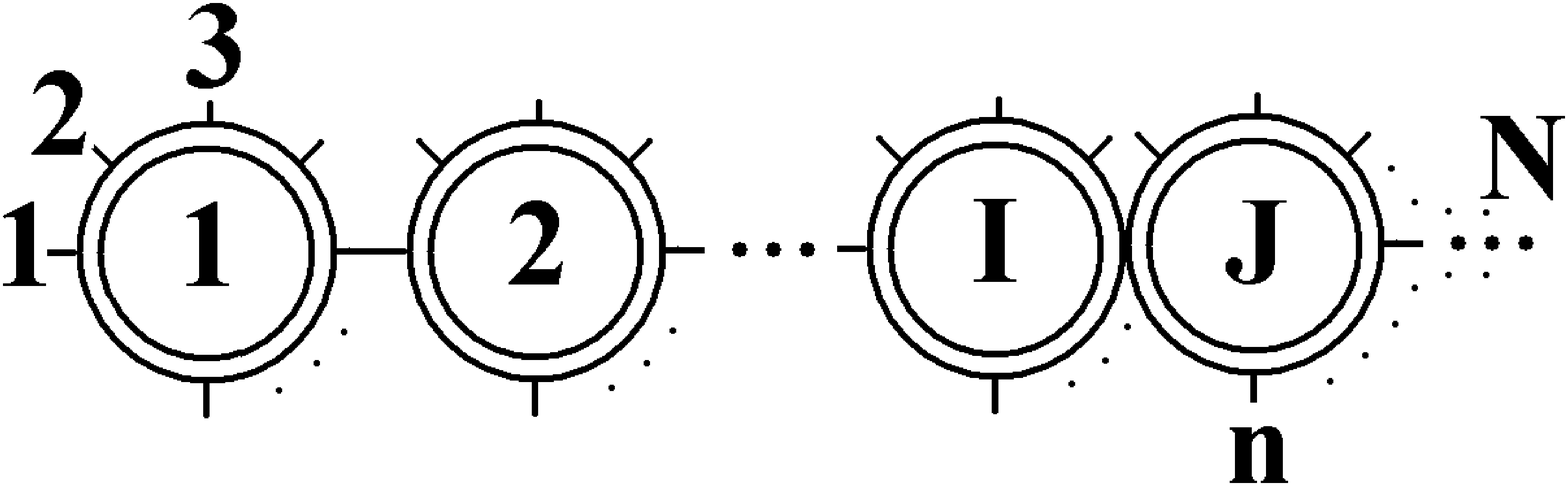}}
\parbox{2cm}{\includegraphics[height=1.8cm]{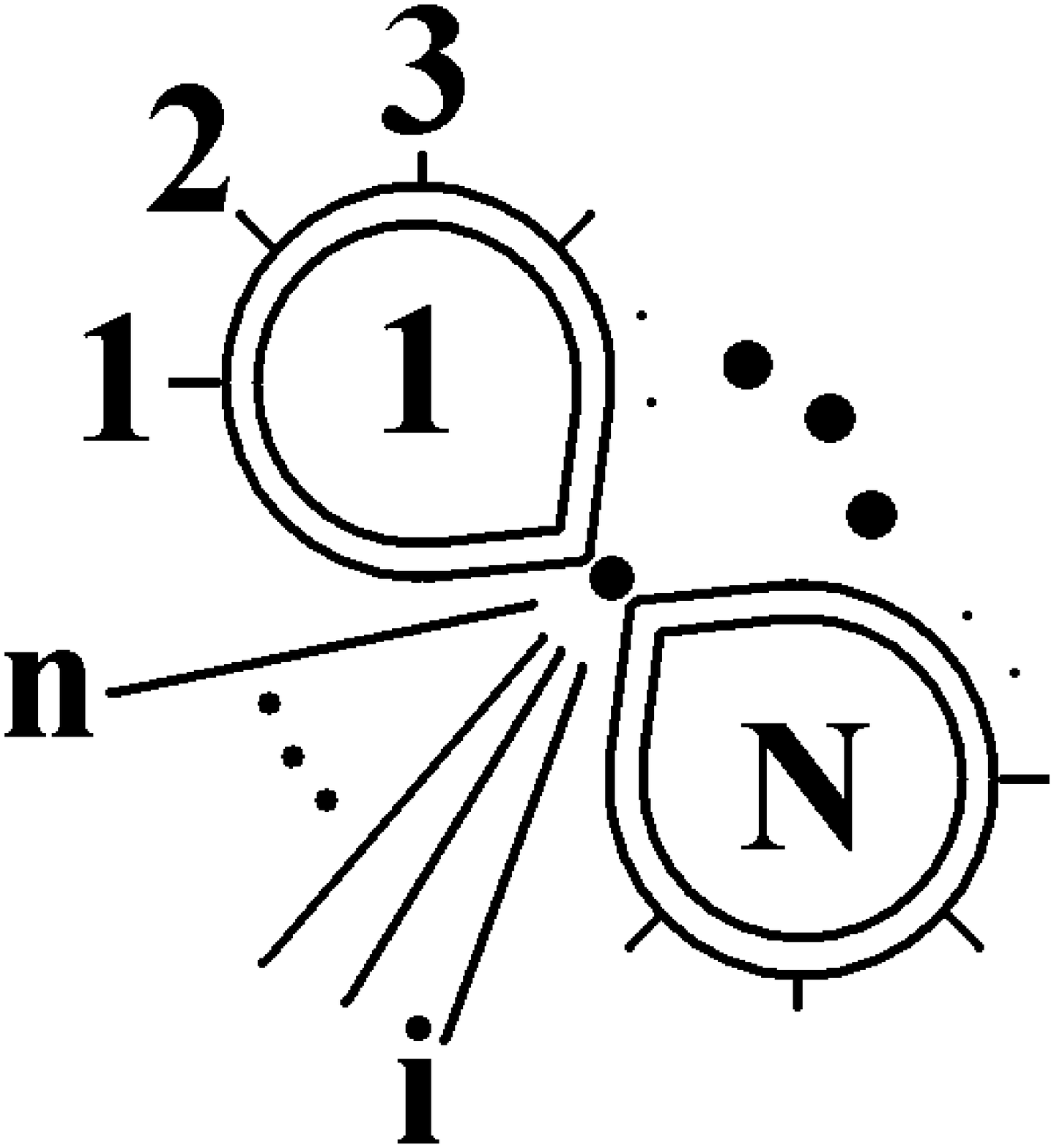}}
\caption{Some generic examples of graphs with leading
contributions in the gravity large-$D$ limit. The external lines
originating from a loop can only have two leading index
structures, ($i.e.$, they can be of the same type as the external
lines originating from a (3B) or a (3C)) loop contribution. Tree
external lines are not restricted to any particular structure,
here the full vertex factor will contribute.\label{larged}}
\end{figure}

More generically the large-$D$, $N$-loop corrections to the
$n$-point function will consist of combinations of leading bubble
and vertex-loop contributions, such as shown in figure
\ref{larged}.

For the bubble contributions it is seen that all loops are
completely separated. Therefore everything is known once the
1-loop contributions are calculated. That is, in order to derive
the full $n$-point sole bubble contribution it is seen that if we
know all $j$-point tree graphs, as well as all $i$-point 1-loop
graphs (where, $i,j \leq n$) are known, the derivation of the full
bubble contribution is a matter of contractions of diagrams and
combinatorics.

For the vertex-loop combinations the structure of the
contributions are a bit more complicated to analyze. The loops
will in this case be connected through the vertices, not via
propagators, and the integrals can therefore be joined together by
contractions of their loop momenta. However, this does in fact
not matter; we can still do the diagrams if we know all integrals
which occur in the generic 1-loop $n$-point function. The
integrals in the vertex-loop diagrams will namely be of completely
the same type as the bubble diagram integrals. There will be no
shared propagator lines which will combine the denominators of the
integrals. The algebraic contractions of indices will however be
more complicated to do for vertex-loop diagrams than for sole
bubble loop diagrams. Multiple index contractions from products of
various integrals will have to be carried out in order to do these
types of diagrams.

Computer algebraic manipulations can be employed to do the algebra
in the diagrams, so we will not focus on that here. The real
problem lays in the mathematical problem calculating generic
$n$-point integrals.

In the Einstein-Hilbert case, where each vertex can add only two
powers of momentum, we see that problem of this diagram is about
doing integrals such as:
\begin{equation}\begin{split}
I_n^{(\mu_1\nu_1\mu_2\nu_2\cdots\mu_n\nu_n)}= \int d^D l
\frac{l^{\mu_1}l^{\nu_1}l^{\mu_2}l^{\nu_2}\cdots
l^{\mu_n}l^{\nu_n}}{l^2
(l+p_1)^2(l+p_1+p_2)^2\cdots(l+p_1+p_2+\ldots p_n)^2},
\end{split}\end{equation}
\begin{equation}\begin{split}
I_n^{(\nu_1\mu_2\nu_2\cdots\mu_n\nu_n)}= \int d^D l
\frac{l^{\nu_1}l^{\mu_2}l^{\nu_2}\cdots l^{\mu_n}l^{\nu_n}}{l^2
(l+p_1)^2(l+p_1+p_2)^2\cdots(l+p_1+p_2+\ldots p_n)^2},
\end{split}\end{equation}
$$\vdots$$
\begin{equation}\begin{split}
I_n= \int d^D l \frac{1}{l^2
(l+p_1)^2(l+p_1+p_2)^2\cdots(l+p_1+p_2+\ldots p_n)^2},
\end{split}\end{equation}
Because the graviton is a massless particle such integrals will be
rather badly defined. The denominators in the integrals are close
to zero, and this makes the integrals divergent. One way to deal
with these integrals is to use the following parametrization of
the propagator:
\begin{equation}
1/q^2 = \int^\infty_0 dx \exp (-x q^2), {\rm for} \ \ \ q^2>0,
\end{equation}
and define the momentum integral over a complex $(2w)$-dimensional
Euclidian space-time. Thus, what is left to do are gaussian
integrals in a $(2w)$-dimensional complex space-time. Methods to
do such types of integrals are investigated
in refs.~\cite{Capper:vb,Leibbrandt:1975dj,Capper:pv,Capper:dp}. The
results for ($I_2$), ($I_2^{(\mu)}$), ($I_2^{(\mu\nu)}$) and
($I_3$) are in fact explicitly stated there. It is outside our
scope to actually delve into technicalities about explicit
mathematical manipulations of integrals, but it should be clear
that mathematical methods to deal with the 1-loop $n$-point
integral types exist and that this could be a working area for
further research. Effective field theory will add more derivatives
to the loops, thus the nominator will carry additional momentum
contributions. Additional integrals will hence have to be carried
out in order to do explicit diagrams in an effective field theory
framework.

\label{appendix1}

%\end{fmffile}
\end{document}